\documentclass[10pt, a4paper]{article}
\usepackage{amsmath}
\usepackage{amssymb}
\usepackage{amsfonts}
\usepackage{setspace}
\usepackage[]{graphicx}
\usepackage[pdfauthor={P. P. Cook},
pdftitle={Connections between Kac-Moody algebras and M-theory}]{hyperref}
\usepackage{oldgerm}
\usepackage{fancyhdr} 
\usepackage{vmargin}
\usepackage{longtable}
\usepackage{lscape}
\numberwithin{equation}{section}

\begin{document}
\setpapersize{A4}
\setmarginsrb{40mm}{20mm}{20mm}{20mm}{12pt}{11mm}{0pt}{11mm}

\pagestyle{fancy}
\rhead{}
\chead{}
\cfoot{\thepage}
\title{Connections between Kac-Moody algebras and M-theory}
\author{Paul P. Cook\footnote{email: p.cook@sns.it}\\  \\{\itshape Department of Mathematics, King's College London,\/}\\{\itshape Strand, London WC2R 2LS, U.K.\/}}
\begin{titlepage}
\begin{flushright}
\end{flushright}
\begin{center}
\vspace{10pt}
\resizebox{5cm}{!}{
\includegraphics{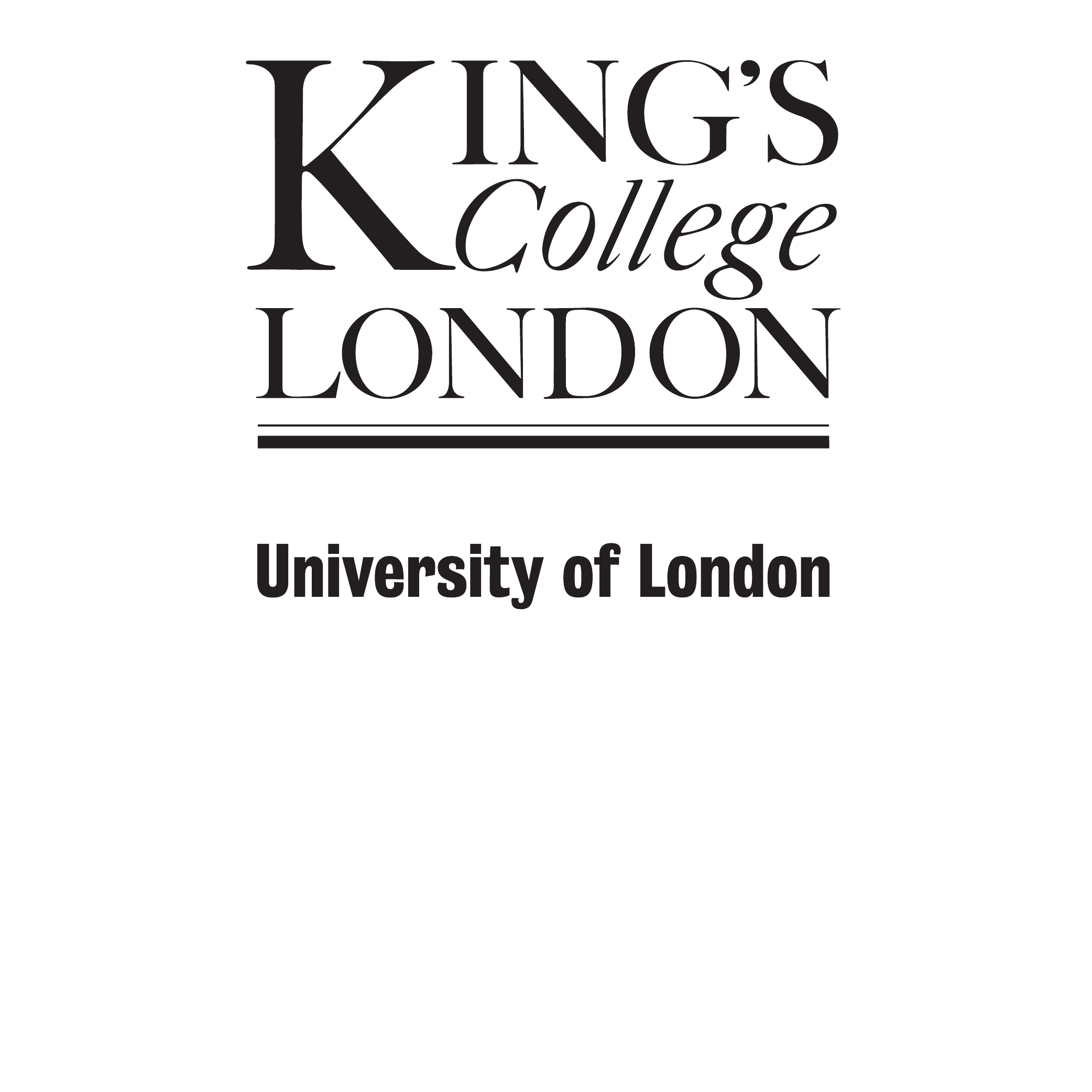}}
\end{center}
\vspace{60pt}
\centering{\LARGE{\bfseries Connections between Kac-Moody algebras and M-theory}}\\
\vspace{30pt}
\def\thefootnote{\fnsymbol{footnote}}
Paul P. Cook\footnote{\href{mailto:p.cook@sns.it}{email: p.cook@sns.it}} \\
\setcounter{footnote}{0}
\vspace{10pt}
{\itshape Department of Mathematics, King's College London,\/}\\
{\itshape Strand, London WC2R 2LS, U.K.\/}\\
\vspace{200pt}
\normalsize{Ph.D. Thesis}\\
\normalsize{Supervisor: Professor Peter West}\\
\normalsize{Submitted in September, 2006}\\
\end{titlepage}
\begin{spacing}{1.5}
\clearpage
\newpage

\newpage
\begin{center}
\LARGE{Abstract} \\ \ \\
%
%
\end{center}
\normalsize
\begin{slshape}
We investigate the motivations and consequences of the conjecture that the Kac-Moody algebra $E_{11}$ is the symmetry algebra of M-theory, and we develop methods to aid the further investigation of this idea.

The definitions required to work with abstract root systems of Lie algebras are given in review leading up to the definition of a Kac-Moody algebra. The motivations for the $E_{11}$ conjecture are reviewed and the nonlinear realisation of gravity relevant to the conjecture is explicitly described.

The algebras of $E_{11}$, relevant to $M$-theory, and $K_{27}$, relevant to the bosonic string theory, along with their $l_1$ representations are constructed. Tables of low level roots are produced for both the adjoint and $l_1$ representations of these algebras.

A solution generating element of the Kac-Moody algebra is given, and it is shown by construction that it encodes all the known half-BPS brane solutions of the maximally oxidised supergravity theories. It is then used to look for higher level branes associated to the roots of the Kac-Moody algebra associated to the oxidised theory.

The nature of how spacetime signature is encoded within the $E_{11}$ formulation of $M$-theory is analysed. The effect of the multiple signatures that arise from the Weyl reflections of $E_{11}$ on the solution generating group element is found and precise conditions of when an electric brane solution exists and in which signatures are given. As a corollary to these investigations the spacelike branes of $M$-theory are found associated to the solution generating element.

The U-duality multiplets of exotic brane charges are shown to have a natural $E_{11}$ origin. General formulae for finding the content of arbitrary brane charge multiplets are given and the exact content of the particle and string multiplets in dimensions $4,5,6,7,8$ is related to the $l_1$ representation of $E_{11}$.

\end{slshape}

\newpage

\begin{center}
{\Large Acknowledgements}\\
\end{center}
\vspace{10pt}
It is a great pleasure to thank Peter West for his impressive insight, expert advice and humour throughout the course of my studies. His patience with my timekeeping was also laudable. 

Throughout my studies I have been supported by PPARC grant PPA/S/S/2003/03644, and I want to thank the "powers that be" for continuing to fund such research as mine, which I believe remains an important pursuit in the modern world.

Studying an esoteric subject can be a lonely business, the past three years have been considerably more enjoyable than one might have imagined due to my incredible friends. In particular the occupants of Room 102 (and formerly 32F): Ant, Manuel, Carmen (who taught me that everything I begin in Latex I must also end), Sven, Vid, Alexis, Laura, Dan, Philipp, Peter, Our Mighty Emperor, Nick, Arnold, Jack and the hairy little man. I have also spent time with people from outside of my office, in particular I would like to thank the following for their company, support and good humour: Joe, Nima, Geoff, Sarah, Rosie, Alfie, George, Dennis, Lisa, My Long, Tam, Kevvy, the Andys, Katie, James, Trevor, Gina, Ben, Phil, Nadia, Dave, Nikos and all the hippos of the world.

I would like to particularly thank Sam for always being able to cross every line anyone ever put down for him, and for making me laugh. Sometimes so hard I had to clean my computer afterwards.

Finally this thesis would just be a twinkling in my eye if it weren't for the help, financial support, occasional laundry, hot meals and love of my parents. I owe them more than a thesis. But for now, with my thanks, that's all I have to offer them.
\newpage
\begin{center}
\ \\
\vspace{10cm}
{\LARGE FOR MUM AND DAD}
\end{center}
\normalfont
\newpage

\tableofcontents
\listoftables
\listoffigures

\newpage

\section{Introduction}
\begin{quotation}
"Physical laws should have mathematical beauty and simplicity"
\end{quotation}
\begin{flushright}
\em{Paul Dirac}
\end{flushright}
It is the great hope of theoretical physics that the vast complexity of the world we see around us may be described by a manageable mathematical formalism. That is not to say that the theoretical physicist believes that the natural laws of the universe adhere to a mundane mathematical model, but rather that he or she frequently hopes that there exists some elegant mathematical means of encoding and enhancing our appreciation of the universe. The most successful of modern theoretical models have all been based around the identification of symmetry groups. In this thesis we examine the conjectured symmetry algebra of $M$-theory, a candidate for the unifying theory of the observed forces of nature.

Symmetry groups underpin modern particle physics. Indeed it may be argued that theoretical physics {\em is} the identification and exploitation of symmetry groups in natural systems. From the identification of the Poincar{\'e} invariance of relativity, through the unitary operators and Lie algebras of quantum mechanics to the construction of the standard model, the application of group theory has accompanied the greatest theoretical successes of the last century. These accomplishments have ushered in the modern era of high-energy particle physics in which theory anticipates the results of experiments that lie beyond the energy limit of current laboratories and accelerators. It is the goal of theoretical physics to unify the laws of nature in these high-energy domains.

A priori there is no reason to be certain that such a unifying framework exists. The widespread belief that there is a high-energy symmetry is based upon the previous successful unifications of the strongest forces of nature. This process culminated in the standard model's unification of electromagnetism, the strong and the weak nuclear forces. Only gravity stands apart. Attempts to incorporate gravity in a unified quantum field theory have motivated astonishing changes of theoretical perspective in the last thirty years, in particular the emergence of supergravity and string theory as the leading candidates for a theory of quantum gravity. One serious stumbling block was that there were discovered five different string theories with the potential to be the unique unifying theory. This problem was surmounted when it was realised that regions of different string coupling strength in the separate theories could be identified by a series of $S$ and $T$-dualities and in this way the five disparate theories were identified as branches of a single higher-energy theory that became known as $M$-theory \cite{Townsend,Witten}. However, $M$-theory contains such a rich structure that a new insight or approach may be needed before rapid progress will be made. The main topic of this thesis will be the conjecture that the Kac-Moody algebra $E_{11}$ encodes the symmetries of $M$-theory \cite{West1} and it is hoped it may offer an expediting algebraic approach to learn more about $M$-theory.

In March of 1978, Cremmer, Julia and Scherk wrote down the unique supergravity theory in eleven dimensions \cite{CremmerJuliaScherk} whose Lagrangian is,
\begin{align}
\nonumber {\cal L}_{11}&=\frac{e}{4\kappa^2}R-\frac{e}{2!4!}F_{\mu_1\ldots\mu_4}F^{\mu_1\ldots\mu_4}-\frac{e}{2}\bar{\psi}_\mu\Gamma^{\mu\nu\rho}D_\nu(\frac{1}{2}(\tilde{\Omega}+\hat{\Omega}))\psi_\rho \\
\nonumber&\quad -\frac{e\kappa}{8.4!}(\bar{\psi}_{\mu_1}\Gamma^{\mu_1\ldots  \mu_6}\psi_{\mu_2}+12\bar{\psi}^{\mu_3}\Gamma^{\mu_4\mu_5}\psi^{\mu_6})(F_{\mu_3\ldots\mu_6}+\hat{F}_{\mu_3\ldots\mu_6}) \\
&\quad +\frac{\kappa}{8(3!)^4}\epsilon^{\mu_1\ldots \mu_{11}}F_{\mu_1\ldots \mu_4}F_{\mu_5\ldots \mu_8}A_{\mu_9\mu_{10}\mu_{11}}
\end{align}
Where,
\begin{align}
\nonumber F_{\mu_1\ldots \mu_4}&=4\partial_{[\mu_1}A_{\mu_2\mu_3\mu_4]}\\
\nonumber \hat{F}_{\mu_1\ldots \mu_4}&=F_{\mu_1\ldots\mu_4}+3\bar{\psi}_{[\mu_1}\Gamma_{\mu_2\mu_3}\psi_{\mu_4]}\\
\nonumber \hat{\Omega}_{\mu bc}&=\omega_{\mu bc}-\frac{1}{2}(\bar{\psi}_\nu\Gamma_c\psi_b-\bar{\psi}_\nu\Gamma_b\psi_c+\bar{\psi}_c\Gamma_\mu\psi_b) \\
\nonumber \tilde{\Omega}_{\mu bc}&=\hat{\Omega}_{\mu bc}+\frac{1}{4}\bar{\psi}_\nu\Gamma_{\mu bc}^{\nu\lambda}\psi_\lambda\\
\omega_{\mu bc}&=\frac{1}{2}({e_b}^\rho\partial_\mu e_{\rho c}-{e_c}^\rho\partial_\mu e_{\rho b})-\frac{1}{2}({e_b}^\rho\partial_\rho e_{\mu c}-{e_c}^\rho\partial_\rho e_{\mu b})-\frac{1}{2}({e_b}^\lambda{e_c}^\rho\partial_\lambda e_{\rho a}-{e_c}^\lambda{e_b}^\rho\partial_\lambda e_{\rho a}){e_\mu}^a
\end{align}
Where ${e_\mu}^a$ is the vielbein and $e$ is its determinant. The equations of motion for the metric, $g_{\mu\nu}$, the gravitino, $\psi_\mu$ and the three form potential $A_{\mu_2\mu_3\mu_4}$ are,
\begin{align}
\nonumber \frac{1}{2!4!}(F_{\mu\alpha_1\ldots \alpha_3}{F_{\nu}}^{\alpha_1\ldots \alpha_3}-\frac{1}{12}g_{\mu\nu}F_{\alpha_1\ldots \alpha_4}F^{\alpha_1\ldots \alpha_4})&=R_{\mu\nu} \\
\nonumber \Gamma^{\mu_1\mu_2\mu_3}\{ D_{\mu_2}\psi_{\mu_3} - \frac{1}{12^2}({\Gamma^{\nu_1\ldots \nu_4}}_{\mu_2}-8\Gamma^{\nu_2\ldots \nu_4}{\delta^{\nu_1}}_{\mu_2})\psi_{\mu_3}\hat{F}_{\nu_1\ldots \nu_4}\} &=0\\
\partial_{\mu_1}F^{\mu_1\ldots \mu_4}+\frac{e^{-1}}{4.12^2} \epsilon^{\mu_2\ldots \mu_{11}\mu_1}F_{\mu_5\ldots \mu_8}F_{\mu_9\ldots \mu_{11}\mu_1}&=0
\end{align}
Where we have set $\kappa=1$. The content of the theory can be understood as the result of trying to build an eleven dimensional gravitational theory possessing supersymmetry. The massless fields of the theory form representations of the little group, $SO(D-2)$, which in eleven dimensions is $SO(9)$, the metric being symmetric has $\frac{9(10)}{2}=45$ bosonic degrees of freedom. The superpartner of the metric is the gravitino field, $\psi_\mu$ which transforms as a vector under the little group and also carries a spinor representation, it transforms under an irreducible representation having $128$ fermionic degrees of freedom. For the theory to be supersymmetric the number of bosonic and fermionic degrees of freedom should match, the shortfall is removed by introducing an extra field $A_{\mu_1\mu_2\mu_3}$ having $\frac{9!}{3!6!}=84$ bosonic degrees of freedom. For a modern review of eleven-dimensional supergravity, containing explicit calculations, the reader is encouraged to refer to \cite{MiemecSchnakenburg}.
The equation of motion for the three form can be rewritten in a linear form,
\begin{equation}
F_{\mu_1\ldots \mu_4}=\frac{e}{7!}\epsilon_{\mu_1\ldots \mu_{11}}\tilde{F}^{\mu_5\ldots\mu_{11}} \label{eomA3}
\end{equation}
Where we have defined,
\begin{equation}
\tilde{F}_{\mu_1\ldots \mu_7}\equiv 7(\partial_{[\mu_1}A_{\mu_2\ldots\mu_7]}+5A_{[\mu_1\ldots \mu_3}F_{\mu_4\ldots\mu_7]})
\end{equation}
The field $A_{\mu_1 \ldots \mu_6}$ is the dual of the three form, defined via the Hodge dual of the four-form field strength.

It is, at first sight, quite an ugly theory. However it was soon realised that the theory had much more symmetry than was apparent in its eleven dimensional formulation. Indeed by September of 1978 Cremmer and Julia reported the first observation of a hidden symmetry in the eleven dimensional theory where the emergence of an $SU(8)$ local symmetry and an $E_{7(7)}$ global symmetry was demonstrated upon dimensional reduction to four dimensions\cite{CremmerJulia,CremmerJulia1}. The notation $E_{d(d)}$ is used to indicate the normal real form of the complex exceptional algebra $E_d({\mathbb C})$, meaning that its generators are maximally non-compact \cite{Gilmore,Helgason}. The number in brackets is the character of the algebra, giving the difference between the number of noncompact and compact generators. For the maximally non-compact algebras the character is the rank of the algebra. Throughout this thesis we will not emphasise the choice of the normal real form and will use the simple notation $E_d$ throughout. Dimensional reduction of the maximal theory to other dimensions revealed a series of hidden global symmetry group, $\cal G$, with a local symmetry given by the maximal compact group, $\cal H$ \cite{Cremmer,Julia,FerraraScherkZumino, CremmerFerraraScherk}. These symmetry groups were realised as scalars parameterising the coset spaces of $\frac{\cal G}{\cal H}$ and are listed in table \ref{tab:TheHiddenSymmetriesEmergentUponDimensionalReductionOfD11Supergravity}.
\begin{table}[htbp]
	\centering
		\begin{tabular}{|c|c|c|}
			\hline
			Dimension & $\cal{G}$ &$\cal{H}$\\
			\hline
			$10$ & $\mathbb R$ &$1$\\
			\hline
			$9$ & $Sl(2,{\mathbb R})\times {\mathbb R}$ &$U(1)$\\
			\hline
			$8$ & $Sl(3,{\mathbb R})\times Sl(2,{\mathbb R})$ &$SO(3)\times SO(2)$\\
			\hline
			$7$ & $Sl(5,{\mathbb R})$ &$SO(5)$\\
			\hline
			$6$ & $SO(5,5,{\mathbb R})$ &$SO(5)\times SO(5)$\\
			\hline
			$5$ & $E_6$ &$USp(8)$\\
			\hline
			$4$ & $E_7$ &$SU(8)$\\
			\hline 
			$3$ & $E_8$ &$SO(16)$\\
			\hline
		\end{tabular}
	\caption{The hidden symmetries emergent upon dimensional reduction of D=11 supergravity}
	\label{tab:TheHiddenSymmetriesEmergentUponDimensionalReductionOfD11Supergravity}
\end{table}
Hidden symmetries had also been uncovered in related IIB supergravity theories. $Sl(2,\mathbb{R})$, hidden symmetries were uncovered by Schwarz and West in the ten-dimensional IIB supergravity theories\cite{SchwarzWest}. Being a chiral theory the IIB theory is not  not straightforwardly related to the eleven dimensional supergravity.
It has been proposed that discrete subgroups of these continuous hidden symmetries are symmetries of $M$-theory \cite{HullTownsend}. 

Discrete symmetries occur naturally in string theory where symmetries are defined by sets S and T-dualities applied so as to relate a single string theory to itself. S-duality maps a theory of strong string coupling to one of weak coupling. While T-duality interchanges winding modes of wrapped strings around compact dimensions with the momentum carried by the string, as well as the radius of the compact dimension with its reciprocal. T-duality generates the discrete symmetry group $SO(d,d,\mathbb{Z})$ on the parameters of a string theory compactified on a torus $T^d$. In terms of a physical theory these dualities are not only astounding, but are incredibly useful. These are the dualities that led to the conjecture of $M$-theory \cite{Townsend,Witten}, and encode what is known about $M$-theory. The exact nature of which theories are related to each other by these dualites may be found in the literature \cite{DaiLeighPolchinski,DineHuetSeiberg,Townsend,Witten,Schwarz,Schwarz1}.
The discrete symmetry groups of string theory are unified in the U-duality group \cite{ObersPioline}, which turns out to be the discrete version of the $E_d$ hidden symmetries of supergravity. 

It has been argued that eleven-dimensional supergravity may be expressed as a non-linear realisation of symmetry group $E_{11}$\cite{West1}. In this thesis we take up this idea and attempt to use it to discover tools that may be useful in future work to understand the precise connections between Kac-Moody algebras and M-theory. The advantage of such an $E_{11}$ symmetry is that the hidden symmetries of table \ref{tab:TheHiddenSymmetriesEmergentUponDimensionalReductionOfD11Supergravity} are made manifest. Such a symmetry group would, of course, contain within it the U-duality groups in various dimensions.

Unfortunately, Kac-Moody algebras are not very well understood. In fact it may be argued that the murky nature of M-theory may just as well be used to inform mathematical insights about special classes of Kac-Moody algebras as vice versa. $E_{11}$ is just one of the very-extended Lie groups, $G^{+++}$. The other $G^{+++}$ are also associated with the conjectured symmetries of alternative supergravity theories, known as the oxidised supergravity theories. These oxidised supergravity theories are the result of commencing in a three-dimensional theory with the scalars parameterising a particular coset group, and then applying the reverse process to dimensional reduction to find the oxidised theory \cite{CremmerJuliaLuPope}. Consequently it may be hoped that the study of these alternative theories may prove useful for discovering new properties concerning the relation of $E_{11}$ to $M$-theory.

The particular aspects of Kac-Moody algebras in modern high energy physics as studied in this thesis present just a small window onto an expanding and exciting area of study. There is an abundance of elegant mathematical literature associated to Kac-Moody algebras and there is a growing interest in such algebras as symmetries of $M$-theory in theoretical physics for the reasons highlighted above. Consequently there are very many other interesting and exciting approaches that will not be touched on in this thesis. By way of apology we mention one notable example concerning the algebra $E_{10}$ in M-theory, in particular the cosmological billiards programme \cite{DamourHenneauxNicolai} which analyses dynamics near a spacelike symmetry, and uncovers a role for the $E_{10}$ algebra. For a recent review of much work not discussed in this thesis see \cite{deBuyl} and the references therein. 

In this thesis we present the basic tools and definitions required to comprehend the $E_{11}$ conjecture. In chapter two we present some basic group theory, in order to familiarise the reader with the identities and definitions that allow us to give the definition of a Kac-Moody group at the end of the chapter. Chapter three concentrates on motivating the $E_{11}$ conjecture; we describe the nonlinear realisation of a theory on a coset space, and how this relates to an extended formulation of eleven dimensional supergravity having a manifest $E_8$ symmetry. At the end of chapter three we review the general procedure for finding coset symmetries from the scalars that emerge in a dimensional reduction, in particular we review the work of \cite{LambertWest} and recover the $E_8$ coset symmetry in the reduction to three dimensions. In chapter four, we analyse the Kac-Moody algebra of $E_{11}$ which may be constructed by extending the Dynkin diagram of the semisimple Lie group $E_8$ by adding to it three appropriately placed nodes. The corresponding Cartan matrix may be used to construct a unique Kac-Moody algebra from commutators of the generators of the positive and negative roots of $E_{11}$ and its Cartan sub-algebra, subject to the Serre relations. The resulting algebra may be studied using weights of a more manageable finite dimensional $A_{10}$ sub-algebra of $E_{11}$, obtained by deletion of the orthogonal node in the Dynkin diagram. The number of multiples of the deleted root, or level, is used to classify the resulting infinite dimensional set of generators. In chapter four we carry out the construction of the algebra in explicit detail. The low level generators from this construction exactly reproduce the fields of the bosonic sector of eleven dimensional supergravity and, by extension, M-theory. We also construct the algebra of the conjectured Kac-Moody symmetry of the bosonic string. The formulae of chapter four are used to produce the tables of roots given in appendix \ref{roottables}, which are used throughout the thesis. In chapter five we present a group element of a general very-extended Lie group, $\cal G^{+++}$, and show by construction that it encodes the brane solutions of the maximally oxidised supergravity theories. By applying this group element to so-called high-level roots that occur in the Kac-Moody algebras we begin to probe parts of the oxidised theories that are not so well understood and we find evidence that the higher level roots of the Kac-Moody algebras may play an important role in the theory. The embedding of spacetime signature in the algebra is the focus of chapter six, where we review the work of Keurentjes which found that M-theory as derived from $E_{11}$ exists in multiple signatures, and we look at the effect of signature changes on the solutions of $M$-theory. As a corollary to this investigation, the spacelike S-brane solutions of M-theory arise naturally from the solution generating group element discussed in chapter five. We finish in chapter seven by demonstrating that a conjectured $E_{11}$ symmetry gives rise naturally to the exotic brane charge multiplets of \cite{ObersPioline} arising from the U-duality group, in various dimensions.  

A conscious effort has been made while writing this thesis not to assume much prior knowledge of the field of study on the part of the reader. The majority of this work is devoted to building a bridge between the study of Kac-Moody algebras and theoretical physics and we hope the gentle introduction to our research presented here proves useful in this exciting endeavour. 

Much of the original work in this thesis is also contained in \cite{CookWest,CookWest2}.

\newpage

\section{Some Group Theory}
The subject matter of this thesis has a dual nature, being in part group theoretical and in part based upon modern theoretical physics. Given that M-theory, to which this work pertains, is an unknown entity much of this thesis will focus on group theoretical approaches and seek to apply this knowledge to the theoretical physics. In this chapter we present some fundamental results and techniques in the representation theory of Lie groups and it represents a lightning introduction to a beautiful subject, the classification of the semisimple Lie groups, which deserves a longer treatment. Our main reference text is the excellent introduction by Hall \cite{Hall} from which we have taken or based the majority of our definitions in this chapter, but we have also benefited greatly from the freely available and concise text by Cahn \cite{Cahn}. Readers interested in studying the subjects mentioned in this chapter in greater detail may benefit from the standard references \cite{Humphreys, FultonHarris, BrockertomDieck, Cornwell, Georgi, Varadarajan, Jacobsen, Curtis, Baker}. Due to the long-established nature of the material presented here we refer the reader to the literature previously mentioned to deduce the appropriate historical citations.

It is possible to approach the theory of semisimple Lie groups with the great generality of a pure mathematician, however we opt to dirty our hands in a number of places in order to get a firm appreciation of the concepts we are defining. The definitions of this chapter will be used frequently throughout the thesis and are motivated by gaining just the minimal insight into the subject to be able to understand the definition of a Kac-Moody algebra, which is given at the end of the chapter. For a thorough review of Kac-Moody algebras see \cite{Kac}.

\subsection{What's in a group?}
\begin{description}\item[Definition 2.1]
A \emph{group} is defined as a set, $G$, endowed with an action mapping $G\times G \rightarrow G$ subject to the following conditions:
\begin{enumerate}
	\item For all $g,h,k\in G$, $g(hk)=(gh)k$    (\emph{Associativity})
	\item There exists an identity element, e, in the set, that acts in the following way under the group action,$ge=eg=g$ (\emph{Identity element})
  \item For each element, $g$, there exists an inverse element, $g^{-1}\in G$ such that, $gg^{-1}=g^{-1}g=e$ (\emph{Inverse element})
\end{enumerate}
\end{description}
Groups consist of objects, in a set, which act upon each other so as to form another member of the set. The study of groups began in earnest with Evariste Galois (1811-1832) who considered permutation groups, which transformed the solutions of algebraic equations into each other, in order to discover that the symmetry properties of the solutions to a quintic equation implied that there could be no general solution to the quintic, as there are for equations of degree two, three or four. The work of Galois was motivated by the study of algebraic equations which have finitely many solutions\footnote{For a mathematical introduction to Galois theory see \cite{Stewart}, and for a popular account of the problem that inspired Galois' groups see \cite{Livio}.}; the Norwegian mathematician Sophus Lie (1842-1899) aimed to apply a similar method to analyse the solutions of differential equations \cite{Ronan}, which have infinitely many solutions (as embodied by the familiar refrain "up to a constant").
\subsection{Lie Groups}
While Galois' groups permuted finitely many solutions between themselves, Lie's groups morphed a continuous set of solutions into each other. That is the action of objects in Lie's groups resulted in a continuous symmetry, and operations could be continuously transformed into each other. The essential characteristic of a Lie group that differentiates it from other groups is that it acts upon an infinite set; a good example is the group of rotations in a two dimensional plane, $SO(2,\mathbb{R})$, where the generic element is a rotation by an angle $\theta$ but where $\theta$ is a continuous parameter - it is in this sense that Lie groups are "infinite". It is however a very natural infinity that arises when we consider Lie groups, the kind that physical finite systems appear to exhibit at the macroscopic level. Of course our microscopic picture of the universe is that it is not continuous, but that it is quantum mechanical in nature and even spacetime is quantised so that certain continuous symmetries must break down. However Lie groups have a very important role to play at the quantum level, where the Schrodinger wave equation, being a differential equation has solutions which are continuously related by the action of a Lie group. In short Lie groups are par for the course in modern theoretical physics, see for example \cite{Georgi}. The interplay between continuous mathematics and discrete mathematics demonstrated by Lie groups and Galois' permutation groups has parallels with the continuous properties of spacetime macroscopically and the discrete quantised spacetime expected at the microscopic level. While the route between the infinite of the Lie group and the finiteness of the simple groups has been bridged in mathematics via an analogue of modular arithmetic, the infinite symmetries of macroscopic spacetime have yet to be captured in a finite mathematical picture that neatly interpolates between quantum mechanics and general relativity. For an enjoyable popular account of the classification of the finite simple groups see \cite{Ronan}.

Let us state the formal definition of a Lie group, before we specialise to the case of a matrix Lie group.

\begin{description}
	\item[{\bf Definition 2.2}]A \emph{Lie Group} is a smooth manifold $\cal{M}$ endowed with a smooth action, and its inverse, mapping $\cal{M}\times \cal{M}\rightarrow \cal{M}$. 
\end{description}

The fact that a Lie group acts on a manifold encodes all the necessary properties of continuous maps. The existence of a group action, and its inverse, make the manifold into a group. A straightforward example is ${\cal M}={\mathbb R}^n$ where the group action is vector addition, defined as $x \times y\rightarrow x+y$. More interesting examples include the classical groups (the special linear, orthogonal, unitary and symplectic groups), the exceptional groups ($G_2$,$F_4$,$E_6$,$E_7$,$E_8$), the Heisenberg group, the spin group, the Lorentz group and the Poincare group.

An important subset of Lie groups are matrix Lie groups, which are closed subgroups of the linear transformation group $GL(n,\mathbb{C})$ and so may be represented by matrices. A more technical definition is,

\begin{description}
	\item[{\bf Definition 2.3}]A \emph{matrix Lie Group} is a subgroup, $\cal{G}$, of $GL(n,\mathbb{C})$ such that any convergent sequence of matrices $A_m\in\cal{G}$ converges as $m\rightarrow\infty$ to a matrix that is either in $\cal{G}$ or is not invertible.	
\end{description}

A sequence of matrices is a set of matrices which transform into each other as some parameter is varied. Convergence of such a sequence, $A_m$ to a matrix $B$ implies that the components $(A_m)_{ij}\rightarrow B_{ij}$ as $m\rightarrow\infty$. For example, consider the subgroup $SL(2,\mathbb{C})$, which contains a convergent sequence of matrices,
\[A_m= \left( \begin{array}{cc}
\frac{1}{m} & -1  \\
1 & 0  \end{array} \right)\] 
It converges to an element of $SL(2,\mathbb{C})$. In fact $SL(2,\mathbb{C})$ is a matrix Lie group since it is the subgroup of $GL(2,\mathbb{C})$ with determinant +1, so any sequence of matrices in $SL(2,\mathbb{C})$ all have determinant +1. Since the determinant is a continuous function any convergent sequence will converge to a matrix in $SL(2,\mathbb{C})$. 

All matrix Lie groups are also Lie groups but the converse is not true. Examples of matrix Lie groups are the general linear groups, the unitary groups, the orthogonal groups, the symplectic groups, the Heisenberg group, the Lorentz group, the Poincare group and the exceptional groups. The universal cover of $SL(n,\mathbb{R})$ where $n>1$ is not a matrix Lie group and neither is the spin group, the double cover of $SO(n,\mathbb{C})$ where $n>2$. The proof of both these statements is achieved by demonstrating that these groups do not have any faithful finite dimensional representations since the kernel of the covering homomorphism, the Lie group homomorphism that projects the simply-connected covering group down to the connected group, is non-trivial in each case. However, both the universal cover of $SL(n,\mathbb{R})$ where $n>1$ and the spin group are Lie groups.

\subsection{Compact, Connected and Simply-Connected Groups}
All successful classification programmes rest between the entirely general and the esoterically precise, so it is with the classification of the classical and exceptional Lie groups. The classification relevant to this thesis was worked on by Cartan and Killing and completed by Chevalley in the 1950s and classifies the semisimple matrix Lie groups by their Dynkin diagrams\footnote{Historically, Dynkin diagrams were first used by H. S. M. Coxeter at least four years before Dynkin \cite{Coxeter} as a notation for recording the angles between mirrors in a kaleidoscope arranged to give rise to regular geometric images of an object in its interior. However, the term Dynkin diagram is more commonly used than Coxeter diagram or even Coxeter-Dynkin diagram and we use it in this thesis for the purpose of common clarity and hope that the role of Coxeter in creating these diagrams is recognised by the reader.}. In order to define and to appreciate some of the qualities of the semisimple matrix Lie groups we need a technical understanding of when a matrix Lie group is compact, connected and simply-connected.

\begin{description}
	\item[{\bf Definition 2.4}]A matrix Lie group, $\cal{G}$, is \emph{compact} if there exists an upper bound, $K$, not exceeded by the modulus of any of the components $|A_{ij}|$ for any matrix $A\in\cal{G}$. Additionally any convergent sequence of matrices in $\cal{G}$ converges to a matrix in $\cal{G}$.
\end{description}

These conditions mean that the matrix Lie group is closed and bounded, and we give some examples of matrix Lie groups which are compact and those which are not in Table (\ref{tab:CompactMatrixLieGroups}).
\begin{table}[h]
	\centering
		\begin{tabular}{c|c}
			Compact & Non-compact \\\hline
			$O(n,\mathbb{R})$&$GL(n,\mathbb{R})$\\
			$SO(n,\mathbb{R})$&$GL(n,\mathbb{C})$\\
			$U(n,\mathbb{R})$&$SL(n,\mathbb{R})$\\
			$SU(n,\mathbb{R})$&$SL(n,\mathbb{C})$\\
			$Sp(n,\mathbb{C})\cap U(2n,\mathbb{C})$&$O(n,\mathbb{C})$\\
			$U(n,\mathbb{C})$&$SO(n,\mathbb{C})$\\
			$SU(n,\mathbb{C})$&$Sp(n,\mathbb{R})$\\
			&$Sp(n,\mathbb{C})$\\
			&the Lorentz group\\
			&the Heisenberg group\\
			&the Poincare group\\
		\end{tabular}
	\caption{Examples of Compact Matrix Lie Groups}
	\label{tab:CompactMatrixLieGroups}
\end{table}

\begin{description}
	\item[{\bf Definition 2.5}]A matrix Lie group, $\cal{G}$, is \emph{path-connected} if there exists a continuous path $A(t)\in \cal{G}$ connecting any two matrices, $A_1,A_2\in\cal{G}$ such that $A(0)=A_1$ and $A(1)=A_2$. 	
\end{description}
If a matrix Lie group is path-connected then it is necessarily connected. We give examples of connected and not connected matrix Lie groups in table (\ref{tab:ConnectedLieGroups}).
\begin{table}[h]
	\centering
		\begin{tabular}{c|c}
			Connected & Not connected \\\hline
			$GL(n,\mathbb{C})$&$GL(n,\mathbb{R})$\\
			$SL(n,\mathbb{C})$&$O(n,\mathbb{R})$\\
			$SL(n,\mathbb{R})$&$O(n,\mathbb{C})$\\
			$SO(n,\mathbb{C})$&the Poincare group\\
			$SO(n,\mathbb{R})$&the Lorentz group\\
			$U(n,\mathbb{R})$&\\
			$U(n,\mathbb{C})$&\\
			$SU(n,\mathbb{R})$&\\
			$SU(n,\mathbb{C})$&\\
			$Sp(n,\mathbb{C})\cap U(2n,\mathbb{C})$&\\
			$Sp(n,\mathbb{C})$&\\
			the Heisenberg group&\\
		\end{tabular}
	\caption{Examples of connected matrix Lie groups}
	\label{tab:ConnectedLieGroups}
\end{table}

\begin{description}
	\item[{\bf Definition 2.6}]A matrix Lie group is \emph{simply-connected} if it is connected and if every loop, $A(t)\in \cal{G}$ can be continuously shrunk to a matrix $A\in\cal{G}$.	
\end{description}
Specifically one considers a continuous loop $A(t)\in\cal{G}$ such that $A(0)=A_0$, $A(1)=A_1$ and $A_0=A_1$ and asks whether there exists a continuous function $B(s,t)$ that morphs the loop $A(t)$ into the matrix $B\in\cal{G}$, or algebraically, such that $B(0,t)=A(t)$, $B(s,0)=B(s,1)$ and $B(1,t)=B$. If such a function exists for all $A_0,B\in\cal{G}$ then the group $\cal{G}$ is simply-connected. We give some examples of simply-connected matrix Lie groups in table (\ref{tab:SimplyConnectedLieGroups}) below.
\begin{table}[h]
	\centering
		\begin{tabular}{c|c}
			Simply-Connected & Connected but not Simply-Connected \\\hline
			$SL(n,\mathbb{C})$&$SO(n,\mathbb{R})$ $n\geq2$\\
			$SO(n,\mathbb{R})$&$SO(n,\mathbb{C})$ $n\geq2$\\
			$SU(n,\mathbb{R})$&$U(n,\mathbb{R})$\\
			$SU(n,\mathbb{C})$&$U(n,\mathbb{C})$\\
			$Sp(n,\mathbb{C})\cap U(2n,\mathbb{C})$&$GL(n,\mathbb{C})$\\
			$Sp(n,\mathbb{C})$&$SL(n,\mathbb{R})$ $n\geq2$\\
			&$SO(n,\mathbb{C})$\\
		\end{tabular}
	\caption{Examples of simply-connected matrix Lie groups}
	\label{tab:SimplyConnectedLieGroups}
\end{table}
\subsection{The Exponential Map and the Lie Algebra}
In practise when working with Lie groups it is often simpler to work with the Lie algebra. The Lie algebra contains much of the information about the group but operations on the algebra are linear and so it is often simpler to use the algebra for calculations. The group and its algebra are related by the exponential map, which we now define for a matrix.
\begin{description}
	\item[{\bf Definition 2.7}]The \emph{exponential map}, $e^A$ or $\exp{A}$, of a matrix, $A$, is defined by:
	\begin{equation}e^A=\sum_{n=0}^{\infty}\frac{A^n}{n!}\end{equation}	
\end{description}
The exponential map always converges. This can be seen if we know that any matrix, $A$, may be decomposed into a diagonal part, $D$, and a nilpotent part, $N$, with the two parts commuting, such that $A=D+N$. Consequently, $e^A=e^De^N$ and $e^D$ converges since it becomes a diagonal matrix of exponentials, and $e^N$ terminates as a sum after a finite number of terms ($N$ is nilpotent $\implies$ $N^k=0$ for some integer $k$).
\begin{description}
	\item[{\bf Definition 2.8}]The \emph{Lie algebra}, $\textfrak{g}$, of a matrix Lie group $\cal{G}$ is the set of all matrices, $X$, such that $e^{tX}\in\cal{G}$ for all $t\in \mathbb{R}$.
\end{description}
There is a unique correspondence between paths $A(t)\in\cal{G}$ and the exponentiation of an element $X$ of the Lie algebra, $\textgoth g$. Differentiating the one-parameter subgroup $e^{tX}$ with respect to $t$ and evaluating at $t=0$ gives $X$ so the unique correspondence is observed. It is worth observing that as $t$ runs from $0$ to $1$, $e^{tX}$ is a path in $\cal{G}$ connecting the identity element to $e^X$.  Furthermore, $\left.\frac{d}{dt}\right|_{t=0}$ acting upon $e^{tX}$ gives $X$ as the tangent vector to $\cal{G}$ at the identity, so that the Lie algebra is the tangent space of the Lie group at the identity.

When $t=1$ the one-parameter subgroup $e^{tX}$ is called the exponential mapping,
\begin{description}
	\item[{\bf Definition 2.9}]The \emph{exponential mapping} of a matrix Lie group $\cal{G}$ with Lie algebra, $\textgoth{g}$ is given by, 
	\begin{equation}\exp : \textgoth{g}\rightarrow \cal{G}\end{equation}
\end{description}
Note that this does not imply that every element of $\cal{G}$ may be expressed as a single exponential of an element of its Lie algebra. Generally the exponential map is not an isomorphism between the algebra and the group, in fact the exponential mapping is neither necessarily onto nor one-to-one. However under certain conditions which we will describe presently the exponential mapping is a bijection, an inverse map exists and both are continuous, so that the exponential mapping becomes a homeomorphism. The existence of the homeomorphism rests upon the existence of an inverse map to the exponential map, the matrix logarithm.
\begin{description}
	\item[{\bf Definition 2.10}]The \emph{matrix logarithm} of a matrix $A$ is defined, when it is a convergent series, as, \begin{equation}\log{A}=\sum_{n=1}^{\infty}(-1)^{n+1}\frac{(A-I)^n}{n}\end{equation}
\end{description}
This is the natural generalisation of the Mercator series to a matrix variable. The Mercator series for the natural logarithm of a variable $x$ converges when $|x|<1$, similarly the series expansion for the matrix logarithm only converges if $\left\|A-I\right\|<1$. Where we define $\left\|\cdot\right\|$ to be a norm on the space of matrices, for example, the Hilbert-Schmidt norm,
\begin{equation}\left\|A\right\|\equiv\sqrt{\sum_{i,j=1}^{n}(A_{ij})^2}\end{equation}
Where $A$ is an $n$-by-$n$ matrix. Crucially we can define a radius of convergence in terms of the matrix norm classifying an open set of matrices, $X\in U$, about the identity matrix such that $\log{(e^X)}=X$ where, 
\begin{equation}U=\{X\in\textgoth{g}|\left\|X\right\|<\ln{2}\}\end{equation} so that $e^X\in\cal{G}$. So long as we consider the open set $U$ about the identity matrix in the Lie algebra then we can treat the exponential mapping as a homeomorphism between the algebra and the matrix Lie group. We only capture the group structure local to the group identity from the algebra using the exponential mapping of a single element of the Lie algebra. If $\cal{G}$ is a connected matrix Lie group, then we can find a path from the identity to any element in the group. We find in fact,
\begin{description}
	\item[{\bf Theorem 2.12}]If $\cal{G}$ is a connected matrix Lie group, then every element $A\in\cal{G}$ may be expressed in the form:
\begin{equation}A=e^{X_1}e^{X_2}\ldots e^{X_n}\end{equation} where $X_1, X_2 \ldots X_n\in\textgoth{g}$.
\end{description}
\emph{Proof} Let $A(t)$ be a path in $\cal{G}$ connecting the identity, $I$, to an arbitrary element, $A\in\cal{G}$, such that $A(0)=I$ and $A(1)=A$. Taking $U$ as defined above and defining $V=\exp{U}$ to be the associated open set in the group. Note that every element of $U$ under the exponential mapping goes to an element of $\cal{G}\in V$. We are interested in finding the map for arbitrary elements of $\cal{G}$, in particular those not in the open set $V$.

\begin{figure}[cth]
\hspace{40pt}\includegraphics[scale=0.9,angle=90]{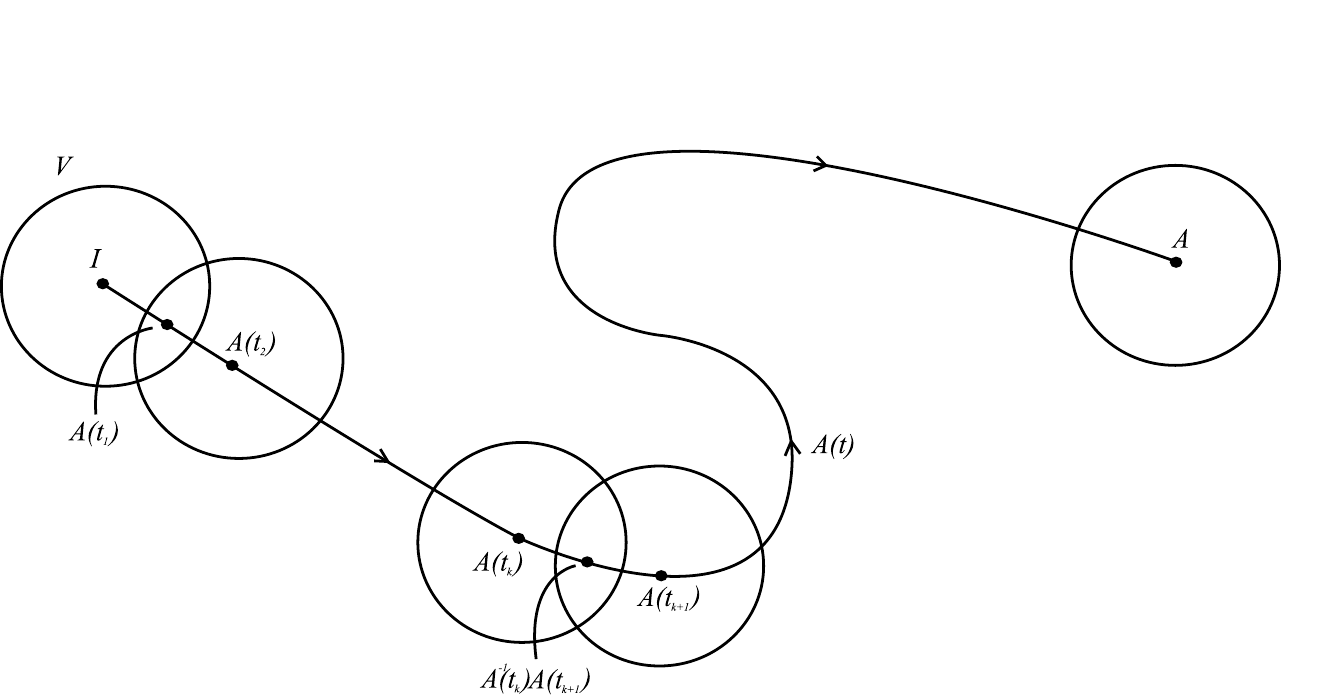}
\caption{An arbitrary path, $A(t)$, through a connected matrix Lie group.} \label{grouppath}
\end{figure}

An arbitrary path, $A(t)$, in the group is shown in figure \ref{grouppath}. In order to move about the group one acts upon the present group element with another element, so that the infinitesimal part of the path $A(t)$ running from $A(t_k)$ to $A(t_{k+1})$ corresponds to the action of the group element $A^{-1}(t_k)A(t_{k+1})$ under a right-action. Now around each group element in the path there is an open set where the matrix logarithm is well-defined, and since the path $A(t)$ is continuous one can always find consecutive group elements $A{(t_1)},A{(t_2)}\ldots A{(t_n)}$ such that the open sets, $V$, around each element overlap, and within that overlap there exists a transition element, $A^{-1}(t_k)A(t_{k+1})$ taking $A(t_k)$ to $A(t_{k+1})$. Using the well-defined matrix logarithm we can find the corresponding element of the Lie algebra, let us label the Lie algebra element for the transition element $A^{-1}(t_k)A(t_{k+1})$ by $X(k+1)$. Then $e^{X(k+1)}=A^{-1}(t_k)A(t_{k+1})$ and the group element $A$ is expressed as,
\begin{align}
\nonumber A&=IA_1(A_1^{-1}A_2)(A_2^{-1}A_3)\ldots(A_{n-1}^{-1}A) \\
&=e^{X_1}e^{X_2}\ldots e^{X_n}
\end{align}
\subsection{The Adjoint and the Lie Bracket}
There is a natural representation of the Lie group which connects it to the Lie algebra, it is called the adjoint representation. Let us describe the adjoint transformation,
\begin{description}
	\item[{\bf Definition 2.13}]Let $A\in\cal{G}$ and $X\in\textgoth{g}$ where $\textgoth{g}$ is the Lie algebra of $\cal{G}$. Then the \emph{adjoint transformation} of $X$ is 
\begin{equation}Ad_A(X)=AXA^{-1}\end{equation}
\end{description}
The adjoint map is an automorphism of the Lie algebra, since,
\begin{equation}e^{tAXA^{-1}}=Ae^{tX}A^{-1}\in{\cal{G}}\quad \forall \quad t\in\mathbb{R}\end{equation}
Hence, $AXA^{-1}\in\textgoth{g}$. One can also view the adjoint transformation as an action on the Lie algebra that maps an element of the Lie group, $A$ into a linear transformation of the Lie algebra, $GL(\textgoth{g})$. Let us recall the notion of a representation.
\begin{description}
	\item[{\bf Definition 2.14}]A \emph{representation} of a group $\cal{G}$ acting on a vector space $V$ is a homomorphism, $\pi$, 
\begin{equation}\pi : {\cal{G}}\rightarrow GL(V)\end{equation}
\end{description}
A representation turns a group action into a matrix action over a suitable vector space. Now,
\begin{equation}Ad_A\cdot Ad_B (X)=ABXB^{-1}A^{-1}=Ad_{AB}(X)\end{equation} 
so the adjoint is a homomorphism and since, 
\begin{equation}Ad_{\cal{G}}:{\cal{G}}\rightarrow GL(\textgoth{g})\end{equation}
then the adjoint transformation is a representation of $\cal{G}$ acting upon its own Lie algebra.
From the group representation we can derive a representation of the algebra by acting with, $\left.\frac{d}{dt}\right|_{t=0}$ on the adjoint action,
\begin{equation}\left.\frac{d}{dt} Ad_A(X)\right|_{t=0}=\left.\frac{d}{dt} e^{tY}(X)e^{-tY}\right|_{t=0}=YX-XY\equiv[Y,X]\end{equation}
That is the adjoint action on the Lie algebra is, 
\begin{equation}ad_Y(X)=[Y,X]\end{equation} This is also a homomorphism,
\begin{align}
\nonumber [ad_Z,ad_Y](X)&=ad_Zad_Y(X)-ad_Yad_Z(X)\\
\nonumber &=[Z,[Y,X]]-[Y,[Z,X]]\\
\nonumber &=-[X,[Z,Y]]\\
&=ad_{[Z,Y]}(X)
\end{align}
This homomorphism is the Jacobi identity. In showing it we implicitly assumed that the multiplication operator in the Lie algebra is the bracket $[,]$, which is also called the Lie bracket. Recall we demonstrated that the adjoint action $Ad_A$, for $A\in\cal{G}$ acted as an automorphism of the Lie algebra, so that $AXA^{-1}\in\textgoth{g}$ if $X\in\textgoth{g}$. Now the Lie algebra is a vector space closed under addition, that is, $X,Y\in\textgoth{g} \implies X+Y\in\textgoth{g}$ since $e^{t(X+Y)}\in\cal{G}$. Then $e^{tY}Xe^{-tY}-X\in\textgoth{g}$, and, noting that $X\in\textgoth{g} \implies tX\in\textgoth{g}$, then $\frac{e^{tY}Xe^{-tY}-X}{t}\in\textgoth{g}$ for all $t\in\mathbb{R}$. In particular,
\begin{equation}\lim_{t\rightarrow0}[\frac{e^{tY}Xe^{-tY}-X}{t}]\in\textgoth{g}\end{equation}
This is simply the derivative evaluated at $X$ along the path $e^{tY}Xe^{-tY}$ from $t=0$ to $t=1$,
\begin{equation}\left.\frac{d}{dt}(e^{tY}Xe^{-tY})\right|_{t=0}=YX-XY=[Y,X]\in\textgoth{g}\end{equation}
We have shown that if $X,Y\in\textgoth{g}$ then $[X,Y]\in\textgoth{g}$ in order to convince ourselves that $ad_X$ is a representation of the Lie algebra.
\subsection{Campbell-Baker-Hausdorff and other identities} \label{identities}
It will be necessary for us to manipulate terms in the connected group element, in particular to move exponentials of different elements of the algebra past each other. In this section we find an identity for achieving this end in special circumstances and also give the Campbell-Baker-Hausdorff formula for the more general case.

Consider a path in the algebra,
\begin{equation}
a(t)=e^{tX}Ye^{-tX} \label{expadj}
\end{equation}
Where $X,Y\in\cal{G}$ and $t\in\mathbb{R}$. Differentiating with respect to the parameter $t$, we find,
\begin{equation}
\frac{da(t)}{dt}=e^{tX}[X,Y]e^{-tX}=[X,a(t)] 
\end{equation}
Since $a(0)=Y$ we find,
\begin{equation}
a(t)=e^{tad_X}Y \label{expadj1}
\end{equation}
Now consider a path,
\begin{equation}b(t)=e^{tY}e^{tX}e^{-t(Y+X)}\end{equation}
Differentiating, we find,
\begin{equation}\frac{db(t)}{dt}=e^{tY}e^{tX}(e^{-tX}Ye^{tX}-Y)e^{t(Y+X)}=e^{tY}e^{tX}(a(-t)-Y)e^{t(Y+X)}\end{equation}
Now we restrict ourselves to the special case where the commutator, $[X,Y]$, itself commutes with the two generators $X$ and $Y$, i.e. $[X,[X,Y]]=[Y,[X,Y]]=0$. In particular this means that the expression for $a(-t)$ terminates quickly,
\begin{equation}a(-t)=Y-t[X,Y]\end{equation}
So that,
\begin{equation}\frac{db(t)}{dt}=e^{tY}e^{tX}t[Y,X]e^{t(Y+X)}=t[Y,X]b(t)\end{equation}
And,
\begin{equation}b(t)=e^{\frac{1}{2}t^2[Y,X]}\end{equation}
Since $[Y,X]$ commutes with $Y$ and $X$ we may rearrange this expression for $b(t)$ for the case when $t=1$ to find,
\begin{equation}e^Ye^X=e^{(Y+X)+\frac{1}{2}[Y,X]}=e^{Y+X}e^{\frac{1}{2}[Y,X]}\end{equation}
Swapping $X$ and $Y$ gives,
\begin{equation}e^Xe^Y=e^{X+Y}e^{-\frac{1}{2}[Y,X]}\end{equation}
And combining these two we find,
\begin{equation}e^Ye^X=e^Xe^Ye^{[Y,X]}\end{equation}
This is an expression we will use frequently throughout this thesis.
If we had not insisted on $[X,Y]$ commuting with $X$ and $Y$ and had taken a lengthier route we would have found a more complex expression for combining multiple exponentials into a single exponential. This expression is called the Baker-Campbell-Hausdorff formula,
\begin{description}
	\item[{\bf Definition 2.15}]The \emph{Campbell-Baker-Hausdorff} formula for combining the product of two matrix exponentials, $e^X$ and $e^Y$, into a single exponential is	
	\begin{equation}e^Xe^Y=e^{X+\int_0^1dtf(e^{ad_X}e^{tad_Y})(Y)}\end{equation} 
	Where, 
	\begin{equation}f(X)=I+\sum_{i=1}^\infty\frac{(-1)^{i+1}}{i(i+1)}(X-I)^i\end{equation}
\end{description}
In particular the first few terms in the series expansion are,
\begin{equation}e^Xe^Y=e^{X+Y+\frac{1}{2}[X,Y]+\frac{1}{12}[X,[X,Y]]-\frac{1}{12}[Y,[X,Y]]+\ldots}\end{equation}
Where no pattern in the series is inferred by the dots, just missing higher order terms.  
\subsection{Representations of SU(2)}
The algebra of the two-dimensional special unitary group will be the prototype group used for the classification of the semisimple Lie algebras. Let us spend some time looking at the representations of $SU(2)$ and its algebra, $su(2)$ and their classification. A particularly illuminating set of representations of $SU(2)$ acts on the space of homogeneous polynomials of degree $m$ in two complex variables, $z_1$, $z_2$, which we denote, $V_{m+1}$ and whose prototype "vector" is,
\begin{equation}f=a_0z_1^m+a_1z_1^{m-1}z_2+\ldots+a_mz_2^m\end{equation}
Now elements $U$ of $SU(2)$ are automorphisms of $\mathbb{C}^2$. So we may define a linear transformation acting on the space of polynomials described above by the action of $SU(2)$ upon the two dimensional complex space spanned by $\{z_1,z_2\}$. The representation $\Pi$ of $SU(2)$ is explicitly,
\begin{equation}[\Pi(U)f](z)=f(U^{-1}z)\end{equation}
Where $z=(z_1,z_2)$. This is indeed a homomorphism since,
\begin{equation}\Pi(U)[\Pi(V)f](z)=\Pi(V)f(U^{-1}z)=f(V^{-1}U^{-1}z)=[\Pi(UV)f](z)\end{equation}
The description of the representation is in no way dependent on the degree $m$ of the polynomial and in fact we could pick any value of $m$ and we would find a new representation of $SU(2)$ which we denote $\Pi_m$. None of these representations for different $m$ are equivalent since they act on vector spaces of different dimension, $m+1$, but each one is classified uniquely by the integer $m$. We can derive a representation of the algebra from a representation of the group. If we denote by $\pi_m$ a representation of the algebra $su(2)$, then,
\begin{equation}\pi_m(X)=\left.\frac{d}{dt}\Pi_m(e^{tX})\right|_{t=0}\end{equation}
The representation $\pi_m$ acts on the same space as $\Pi_m$. We will need an explicit basis for $su(2)$, we recall that an element $X$ of $su(2)$ must satisfy the following conditions, 
\begin{equation}(e^{X})(e^{X})^\dagger=I \qquad \det(e^{X})=1\end{equation}
These turn into constraints on the Lie algebra,
\begin{equation}X=-X^\dagger \qquad Tr(X)=0\end{equation}
Where $Tr$ is the trace. A suitable basis is,
\begin{equation}H= \left( \begin{array}{cc}
i & 0  \\
0 & -i  \end{array} \right),
\quad X= \left( \begin{array}{cc}
0 & i  \\
i & 0  \end{array} \right),
\quad Y= \left( \begin{array}{cc}
0 & 1  \\
-1 & 0  \end{array} \right)\end{equation}
We are now in a position to explicitly write down our representation of $su(2)$ on the space of complex polynomials of degree $m$. Let $W\in\{X,Y,H\}$, then,
\begin{align}[\nonumber \pi_m(W)f](z)&=\left.\frac{d}{dt}([\Pi(e^{tW})f](z))\right|_{t=0}\\
\nonumber &=\left.\frac{d}{dt}(f(e^{-tW}z))\right|_{t=0}\\
\nonumber &=\left.(f'\frac{d}{dt}(e^{-tW})z)\right|_{t=0}\\
&=-\frac{\partial f}{\partial z_1}(W_{11}z_1+W_{12}z_2)-\frac{\partial f}{\partial z_2}(W_{21}z_1+W_{22}z_2)
\end{align}
And,
\begin{align}
\nonumber \pi_m(H)&=-iz_1\frac{\partial }{\partial z_1}+iz_2\frac{\partial }{\partial z_2}\\
\nonumber \pi_m(X)&=-iz_2\frac{\partial }{\partial z_1}-iz_1\frac{\partial }{\partial z_2}\\
\pi_m(Y)&=-z_2\frac{\partial }{\partial z_1}+z_1\frac{\partial }{\partial z_2}
\end{align}
Let us see how this representation acts on a generic term of $f$ which takes the form $a_kz_1^mz_2^{k-m}$,
\begin{align}
\nonumber \pi_m(H)a_kz_1^mz_2^{k-m}&=ia_k(m-2k)z_1^kz_2^{m-k}\\
\nonumber \pi_m(X)a_kz_1^mz_2^{k-m}&=ia_k(-kz_1^{k-1}z_2^{m-(k-1)}-(m-k)z_1^{k+1}z_2^{m-(k+1)})\\
\pi_m(Y)a_kz_1^mz_2^{k-m}&=a_k(-kz_1^{k-1}z_2^{m-(k-1)}+(m-k)z_1^{k+1}z_2^{m-(k+1)})
\end{align}
From looking at how the representations of $X$ and $Y$ act on $f$ we deduce that a more convenient set of operators to work with are $\pi_m(X)+i\pi_m(Y)=\pi_m(X+iY)\equiv\pi_m(\hat{X})$ and $\pi_m(X)-i\pi_m(Y)=\pi_m(X-iY)\equiv\pi_m(\hat{Y})$ since,
\begin{align}
\nonumber \pi_m(\hat{X})a_kz_1^mz_2^{k-m}&=-2ika_kz_1^{k-1}z_2^{m-(k-1)}\\
\pi_m(\hat{Y})a_kz_1^mz_2^{k-m}&=2(m-k)a_kz_1^{k+1}z_2^{m-(k+1)}
\end{align}
Now we see that a single term in $f$, $z_1^kz_2^{m-k}$ acts as an eigenvector for $\pi_m(H)$ in the space $V_{m+1}$, and that the action of $\pi_m(\hat{X})$ and $\pi_m(\hat{Y})$ transform $f$ into an eigenfunction of $\pi_m(H)$. These are results that we are very familiar with, although we are, perhaps, more used to working with the bare Lie algebra in order to see them. Having looked at a concrete example of a representation of $su(2)$ let us now look at the more abstract, yet more familiar, version of the same situation using the commutator relations of the Lie algebra. We will need the commutators,
\begin{equation}[H,X]=-2Y, \quad [H,Y]=2X, \quad [X,Y]=-2H\end{equation}
Let us make the change of variables that we made above, i.e. $\hat{X}=X+iY$, $\hat{Y}=X-iY$, we find the new commutators,
\begin{align}
\nonumber [H,\hat{X}]&=[H,X+iY]=-2Y+i2X=2i\hat{X}\\
\nonumber [H,\hat{Y}]&=[H,X-iY]=-2Y-i2X=-2i\hat{Y}\\
[\hat{X},\hat{Y}]&=[X+iY,X-iY]=-2i[X,Y]=4iH
\end{align}
After rescaling $H'=iH$, $\hat{X}'=\frac{1}{2}\hat{X}$ and $\hat{Y}'=\frac{1}{2}\hat{Y}$, we find the familiar set of commutators,
\begin{equation}[H',\hat{X}']=-2\hat{X}', \quad [H',\hat{Y}']=2\hat{Y}', \quad [\hat{X}',\hat{Y}']=H'\end{equation}
We assert that every set of generators whose commutators take this form carry an irreducible representation of $su(2)$, that is uniquely characterised by its dimension $m+1$. Let us see how this comes about. Every representation of $H$, $\pi(H)$ at least has some eigenvalue $\lambda\in\mathbb{C}$ and a corresponding eigenvector $v$ such that,
\begin{equation}\pi(H)v=\lambda v\end{equation}
Just from the commutator relations we find,
\begin{align}
\nonumber \pi(H')\pi(\hat{X}')v&=\pi(\hat{X}')\pi(H')v+[\pi(H'),\pi(\hat{X}')]v=(\lambda-2)\pi(\hat{X}')v \\
\nonumber \pi(H')\pi(\hat{Y}')v&=\pi(\hat{Y}')\pi(H')v+[\pi(H'),\pi(\hat{Y}')]v=(\lambda+2)\pi(\hat{Y}')v
\end{align}
We see that $\pi(\hat{X}')$ is the lowering operator and $\pi(\hat{Y}')$ is the raising operator. And, in general,
\begin{align}
\nonumber \pi(H')\pi(\hat{X}')^nv&=(\lambda-2n)\pi(\hat{X}')^nv \\
\nonumber \pi(H')\pi(\hat{Y}')^nv&=(\lambda+2n)\pi(\hat{Y}')^nv
\end{align}
At this point $\lambda$ may well be a complex number and we have an infinite string of eigenvectors. Let us return to our particular representation, $\pi_m$ on the space $V_{m+1}$ and see what more we can learn. We have already observed that an eigenvector exists for $\pi(H)$, that is, $v=z_1^kz_2^{m-k}$, and we find,
\begin{align}
\nonumber \pi_m(H')z_1^kz_2^{m-k}&=-(m-2k)z_1^kz_2^{m-k}\equiv\lambda v\\
\nonumber \pi_m(H')\pi_m(\hat{X}')z_1^kz_2^{m-k}&=ik(m-2(k-1))z_1^{k-1}z_2^{m-(k-1)}\\
\nonumber &=-(m-2(k-1))\pi_m(\hat{X}')z_1^kz_2^{m-k}=(\lambda-2)\pi_m(\hat{X}') v\\
\nonumber \pi_m(H')\pi_m(\hat{Y}')z_1^kz_2^{m-k}&=i(m-k)(m-2(k+1))z_1^{k+1}z_2^{m-(k+1)}\\
&=-(m-2(k+1))\pi_m(\hat{Y}')z_1^kz_2^{m-k}=(\lambda+2)\pi_m(\hat{Y}') v
\end{align}
We see the expected pattern of eigenvectors but now we are working on an $m+1$ dimensional space and the action of the raising and lowering generators is to shift the eigenvector as indicated here,
\begin{align}
\nonumber &\stackrel{\pi_m(\hat{X}')}{\longrightarrow}\\
\nonumber z_1^m\quad z_1^{m-1}z_2\quad &\ldots \ldots \quad z_1z_2^{m-1}\quad z_2^m\\
&\stackrel{\longleftarrow}{\pi_m(\hat{Y}')}
\end{align}
We have $m+1$ eigenvectors which span the space $V_{m+1}$. Consequently the only invariant subspace of $V_{m+1}$ under the action of $\pi_m(H')$, $\pi_m(\hat{X}')$ and $\pi_m(\hat{Y}')$ is the whole of $V_{m+1}$, hence the representation is irreducible. Furthermore we have found that $\lambda=m-2k\in\mathbb{R}$ and that the sequence of eigenvectors terminates when $k=0$, with eigenvector $z_1^m$ ($\lambda=m$), or $k=m$, with eigenvector $z_2^m$ ($\lambda=-m$). Therefore irreducible representations of $su(2)$ are uniquely characterised by their highest eigenvalue, $m$. This is the cornerstone of classifying representations of larger algebras, such as $su(n)$ representations which will be classified by $n-1$ such highest eigenvalues.
\subsection{Weights and Roots}
The classification of the semisimple Lie groups was completed over half a century ago and the accomplishment itself was achieved within about half a century of its inception. The problem was worked on by many mathematicians, but notable contributors include Leonard Eugene Dickson (1874-1954), Elie Cartan (1869-1951), Wilhelm Killing (1847-1923) and Claude Chevalley (1909-1984), who completed the most satisfying version of the classification. The modern approach classifies the groups by their Dynkin diagrams, an approach that requires us to generalise our notion of classifying a representation of $su(2)$ by its highest eigenvector to larger groups. The generalisation of the eigenvector is called the weight and a special case of the weight is called a root, and it is the properties of the simple roots of a semisimple Lie algebra that are encoded in its Dynkin diagram.

The strategy for $su(2)$ began by finding an eigenvalue and eigenvector for a representation of its commuting element $H$. When we work with larger algebras, we will find that the generators are for the mostpart multiple copies of the generators of $su(2)$. Importantly there may be more than one such element that commutes with the rest of the algebra. The largest set of commuting generators forms the Cartan sub-algebra.
\begin{description}
	\item[{\bf Definition 2.16}]The \emph{Cartan sub-algebra} is an abelian sub-algebra, $\textgoth{h}\subset\textgoth{g}$, whose adjoint representation has as its eigenvectors the rest of the generators of the algebra, $\textgoth{g}$.
\end{description}
The Cartan sub-algebra of $su(2)$ only contained the element $H'$. Any commutating set of matrices on a complex vector space has at least one simultaneous eigenvector. Suppose that $\pi(H_1)$ and $\pi(H_2)$ are representations of elements of a Cartan sub-algebra. Certainly $\pi(H_1)$ has an eigenvalue, $\lambda$, (since we are working over the complex numbers) and an eigenvector, $v$. So we find,
\begin{equation}\pi(H_2)\pi(H_1)v=\lambda[\pi(H_2) v]=\pi(H_1)[\pi(H_2)v]\end{equation}
So that $\pi(H_2)v$ is another eigenvector, with the same eigenvalue, for $\pi(H_1)$ - i.e. it acts as an endomorphism of the eigenspace of $\pi(H_1)$ the subspace spanned by the eigenvectors having eigenvalue $\lambda$ for $\pi(H_1)$. If we now look at the action of $\pi(H_2)$ on this eigenspace we can find a new eigenvector, $u$, of $\pi(H_2)$, simply because we are working with a complex space. This new eigenvector, $u$, of $\pi(H_2)$ by construction is in the eigenspace of $\pi(H_1)$ and hence is a simultaneous eigenvector of both $\pi(H_1)$ and $\pi(H_2)$. For any set of commuting generators we can always find at least one simultaneous eigenvector. Now let $H_i\in\textgoth{h}$ and let $u$ be the simultaneous eigenvector for the Cartan sub-algebra then we find,
\begin{equation}\pi(H_i)u=\lambda_iu\end{equation}
The eigenvalue $\lambda_i$ is a function of the Cartan sub-algebra element acting on $u$ in its representation; $\lambda_i$ is a weight of $\pi$.
\begin{description}
	\item[{\bf Definition 2.17}]A \emph{weight}, $\lambda_i$, is a set of simultaneous eigenvalues for the representation $\pi$ of a commuting set of generators such that there exists an associated non-zero simultaneous eigenvector, $u$, called the \emph{weight vector},
	\begin{equation}\pi(H_i)u=\lambda_iu\end{equation}
\end{description}
A special case of a weight applied to the adjoint representation of the Cartan sub-algebra is called a root.
\begin{description}
	\item[{\bf Definition 2.18}]A \emph{root}, $\alpha_i$, is a set of simultaneous eigenvalues for the adjoint representation $ad_H$ of the Cartan sub-algebra, $H\in\textgoth{h}$, such that there exists an associated non-zero element, $X\in\textgoth{g}$, called the \emph{root vector},
	\begin{equation}[H_i,X]=\alpha_iX\end{equation}
\end{description}
For example, a root of $su(2)$ is two (it is single valued since there is only one element in the Cartan sub-algebra of $su(2)$, namely $H'=\textgoth{h}$) and the corresponding root vector is $\hat{Y}'$. 
The root, $\alpha_i$, carries an index, $i$, that runs over the elements of the Cartan sub-algebra. If we specify a particular value for the index, say $i=H_1$, then the root gives us a number, $\alpha_{H_1}$. In other words a root, or a weight, is an element of the dual space, $\textgoth{h}^*$, of the Cartan sub-algebra (i.e. a map whose domain is the Cartan sub-algebra and whose image is $\mathbb{R}$). A familiar map with these properties is the inner product on the Cartan sub-algebra, which we denote $<\cdot,\cdot>$, which identifies elements of $\textgoth{h}^*$ with elements in the space spanned by the Cartan sub-algebra, $\textgoth{h}$. With this in mind we rewrite our definitions for the weight and the root as,
\begin{equation}\pi(H)u=<\lambda,H>u, \qquad [H,X]=<\alpha,H>X\end{equation}
Where $\alpha$ and $\lambda$ are vectors in the space spanned by the Cartan sub-algebra. It is useful to work through an example, we will look at the algebra $so(5)$.
\subsubsection{The roots of $so(5)$}
\begin{figure}[cth]
\hspace{80pt}\includegraphics[viewport=0 150 135 200,angle=90]{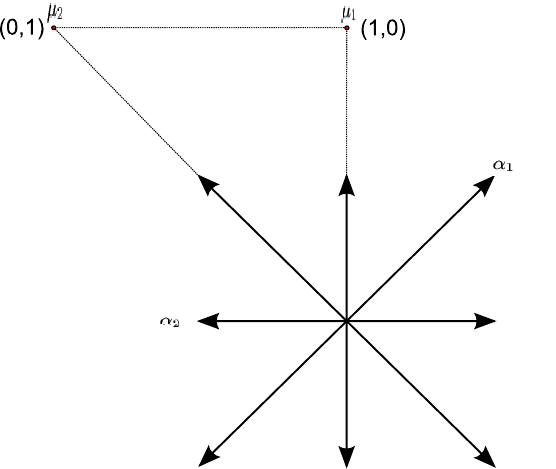}
\caption{The root diagram of $so(5)$ with the fundamental weights superposed} \label{so5roots}
\end{figure}

We find a basis for the algebra $so(5)$, $\textgoth{g}$, by applying the defining relation of the group as follows,
\begin{equation}
e^X(e^X)^T=I \implies X=-X^T
\end{equation}
The determinant being equal to one implies that the matrices representing the algebra are traceless. A basis of $so(5)$ is provided by the skew-symmetric 5-by-5 matrices with real components, these have zeroes along the leading diagonal and so are automatically traceless. We denote our basis,
\begin{align}
\nonumber H_1&= \left( \begin{array}{ccccc}
0 & 1 & 0 & 0 & 0  \\
-1 & 0 & 0 & 0 & 0 \\
0 & 0 & 0 & 0 & 0 \\
0 & 0 & 0 & 0 & 0 \\
0 & 0 & 0 & 0 & 0 \\  \end{array} \right),\quad
H_2= \left( \begin{array}{ccccc}
0 & 0 & 0 & 0 & 0  \\
0 & 0 & 0 & 0 & 0 \\
0 & 0 & 0 & 1 & 0 \\
0 & 0 & -1 & 0 & 0 \\
0 & 0 & 0 & 0 & 0 \\  \end{array} \right),\\
\nonumber S&= \left( \begin{array}{ccccc}
0 & 0 & 1 & 0 & 0  \\
0 & 0 & 0 & 0 & 0 \\
-1 & 0 & 0 & 0 & 0 \\
0 & 0 & 0 & 0 & 0 \\
0 & 0 & 0 & 0 & 0 \\  \end{array} \right),\quad
T= \left( \begin{array}{ccccc}
0 & 0 & 0 & 1 & 0  \\
0 & 0 & 0 & 0 & 0 \\
0 & 0 & 0 & 0 & 0 \\
-1 & 0 & 0 & 0 & 0 \\
0 & 0 & 0 & 0 & 0 \\  \end{array} \right),\\
\nonumber U&= \left( \begin{array}{ccccc}
0 & 0 & 0 & 0 & 1  \\
0 & 0 & 0 & 0 & 0 \\
0 & 0 & 0 & 0 & 0 \\
0 & 0 & 0 & 0 & 0 \\
-1 & 0 & 0 & 0 & 0 \\  \end{array} \right),\quad
V= \left( \begin{array}{ccccc}
0 & 0 & 0 & 0 & 0  \\
0 & 0 & 1 & 0 & 0 \\
0 & -1 & 0 & 0 & 0 \\
0 & 0 & 0 & 0 & 0 \\
0 & 0 & 0 & 0 & 0 \\  \end{array} \right), \\
\nonumber W&= \left( \begin{array}{ccccc}
0 & 0 & 0 & 0 & 0  \\
0 & 0 & 0 & 1 & 0 \\
0 & 0 & 0 & 0 & 0 \\
0 & -1 & 0 & 0 & 0 \\
0 & 0 & 0 & 0 & 0 \\  \end{array} \right),\quad
X= \left( \begin{array}{ccccc}
0 & 0 & 0 & 0 & 0  \\
0 & 0 & 0 & 0 & 1 \\
0 & 0 & 0 & 0 & 0 \\
0 & 0 & 0 & 0 & 0 \\
0 & -1 & 0 & 0 & 0 \\  \end{array} \right), \\
Y&= \left( \begin{array}{ccccc}
0 & 0 & 0 & 0 & 0  \\
0 & 0 & 0 & 0 & 0 \\
0 & 0 & 0 & 0 & 1 \\
0 & 0 & 0 & 0 & 0 \\
0 & 0 & -1 & 0 & 0 \\  \end{array} \right),\quad
Z= \left( \begin{array}{ccccc}
0 & 0 & 0 & 0 & 0  \\
0 & 0 & 0 & 0 & 0 \\
0 & 0 & 0 & 0 & 0 \\
0 & 0 & 0 & 0 & 1 \\
0 & 0 & 0 & -1 & 0 \\  \end{array} \right)
\end{align}
We compute the following commutator relations,
\begin{align}
\nonumber [H_1,H_2]&=0 \\
\nonumber [H_1,S]&=-V,\quad [H_1,T]=-W,\quad [H_1,U]=-X,\quad [H_1,V]=S \\
\nonumber [H_1,W]&=T,\quad [H_1,X]=U,\quad [H_1,Y]=0,\quad [H_1,Z]=0 \\
\nonumber [H_2,S]&=-T,\quad [H_2,T]=S,\quad [H_2,U]=0,\quad [H_2,V]=-W \\
[H_2,W]&=V,\quad [H_2,X]=0,\quad [H_2,Y]=-Z,\quad [H_2,Z]=Y 
\end{align}
After much thought we find a set of eight Lie algebra elements which act as eigenvectors under the adjoint action of $iH_1$ and $iH_2$. In full, we compute,
\begin{align}
\nonumber [iH_1,V+T+i(S-W)]&=V+T+i(S-W)\\
\nonumber [iH_2,V+T+i(S-W)]&=V+T+i(S-W)\\
\nonumber [iH_1,V+T-i(S-W)]&=-(V+T-i(S-W))\\
\nonumber [iH_2,V+T-i(S-W)]&=-(V+T-i(S-W))\\
\nonumber [iH_1,V-T-i(S+W)]&=-(V-T-i(S+W))\\
\nonumber [iH_2,V-T-i(S+W)]&=(V-T-i(S+W))\\
\nonumber [iH_1,V-T+i(S+W)]&=V-T+i(S+W)\\
\nonumber [iH_2,V-T+i(S+W)]&=-(V-T+i(S+W))\\
\nonumber [iH_1,X+iU]&=X+iU\\
\nonumber [iH_2,X+iU]&=0\\
\nonumber [iH_1,X-iU]&=-(X-iU)\\
\nonumber [iH_2,X-iU]&=0\\
\nonumber [iH_1,Z+iY]&=0\\
\nonumber [iH_2,Z+iY]&=Z+iY\\
\nonumber [iH_1,Z-iY]&=0\\
[iH_2,Z-iY]&=-(Z+iY)
\end{align}
We read off the roots from the above relations, and they are listed for reference in table \ref{tab:Roots and vectors of so(5)}.
\begin{table}[h]
	\centering
		\begin{tabular}{c|c}
			Root, $\alpha$ & Root vector \\\hline
			$(1,1)$&$V+T+i(S-W)$\\
			$(-1,-1)$&$V+T-i(S-W)$\\
			$(-1,1)$&$V-T-i(S+W)$\\
			$(1,-1)$&$V-T+i(S-W)$\\
			$(1,0)$&$X+iU$\\
			$(-1,0)$&$X-iU$\\
			$(0,1)$&$Z+iY$\\
			$(0,-1)$&$Z-iY$\\
		\end{tabular}
	\caption{The roots and root vectors of $so(5)$}
	\label{tab:Roots and vectors of so(5)}
\end{table}
We pick two roots to act as basis for the root space, and the convention is to pick the simple positive roots, which have the property that the remaining roots are either all negative or all positive multiples of the simple positive roots. We pick as basis the roots $\alpha_1=(1,1)$ and $\alpha_2=(-1,0)$. At this juncture it is worth noting, in passing, a general trend exhibited by our example: There is an element of the algebra, or generator, associated to the positive and the negative of the simple roots. We will, later, in this thesis denote the generators of the positive simple roots by $E_{\alpha}$ and the negative simple roots by $F_\alpha$, so that for every simple root of an algebra we will find a set of generators, $\{H_\alpha,E_\alpha,F_\alpha\}$, of $su(2)$. In the $so(5)$ case we have 
\begin{align}
\nonumber E_{\alpha_1}&=V+T+i(S-W),\qquad F_{\alpha_1}=V+T-i(S-W)\\
E_{\alpha_2}&=X-iU, \qquad F_{\alpha_2}=X+iU
\end{align}
With our choice we have the roots given in table \ref{tab:Roots of so(5)}.
\begin{table}[h]
	\centering
		\begin{tabular}{c|c}
			Root, $\alpha$ & Expansion in $\alpha_1$ and $\alpha_2$ basis \\\hline
			$(1,1)$&$\alpha_1$\\
			$(-1,-1)$&$-\alpha_1$\\
			$(-1,1)$&$\alpha_1+2\alpha_2$\\
			$(1,-1)$&$-\alpha_1-2\alpha_2$\\
			$(1,0)$&$-\alpha_2$\\
			$(-1,0)$&$\alpha_2$\\
			$(0,1)$&$\alpha_1+\alpha_2$\\
			$(0,-1)$&$-\alpha_1-\alpha_2$\\
		\end{tabular}
	\caption{The roots of $so(5)$}
	\label{tab:Roots of so(5)}
\end{table}
Our next step is to define an appropriate inner product and express the roots as elements of the space spanned by the Cartan sub-algebra. Even though we are working with traceless matrices we can make use of the Hilbert-Schmidt inner product whose sum is restricted to the diagonal elements of our matrix space, that is,
\begin{equation}<A,B>\equiv Trace(AB)\end{equation}
Using this inner product we can find two 5-by-5 matrices in the Cartan sub-algebra to represent the simple roots, they are,
\begin{equation}\nonumber \alpha_1= \frac{1}{2}\left( \begin{array}{ccccc}
0 & i & 0 & 0 & 0  \\
-i & 0 & 0 & 0 & 0 \\
0 & 0 & 0 & i & 0 \\
0 & 0 & -i & 0 & 0 \\
0 & 0 & 0 & 0 & 0 \\  \end{array} \right),\quad
\alpha_2= \frac{1}{2}\left( \begin{array}{ccccc}
0 & -i & 0 & 0 & 0  \\
i & 0 & 0 & 0 & 0 \\
0 & 0 & 0 & 0 & 0 \\
0 & 0 & 0 & 0 & 0 \\
0 & 0 & 0 & 0 & 0 \\  \end{array} \right)
\end{equation}
The inner product gives us a way of directly evaluating the root lengths and the orientation of the roots - in more complicated scenarios this may not be so obvious. For example,
\begin{align}
\nonumber <\alpha_1,\alpha_1>&=1=\left|\alpha_1\right|^2\\
\nonumber <\alpha_2,\alpha_2>&=\frac{1}{2}=\left|\alpha_2\right|^2\\
<\alpha_1,\alpha_2>&=-\frac{1}{2}
\end{align}
We deduce that the positive simple roots are at an angle of $\frac{3\pi}{4}$, or, 145 degrees, to each other. We draw the root diagram of $so(5)$ in figure \ref{so5roots}.
Also indicated on the diagram are the fundamental weights of $so(5)$. The symmetry of the root diagram is an important property and we will later make use of the fact that if we know the simple roots of a group we will be able to generate the full set of roots by insisting that the skeleton root diagram of the simple roots generates the full root diagram under reflections in the planes perpendicular to the simple roots. Such a reflection is called a Weyl reflection and we will define it explicitly in chapter 3.
\begin{description}
	\item[{\bf Definition 2.19}]The \emph{fundamental weights}, $\mu_i$, of an algebra are the objects in the root space satisfying,
	\begin{equation}<\mu_i,\alpha_j>=\delta_{ij}\end{equation} 
	Where $\alpha_i$ are the positive simple roots of the algebra.
\end{description}
Hence for $so(5)$ we find,
\begin{equation}\mu_1=2\alpha_1+2\alpha_2\qquad\mu_2=2\alpha_1+4\alpha_2\end{equation}
In the $su(2)$ case we considered earlier we were able to classify representations uniquely by specifying a positive eigenvalue for its single Cartan sub-algebra element. For more general cases we are able to classify a representation in the same way if we specify a highest weight, for irreducible representations the highest weight will be a positive sum of the fundamental weights. In figure \ref{so5roots} we have indicated the fundamental weights, $\mu_1$ and $\mu_2$. For $so(5)$ every irreducible representation has a highest weight that occurs inside or on the triangle spanned by $\mu_1$ and $\mu_2$. The action of the generators on the root vectors is almost identical to the $su(2)$ prototype. Specifically if $u$ is a weight vector with weight $\lambda_i$ then,
\begin{equation}H_iu=\lambda_iu\end{equation}
So that if $E_\alpha$ denotes a generator of a positive simple root and $F_\alpha$ denotes a generator of a negative simple root then,
\begin{align}
\nonumber H_iE_\alpha u&=E_\alpha H_i u+[H_i,E_\alpha]u=(\lambda_i+\alpha)E_\alpha u \\
H_iF_\alpha u&=F_\alpha H_i u+[H_i,F_\alpha]u=(\lambda_i-\alpha)F_\alpha u
\end{align}
The major difference to the $su(2)$ case is that we have multiple roots, and given a weight vector, $u$, we are able to "raise" and "lower" the weight in a number of different directions in the root space. There is another useful fact that we can see in this example and this is that the simple roots are multiples of the corresponding elements of the Cartan sub-algebra, we find, in fact,
\begin{equation}H_\alpha=2\frac{\alpha}{<\alpha,\alpha>}\end{equation}
which allows us to write, for future use, that,
\begin{align}
\nonumber [H_\alpha,E_\beta]&=2\frac{<\alpha,\beta>}{<\alpha,\alpha>}E_\beta\\
[H_\alpha,F_\beta]&=-2\frac{<\alpha,\beta>}{<\alpha,\alpha>}F_\beta
\end{align}

\subsection{Semisimple Lie Algebras}
In this section we will give a summary of the classification of the semisimple Lie algebras. But we need a few more definitions before we list all the classes of semisimple Lie algebras.
\begin{description}
	\item[{\bf Definition 2.20}]The \emph{Cartan matrix}, $A_{ab}$, of an algebra is defined to be,
	\begin{equation}A_{ab}=2\frac{<\alpha_a,\alpha_b>}{<\alpha_a,\alpha_a>}\end{equation}
	Where $\alpha_a$ are the positive simple roots of the algebra.
\end{description}
 The diagonal elements of $A_{ab}$ are all 2 and the off-diagonal elements are negative integers, with the zeroes being symmetrically distributed. For our example of $so(5)$ we have,
\begin{equation}A_{ab}(so(5))=\left(\begin{array}{cc}
2 & -2\\
-1 & 2\\
\end{array}\right)\end{equation}
We can now write the commutators of the generators $\{H_a,E_a,F_a\}$ in the compact notation,
\begin{align}
\nonumber [E_a, F_b]&=\delta_{ab}H_b\\
[H_a, E_b]=A_{ab}E_b \qquad & \qquad [H_a, F_b]=-A_{ab}F_b 
\end{align}
It is now clear that when $a=b$ these generators span a sub-algebra isomorphic to $su(2)$.

\begin{figure}[cth]
\hspace{40pt} \includegraphics[viewport=0 50 35 200,angle=90]{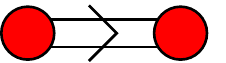}
\caption{The Dynkin diagram of $so(5)$} \label{dynkinso(5)}
\end{figure}

\begin{figure}[cth]
\hspace{170pt}\includegraphics[viewport=0 150 135 200,angle=90]{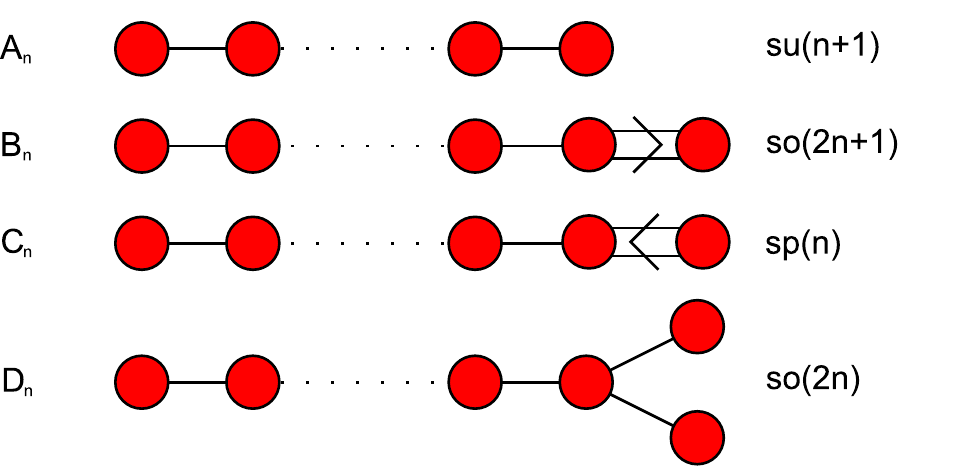}
\caption{The classical Lie groups} \label{classicalgroups}
\end{figure}

\begin{figure}[cth]
\hspace{145pt}\includegraphics[viewport=0 150 200 200,angle=90]{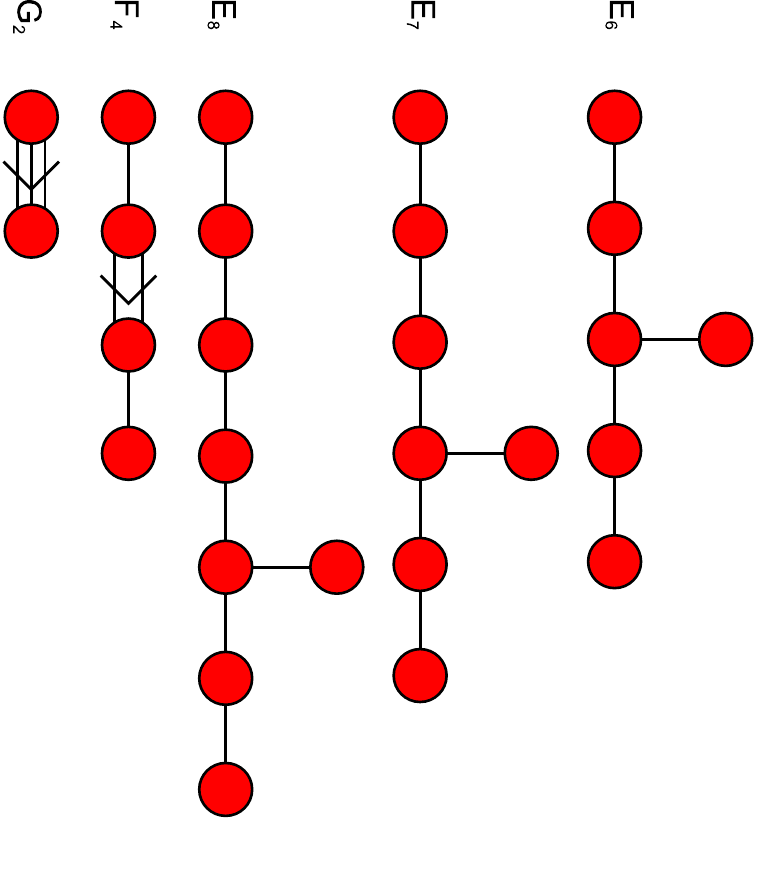}
\caption{The exceptional Lie groups} \label{exceptionalgroups}
\end{figure}

The ratio of off-diagonal components is the ratio of the length-squared of the simple roots,
\begin{equation}\frac{A_{ab}}{A_{ba}}=\frac{<\alpha_a,\alpha_b><\alpha_b,\alpha_b>}{<\alpha_a,\alpha_a><\alpha_b,\alpha_a>}=\frac{\alpha_b^2}{\alpha_a^2}\end{equation}
The semisimple Lie algebras are classified by their Dynkin diagrams, which are a useful way of encoding the properties of the roots and their relations to each other. Let us now set down the rules for drawing Dynkin diagrams from the Cartan matrix.
\begin{enumerate}
	\item Draw $r$ nodes, where $r$ is the rank of the Cartan matrix.
	\item Connect node $a$ to node $b$ by $A_{ab}A_{ba}$ lines.
	\item If $\frac{A_{ab}}{A_{ba}}<1$ then indicate that $\alpha_a$ is longer than $\alpha_b$ by drawing a larger than, $>$, symbol along the lines connecting node $a$ and node $b$.
\end{enumerate}
The Dynkin diagram of $so(5)$ is shown in figure \ref{dynkinso(5)}.
\begin{description}
	\item[{\bf Definition 2.21}]A Lie algebra is \emph{simple} if it has no non-trivial ideals and has dimension greater than or equal to two.
\end{description}
We recall that an ideal is a sub-algebra $\textgoth{h}\in\textgoth{g}$ with the property that if $H\in \textgoth{h}$ then for all $X\in\textgoth{g}$, $[X,H]\in\textgoth{h}$, and that every algebra has two trivial ideals, $\{0\}$ and $\textgoth{g}$.
\begin{description}
	\item[{\bf Definition 2.22}]A Lie algebra is \emph{semisimple} if it is isomorphic to a direct sum of simple Lie algebras.
\end{description}
There are a number of equivalent definitions of a semisimple Lie algebra, a useful alternative definition is the following.
\begin{description}
	\item[{\bf Definition 2.23}]A complex Lie algebra is \emph{semisimple} if and only if it is isomorphic to the complexification of the Lie algebra of a simply-connected, compact matrix Lie group.
\end{description}
Examples of semisimple Lie algebras are $so(n)$, $sl(n)$ and $sp(n)$. Returning to the Dynkin diagrams we observe that 
\begin{equation}A_{ab}A_{ba}=4\frac{<\alpha_a,\alpha_b>^2}{<\alpha_a,\alpha_a><\alpha_b,\alpha_b>}=4\cos^2{\theta_{ab}}\end{equation}
Therefore the number of lines joining any two nodes in a Dynkin diagram must be less than, or equal to, four. This immediately limits the possible Dynkin diagrams that may exist. It will not be beneficial to list further restrictions on possible allowed Dynkin diagrams here, the reader can read about the classification elsewhere \cite{Cahn, Jacobsen, Humphreys}. The result, which should not appear too surprising at this stage, is that there are only a finite class of Dynkin diagrams, or root systems that exist. They are classified $A_n$, $B_n$, $C_n$, $D_n$, $E_6$, $E_7$, $E_8$, $F_4$ and $G_2$; we give their Dynkin diagrams in figures \ref{classicalgroups} and \ref{exceptionalgroups}. The Dynkin diagrams of arbitrary length are called the classical groups and the remainder, which were discovered later, are the exceptional groups. From the Dynkin diagrams we notice that the example algebra we have been considering, $so(5)$ has the same Dynkin diagram as $sp(2)$, and indeed $so(5)\cong sp(2)$.
Without proof, the classification of the semisimple Lie algebras is equivalent to the classification of the Dynkin diagrams shown in figures \ref{classicalgroups} and \ref{exceptionalgroups}, which is a remarkable result made clear by the work of Chevalley.
\subsection{From Semisimple Lie Groups to Kac-Moody Algebras}
Finally we arrive at the central purpose of this chapter: the defining properties of a Kac-Moody algebra. A Kac-Moody algebra may be thought of as a generalisation of the semisimple Lie algebras. For a thorough review and references to the original literature see \cite{Kac}. It is defined in terms of a Cartan matrix and the generators of the positive $E_a$ and negative $F_a$ simple roots. Here the index $a$ runs over the simple roots of the algebra. A Kac-Moody algebra is uniquely constructed from the commutators of $E_a$, $F_a$ and the Cartan sub-algebra generators, $H_a$, subject to the following relations
\begin{align}
\nonumber [H_a, H_b]&=0\\
\nonumber [H_a, E_b]=A_{ab}E_b \qquad & \qquad [H_a, F_b]=-A_{ab}F_b \\
\nonumber [E_a, F_b]&= \delta_{ab}H_b\\
[E_a, \ldots [E_a,E_b]]=0 \qquad & \qquad [F_a, \ldots [F_a,F_b]]=0 \label{Serrerelations}
\end{align}
In the final line there are $(1-A_{ab})$ $E_a$'s in the first equation and the same number of $F_a$'s in the second equation. These relations are called the Serre relations.

Any set of generators satisfying these conditions is a Kac-Moody algebra. Most such algebras are infinite dimensional, in the sense that the string of weights associated to a particular representation will not generally terminate in both "up" and "down" directions and also in the sense that the set of generators do not close in on themselves after a finite number of applications of the adjoint action. In fact they are not comparatively well understood, and one of the aims of this thesis is to get some insight into a particular class of Kac-Moody algebras which are thought to be relevant to physics. In the next chapter we shall see why these algebras might turn up in an extension of eleven dimensional supergravity (or M-Theory) and string theory.
\newpage

\section{Symmetries of M-theory}
In this chapter we look at the motivations for the conjecture that the Kac-Moody group $E_{11}$ is a global symmetry of M-theory. Much of the material reviewed in this chapter has also been reviewed in detail in the excellent masters thesis of Ling Bao \cite{Bao}, which is recommended as a complementary resource when reading the original literature and this chapter.

We review the steps in the literature that lead up to the conjecture that the Kac-Moody algebra $E_{11}$ encodes the symmetries of M-theory commencing by describing nonlinear realisations on coset spaces \cite{ColemanWessZumino,Volkov,SalamStrathdee}, a technique used to encode group symmetries in a Lagrangian formulation, and which is the cornerstone of connecting any group symmetries with a physical formulation of a theory. Having familiarised ourselves with the main technique, we describe an old observation \cite{Ogievetsky} for generating the infinite dimensional diffeomorphism group as the closure of an affine group with the conformal group. We then review essential parts of \cite{West2} and find the closure of a group $G_{11}$, which is an enlargement of the affine group $IGL(11)$ to include two generators whose associated gauge fields are those of the M2 and the M5 branes, with the conformal group. As a result of the nonlinear realisation we find the field equations of the bosonic sector of eleven dimensional supergravity. Having gained confidence in the nonlinear realisation of eleven dimensional supergravity, we review \cite{West1} where the argument that an $E_8$ symmetry of a nonlinear realisation of supergravity ought to be manifest before compactification to three dimensions encourages the extension of the group $G_{11}$ by including a generator associated to the dual gravity field, and the subsequent conjecture of an $E_{11}$ symmetry underlying M-theory \cite{West1}. We finish the chapter by reviewing the elegant work of \cite{LambertWest} associating a coset symmetry in three dimensions with terms appearing directly in the dimensional reduction. By reviewing this topic we gain an understanding of when it is possible to formulate a coset symmetry of a theory, as well as observing the hidden $E_8$ symmetry in eleven-dimensional supergravity, two topics which are pillars of the $E_{11}$ conjecture, and we begin to see the outline of a direct bridge connecting group theory and M-theory.

\subsection{Nonlinear Realisations on a Coset Space}
A nonlinear realisation \cite{ColemanWessZumino,Volkov,SalamStrathdee} is a means of expressing a theory such that its Lagrangian and equations of motion are manifestly invariant under the action of a specified group. Crucial to writing a nonlinear realisation of a theory is the Maurer-Cartan form, ${\cal \nu}_g$, which is a map that takes the tangent space at $g$ for some $\cal{G}$ to the tangent space at the identity, which is the Lie algebra. For matrix Lie groups we may write the Maurer-Cartan form explicitly as,
\begin{equation}{\cal \nu}_g=g^{-1}dg\end{equation}
Where $g$ is an element of $\cal{G}$ and we have used the exterior derivative $d\equiv dx^\mu\wedge \partial_\mu$, and in particular we have introduced space-time coordinates, let us now define some space-time dependent group elements. Recall that we demonstrated in chapter two that any connected Lie algebra we may express any element of the group by
\begin{equation}g_0=e^{X_1}e^{X_2}\ldots e^{X_r}\end{equation}
Where $X_i$ are the generators of the group. This is an example of a rigid group element, since it does not depend upon the space-time coordinates. We may express a general path through a connected Lie group by the group element,
\begin{equation}g=e^{t_1X_1}e^{t_2X_2}\ldots e^{t_rX_r}\end{equation}
Where $t_i$ parameterise the path through the group. Physically the $t_i$ will be functions dependent on the space-time coordinates of our theory - they will become the gauge fields, and in general will have a more complex index structure than indicated here, as will the associated generators. This will be our prototype group element. We shall use the left-action of the group $\cal{G}$ upon itself so that the action of a rigid group element, $g_0$ on a general group element, $g$ is given by,
\begin{equation}g_0g=e^{\hat{t}_1X_1}e^{\hat{t}_2X_2}\ldots e^{\hat{t}_rX_r}=g'\end{equation}
Under this action of a rigid group element, $g_0$, the Maurer-Cartan form is invariant,
\begin{equation}{\cal \nu}_g\rightarrow g^{-1}g_0^{-1}g_0dg={\cal \nu}_g\end{equation}
A Lagrangian constructed from the Maurer-Cartan one-form is,
\begin{equation}{\cal L}= -Tr(g^{-1}\partial_\mu g) (g^{-1}\partial^\mu g)\end{equation}
By constructing a Lagrangian from the Maurer-Cartan one-forms we have arrived at a formulation of a theory invariant under the rigid action of the group $\cal{G}$. In practise we also wish for some symmetries of the action to be local, i.e. non-rigid, for example the action may be Lorentz invariant. We can incorporate such a symmetry by modifying the Maurer-Cartan form and letting our nonlinear realisation act on a coset space. 

A coset model is a special type of nonlinear theory, where the scalars in the theory parameterise elements of a coset space $\frac{\cal{G}}{H}$, where $\cal{G}$ is a Lie group and $H$ is a sub-group of $\cal{G}$. One may think of this as a fibre, $H$, over a base manifold, $\frac{\cal{G}}{H}$, so that the total space is $\cal{G}$. An element of $\cal{G}$, that is neither the identity map nor a member of $H$, uniquely specifies each coset. The group $H$ is a local symmetry of the model and the group $\cal{G}$ is a global symmetry. It is convenient to take the $H$ action on $\cal{G}$ to be the adjoint action:
\begin{equation}g\rightarrow hgh^{-1}\end{equation}
Our group elements are now the representatives of the coset space, we have one such element defining each unique orbit under the action of $H$. Consequently our local group elements are now modulo the subgroup $H$, and the action of a generic rigid group element will incorporate the generators of $\cal{H}\subset\cal{G}$,
\begin{equation}g_0g=e^{\hat{t}_1X_1}e^{\hat{t}_2X_2}\ldots e^{\hat{t}_rX_r}=g'h\end{equation}
We define the action of the group on the coset representative to be given by,
\begin{equation}g\rightarrow g_0gh^{-1}=g'\end{equation}
Under this coset action the Maurer-Cartan form transforms as,
\begin{equation}{\cal \nu}_g\rightarrow (g_0gh^{-1})^{-1}d(g_0gh^{-1})=hg^{-1}dg h^{-1}+hdh^{-1}=h{\cal \nu}_gh^{-1}+{\cal \nu}_h\end{equation}
The Maurer-Cartan form is no longer invariant, but its transformation is determined entirely by $h\in\cal H$. If we extend the Maurer-Cartan form by subtracting a term $w$ which transforms under the full group as,
\begin{equation}w\rightarrow hwh^{-1}+\nu_h\end{equation}
Then we find that the extended form ${\cal V}_g\equiv {\cal \nu}_g-w$ is invariant modulo an adjoint transformation under an element of the local group $\cal{H}$,
\begin{equation}{\cal V}_g\rightarrow h{\cal \nu}_gh^{-1}-hwh^{-1}=h{\cal V}_gh^{-1}\end{equation}
Since we are now considering a theory on a coset space we would like to construct a Lagrangian which is invariant under the rigid group transformation but which may transform under the local group. Using the extended Maurer-Cartan form this is exactly what we have achieved with the Lagrangian,
\begin{equation}{\cal L} = -Tr(g^{-1}\partial_\mu g-w_\mu)(g^{-1}\partial^\mu g-w^\mu)\rightarrow h{\cal L}h^{-1}\end{equation}
The subtraction of a term $w$ is simply compensating for part of the rigid transformation of the full group that occurs in the local group, and may be interpreted as the promotion of the partial derivative in the original Maurer-Cartan form to a covariant derivative $D_\mu$ where $D_\mu g = \partial_\mu g +gw_\mu$ and,
\begin{equation}{\cal L}=-Tr(g^{-1}D_\mu g)(g^{-1}D^\mu g)\end{equation}
If the local group $\cal H$ is the Lorentz group then $w$ is the Lorentz connection. If there were only a single generator for the local group, i.e. $h=e^{hX}$, then $w\equiv dx^\mu(\partial_\mu h)\equiv dx^\mu h_\mu$.

\subsection{The Diffeomorphism Group and Bosonic Supergravity}
Let us consider the arguments for a hidden superconformal symmetry being encoded in the eleven dimensional supergravity algebra given in \cite{West2}. The argument rests upon the observation that in the bosonic case, the closure of the conformal algebra and the affine algebra results in the full diffeomorphism group \cite{Ogievetsky} in eleven dimensions which we require for general relativity. The affine group and conformal group have three generators in common (the dilation, Lorentz and translation generators) and two generators not in common, namely the symmetric and traceless part of ${K^a}_b$ in the affine group and the generator of special conformal transformations. It is the commutator of these two generators that forms the generators of the diffeomorphism group \cite{Bao}.

In fact in \cite{West2} the closure of a group larger than the affine group was considered, which includes the affine algebra $\{{K^a}_b,P_a\}$ as a subgroup as well as generators of rank three $R^{c_1\ldots c_3}$ and rank six $R^{c_1\ldots c_6}$ whose associated gauge fields are those of the membrane and the fivebrane, this group is denoted $G_{11}$. The motivation for working with this group is given in \cite{BarwaldWest} where it was shown that the sub-algebra formed by $R^{c_1\ldots c_3}$ and $R^{c_1\ldots c_6}$ are a symmetry of the fivebrane equations of motion. This same sub-algebra has also been used in the nonlinear realisation of the fivebrane \cite{West5}, where it is argued that such an enlargement of the algebra is to be expected in a theory that includes superbranes alongside the superparticle - it is pointed out that the full automorphism group of the supercharges (i.e. the general linear group mixing the supercharges, which is $GL(32,{\mathbb R})$ in eleven dimensions) is the natural generalisation to include branes of the spin group for the point particle. The enlargement of the Poincare group by the three-form and the six-form involves a restriction of the full automorphism group, which has previously appeared as a symmetry of the covariant fivebrane equations of motion \cite{BarwaldWest}. Hence this group generated by this algebra is a reasonable starting point to begin the search for an enlarged symmetry group relevant to M-theory. $G_{11}$ has non-vanishing commutators,
\begin{align}
\nonumber [{K^a}_b,{K^c}_d]&=\delta^c_b{K^a}_d-\delta^a_d{K^c}_b \\
\nonumber [{K^a}_b,P_c]&=-\delta^a_cP_b\\
\nonumber [{K^a}_b,R^{c_1c_2c_3}]&=\delta^{c_1}_bR^{ac_2c_3}+\ldots\\
\nonumber [{K^a}_b,R^{c_1\ldots c_6}]&=\delta^{c_1}_bR^{ac_2\ldots c_6}+\ldots\\
[R^{c_1c_2c_3},R^{c_4c_5c_6}]&=2R^{c_1\ldots c_6} \label{G11commutators}
\end{align}
Where $+\ldots$ denotes the appropriate permutation of the delta function on the indices of the relevant tensor in each case. We will consider the closure of this finite dimensional algebra with the conformal group. Let us recall that the conformal group has the defining property that the action of its generators preserve the metric up to a scale factor:
\begin{equation}
g'_{\mu\nu}=e^\phi g_{\mu\nu} \label{conformal}
\end{equation}
To give a sense of the conformal group we outline how one finds the generators of the conformal group from this relation. Following the review in \cite{Ginsparg}, we apply a change of coordinates $x^\mu\rightarrow x'^\mu=x^\mu+\epsilon^\mu$ where $\epsilon^\mu$ is an infinitesimal translation, which alters the metrics accordingly,
\begin{align}
\nonumber g_{\mu\nu}&=\frac{\partial x'^\mu}{\partial x^\mu}\frac{\partial x'^\nu}{\partial x^\nu}g'_{\mu\nu}\\
\nonumber &=(1+\partial_\mu\epsilon^\mu)(1+\partial_\nu\epsilon^\nu)g'_{\mu\nu}\\
&=(g_{\mu\nu}+\partial_\mu\epsilon_\nu+\partial_\nu\epsilon_\mu)e^\phi+{\cal{O}}(\epsilon^2) \label{conformal1}
\end{align}
Where we have used (\ref{conformal}) in the final line. Herein we drop the terms of order $\epsilon^2$. We are seeking a constraint on the form of $\epsilon^\mu$, so we premultiply by $g^{\mu\nu}$ to find,
\begin{equation}(e^{-\phi}-1)D=2\partial\cdot\epsilon\end{equation}
Where we have used the notation $\partial\cdot\epsilon\equiv\partial_\mu\epsilon^\mu$. If we now use this expression to eliminate $e^\phi$ from (\ref{conformal1}) we obtain,
\begin{equation}\frac{2}{D}\partial\cdot\epsilon g_{\mu\nu}=\partial_\mu\epsilon_\nu+\partial_\nu\epsilon_\mu\end{equation} Finally if we act on this equation with $\partial^\nu\partial^\mu$ we have,
\begin{equation}(\frac{2}{D}-2)\partial^2(\partial\cdot\epsilon)=0\end{equation}
That is, $\epsilon^\mu$ is at most quadratic in $x^\mu$,
\begin{equation}\epsilon^\mu=a^\mu+\lambda x^\mu+{b^\mu}_\nu x^\nu-c^\mu x^2+2x^\mu c\cdot x\end{equation}
These may be classified as four types of transformation:
\begin{enumerate}
	\item Translations \begin{equation}x'^\mu=x^\mu+a^\mu \end{equation}
	\item Dilations \begin{equation}x'^\mu=\lambda x^\mu\end{equation}
	\item Lorentz transformations \begin{equation}x'^\mu={b^\mu}_\nu x^\nu\end{equation}
	\item Special conformal transformations \begin{equation}x'^\mu=2x^\mu c\cdot x-c^\mu x^2\end{equation}
\end{enumerate}
The generators of the conformal group, $X_\nu$ are defined by $x'^\mu=x^\mu+\epsilon^\nu X_\nu x^\mu$, and so for the conformal group transformations classified above we have,
\begin{enumerate}
	\item Translations \begin{equation}P_\mu=\partial_\mu \end{equation}
	\item Dilations \begin{equation}D=x^\mu\partial_\mu\end{equation}
	\item Lorentz transformations \begin{equation}J_{\mu\nu}=x_\mu\partial_\nu-x_\nu\partial_\mu\end{equation}
	\item Special conformal transformations \begin{equation}K_\mu=2x_\mu x^\nu\partial_\nu-x^2\partial_\mu \end{equation}
\end{enumerate}
Therefore we have the non-vanishing commutators for the conformal group,
\begin{align}
\nonumber [P_\mu,D]&=P_\mu \\
\nonumber [P_\mu,J_{\nu\rho}]&=\eta_{\mu\nu}P_\rho-\eta_{\mu\rho}P_\nu\\
\nonumber [P_\mu,K_\nu]&=2\eta_{\mu\nu}D-2J_{\mu\nu}\\
\nonumber [D,K_\mu]&=K_\mu\\
\nonumber [J_{\mu\nu},K_\rho]&=\eta_{\nu\rho}K_\mu-\eta_{\rho\mu}K_\nu\\
[J_{\mu\nu},J_{\rho\sigma}]&=\eta_{\mu\sigma}J_{\nu\rho}+\eta_{\nu\rho}J_{\mu\sigma}-\eta_{\mu\rho}J_{\nu\sigma}-\eta_{\nu\sigma}J_{\mu\rho}
\end{align}
The self-evident way to find the closure of $G_{11}$ and the conformal algebra is to write down all the possible permutations of commutators for the full set of generators and then find the nonlinear realisation of the infinite set of generators. However since it will be useful to familiarise ourselves with the nonlinear realisation we shall repeat the efficient method of \cite{West2} which was to write down the nonlinear realisation for coset models of $G_{11}$ and the conformal group independently and then construct a theory that contains forms which are invariant under each realisation. 

For $G_{11}$ we consider the the local group $\cal{H}$ to be the Lorentz group, whose generators are $J_{ab}=\eta_{ac}{K^c}_b-\eta_{bc}{K^c}_a=K_{ab}-K_{ba}$ so that $h=e^{\frac{1}{2}{w_a}^b{J^a}_b}$ and so $w=\frac{1}{2}dx^\mu{\omega_{\mu a}}^b{J^a}_b$. The local group element takes the generic form,
\begin{equation}g=e^{x^\mu P_\mu}e^{{h_a}^b{K^a}_b}e^{\frac{1}{3!}A_{c_1c_2c_3}R^{c_1c_2c_3}}e^{\frac{1}{6!}A_{d_1\ldots d_6}R^{d_1\ldots d_6}}\end{equation}
Where $x^\mu$, ${h_a}^b$, $A_{c_1c_2c_3}$ and $A_{d_1\ldots d_6}$ are the generalisation of our $t_i$ which parameterise the path through our group and are dependent upon the space-time coordinates which we choose to be $x^\mu$ due to their association with the generator of translations, although the natural choice would be to generalise the notion of coordinate to apply equally to all the gauge fields. It is suggested in \cite{West2} that this would be achieved by allowing the fields to be specified by $D$ parameters of space-time, but we can proceed equally well by choosing the $D$ parameters $x^\mu$ to specify the coordinates. We can now write down the Maurer-Cartan form,
\begin{align}
\nonumber {\cal V}_g&=dx^\mu[(g^{-1}\partial_\mu g)-w_\mu]\\
\nonumber &=dx^\mu[g^{-1}(P_\mu g + e^{x^\mu P_\mu}(\partial_\mu {h_a}^b){K^a}_be^{\frac{1}{3!}A_{c_1c_2c_3}R^{c_1c_2c_3}}e^{\frac{1}{6!}A_{d_1\ldots d_6}R^{d_1\ldots d_6}}\\
\nonumber & \quad + e^{x^\mu P_\mu}e^{{h_a}^b{K^a}_b}\frac{1}{3!}(\partial_\mu A_{c_1c_2c_3})R^{c_1c_2c_3} e^{\frac{1}{3!}A_{c_1c_2c_3}R^{c_1c_2c_3}}e^{\frac{1}{6!}A_{d_1\ldots d_6}R^{d_1\ldots d_6}}
\\
& \quad +e^{x^\mu P_\mu}e^{{h_a}^b{K^a}_b}e^{\frac{1}{3!}A_{c_1c_2c_3}R^{c_1c_2c_3}}\frac{1}{6!}(\partial_\mu A_{d_1\ldots d_6})R^{d_1\ldots d_6}e^{\frac{1}{6!}A_{d_1\ldots d_6}R^{d_1\ldots d_6}})
-w_\mu]
\end {align}
We may rearrange and simplify this equation using equations (\ref{expadj}) and (\ref{expadj1}) from chapter two. As an example, let us concentrate on simplifying the first term above,
\begin{align}
\nonumber dx^\mu g^{-1}P_\mu g&= dx^\mu e^{-{h_a}^b{K^a}_b}P_\mu e^{{h_a}^b{K^a}_b} \\
\nonumber &= dx^\mu e^{-{h_a}^b[{K^a}_b,\cdot]}P_\mu \\
\nonumber &= dx^\mu (1+{h_a}^b\delta_\mu^aP_b+\frac{1}{2!}{h_c}^d{h_a}^b\delta^c_b\delta_\mu^aP_d+\ldots) \\
&= dx^\mu {(e^{h})_\mu}^aP_a
\end{align}
The naturally appearing object ${(e^{h})_\mu}^a$ is the vielbein, which we denote ${e_\mu}^a$. Carrying out the same simplification for the other generators we find,
\begin{align}
\nonumber e^{-\frac{1}{6!}A_{d_1\ldots d_6}R^{d_1\ldots d_6}}&e^{-\frac{1}{3!}A_{c_1c_2c_3}R^{c_1c_2c_3}}\partial_\mu({h_a}^b){K^a}_be^{\frac{1}{3!}A_{c_1c_2c_3}R^{c_1c_2c_3}}e^{\frac{1}{6!}A_{d_1\ldots d_6}R^{d_1\ldots d_6}} \\
\nonumber &=e^{-\frac{1}{6!}A_{d_1\ldots d_6}R^{d_1\ldots d_6}}\partial_\mu({h_a}^b)[{K^a}_b+\frac{1}{3!}A_{c_1c_2c_3}(\delta_b^{c_1}R^{ac_2c_3}+\ldots)\\
\nonumber \quad & -\frac{1}{3!}A_{c_1c_2c_3}A_{c_4c_5c_6}(\delta_b^{c_1}R^{ac_2c_3c_4c_5c_6}+\ldots)]e^{\frac{1}{6!}A_{d_1\ldots d_6}R^{d_1\ldots d_6}} \\
\nonumber &= \partial_\mu({h_a}^b)[{K^a}_b+\frac{1}{3!}A_{c_1c_2c_3}(\delta_b^{c_1}R^{ac_2c_3}+\ldots) \\
\quad & -\frac{1}{(3!)^2}A_{c_1c_2c_3}A_{c_4c_5c_6}(\delta_b^{c_1}R^{ac_2c_3c_4c_5c_6}+\ldots)+\frac{1}{6!}A_{d_1\ldots d_6}(\delta_b^{d_1}R^{ad_2\ldots d_6}+\ldots)]
\end{align}
\begin{align}
\nonumber e^{-\frac{1}{6!}A_{d_1\ldots d_6}R^{d_1\ldots d_6}}&e^{-\frac{1}{3!}A_{c_4c_5c_6}R^{c_4c_5c_6}}\frac{1}{3!}(\partial_\mu A_{c_1c_2c_3})R^{c_1c_2c_3}e^{\frac{1}{3!}A_{c_4c_5c_6}R^{c_4c_5c_6}}e^{\frac{1}{6!}A_{d_1\ldots d_6}R^{d_1\ldots d_6}} \\
&= \frac{1}{3!}(\partial_\mu A_{c_1c_2c_3})[R^{c_1c_2c_3}+\frac{2}{3!}A_{c_4c_5c_6}R^{c_1\ldots c_6}]
\end{align}
We may write, using the notation of \cite{West2},
\begin{equation}{\cal V}_g =dx^\mu[{e_\mu}^aP_a+{\Omega_{\mu a}}^b{K^a}_b+\frac{1}{3!}(\tilde{D}_\mu A_{c_1c_2c_3})R^{c_1c_2c_3}+\frac{1}{6!}(\tilde{D}_\mu A_{d_1\ldots d_6})R^{d_1\ldots d_6}]\end{equation}
Where,
\begin{align}
\nonumber {e_\mu}^a&={(e^h)_\mu}^a \\
\nonumber {\Omega_{\mu a}}^b&=\partial_\mu({h_a}^b)-{w_{\mu a}}^b\\
\nonumber \tilde{D}_\mu A_{c_1c_2c_3}&=(\partial_\mu A_{c_1c_2c_3})+\partial_\mu({h_{c_1}}^b)(A_{bc_2c_3}+\ldots)\\
\nonumber \tilde{D}_\mu A_{d_1\ldots d_6}&=(\partial_\mu A_{d_1\ldots d_6})+\partial_\mu({h_{d_1}}^b)(A_{bd_2\ldots d_6}+\ldots)\\
\nonumber &\quad -20\partial_\mu({h_{d_1}}^b)(A_{bd_2d_3}+\ldots)A_{d_4d_5d_6}-20A_{d_1d_2d_3}(\partial_\mu A_{d_4d_5d_6})\\
&=(\partial_\mu A_{d_1\ldots d_6})+\partial_\mu({h_{d_1}}^b)(A_{bd_2\ldots d_6}+\ldots)-A_{[d_1d_2d_3}(\tilde{D}_{|\mu |}A_{d_4d_5d_6]})
\end{align}
Where the $"+\ldots"$ denote the symmetrised application of $\partial_\mu (h_{c_1})^b$ to all the indices of the associated field, e.g. $A_{bc_2c_3}$. We note that we have neglected terms of order $h^2$ and higher, and for comparison with the literature we observe that $\partial_\mu {h_a}^b={(e^{-1}\partial_\mu e)_a}^b$ where $e\equiv (e^h)$. To find the covariant derivatives of the fields we convert the spacetime index to a tangent space index by applying ${(e^{-1})_a}^\mu$ to each of the derivatives above.

We now do the same for the conformal algebra, where $\cal{H}$ is taken to be the Lorentz group and for the conformal symmetry we take the Lorentz symmetries to be rigid, so that ${\cal V}_g=g^{-1}dg$. The representative element of the coset is formed from the remaining generators,
\begin{equation}g=e^{x^\mu P_\mu}e^{\phi^bK_b}e^{\sigma D}\end{equation}
Only one term in the Maurer-Cartan form takes any time to simplify, and we calculate it explicitly here,
\begin{align}
\nonumber dx^\mu g^{-1}P_\mu g &= e^{-\sigma D}[P_\mu -\phi^a[K_a,P_\mu]+\frac{1}{2!}[K_c,[K_a,P_\mu]]]e^{\sigma D}\\
\nonumber &= dx^\mu e^{-\sigma D}[P_\mu -\phi^a\delta^b_\mu(2\eta_{ab} D-2J_{ba})-\phi^c\phi^a\delta^b_\mu[K_c,(2\eta_{ab}D-2J_{ba})]]e^{\sigma D}\\
\nonumber &= dx^\mu e^{-\sigma D}[P_\mu -\phi^a\delta^b_\mu(2\eta_{ab} D-2J_{ba})+\phi^c\phi^a\delta^b_\mu(\eta_{ab}K_c+\eta_{bc}K_a-\eta_{ac}K_b)]e^{\sigma D}\\
\nonumber &= dx^\mu e^{-\sigma D}[P_b -2\phi_bD-2\phi^aJ_{ba}+2\phi_b\phi^a K_a-\phi_a\phi^aK_b]\delta^b_\mu e^{\sigma D}\\
&= dx^\mu [e^\sigma P_b -2\phi_bD-2\phi^aJ_{ba}+2\phi_b\phi^a e^{-\sigma}K_a-\phi_a\phi^ae^{-\sigma}K_b]\delta^b_\mu
\end{align}
The Maurer-Cartan form is then,
\begin{equation}{\cal V}_g=dx^\mu [e^\sigma P_b+ e^{-\sigma}(2\phi_b\phi^a-\phi^c\phi_c\delta^a_b+\partial_b\phi^a)K_a+(2\phi_b+\partial_b\sigma)D+2\phi^aJ_{ab}]\delta_\mu^b\end{equation}
As we did for $G_{11}$ we may find the covariant derivatives of the fields by applying the inverse coefficient of the translation generator in the Maurer-Cartan form to the coefficients of the other generators, the term that takes the expressions back to the tangent space is $e^{-\sigma}$ for the conformal group. The covariant derivatives for the fields of the conformal group are explicitly,
\begin{align}
\nonumber \Delta_\mu \phi^a&=e^{-2\sigma}(2\phi_b\phi^a-\phi^c\phi_c\delta^a_b+\partial_b\phi^a)\delta_\mu^b\\
\Delta_\mu \sigma &= e^{-\sigma}(2\phi_b+\partial_b\sigma)\delta_\mu^b
\end{align}
The covariant derivative of the dilaton field $\sigma$ is set to zero yielding the identity,
\begin{equation}\partial_a\sigma=-2\phi_a\end{equation}
With this identity we may eliminate $\phi_a$ as an independent field, leaving only $\sigma$ as an independent field of the conformal group.

Now we wish to find the closure of the groups $G_{11}$ and the conformal group, and we achieve this by placing conditions on the fields so that the $G_{11}$ covariant derivatives are identical to the conformal covariant derivatives. We start by relating the fundamental fields ${h_a}^b$ and $\sigma$, since the dilation is a scaling of the diagonal of the metric we write,
\begin{equation}{h_a}^b={\bar{h}_a}^b+\sigma{\delta_a}^b\end{equation}
Where ${\bar{h}_a}^b$ is the off-diagonal part of ${h_b}^a$ and $\sigma{\delta_a}^b$ is the diagonal. So that, 
\begin{equation}{(e^h)_\mu}^a={(e^{\bar{h}})_\mu}^ae^\sigma\equiv {(\bar{e})_\mu}^a e^\sigma\end{equation}
With this new notation we write out the $G_{11}$ covariant derivative for the field $A_{a_1a_2a_3}$,
\begin{align}
\nonumber \tilde{D}_a A_{c_1c_2c_3}&={(e^{-h})_a}^\mu[(\partial_\mu A_{c_1c_2c_3})+(\partial_\mu ({{\bar h}_{c_1}}^{\hspace{7pt} b}+{\delta_{c_1}}^b\sigma)A_{bc_2c_3}+\ldots)]\\
&={(\bar{e})_a}^\mu[e^{-\sigma}\partial_\mu A_{c_1c_2c_3}+e^{-\sigma}((\partial_\mu({\bar{h}_{c_1}}^{\hspace{7pt}b})+\delta_{c_1}^{b}\partial_\mu\sigma)A_{bc_2c_3}+\ldots)] \label{cdA3}
\end{align}
How does the conformal covariant derivative act on the field $A_{a_1a_2a_3}$? From the Maurer-Cartan form we see that representations of the Lorentz group pick up an extra contribution, which must be taken account of when writing down the covariant derivative with respect to the conformal transformations of an arbitrary field, $A$, carrying a Lorentz representation, $\Sigma_{ab}$\footnote{By $\Sigma_{ab}$ we effectively mean $2\eta_{[a|\mu_1|}A_{b]\mu_2\ldots \mu_n}+\ldots$},
\begin{align}
\nonumber \Delta_a A&= e^{-\sigma}[\partial_a - 2\phi^b\Sigma_{ab}]A \\
&= e^{-\sigma}[\partial_a + \partial^b \sigma\Sigma_{ab}]A
\end{align}
In particular for $A_{b_1b_2b_3}$
\begin{equation}\Delta_a A_{b_1b_2b_3}=e^{-\sigma}[\partial_aA_{b_1b_2b_3} + (\eta_{ab_1}\partial^{c}\sigma A_{cb_2b_3}+\ldots)-(\partial_{b_1}\sigma A_{ab_2b_3}+\ldots) ]\end{equation}
We now substitute this expression into equation (\ref{cdA3}) and can now relate the $G_{11}$ and the conformal covariant derivatives,
\begin{align}
\nonumber \tilde{D}_a A_{b_1b_2b_3}&={(\bar{e})_a}^\mu[\Delta_\mu A_{b_1b_2b_3}-e^{-\sigma}\{(\eta_{\mu b_1}\partial^{c}\sigma A_{cb_2b_3}+\ldots)-(\partial_{b_1}\sigma A_{\mu b_2b_3}+\ldots)\\  
&\quad -(\partial_\mu({\bar{h}_{b_1}}^c)A_{cb_2b_3}+\ldots)-(\delta_{b_1}^c\partial_\mu\sigma A_{cb_2b_3}+\ldots)\}]
\end{align}
Our aim is to equate the two covariant derivatives by means of a constraint on the field $A_{a_1a_2a_3}$. At level $\bar{h}^0$ we have only three sets of terms that remain and distinguish the two covariant derivatives, we would like these terms to vanish, they are,
\begin{equation}e^{-\sigma}\{(\eta_{\mu b_1}\partial^{c}\sigma A_{cb_2b_3}+\ldots)-(\partial_{b_1}\sigma A_{\mu b_2b_3}+\ldots)-(\delta_{b_1}^c\partial_\mu\sigma A_{cb_2b_3}+\ldots)\}\end{equation}
If we were to antisymmetrise the indices $\{\mu,b_1,b_2,b_3\}$ then we would find that the first term vanishes identically, since $\eta_{ab}$ is symmetric, and the remaining two terms cancel each other out due to the ordering of their indices, so that all these terms vanish as desired. We find,
\begin{equation}\tilde{D}_{[a} A_{b_1b_2b_3]}=\Delta_{[a} A_{b_1b_2b_3]}\end{equation}
The object that is covariant under both the conformal and the $G_{11}$ groups is,
\begin{equation}\tilde{F}_{a_1\ldots a_4}\equiv 4({e_{[a_1}}^\mu\partial_\mu A_{a_2\ldots a_4]}+3{e_{[a_1}}^\mu{(\partial_{|\mu|} h)_{a_2}}^bA_{ba_3a_4]})\end{equation}
Following the same procedure for the field $A_{a_1\ldots a_6}$ we find,
\begin{equation} \tilde{F}_{a_1\ldots a_7}\equiv 7{e_{[a_1}}^\mu(\partial_{|\mu|} A_{a_2\ldots a_7]})+6{e_{[a_1}}^\mu\partial_\mu({h_{[a_2}}^b)(A_{|b|a_3\ldots a_7]})-{e_{[a_1}}^\mu A_{[a_2a_3a_4}(\tilde{D}_{|\mu |}A_{a_5a_6a_7]})\end{equation}
The closure of the conformal group with the extension of the affine group $G_{11}$ is enough to ensure the antisymmetrisation of the field strengths formed from the potentials and as a result of the constraint we find \cite{West2},
\begin{equation}\tilde{F}_{a_1\ldots a_4}=\frac{1}{7!}\epsilon_{a_1\ldots a_{11}}\tilde{F}^{a_5\ldots a_{11}}\end{equation}
This is the equation of motion of the field $A_{a_1\ldots a_3}$ in eleven dimensional supergravity as given in equation (\ref{eomA3}) and we see that it occurs in the nonlinear realisation as an identity between fields which are simultaneously covariant under the conformal group and the extended affine group, $G_{11}$. 

Following the literature \cite{West2} we have looked at just the bosonic part of eleven dimensional supergravity, but have not paid any attention to the fermionic sector. The reason for this is that the treatment for the supersymmetric theory would involve the closure of $G_{11}$ and the superconformal group. Since there exists a unique extension of the conformal group that also contains the superalgebra \cite{HoltenProeyen}, being the group denoted $Osp(1/64)$, we are able to conclude that the supersymmetric version of the nonlinear realisation involves the closure of $Osp(1/64)$, or a group which contains it as a subgroup, with $G_{11}$, or again some group containing it as a subgroup. This expectation, in part, motivates us to concentrate on the bosonic sector of the theory and consider some very large algebraic constructions, under which the full theory is invariant.

\subsection{An Awfully Big Conjecture}
In the introduction we reviewed the various hidden symmetries that emerge upon a dimensional reduction of eleven dimensional supergravity. Such hidden symmetries do not have an obvious higher-dimensional origin. The approach of \cite{West2} that we have described in the previous section, while it demonstrates the success of the nonlinear realisation does not involve a group which contains the the hidden symmetry groups as subgroups. The cosets, $\frac{\cal G}{\cal H}$, which arise in the dimensional reduction and reveal the hidden symmetries are such that $\cal H$ is the maximal compact subgroup of $\cal G$ and that $\cal G$ and $\cal H$ have the same rank. The coset representatives, which correspond to the scalar fields of the theory, may be chosen to be in the exponentiation of the Borel sub-algebra of $\cal G$. 
\begin{description}
	\item[{\bf Definition 3.1}]The \emph{Borel sub-algebra} is the set of generators of positive roots and the Cartan sub-algebra.
\end{description}
There are as many scalar fields as there are generators of the Borel sub-algebra. 

The coset model of \cite{West2} does not take the same form as that used to reveal hidden symmetries \cite{CremmerJulia,Julia,CremmerJulia1}. Consequently in \cite{West1} the approach was altered by enlarging the group $G_{11}$. The additional generators may occur in the local subgroup $\cal H$ of the coset model, in which case their effect on the resulting theory is limited, as the form of Maurer-Cartan form is unaltered, but the local transformations of the theory are of course enlarged and the field equations may no longer be invariant under the local group. Our aim in this section is to follow \cite{West1} and see if we are able to enlarge $G_{11}$ to manifestly contain the exceptional Lie algebras, $E_n$, thereby recreating the known coset symmetries that appear in the reduction to four and three dimensions. 

We commence by looking for an $E_7$ coset symmetry and our first step is to split the generators of $G_{11}$ into distinguished sets. The first set contains the generators of the Cartan sub-algebra, and the second set the generators of the positive roots of $SL(11)$ as well as the three form and the six form from $G_{11}$, and the final set contains the remaining negative root generators of $SL(11)$:
\begin{align}
\nonumber G^0_{11}&\equiv \{H_a={K^{a}}_a-{K^{a+1}}_{a+1}, D=\sum_{b=1}^{11}{K^b}_b \left| \right. a=1\ldots 10\} \\
\nonumber G^+_{11}&\equiv \{{K^a}_b, R^{c_1c_2c_3}, R^{c_1\ldots c_6} \left| \right. b>a; a,b=1\ldots 11\} \\
G^-_{11}&\equiv \{{K^a}_b \left| \right. b<a; a,b=1\ldots 11\} 
\end{align}
The Borel sub-algebra is contained in the union $G^0_{11}\cup G^+_{11}\equiv G^{0+}_{11}$. Our hope is that we can construct a group containing $G_{11}$ that can be written as a coset in the same format as the cosets found from dimensional reduction by adding generators to $G^-_{11}$, which will correspond to an enlargement of the local group. Or, in other words, we work with the coset space $\frac{G_{11}}{G^-_{11}}$. Our next step is to identify the generators in $G^{0+}_{11}$ that correspond to the Borel sub-algebra of $E_7$, which we may calculate by applying all possible Weyl reflections in the simple roots to the set of simple roots. Doing this we find that the positive root generators of $E_7$ are,
\begin{equation}{K^a}_b , R^{a_1a_2a_3}, R^{a_1\ldots a_6} \quad b>a\end{equation}
There are $21+35+7=63$ positive roots and generators. They satisfy the same commutators as the equivalent objects in $G_{11}$ as given in equation (\ref{G11commutators}). A convenient basis for the Cartan sub-algebra of $E_7$ is,
\begin{align}
\nonumber \hat{H}_i&={K^i}_i-{K^{i+1}}_{i+1} \quad i=5\ldots 10 \\
\hat{H}_{11}&=-\frac{1}{3}({K^5}_5+\ldots+{K^8}_8)+\frac{2}{3}({K^9}_9+\ldots+{K^{11}}_{11})
\end{align}
The dimension of the adjoint representation of the Borel sub-algebra is $7+(7+21+35)=70$. It is possible therefore by restricting the generators of $G^{0+}_{11}$ to a seven-dimensional subspace to find the algebra satisfied by the Borel sub-algebra of $E_7$. We observe that if we number the nodes on the $E_7$ Dynkin diagram $5\ldots 11$ left to right along the $SU(7)$ sub-algebra, the remaining node being labelled $11$, then we find the restricted $G^{0+}_{11}$ content is,
\begin{equation}\{{K^i}_j,R^{k_1k_2k_3},R^{k_1\ldots k_6}, \hat{D}=\sum_{l=5}^{11}{K^l}_l \left| \right. i,j,k = 5\ldots 11, j\geq i \}\end{equation} 
Instead of the generator $\hat{D}$ we make use of the variable $H_{11}\equiv {K^9}_9+{K^{10}}_{10}+{K^{11}}_{11} -\frac{1}{3}\hat{D}$ and then since the generators and the commutators of the restricted subspace of $G^{0+}_{11}$ are identical to those of the Borel sub-algebra of $E_7$ we are content that we have identified the $E_7$ Borel sub-algebra within $G^{0+}_{11}$. But this is not enough to conclude that the full $E_7$ symmetry is present in $G_{11}$ for we have not identified the generators of the negative roots. As proposed above we will simply add in the missing generators as generators of a local symmetry and then check to see if the field equations are satisfied. For notational convenience we will relabel $R^{k_1\ldots k_6}$ by $S_j$ where,
\begin{equation}S_j\equiv \frac{1}{6!}\epsilon_{jk_1\ldots k_6}R^{k_1\ldots k_6}\end{equation} 
By counting the degrees of freedom we can deduce the missing generators, since we know that the adjoint representation of $E_7$ is 133 dimensional, we have,
\begin{equation}133=6(H_i)+1(H_{11})+21({K^i}_j, j>i)+21({K^j}_i, j>i)+35(R^{k_1k_2k_3})+\bar{7}S_j+\bar{35}(R_{k_1k_2k_3})+7S^j\end{equation}
Where to find the full set of $E_7$ generators we have systematically added two extra generators $R_{k_1k_2k_3}$ and $S^j$ associated to the generator of the negative roots, $-\alpha_{11}$ and $-(\alpha_6+2\alpha_7+3\alpha_8+2\alpha_{9}+\alpha_{10}+2\alpha_{11})$\footnote{These roots are those associated to the specific generators $R_{91011}$ and $R_{67891011}$, or. equivalently, $S^5$.}. These generators together with the commutator action form the dual to the generators associated to $\alpha_{11}$. Thus to incorporate the full $E_7$ symmetry in our algebra we are lead to enlarge $G_{11}$ to include these two extra generators, and in particular to enlarge the local symmetry group to include these generators as well as the Lorentz group. Our sole concern with our adhoc approach is that the equations of motion are invariant under the new local subgroup action. As observed in \cite{West1}, the $63$ degrees of freedom of $SU(8)$ decompose naturally into representations of $SO(7)$ as 
\begin{equation}63=21(K_{(ij)})+\bar{35}(R_{k_1k_2k_3})+7(S^j)\end{equation}
So that the field equations of eleven dimensional supergravity are unaltered by the enlargement of $G_{11}$ reviewed above. 

Having suggested a way to include an $E_7$ sub-algebra into supergravity it is now a short step to conjecturing an $E_{11}$ symmetry. In the enlarged $\tilde{G}_{11}$ we have a set of generators for $SL(11)$, ${K^a}_b$, as well as the generators for an $E_7$ sub-algebra. The algorithm for finding the diffeomorphism group requires finding the closure of the set of generators $\tilde{G}_{11}$ with the conformal group, so at the very least we are looking for an infinite-dimensional symmetry having both $E_7$ and $A_{10}$ sub-algebras. The "simplest" algebra of satisfying these conditions and naturally giving rise to a three form field when decomposed with respect to the $A_{10}$ sub-algebra is the very-extended Kac-Moody algebra $E_8^{+++}\equiv E_{11}$, which has Dynkin diagram,
\begin{figure}[cth]
\hspace{160pt} \includegraphics[viewport=0 150 50 200,angle=90]{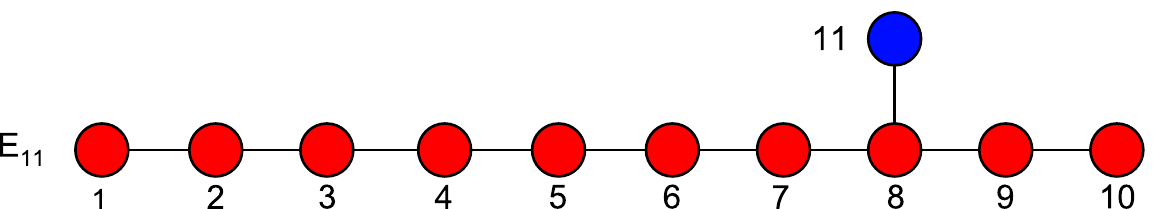}
\caption{The Dynkin Diagram of $E_{11}$} \label{E11}
\end{figure}
$E_{11}$, and falls into a class of Kac-Moody algebras which are called Lorentzian \cite{GaberdielOliveWest}. This means that the deletion of at least one node of the diagram leads to a finite dimensional Lie group. In the case of $E_{11}$ deletion of the node corresponding to $\alpha_{11}$ leads to $A_{10}$ which is given by the string of roots coloured in red in figure \ref{E11}. 

Continuing our review of \cite{West1} we address the question of whether a similar scheme can be employed to enlarge $G_{11}$ and find the Borel sub-algebra of $E_8$ which is the symmetry group that reveals itself upon dimensional reduction to three dimensions. We will find that it cannot. To see that this scheme will not extend to $E_8$ in a similar manner we require a little familiarity with the Borel sub-algebra of $E_8$. The Cartan sub-algebra presents us with no problems, it is trivially extended as,
\begin{align}
\nonumber \hat{H}_i&={K^i}_i-{K^{i+1}}_{i+1} \quad i=4\ldots 10 \\
\hat{H}_{11}&=-\frac{1}{3}({K^4}_4+\ldots+{K^8}_8)+\frac{2}{3}({K^9}_9+\ldots+{K^{11}}_{11})
\end{align}
The generators of the positive positive roots are not so straightforward. As before we can find all these generators by finding all the Weyl reflections of the positive simple root system. It may be worthwhile carrying this process out in detail to convince ourselves that the field content of $E_8$ differs from $E_7$ in one significant way. The interesting set of generators are those associated to the distinguished root, $\alpha_{11}$ since the Weyl reflections of the root system associated to the gravity line $\{\alpha_4, \ldots ,\alpha_{10} \}$ gives the roots associated to the generators ${K^i}_j$ where $j>i$ and $i,j=4\ldots 11$. Commencing with $\alpha_{11}$ the only Weyl reflection that is not inert is the reflection in the root $\alpha_8$, which we denote $S_{8}$. We recall that the Weyl reflection is defined to act as 
\begin{equation}
S_a\beta\equiv\beta-2\frac{<\alpha_a,\beta>}{<\alpha_a,\alpha_a>}\alpha_a \label{Weylreflection}
\end{equation} 
It can be convenient to write the coefficients of a particular sum of simple roots over the Dynkin diagram as a visual aid to following the Weyl reflections, thus $\alpha_{11}$ and its associated generator, decomposed with respect to an $SL(8)$ sub-algebra, are written,
\end{spacing}
\begin{align}
\nonumber &1\\
\nonumber 0 \quad0 \quad0 \quad&0 \quad0 \quad0 \qquad\qquad R^{91011}
\end{align}
\begin{spacing}{1.5}
The use of this approach is that we may quickly write down the Weyl reflections, which translate into the rule that a Weyl reflection in $\alpha_i$ subtracts twice the value of the $\alpha_i$ coefficient and adds the value of all the coefficients that it is attached to on the Dynkin diagram to the coefficient of $\alpha_i$. This may sound cumbersome but it often speeds calculations. The addition of a simple root acts as the adjoint action of the generator of the added simple root upon the generator of the root that we Weyl reflected. In particular by applying all possible Weyl reflections we find, amongst many others, the generators,
\end{spacing}
\begin{align}
\nonumber &2\\
\nonumber 0\quad 0 \quad1 \quad2 \quad&3 \quad2 \quad1 \qquad\qquad R^{67891011}\\
\nonumber \\
\nonumber &3\\
\nonumber 1\quad 2 \quad3 \quad4 \quad&5 \quad4 \quad2 \qquad\qquad R^{4567891011,9}
\end{align}
\begin{spacing}{1.5}
We deduce from the Weyl reflections that the peculiar generator with nine indices must be antisymmetrised over the first eight indices before the comma and symmetrised in the index following the comma. The new object acts like a vector and has dimension $8$. The dimension of $E_8$ is therefore \begin{equation}64({K^i}_j)+56(R^{k_1k_2k_3})+28(R^{k_1\ldots k_6})+8(R^{k_1\ldots k_8,j})+\bar{56}(R_{k_1k_2k_3})+\bar{28}(R_{k_1\ldots k_6})+\bar{8}(R_{k_1\ldots k_8,j})=248\end{equation}
It is the presence of this new object $R^{k_1\ldots k_8,j}$ in the Borel sub-algebra of $E_8$ that is the problem. It is not present in the restriction of $G^{0+}_{11}$ to an appropriate eight-dimensional subspace. In order to find a higher dimensional origin of the hidden $E_8$ symmetry, which we know is present after the dimensional reduction, we must enlarge $G_{11}$ to include $R^{k_1\ldots k_8,j}$ as well as add the missing generators of the negative roots to the local sub-algebra $G^-_{11}$ in our coset formulation. The effect of enlarging the Borel sub-algebra is non-trivial, a new field is introduced into supergravity. As explained in \cite{West1}, the field associated to the generator is $h_{a_1\ldots a_8,b}$. If we were to treat its $b$ index as an internal index and then form a field strength from it we would find,
\begin{equation}F_{a_1\ldots a_9,b}=9\partial_{[a_1}h_{a_2\ldots a_9],b}\end{equation}
This field strength is dual to 
\begin{equation}{F^{a_{10}a_{11}}}_b\equiv\frac{1}{9!}\epsilon^{a_1\ldots a_{11}}F_{a_1\ldots a_9,b}\end{equation}
Which is the field strength we would find if we treated ${h^a}_b$, the field associated to the vielbein, as a gauge potential with an internal index $b$. This observation is aesthetically very satisfying, since it results in all the gauge fields of the enlarged $G_{11}$ having a dual counterpart in the algebra.

The incorporation of an $E_8$ symmetry into supergravity again puts us back on the trail of an $E_{11}$ symmetry in M-theory. Let us call our enlarged set of supergravity generators $\hat{G}_{11}$, and record here for completeness the upgraded set of commutators \cite{West1},
\begin{align}
\nonumber [{K^a}_b,{K^c}_d]&=\delta^c_b{K^a}_d-\delta^a_d{K^c}_b \\
\nonumber [{K^a}_b,P_c]&=-\delta^a_cP_b\\
\nonumber [{K^a}_b,R^{c_1c_2c_3}]&=\delta^{c_1}_bR^{ac_2c_3}+\ldots\\
\nonumber [{K^a}_b,R^{c_1\ldots c_6}]&=\delta^{c_1}_bR^{ac_2\ldots c_6}+\ldots\\
\nonumber [{K^a}_b,R^{c_1\ldots c_8,d}]&=\delta^{c_1}_bR^{ac_2\ldots c_8,d}+\ldots +\delta^{d}_bR^{c_1c_2\ldots c_8,a}\\
\nonumber [R^{c_1c_2c_3},R^{c_4c_5c_6}]&=2R^{c_1\ldots c_6}\\
[R^{c_1\ldots c_6},R^{c_7c_8c_9}]&=3R^{c_1\ldots [c_7c_8,c_9]} \label{Dualgravcommutators}
\end{align}
The remaining commutators being zero.

The enlargement of the bosonic part of the supergravity algebra to find subsequently $E_7$ and $E_8$ algebras which are present prior to dimensional reduction and are no longer hidden symmetries emboldens the conjecture discussed earlier, namely that there is an $E_{11}$ symmetry algebra underlying M-theory. The approach in the subsequent chapters of this thesis will be to commence with the algebra of $E_{11}$ and see what top-down connections can be made between this algebraic approach and M-theory that may shed light on the status of this conjecture. But it will be encouraging first of all to look at the role of the scalar fields and their association with the coset symmetries, which were historically the first indicators of hidden symmetries, in three dimensions.

\subsection{A Return to Three Dimensions} \label{scalars}
In three dimensions the entire field content of a theory may be described by the scalar fields. There exist only four types of tensorial object, and all those which carry information about the theory are directly derived from the scalars. The tensors have at most three antisymmetrised indices, these being a scalar, the vector field strength of a scalar, the dualised field strength of a scalar and the completely antisymmetric epsilon tensor. In the coset model the scalars are used to parameterise the cosets and enhanced symmetries are observed.

There is an elegant formulation of the Kac-Moody symmetries which comes from identifying the scalars in a reduced theory with the simple roots of an algebra \cite{LambertWest}. This approach goes some way towards identifying an $E_{11}$ theory, in that it identifies an $E_8$ symmetry amongst the scalars resulting from the dimensional reduction of a pure gravitational theory coupled to a four form. This approach will be illuminating to review briefly. The paper \cite{LambertWest} concentrates on a number of cases including those of pure gravity, the bosonic string, eleven dimensional supergravity. We will concentrate on the pure gravity and eleven dimensional theory dimensional reductions and state the conjectured symmetry, partly uncovered by the same process, of bosonic string theory.

To commence we must be familiar with the notion of dimensional reduction. We shall use the conventions detailed in the excellent lecture notes of Chris Pope \cite{Pope}. Dimensional reduction is a process whereby the metric and all the fields of the theory are made to be independent of the coordinates spanning the reduced dimension. Frequently one may imagine the reduced dimensions to be wrapped into circles, this is a toroidal compactification, whose circumferences are so small that variations in their coordinates have negligible effect upon the fields which they depend upon. The metric becomes decomposed into an infinite number of modes, all bar the zero mode having a very large associated mass. The scale of the massive fields is large due to the radius of compactification being so small, thus the opportunities to excite these modes lie above the energy realms of modern day particle physics. Consequently it is interesting enough simply to concentrate on the massless sector and truncate the infinite set of fields to only the massless sector. This is normally implicit in dimensional reduction. 

Consider the effect of a one-dimensional reduction upon the degrees of freedom of a $D+1$ dimensional metric, $\hat{g}_{\mu\nu}$. We have three different choices, depending upon whether the metric component is dependent upon zero, one or two dimensionally reduced parameters. If we allow $z$ to denote the reduced dimension, then we find a new metric $g_{\mu\nu}$ where the missing hat indicates that $\mu,\nu = 1\ldots D$, a column vector $A_\mu$ coming from the components of the metric of the form $\hat{g}_{\mu z}$ and a scalar, $\phi$ arising from the component $\hat{g}_{zz}$. When we reduce over more than one dimension we repeat this process and find more complex fields arising from the dimensional reduction. To summarise we find that when we dimensionally reduce a pure gravity theory we find field content of a more complex theory plus a lower dimensional gravity theory. This remarkable observation was originally made by Theodore Kaluza and Oskar Klein and \cite{Kaluza, Klein} for the reduction of a five-dimensional gravitational theory on a circle to arrive at a four-dimensional unified theory of gravity and the electromagnetic force. 

As described above the line element is decomposed into 
\begin{equation}d\hat{s}^2=ds^2+2A_\mu dz dx^\mu + \phi dz^2\end{equation}
However for computational convenience we prefer to use the decomposition ansatz of \cite{Pope},
\begin{equation}d\hat{s}^2=e^{2\alpha\phi}ds^2+e^{2\beta\phi}(dz+A_\mu dx^\mu)^2\end{equation}
Using this ansatz and taking $\beta=-(D-2)\alpha$ as well as $\alpha^2=\frac{1}{2(D-2)(D-1)}$, choices which ensure we find a familiar gravity action and scalar kinetic term, we find that a pure gravity action decomposes into,
\begin{equation}\sqrt{-\hat{g}}\hat{R}=\sqrt{-g}\{R-\frac{1}{2}\partial_\mu\phi\partial^\mu\phi-\frac{1}{4}e^{-2(D-1)\alpha\phi}F_{\mu\nu}F^{\mu\nu}\}\end{equation}
The one-dimensional reduction of an m-form field strength $F_{m}=dA_{m-1}$ proceeds similarly. We first note that,
\begin{equation}\hat{F}_{(m)}=dA_{(m-1)}+dA_{(m-2)}\wedge dz\end{equation}
Which we rewrite to fit our line element ansatz as,
\begin{align}
\nonumber \hat{F}_{(m)}&=dA_{(m-1)}+dA_{(m-2)}\wedge dz+dA_{(m-2)}\wedge A_{(1)}-dA_{(m-2)}\wedge A_{(1)} \\
\nonumber &=dA_{(m-1)}-dA_{(m-2)}\wedge A_{(1)}+dA_{(m-2)}\wedge (dz+A_{(1)}) \\
&\equiv F_{(m)}+F_{(m-1)}\wedge (dz+A_{(1)})
\end{align}
Where we have defined arbitrarily,
\begin{equation}F_{(m)}=dA_{(m-1)}-dA_{(m-2)}\wedge A_{(1)}, \quad F_{(m-1)}=dA_{(m-2)}\end{equation}
The vielbein for our ansatz is read off from the line element as ${\hat{e}^a}=e^{\alpha\phi}{e^a}$ and ${\hat{e}^z}=e^{-(D-2)\phi}(dz+A)$ so that upon reduction the form becomes
\begin{align}
\nonumber \hat{F}_{(m)}&=\frac{1}{m!}\hat{F}_{a_1\ldots a_m}(\hat{e}^{a_1}\wedge \ldots \wedge \hat{e}^{a_m})+\frac{1}{(m-1)!}\hat{F}_{a_1 \ldots a_{m-1} z}\wedge (\hat{e}^{a_1}\wedge \ldots \wedge \hat{e}^{a_{m-1}}\wedge \hat{e}^z)\\
&=\frac{e^{m\alpha\phi}}{m!}\hat{F}_{a_1\ldots a_m}(e^{a_1}\wedge \ldots \wedge e^{a_m})+\frac{e^{(m-D+1)\alpha\phi}}{(m-1)!}\hat{F}_{a_1 \ldots a_{m-1} z}\wedge ({e}^{a_1}\wedge \ldots \wedge {e}^{a_{m-1}}\wedge {e}^z)
\end{align}
We therefore find for the reduced components,
\begin{equation}\hat{F}_{a_1\ldots a_m}=e^{-m\alpha\phi}F_{a_1\ldots a_m}, \quad \hat{F}_{a_1\ldots a_(m-1)z}=e^{(D-m-1)\alpha\phi}F_{a_1\ldots a_{m-1}z}\end{equation}
In terms of the reduction of a kinetic term formed from the m-form, we bear in mind that $\sqrt{-\hat{g}}=\sqrt{-ge^{2D\alpha}e^{-2(D-2)\alpha\phi}}=e^{2\alpha\phi}\sqrt{-g}$, we find,
\begin{equation}\frac{\sqrt{-\hat{g}}}{2m!}\hat{F}_{(m)}^2=\sqrt{-g}\{\frac{e^{2(1-m)\alpha\phi}}{2m!}F_{(m)}^2+\frac{e^{2(D-m)\alpha\phi}}{2(m-1)!}F_{(m-1)}^2\}\end{equation}
We are interested in finding the largest coset symmetry in the dimensional reduction, and since the scalars in the theory will parameterise the cosets we must generate as many scalars as possible, this means we must also consider scalars that arise from dualisation of the $m$-forms. In a $D$ dimensional spacetime the dual of an $m$-form is a $(D-m)$-form, which is the exterior derivative of a $(D-m-1)$-form. Consequently when $D=m+1$ we find an extra scalar. 

Let us look at the reduction of pure gravity theory to get an understanding of the appearance of scalars. Commencing in $D+1$ dimensions and applying a reduction on a two-torus we find the the following Lagrangian in $D-1$,
\begin{align}
\nonumber \int d^{D+1}x\sqrt{-g} R&=\int d^Dx \sqrt{-g}\{R^{(D)}-\frac{1}{2}\partial_\mu\phi_{(D)}\partial^\mu\phi_{(D)}-\frac{1}{2.2!}e^{-2(D-1)\alpha_D\phi_D }F^{(D)}_{\mu\nu}F^{(D)\mu\nu}\}\\
\nonumber &=\int d^{(D-1)}x \sqrt{-g}\{R^{(D-1)}-\frac{1}{2}\partial_\mu\phi_{(D-1)}\partial^\mu\phi_{(D-1)}\\
\nonumber &-\frac{1}{2.2!}e^{-2(D-2)\alpha_{(D-1)}\phi_{(D-1)}}F^{(D-1)}_{\mu\nu}F^{(D-1)\mu\nu}-\frac{1}{2}\partial_\mu\phi_{(D)}\partial^\mu\phi_{(D)}\\
\nonumber &-\frac{1}{2.2!}e^{-2(D-1)\alpha_D\phi_D}e^{-2\alpha_{(D-1)}\phi_{(D-1)}}F^{(D)}_{\mu\nu}F^{(D)\mu\nu}\\
&-\frac{1}{2}e^{-2(D-1)\alpha_D\phi_D }e^{2(D-3)\alpha_{(D-1)}\phi_{(D-1)}}F^{(D)}_{\mu}F^{(D)\mu}\} \label{reduction}
\end{align}
Where we indicate the dimension the scalars and forms appeared in brackets, e.g. $F^{(D)}_\mu$ is a one-form reduced from the two-form that appeared upon reduction to $D$ dimensions. We have also used the notation $\alpha_D=\sqrt{\frac{1}{2(D-2)(D-1)}}$. We could continue on but the notation becomes cumbersome. Instead, following \cite{LambertWest}, we observe that upon each reduction we obtain a new $\alpha_d$ and $\phi_d$ which appear together and which we may write as components of a scalar product of vectors $\alpha$ and $\phi$ whose dimension increases by one upon each dimensional reduction. That is, upon compactification on a 2-torus to $D-1$ dimensions we have the natural 2-dimensional vectors, $\alpha_{(j,k)}\equiv(2(D-j-3)\alpha_{D-1},-2(D-k-1)\alpha_{(D)})$ and $\phi\equiv(\phi_{D-1},\phi_{(D)})$.
We may rewrite equation (\ref{reduction}) as,
\begin{align}
\nonumber \int d^{D+1}x\sqrt{-g} R&=\int d^{(D-1)}x\sqrt{-g}\{R^{(D-1)}-\frac{1}{2}\partial_\mu\phi\cdot\partial^\mu\phi-\frac{1}{2.2!}e^{\alpha_{(2D-5,D-1)}\cdot\phi}F^{(D-1)}_{\mu\nu}F^{(D-1)\mu\nu}\\
&-\frac{1}{2.2!}e^{\alpha_{(D-2,0)}\cdot\phi}F^{(D)}_{\mu\nu}F^{(D)\mu\nu}-\frac{1}{2}e^{\alpha_{(0,0)}\cdot\phi}F^{(D)}_{\mu}F^{(D)\mu}\} \label{reduction_roots}
\end{align}
We note that the vectors that arise satisfy the linear relation $\alpha_{(2D-5,D-1)}+\alpha_{(0,0)}=\alpha_{(D-2,0)}$\footnote{We will appreciate this linear relation between the vectors appearing at each stage of the dimensional reduction as being identical to the linear relations between the positive roots of a Lie algebra, which itself may be classified by identifying solely its simple roots, of which here there are two associated to $\alpha_{(0,0)}$ and $\alpha_{(2D-5,D-1)}$.}. Now we note that we commence finding a second class of scalars, besides $\phi$, when we reduce the scalar curvature, $R$, on a 2-torus. In the above reduced Lagrangian the first additional scalar is revealed, where it appears as a one-form field strength, $F_\mu=\partial_\mu\psi$. In general, for a reduction from $D+1$ dimensions to $D+1-r$ dimensions on an r-torus, we will find $r-2$ additional scalars and we will construct vectors, characterising these scalars, in an r-dimensional space such that $\phi\equiv(\phi_{(D+1-r)},\ldots,\phi_{(D)})$ and we will find we are able to write the reduced Lagrangian using the general vectors,
\begin{equation}
\alpha_{(k_1,\ldots k_r)}\equiv-2((D-k_1-1)\alpha_D,(D-k_2-3)\alpha_{D-1},(D-k_3-4)\alpha_{D-2},\ldots,(D-k_r-r-1)\alpha_{D-r+1})
\end{equation}
Using vectors in this format we can write out the reduced Lagrangian, as we did when we reduced on the two-torus above. A suitable basis for these vectors which characterise the scalars is given by the vectors used in the expression for the one-forms which originate in different dimensions, these all take the form,
\begin{equation}
\alpha_{k}\equiv(0,\ldots,0,-2(D-k-1)\alpha_{D-k},2(D-k-3)\alpha_{D-k-1},0,\ldots,0) \label{rootvectors}
\end{equation}
Where there are $k$ zeroes on the left and $r-2-k$ zeroes on the right, and $k=0,\ldots, r-2$. Our expression differs from that given in \cite{LambertWest} because we began our reduction in $D+1$ dimensions, and if we replace $D$ with $D-1$ in our expression we find the same basis vectors as given in \cite{LambertWest}. The vectors satisfy the relations,
\begin{align}
\nonumber \alpha_k\cdot\alpha_k&=4\\
\nonumber \alpha_k\cdot\alpha_{k-1}&=-2\\
\alpha_k\cdot\alpha_{k+n}&=0 \qquad |n|>1
\end{align}
Vectors of this type span $r-1$ dimensions of the r-dimensional vector space, but our experience of the notation in the reduction on the 2-torus leads us to expect that the full Lagrangian is characterised by $\frac{r(r+1)}{2}$ vectors and that in addition to the $r-1$ vectors, $\alpha_k$, we find one further vector associated to the terms formed in the reduction whose scalar association has yet to be made manifest by reduction. For example, in the reduction on the two torus, we would have found one vector of the type $\alpha_k$ which in the notation of equation (\ref{reduction_roots}) is the vector $\alpha_{(0,0)}$, in addition to this vector we found one further linearly independent vector, $\alpha_{(2D-5,D-1)}$, which was associated to the two-form which appeared in the last stage of the reduction to dimension $(D-1)$. Similarly in the general case when we reduce on an r-torus there will arise a two-form $F_{\mu\nu}^{(D+1-r)}$ upon reduction of the r'th coordinate. 

The coefficient of this term is,
\begin{equation}
-\frac{1}{2.2!}e^{-2(D-r)\alpha_{D+1-r}\phi_{(D+1-r)}}
\end{equation}
So that we find an additional vector,
\begin{equation}
\alpha_{r-1}=(0,\ldots,0,-2(D-r)\alpha_{D+1-r})
\end{equation}
Which satisfies the following relations,
\begin{align}
\nonumber \alpha_{r-1}\cdot\alpha_{r-1}&=\frac{2(D-r)}{(D-r-1)}\\
\nonumber \alpha_{r-1}\cdot\alpha_{r}&=-2\\
\alpha_{r-1}\cdot\alpha_{r-n}&=0 \qquad n\geq2
\end{align}
In particular we observe that only when we have reduced to a dimension where it is possible to dualise the two-form to a scalar, namely when we are in three dimensions, do the inner products of this final vector take the same form as the others. Specifically, when we reduce to three dimensions, $D+1-r=3$, so that 
\begin{equation}
\alpha_{r-1}\cdot\alpha_{r-1}=\frac{2(2)}{(1)}=4
\end{equation}
We emphasise that only when the pure gravity theory is reduced to three dimensions do the basis vectors all have the same length, in all higher dimensions the final basis vector is shorter. These vectors that we have directly identified with the scalars of the theory may now be associated to roots of a Dynkin diagram and we find the simple roots of $A_{r}=A_{D-4}$ have an obvious one-to-one correspondence with the scalars. The remaining other roots that occurred in our reduction, which we could have chosen as basis vectors, are associated with the positive roots of the algebra, of which there are $\frac{r(r+1)}{2}$ as anticipated. We note that had we commenced our reduction in $D$ dimensions, instead of $D+1$, we would have ended up with the simple roots of $A_{D-3}$. 

Finally we apply the same approach to uncovering the coset symmetries of a pure gravity action in eleven dimensions and an additional four-form field strength, which is the bosonic sector of M-theory. 
\begin{equation}
\int d^{D}x\sqrt{-g} \{R-\frac{1}{2.4!}F_{\mu_1\ldots \mu_4}F^{\mu_1\ldots \mu_4}\}
\end{equation}
From the pure gravity sector we clearly obtain the same vectors parameterising an $A_{D-3}$ symmetry as before, but we also find an additional vector coming from the reduction of the four-form. For simplicity we consider the effect of a reduction on a 3-torus and find that the one-form term that remains is
\begin{equation}
\int d^{D-3}x\sqrt{-g}\frac{1}{2}e^{2(D-5)\alpha_{D-1}\phi_{D-1}}e^{2(D-5)\alpha_{D-2}\phi_{D-2}}e^{2(D-5)\alpha_{D-3}\phi_{D-3}}F_{\mu}F^{\mu}
\end{equation}
The vector associated with this additional scalar upon reduction to three dimensions is $\alpha_{7}=(2(D-5)\alpha_{D-1},2(D-5)\alpha_{D-2},2(D-5)\alpha_{D-3},0\ldots 0)$, the vector being 8-dimensional. It has the properties,
\begin{align}
\nonumber \alpha_7\cdot\alpha_7&=2(D-5)^2\{\frac{1}{(D-3)(D-2)}+\frac{1}{(D-4)(D-3)}+\frac{1}{(D-5)(D-4)}\}\\
\nonumber &=\frac{6(D-5)}{(D-2)}\\
\nonumber \alpha_7\cdot\alpha_2 &=-2 \\
\alpha_7\cdot \alpha_i &=0 \qquad i\neq 2,7
\end{align}
Where we have set $D\rightarrow D-1$ in equation (\ref{rootvectors}) and now we have a set of vectors $\{\alpha_0, \ldots \alpha_7\}$ which obey the defining relations of the simple roots of the exceptional group $E_8$. 

A similar analysis is applied in \cite{LambertWest} to the bosonic string theory in twenty-six dimension and the symmetry group paramaterised by the scalars in three dimensions is $D_{24}$.
\begin{figure}[h]
\hspace{185pt}\includegraphics[viewport=0 150 130 200,angle=90]{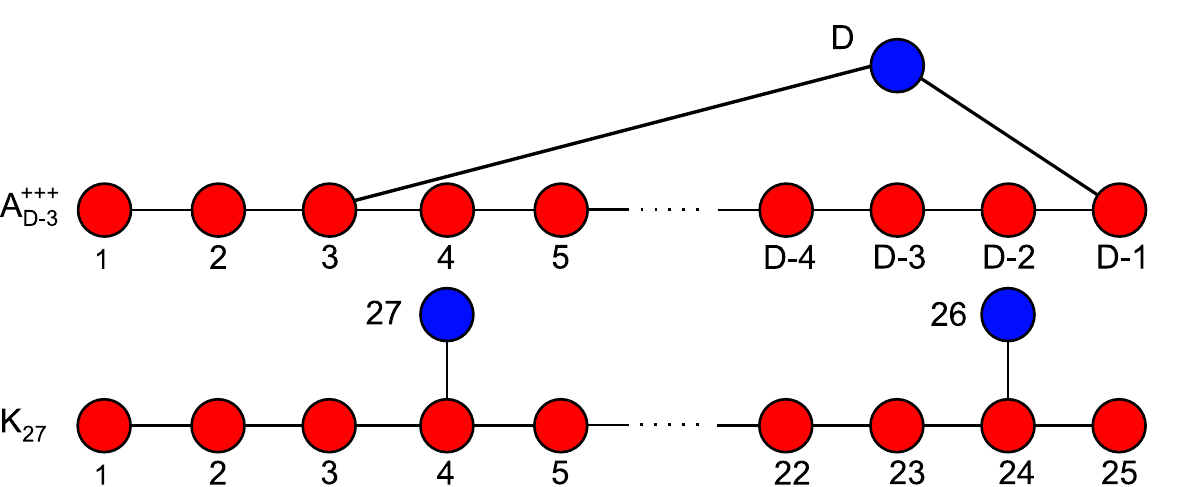}
\caption{The Dynkin diagrams of $A_{D-3}^{+++}$ and $K_{27}$} \label{K27andpuregravity}
\end{figure}

We now have confidence that the dimensional reduction procedure uncovers a hidden $E_8$ symmetry in the bosonic sector of M-theory. Indeed in \cite{LambertWest} it is shown that the reduction of the enlarged algebra, that leads to the $E_{11}$ conjecture, does reduce to the coset symmetries that have been demonstrated above, as well as to an old and well known $SL(2,\mathbb{R})$ symmetry \cite{Ehlers,Geroch}. It may appear that it would be sensible to reduce the theory to two dimensions and see how the symmetry of the gravitational degrees of freedom are encoded in the scalars that then appear, however in two dimensions the only rank two object is the anti-symmetric tensor, $\epsilon^{\mu\nu}$, and so we can no longer study the gravity theory in less than three dimensions. It has however been argued that the expected enlarged symmetry in two dimensions has an affine algebra \cite{JuliaNicolai}. The affine extensions of the coset symmetry groups lead to $A^+_{D-3}$, $E_9$ and $D_{24}^+$, for pure gravity, the bosonic part of M-theory, and the bosonic string respectively. The symmetries are expected to be enlarged in the reduction to zero dimensions to $A_{D-3}^{+++}$, $E_8^{+++}\equiv E_{11}$ and $D_{24}^{+++}\equiv K_{27}$. The Dynkin diagrams for $E_{11}$ was given earlier in figure \ref{E11}, the other Dynkin diagrams are shown in figure \ref{K27andpuregravity}.
\newpage

\section{Constructing the Algebra}
Thus far we have familiarised ourselves with the tools of our work and motivated the study of Kac-Moody algebras within theoretical physics. In this chapter we work with the Kac-Moody algebras directly and produce tables of roots appearing in the algebra. Our method will be to decompose roots in the Kac-Moody algebras into representations of semisimple Lie groups by deleting nodes from the Dynkin diagram. Using the properties of the semisimple Lie groups we will find the algebraic equations that must be solved with integer valued solutions for a root to exist in the Kac-Moody algebra. Since the algebras are Kac-Moody there are an infinite number of such roots and we will label classes of roots by the coefficients of the deleted roots in the decomposition. We will concentrate on the algebras relevant to M-theory and bosonic string theory, these being $E_{11}$ and $K_{27}$ respectively.

\subsection{The Algebra of M-theory} \label{Mtheory}
Although the problem of identifying all the roots existing in an infinite dimensional algebra, is probably best left as a computational problem, there are a number of good reasons for beginning the search by hand. For example one quickly arrives at some governing rules which vastly reduce the number of potential candidate roots which will save computational power at a later stage and secondly one becomes familiar with patterns that proliferate the algebra. We commence with the Dynkin diagram of $E_{11}$ given in figure \ref{E11}.

The setting for our calculations is an eleven-dimensional vector space with basis elements $\{e_1\ldots e_{11}\}$ endowed with an inner product,
\begin{equation}
<a,b>=a\cdot b-{\frac{1}{9}}\sum_i(a^i)\sum_j(b^j) \label{innerproduct1}
\end{equation}
We construct vectors in this space which represent the roots of our algebra such that the inner product relations between the roots as encoded in the Dynkin diagram is carried over to the inner product on our vector space. The largest algebra we will consider is $E_{11}$ and all sub-algebras relevant to M-theory will be formed by deleting the appropriate root vector. The roots of $E_{11}$ in this basis are,
\begin{align}
\nonumber\alpha_i&=e_i-e_{i+1}\quad i=1,2\ldots 10 \\
\alpha_{11}&=e_9+e_{10}+e_{11} \label{E11simpleroots}
\end{align}
A generic root of $E_{11}$ is,
\begin{equation}
\beta=\sum_{i=1}^{11} m_i \alpha_i
\end{equation}
Where $\alpha_i$ are the simple roots of $E_{11}$ and the coefficients $m_i$ are referred to as the Dynkin labels. By deleting the distinguished node, corresponding to $\alpha_{11}$, of the Dynkin diagram we find representations of the $A_{10}$ sub-algebra. In particular we may write the deleted root as being composed of a part orthogonal to the ten dimensional vector space of $A_{10}$ and a vector in the $A_{10}$ lattice. It is clear from the inner product relations of $\alpha_{11}$ with the roots of $A_{10}$ namely,
\begin{equation}
<\alpha_{11},\alpha_i>=-\delta_{i8} \qquad 1<i\leq 10
\end{equation}
that the part of $\alpha_{11}$ lying in the $A_{10}$ lattice is the negative of the fundamental weight, $\nu_8$ of $A_{10}$. Where $\nu_i$, $1<i\leq 10$, are the fundamental weights of $A_{10}$ which we recall have the defining relation,
\begin{equation}
<\nu_i,\alpha_j>=\delta_{ij}
\end{equation}
So we may write,
\begin{equation}
\alpha_{11}=z-\nu_8 \label{root11}
\end{equation}
Where $z$ is some vector orthogonal to the $A_{10}$ lattice. We may determine $z$, an eleven dimensional vector, explicitly by imposing the condition that it is by design orthogonal to the roots of $A_{10}$ and that $\alpha_{11}^2=2$. Let us carry out this process once in detail. Let,
\begin{equation}
z\equiv\sum_{i=1}^{11}a_ie_i
\end{equation}
Where $a_i$ are real coefficients. Now,
\begin{equation}
<z,\alpha_i>=<\sum_{i=1}^{11}a_ie_i,e_i-e_{i+1}>=0 \Rightarrow z=a_{11}(e_1+\ldots +e_{11}) \label{zvector}
\end{equation}
To make use of the condition on the length of $\alpha_{11}$ we need the fact that the fundamental weights, $\nu_i$ of any Lie algebra are related to its Cartan matrix, $A$, by
\begin{equation}
<\nu_i,\nu_j>=A^{-1}_{ij} \label{fundamentalweights}
\end{equation}
Inverting the Cartan matrix may seen like a cumbersome task but we are able to make use of several very nice formulae calculated in \cite{GaberdielOliveWest} giving closed expressions for the inverted matrix. For the case of the $A_n$ Lie groups we have the simple result,
\begin{equation}
A_{ij}^{-1}=\frac{i(n+1-j)}{n+1} \quad i \leq j \label{inverseAn}
\end{equation}
Consequently we find from equation (\ref{root11}) that,
\begin{equation}
2=z^2+\frac{24}{11} \Rightarrow z^2=-\frac{2}{11}
\end{equation}
Using equation (\ref{zvector}) this gives $a_{11}=\pm\frac{3}{11}$. To decide whether the coefficient is plus or minus we may repeat the process and find the weight $\nu_8$ in our vector space basis. In fact for an $A_n$ algebra we find the fundamental weights are,
\begin{equation}
\nu_i=\frac{n+1-i}{n+1}(e_1+\ldots +e_i)-\frac{i}{n+1}(e_{i+1}+\ldots +e_{n+1}) \label{Anweights}
\end{equation}
So that,
\begin{equation}
\nu_8=\frac{3}{11}(e_1+\ldots +e_8)-\frac{8}{11}(e_9+\ldots+e_{11})
\end{equation}
Then from equation (\ref{root11}) we find that,
\begin{equation}
z=\frac{3}{11}(e_1+\ldots +e_{11})
\end{equation}
Finally we are now able to write $\beta$ a general root of $E_{11}$ in terms of weights of its $A_{10}$ sub-algebra and an orthogonal part, $z$,
\begin{equation}
\beta=m_{11}(z-\nu_8)+\sum_{i=1}^{10}m_i\alpha_i\equiv m_{11}z-\Lambda
\end{equation}
Where,
\begin{equation}
\Lambda=m_{11}\nu_8-\sum_{i=1}^{10}m_i\alpha_i\equiv \sum_{i=1}^{10}p_i\nu_i \label{A10weights}
\end{equation}
$\Lambda$ is a weight of $A_{10}$ which we have written in terms of the fundamental weights of $A_{10}$ which span the weight space. The important and useful point is that $m_{11}$ is related to the weights of the $A_{10}$ sub-algebra, and although we have ten unknowns on the right hand side of the equation and eleven on the right hand side, we will be able to find a relation that gives strict conditions that the $m_i$ and must satisfy once $m_{11}$ is specified. The important role played by the coefficient of the deleted root has led to it being called the level in the literature.

Taking the inner product of equation (\ref{A10weights}) with a fundamental weight of $A_{10}$, $\nu_j$, we find,
\begin{equation}
m_j=m_{11}<\nu_8,\nu_j>-\sum_{i=1}^{10}p_i<\nu_i,\nu_j> \qquad j\leq 10
\end{equation}
Using equations (\ref{inverseAn}) and (\ref{fundamentalweights}) we simplify this to,
\begin{equation}
m_j=\begin{cases}m_{11}\frac{3j}{11}-\sum_{i=1}^{j}ip_i-j\sum_{i=j+1}^{10}p_i+\frac{j}{11}\sum_{i=1}^{10}ip_i\equiv \frac{j}{11}(3m_{11}+A)-B_j,  \qquad j\leq 8 \\
m_{11}\frac{8(11-j)}{11}-\sum_{i=1}^{j}ip_i-j\sum_{i=j+1}^{10}p_i+\frac{j}{11}\sum_{i=1}^{10}ip_i\equiv \frac{j}{11}(-8m_{11}+A)-B_j+8m_{11},\hspace{4pt} j>8\end{cases} \label{mj}
\end{equation}
Where we have defined the useful, positive, integer quantities,
\begin{align}
\nonumber A&\equiv\sum_{i=1}^{10}ip_i \\
B_j&\equiv\sum_{i=1}^{j}ip_i+j\sum_{i=j+1}^{10}p_i \label{E11parameters}
\end{align}
Observe that $B$ as a function of $j$ satisfies $B_{10}=A$ and $B_1=\sum_{i=1}^{10}p_i\leq A$. We are motivated to express $m_j$ in the form given in equation (\ref{mj}) by some group theoretical conditions. In chapter two we observed the symmetry of the root diagrams of the semisimple Lie algebras; for every positive root there exists a negative root, which is its negative. Consequently by finding simply the positive roots we also identify the negative roots of the algebra. The triangle decomposition of an algebra tells us we can specify the entire algebra from the generators of the positive and negative roots and the Cartan sub-algebra. We may deduce the Cartan sub-algebra from its Dynkin diagram, and hence if we are able to find all the positive roots we can deduce the rest of the algebraic content. The positive roots have integer valued non-negative coefficients $m_i$, this means equation (\ref{mj}) places strong restrictions on the values $A$ and $B$ may take such that $m_j\in{\mathbb Z}^{0+}$.

From (\ref{mj}) we see that a general solution is given by,
\begin{equation}
m_{11}=\frac{1}{3}(-A+11k) \label{m11}
\end{equation}
Where $k\in\mathbb{Z}$ and satisfies $k\geq \frac{A}{11}$. At this stage we ought to be concerned that we have discarded solutions such as $m_{11}=-4A+\frac{11}{3}k$ or $m_{11}=7A+\frac{11}{3}k$, but it is obvious that we can relate these solutions by adjusting the constant k since they differ by some multiple of eleven. We may always rescale $A$ upwards by multiplying the coefficients $p_i$ by some integer positive constant. The solutions that we have singled out above may thus be rescaled to coincide with the chain of solutions are discussing here, e.g. rescaling $p_i\rightarrow 12p_i$ gives $m_{11}=-4A+\frac{11}{3}k$. Since $p_i$ are positive integers to find the most general set of solutions means minimising the magnitude of the coefficient of $A$ in our solution, which is why we have singled out the solution in (\ref{m11}). We must of course be worried that unless $-A+11k$ is a multiple of three $m_{11}$ will not be an integer. As we will see $-A+11k$ will be forced to be a multiple of three by the constraints on the root length in a Kac-Moody algebra.

Plugging the solution in equation (\ref{m11}) into equation (\ref{mj}) gives,
\begin{equation}
m_j=\begin{cases} jk-B_j,  \qquad j\leq 8 \\
\frac{j}{3}(A-8k)+\frac{8}{3}(-A+11k)-B_j,\qquad j>8\end{cases}
\end{equation}
These are integers again if $m_{11}$ is an integer. To ensure positivity of the coefficients we find from $m_{10}\geq0$ that
\begin{equation}
m_{10}=m_{11}-k \Rightarrow k\leq m_{11}
\end{equation}
Let us now write out the root corresponding to this solution in our vector space basis,
\begin{align}
\nonumber \beta &=\sum_{i=1}^{8}(ik-B_i)\alpha_i+\sum_{i=9}^{10}\{\frac{i}{3}(A-8k)+\frac{8}{3}(-A+11k)-B_i\}\alpha_i+\frac{1}{3}(-A+11k)\alpha_{11}\\
&=\sum_{n=1}^{10}(k-\sum_{i=n}^{10}p_i)e_n+ke_{11}
\end{align}
A root which has connected support on a Dynkin diagram is one whose non-zero Dynkin coefficients are all connected if we write them out in their appropriate place on the Dynkin diagram. Roots with connected support are those which give rise to irreducible representations of the full algebra, for imagine a root of $E_{11}$ of the form $\alpha_1+\alpha_{11}+2\alpha_8+\alpha_7+\alpha_9$ the root $\alpha_1$ is disconnected from the rest of the root, so it forms an invariant subspace carrying a representation of $SU(2)$. We are interested in irreducible representation of $E_{11}$, and consequently the roots all have length squared less than or equal to two \cite{Kac}. This is a necessary condition for a root that comes from the Serre relations. Consider the simple case that $\beta\equiv \alpha_j+\alpha_k$ is a root, where $\{\alpha_i\}$ are positive simple roots. Then we note that,
\begin{equation}
[E_j,E_k]\equiv E_{j+k}\neq 0
\end{equation}
Where $E_i$ are the generators associated to the simple roots $\alpha_i$. The Serre relations tell us that $(1-A_{jk})>1\Rightarrow A_{ij}\leq -1$. This places restrictions on the length squared of $\beta$,
\begin{equation}
\beta^2=(\alpha_j+\alpha_k)^2=2+2(1+A_{jk})\leq 2
\end{equation}
The example for three roots better illustrates the more general case. Suppose $\gamma\equiv \alpha_i+\alpha_j+\alpha_k$ is also a root, then,
\begin{equation}
[E_i,[E_j,E_k]]=[[E_i,E_j],E_k]+[E_j,[E_i,E_k]]\equiv E_{i+j+k}\neq 0
\end{equation}
Where we have permuted the commutators using the Jacobi identity. The Serre relations tell us that $A_{ij}\leq -1$ or $A_{ik}\leq -1$ or even both, and also that $A_{jk}\leq -1$ since $\beta$ is also a root. Consequently, the root length squared of $\gamma$ satisfies,
\begin{equation}
\gamma^2=2(3+A_{jk}+A_{ik}+A_{ij})\leq 2
\end{equation}
Now let us generalise this observation to a more general example. Consider a string of roots containing $\beta$ and $\gamma=\beta+\alpha_i$. Let $\beta=\sum_i m_i \alpha_i$, having a generator formed of the appropriate multiple commutators of the simple roots, which we denote, $E_{\beta}=[E_{n_1},[E_{n_2},\ldots ,[E_{n_{j-1}},E_{n_j}]\ldots ]]$. Since $\gamma$ is a root then, 
\begin{equation}
[E_i,[E_{n_1},[E_{n_2},\ldots ,[E_{n_{j-1}},E_{n_j}]\ldots ]]]\neq 0
\end{equation}
For this general case we can play the same trick as with the three root case and using the Jacobi identity and the Serre relations find that at least one element of the Cartan matrix, $A_{in_k}$ is less than or equal to $-1$. Now,
\begin{equation}
\gamma^2=2+\beta^2+2\sum_j m_j A_{ij}\leq \beta^2
\end{equation}
Of course somewhere in our string of roots is a simple root, which is normalised to two, so by induction we find that for a general root $\gamma$ in our algebra the Serre relations imply that,
\begin{equation}
\gamma^2 \leq 2
\end{equation}
This condition on the root length is necessary but not sufficient to find roots of the algebra, but it is as much as we will use in this thesis. Explicit computation of the Serre relations, and knowledge of the higher level commutators is required to exactly determine the roots of an algebra. For a simply-laced algebra, one whose simple roots all have the same length, such as $E_{11}$ the length squared, $\beta^2$ is always even. This condition on the length of the roots appearing in our algebra is really the second criterion that we will use to generate the algebra, the first condition being that the Dynkin coefficients are non-negative integers. Let us square our root, 
\begin{align}
\nonumber \beta^2&=\sum_{n=1}^{10}(k-\sum_{n}^{10}p_i)^2+k^2-\frac{1}{9}(11k-A)^2\\
&=\sum_{n=1}^{10}(k-\sum_{n}^{10}p_i)^2+k^2-(m_{11})^2 \label{rootlengthsquared}
\end{align}
We see we are bound by our solution to restrict $-A+11k=3m_{11}$ to multiples of $3$ if $\beta^2$ is to be an integer. It is a little more involved to realise that our constraint of the root length squared being even leads to no further restrictions on our variables. In order to see that $\beta^2$ is even, we write,
\begin{align}
\nonumber \beta^2&=\sum_{n=1}^{10}(k^2-2k\sum_{n}^{10}p_i+(\sum_{n}^{10}p_i)^2)+k^2-m_{11}^2 \\
&=20kA+8m_{11}^2-110k^2+[\sum_{n=1}^{10}(\sum_{i=n}^{10}p_i)^2-A^2]
\end{align}
In the last line it is clear that all terms apart from those in the square brackets are even. Expanding, we note that, 
\begin{equation}
A^2=\sum_ii^2p_i^2+2\sum_{j>i}ijp_ip_j
\end{equation}
Similarly,
\begin{equation}
\sum_{n=1}^{10}(\sum_{i=n}^{10}p_i)^2=\sum_iip_i^2+2\sum_{j>i}jp_ip_j
\end{equation}
Therefore,
\begin{equation}
\beta^2=20kA+8m_{11}^2-110k^2+\sum_ii(i-1)p_i^2+2\sum_{j>i}j(i-1)p_ip_j
\end{equation}
So we see explicitly that each term in the expression is even. From requiring that $m_{11}\geq0$ we have the requirement that $A\leq11k$, and we noted earlier that $k\leq m_{11}$. We are now ready to construct the algebra of $E_{11}$. Really this is a computational problem, we have to find roots that satisfy the bounds on $k$ for a given $m_{11}$ such that $\beta^2\leq2$, but we shall go through the first few solutions by hand so as to familiarise ourselves with the algorithm before we use a computer. 

At level $m_{11}=0$ we find that $k=0$ and so we must solve the equation $A=0$ which we can only solve trivially such that all $p_i$ are zero, so we do not find a root. At level $m_{11}=1$, we consider $k=0,1$ in the first instance we find that $A=-3$ and so has no solution, and in the second case we find $A=8$. We have a useful restriction on the number of ways we may solve this equation coming from requiring non-negativity of the Dynkin coefficient $m_1\geq0$, which implies that,
\begin{equation}
\sum_{i=1}^{10}p_i\leq k \label{criterion}
\end{equation}
In this case, therefore, there is only one putative solution where $p_8=1$. We are not yet sure this is a root since we have yet to check that the root length squared is less than or equal to two. Making use of equation (\ref{rootlengthsquared}) we find that $\beta^2=2$ and it is indeed a root of $E_{11}$. Generating the rest of the roots is a tedious process best done by computer. The computation itself is carried out as described in this first example, and, with use of equation (\ref{criterion}), takes very little time to generate the lower level roots. We list the roots below level eight in table \ref{E11roots1} and the roots to level twelve in appendix \ref{roottables}. These low level roots were originally computed in \cite{FischbacherNicolai}. 

The field content associated to the algebra may be interpreted with respect to its spacetime decomposition from the $A_{10}$ weights. We refer the interested reader to \cite{Cahn}, and offer a shorthand rule for beginning to interpret the field content. The fundamental weights of the form $[0,\ldots ,0,1,0,\ldots 0]$, where there are $n-1$ zeroes on the left, are identical to $n$ copies of the vector representation $[1,0,\ldots ,0]$, antisymmetrised under the tensor product. Consequently, the example representation corresponds to a field with $n$ antisymmetrised indices. The same analysis can be carried out by counting from the right, for our example, there are $D-1$ components in our weight vector, where $D$ is the spacetime dimension. If we had counted from the right we would have found a representation with $D-n$ antisymmetric indices. We will use lowered indices on fields to indicate we are counting from the right, and raised indices to indicate a count from the left, and it is natural to associate these different fields using the totally-antisymmetric tensor with $D$ indices,

\begin{equation}
A_{a_1\ldots a_n}=\frac{1}{(D-n)!}\epsilon_{a_1\ldots a_n b_1\ldots b_{D-n}}A^{b_1\ldots b_{D-n}}
\end{equation}

Hence at low levels of $E_{11}$ we find a three-form field, $A_{a_1a_2a_3}$, a six-form field, $A_{a_1\ldots \_6}$, a nine-form field, $A_{a_1\ldots a_9}$ and a field with eight antisymmetrised indices and a vector index, $A_{a_1\ldots a_8,b}$. In our analysis we have not calculated the multiplicities of the root spaces, but this has been carried out in \cite{KleinschmidtSchnakenburgWest} and we see that of this low-level content only the nine-form has multiplicity zero. Consequently the remaining field content at low levels are the bosonic fields of nonlinear realisation of supergravity discussed in chapter three. The algebra has an infinite number of roots, and while the low-level roots do correspond to the fields of bosonic supergravity it is not clear how to interpret the higher level fields nor if they have any role to play in $M$-theory. We postpone the interpretation of the higher-level roots until the next chapter.
\begin{table}[htp]
	\centering
		\begin{tabular}{|c|c|c|c|c|c|}
		\hline
			$k$& $A$&$A_{10}$ Weights& $\beta$ ($\alpha_i$ basis)&$\beta$ ($e_i$ basis)&$\beta^2$ \\
\hline$1$&$8$&$[0,0,0,0,0,0,0,1,0,0]$& $(0,0,0,0,0,0,0,0,0,0,1)$&$(0,0,0,0,0,0,0,0,1,1,1)$&$2$ \\ 
\hline$1$&$5$&$[0,0,0,0,1,0,0,0,0,0]$& $(0,0,0,0,0,1,2,3,2,1,2)$&$(0,0,0,0,0,1,1,1,1,1,1)$&$2$ \\ 
\hline$1$&$2$&$[0,1,0,0,0,0,0,0,0,0]$& $(0,0,1,2,3,4,5,6,4,2,3)$&$(0,0,1,1,1,1,1,1,1,1,1)$&$0$ \\ 
$2$&$13$&$[0,0,1,0,0,0,0,0,0,1]$& $(0,0,0,1,2,3,4,5,3,1,3)$&$(0,0,0,1,1,1,1,1,1,1,2)$&$2$ \\ 
\hline$2$&$10$&$[0,0,0,0,0,0,0,0,0,1]$& $(1,2,3,4,5,6,7,8,5,2,4)$&$(1,1,1,1,1,1,1,1,1,1,2)$&$-2$ \\ 
$2$&$10$&$[0,1,0,0,0,0,0,1,0,0]$& $(0,0,1,2,3,4,5,6,4,2,4)$&$(0,0,1,1,1,1,1,1,2,2,2)$&$2$ \\ 
$2$&$10$&$[1,0,0,0,0,0,0,0,1,0]$& $(0,1,2,3,4,5,6,7,4,2,4)$&$(0,1,1,1,1,1,1,1,1,2,2)$&$0$ \\ 
$3$&$21$&$[1,0,0,0,0,0,0,0,0,2]$& $(0,1,2,3,4,5,6,7,4,1,4)$&$(0,1,1,1,1,1,1,1,1,1,3)$&$2$ \\ 
\hline$2$&$7$&$[0,0,0,0,0,0,1,0,0,0]$& $(1,2,3,4,5,6,7,9,6,3,5)$&$(1,1,1,1,1,1,1,2,2,2,2)$&$-2$ \\ 
$2$&$7$&$[0,1,0,0,1,0,0,0,0,0]$& $(0,0,1,2,3,5,7,9,6,3,5)$&$(0,0,1,1,1,2,2,2,2,2,2)$&$2$ \\ 
$2$&$7$&$[1,0,0,0,0,1,0,0,0,0]$& $(0,1,2,3,4,5,7,9,6,3,5)$&$(0,1,1,1,1,1,2,2,2,2,2)$&$0$ \\ 
$3$&$18$&$[0,0,0,0,0,0,0,0,2,0]$& $(1,2,3,4,5,6,7,8,4,2,5)$&$(1,1,1,1,1,1,1,1,1,3,3)$&$2$ \\ 
$3$&$18$&$[0,0,0,0,0,0,0,1,0,1]$& $(1,2,3,4,5,6,7,8,5,2,5)$&$(1,1,1,1,1,1,1,1,2,2,3)$&$0$ \\ 
$3$&$18$&$[1,0,0,0,0,0,1,0,0,1]$& $(0,1,2,3,4,5,6,8,5,2,5)$&$(0,1,1,1,1,1,1,2,2,2,3)$&$2$ \\ 
\hline$2$&$4$&$[0,0,0,1,0,0,0,0,0,0]$& $(1,2,3,4,6,8,10,12,8,4,6)$&$(1,1,1,1,2,2,2,2,2,2,2)$&$-4$ \\ 
$2$&$4$&$[0,2,0,0,0,0,0,0,0,0]$& $(0,0,2,4,6,8,10,12,8,4,6)$&$(0,0,2,2,2,2,2,2,2,2,2)$&$0$ \\ 
$2$&$4$&$[1,0,1,0,0,0,0,0,0,0]$& $(0,1,2,4,6,8,10,12,8,4,6)$&$(0,1,1,2,2,2,2,2,2,2,2)$&$-2$ \\ 
$3$&$15$&$[0,0,0,0,0,0,1,1,0,0]$& $(1,2,3,4,5,6,7,9,6,3,6)$&$(1,1,1,1,1,1,1,2,3,3,3)$&$2$ \\ 
$3$&$15$&$[0,0,0,0,0,1,0,0,1,0]$& $(1,2,3,4,5,6,8,10,6,3,6)$&$(1,1,1,1,1,1,2,2,2,3,3)$&$0$ \\ 
$3$&$15$&$[0,0,0,0,1,0,0,0,0,1]$& $(1,2,3,4,5,7,9,11,7,3,6)$&$(1,1,1,1,1,2,2,2,2,2,3)$&$-2$ \\ 
$3$&$15$&$[0,1,1,0,0,0,0,0,0,1]$& $(0,0,1,3,5,7,9,11,7,3,6)$&$(0,0,1,2,2,2,2,2,2,2,3)$&$2$ \\ 
$3$&$15$&$[1,0,0,0,1,0,0,0,1,0]$& $(0,1,2,3,4,6,8,10,6,3,6)$&$(0,1,1,1,1,2,2,2,2,3,3)$&$2$ \\ 
$3$&$15$&$[1,0,0,1,0,0,0,0,0,1]$& $(0,1,2,3,5,7,9,11,7,3,6)$&$(0,1,1,1,2,2,2,2,2,2,3)$&$0$ \\ 
$4$&$26$&$[0,0,0,0,0,1,0,0,0,2]$& $(1,2,3,4,5,6,8,10,6,2,6)$&$(1,1,1,1,1,1,2,2,2,2,4)$&$2$ \\ 
\hline$2$&$1$&$[1,0,0,0,0,0,0,0,0,0]$& $(1,3,5,7,9,11,13,15,10,5,7)$&$(1,2,2,2,2,2,2,2,2,2,2)$&$-8$ \\ 
$3$&$12$&$[0,0,0,0,0,2,0,0,0,0]$& $(1,2,3,4,5,6,9,12,8,4,7)$&$(1,1,1,1,1,1,3,3,3,3,3)$&$2$ \\ 
$3$&$12$&$[0,0,0,0,1,0,1,0,0,0]$& $(1,2,3,4,5,7,9,12,8,4,7)$&$(1,1,1,1,1,2,2,3,3,3,3)$&$0$ \\ 
$3$&$12$&$[0,0,0,1,0,0,0,1,0,0]$& $(1,2,3,4,6,8,10,12,8,4,7)$&$(1,1,1,1,2,2,2,2,3,3,3)$&$-2$ \\ 
$3$&$12$&$[0,0,1,0,0,0,0,0,1,0]$& $(1,2,3,5,7,9,11,13,8,4,7)$&$(1,1,1,2,2,2,2,2,2,3,3)$&$-4$ \\ 
$3$&$12$&$[0,1,0,0,0,0,0,0,0,1]$& $(1,2,4,6,8,10,12,14,9,4,7)$&$(1,1,2,2,2,2,2,2,2,2,3)$&$-6$ \\ 
$3$&$12$&$[0,2,0,0,0,0,0,1,0,0]$& $(0,0,2,4,6,8,10,12,8,4,7)$&$(0,0,2,2,2,2,2,2,3,3,3)$&$2$ \\ 
$3$&$12$&$[1,0,0,1,0,0,1,0,0,0]$& $(0,1,2,3,5,7,9,12,8,4,7)$&$(0,1,1,1,2,2,2,3,3,3,3)$&$2$ \\ 
$3$&$12$&$[1,0,1,0,0,0,0,1,0,0]$& $(0,1,2,4,6,8,10,12,8,4,7)$&$(0,1,1,2,2,2,2,2,3,3,3)$&$0$ \\ 
$3$&$12$&$[1,1,0,0,0,0,0,0,1,0]$& $(0,1,3,5,7,9,11,13,8,4,7)$&$(0,1,2,2,2,2,2,2,2,3,3)$&$-2$ \\ 
$3$&$12$&$[2,0,0,0,0,0,0,0,0,1]$& $(0,2,4,6,8,10,12,14,9,4,7)$&$(0,2,2,2,2,2,2,2,2,2,3)$&$-4$ \\ 
$4$&$23$&$[0,0,0,0,1,0,0,1,0,1]$& $(1,2,3,4,5,7,9,11,7,3,7)$&$(1,1,1,1,1,2,2,2,3,3,4)$&$2$ \\ 
$4$&$23$&$[0,0,0,1,0,0,0,0,1,1]$& $(1,2,3,4,6,8,10,12,7,3,7)$&$(1,1,1,1,2,2,2,2,2,3,4)$&$0$ \\ 
$4$&$23$&$[0,0,1,0,0,0,0,0,0,2]$& $(1,2,3,5,7,9,11,13,8,3,7)$&$(1,1,1,2,2,2,2,2,2,2,4)$&$-2$ \\ 
$4$&$23$&$[1,0,1,0,0,0,0,0,1,1]$& $(0,1,2,4,6,8,10,12,7,3,7)$&$(0,1,1,2,2,2,2,2,2,3,4)$&$2$ \\ 
$4$&$23$&$[1,1,0,0,0,0,0,0,0,2]$& $(0,1,3,5,7,9,11,13,8,3,7)$&$(0,1,2,2,2,2,2,2,2,2,4)$&$0$ \\\hline
		\end{tabular}
	\caption{The low level roots of $E_{11}$}
	\label{E11roots1}
\end{table}

\subsection{The Bosonic String Theory Algebra}
In this section we repeat the hands-on approach to constructing the algebra that we applied to $E_{11}$ in the previous section. The Kac-Moody algebra that is conjectured to underpin the theory of the twenty-six dimensional bosonic string is $D_{24}^{+++}\equiv K_{27}$, its Dynkin diagram is shown in figure \ref{K27andpuregravity}. Due to the structure of the Dynkin diagram we will not be able to exactly replicate the process of the previous section. We work in a twenty-seven dimensional vector space with basis elements $e_i$ as before, except we will make use of the inner product,
\begin{equation}
<a,b>\equiv\sum_{i=1}^{26}a_ib_i-a_{27}b_{27} \label{innerproduct2}
\end{equation}
In this basis the simple roots of $K_{27}$ are,
\begin{align}
\nonumber \alpha_i&=e_i-e_{i+1}, \qquad 1\leq i \leq 25\\
\nonumber \alpha_{26}&=e_{25}+e_{26} \\
\alpha_{27}&=-(e_1+\ldots+e_4)+\sqrt{2}e_{27} \label{K27simpleroots}
\end{align}
Our aim is to decompose the algebra into representations of an $A_{25}$ algebra, which will correspond to the twenty-six dimensional spacetime.\footnote{In the literature the singled out $A_n$ sub-algebra into which the Kac-Moody algebra is decomposed is often called the "gravity line" for obvious reasons.} The $A_{25}$ algebra may be chosen in two symmetric ways, we choose to delete nodes $26$ and $27$ as indicated on the diagram of figure \ref{K27andpuregravity}. We delete first node $27$, which leaves the classical algebra $D_{26}$, then we delete node $26$, leaving the $A_{25}$ algebra. Consequently,
\begin{equation}
\alpha_{27}=z-\mu_4=z-2y-\nu_4, \qquad \alpha_{26}=y-\nu_{24}
\end{equation}
Where we use $\mu_i$ to denote the fundamental weights of $D_{26}$ and $\nu_i$ those of $A_{25}$. The vector $z$ is orthogonal to the root lattices of $D_{26}$ and $A_{25}$, while $y$ lies in the root lattice of $D_{26}$ and is orthogonal to the roots of $A_{25}$. In the first expression we have used $\mu_4=\nu_4+2y$. The explicit expression for $z$ and $y$ are,
\begin{align}
\nonumber z&=\sqrt{2}e_{27}\\
y&=\frac{1}{13}(e_1+\ldots+e_{26})
\end{align}
Consequently a general root of $K_{27}$ is given by
\begin{equation}
\beta=m_{26}y+m_{27}(z-2y)-\Lambda
\end{equation}
Where $\Lambda$ lies in the weight lattice of $A_{26}$,
\begin{equation}
\Lambda=-\sum_{i=1}^{25}m_i\alpha_i+m_{26}\nu_{24}+m_{27}\nu_4\equiv \sum_{i=1}^{25}p_i\nu_i
\end{equation}
Therefore we find the following equations which must be integers,
\begin{align}
m_j=\begin{cases}\frac{j}{13}(m_{26}+11m_{27}+\frac{A}{2})-B_j \qquad &j\leq4\\
\frac{j}{13}(m_{26}-2m_{27}+\frac{A}{2})+4m_{27}-B_j \qquad &5\leq j\leq24\\
\frac{1}{13}(12m_{26}+2m_{27}-\frac{A}{2}) &j=25\end{cases}
\end{align}
Where we have used the equivalent expressions of those given in equation (\ref{E11parameters}) for $A$ and $B_j$, namely,
\begin{align}
\nonumber A&\equiv\sum_{i=1}^{25}ip_i \\
B_j&\equiv\sum_{i=1}^{j}ip_i+j\sum_{i=j+1}^{25}p_i \label{K27parameters}
\end{align}
A fine grained set of solutions is given by,
\begin{equation}
m_{26}=-11m_{27}-\frac{A}{2}+13k
\end{equation}
It is clear that we will only find representations such that $A$ is even-valued. This solution corresponds to Dynkin coefficients,
\begin{align}
m_j=\begin{cases}kj-B_j \qquad &j\leq4\\
j(k-m_{27})+4m_{27}-B_j \qquad &5\leq j\leq24\\
-10m_{27}-\frac{A}{2}+12k &j=25\end{cases}
\end{align}
From the condition $m_1\geq 0$ we see that $\sum_{i=1}^{25}p_i\leq k$ and from observing that $m_{25}=m_{26}+m_{27}-k\geq0$ we find an upper bound for our parameter is $k\leq m_{26}+m_{27}$. The prototype root is,
\begin{align}
\nonumber\beta=&\sum_{i=1}^{4}(k-B_i)\alpha_i+\sum_{i=5}^{24}(i(k-m_{27})+4m_{27}-B_i)\alpha_i+(-10m_{27}-\frac{A}{2}+12k)\alpha_{25}\\
\nonumber &+(-11m_{27}-\frac{A}{2}+13k)\alpha_{26}+m_{27}\alpha_{27}\\
=&\sum_{n=1}^{25}(k-m_{27}-\sum_{i=n}^{25}p_i)e_n+(k-m_{27})e_{26}+\sqrt{2}m_{27}e_{27}
\end{align}
We note the root has a relatively simple form in our basis, and has essentially the same structure of the generic $E_{11}$ root. In contrast to the $E_{11}$ case our root is parameterised by two variables, $k$ and $m_{27}$ (corresponding to the two deleted roots) as well as the twenty-five coefficients of the $A_{25}$ fundamental weights. Finally, we find an expression for the square of the root,
\begin{equation}
\beta^2=\sum_{n=1}^{25}(k-m_{27}-\sum_{i=n}^{25}p_i)^2+(k-m_{27})^2-2m_{27}^2
\end{equation}
At this stage we turn to our computational algorithm which simply searches for weights of $A_{25}$ satisfying $\beta^2\leq2$ for specified values of $m_{27}$ and $m_{26}$, or equivalently $k$. Upto level $(m_{26},m_{27})=(1,1)$ we find the following roots,

\begin{table}[htp]
	\centering
		\begin{tabular}{|c|c|c|c|c|}
		\hline
		$(m_{26},m_{27})$& $A_{25}$ Weights& $\beta$ ($\alpha_i$ basis)&$\beta$ ($e_i$ basis)&$\beta^2$ \\
		\hline
		$(1,0)$&$[0,\ldots ,0,1,0]$& $(0,\ldots ,0,1,0)$&$(0,\ldots ,0,1,1,0)$&$2$ \\ 
		$(0,1)$&$[0,0,0,1,0,\ldots ,0]$& $(0,\ldots ,0,1)$&$(-1,-1,-1,-1,0,\ldots ,0,\sqrt{2})$&$2$ \\ 
		$(1,1)$&$[0,1,0,\ldots ,0]$& $(0,0,1,2,\ldots ,2,1,1,1)$&$(-1,-1,0,\ldots ,0,1\sqrt{2})$&$0$ \\ 
		$(1,1)$&$[0,0,1,0,\ldots ,0,1]$& $(0,0,0,1,\ldots ,1,0,1,1)$&$(-1,-1,-1,0,\ldots ,0,1,1\sqrt{2})$&$2$\\
		\hline
		\end{tabular}
	\caption{The first few roots of $K_{27}$}
	\label{K27roots1}
\end{table}
These low level roots and others were found in \cite{KleinschmidtSchnakenburgWest} for a generic $D_{n-3}^{+++}$ algebra. In the appendix, in table \ref{K27roots} the roots of $K_{27}$ up to level $(3,3)$, which to the best of my belief are not given elsewhere in the literature. The gauge field in the space-time decomposition associated to these roots may be deduced from the weights of $A_{25}$. Specifically, we find at low levels a two-form $A_{a_1a_2}$, a four-form $A^{a_1\ldots a_4}$, a second two-form $\hat{A}^{a_1a_2}$, and a field with three antisymmetrised indices and one symmetric index, $A^{a_1a_2a_3,b}$ \cite{KleinschmidtSchnakenburgWest}.
\subsection{The Translation Generator}
Turning our attention back to $E_{11}$ and the motivations given in chapter three for considering it as a symmetry of M-theory, we notice that a crucial part of our nonlinear construction has not appeared in the algebra of $E_{11}$ namely the translation generator $P_a$ is absent. Addressing this point is the main aim of this section, but first we should comment on some other generators that have also not appeared so far in this chapter.

The keen reader will have noted that the ${K^a}_b$ generators occurring in the nonlinear construction have not appeared explicitly and neither has the dilaton field, $\phi$, of string theory. Their absence in the constructions of this chapter has a straightforward explanation, namely, that we have only looked for positive roots associated to multiples of the deleted roots which were used to go between our Kac-Moody algebra and our representations of $A_{10}$ or $A_{25}$ - we have not sought to generate the well-understood finite algebra of these gravity lines themselves. The ${K^a}_b$ are the generators of these algebras and have roots in the $e_i$ basis, $e_a-e_b=(0,\ldots , 0, 1, 0,\ldots ,0, -1,0,\ldots,0)$ where the $+1$ entry occurs as the $a$th component and the $-1$ entry as the $b$th component. Now let us account for the dilaton of string theory. So far, we have been concerned with finding the positive roots of the algebra and their generators, but as we mentioned earlier the full algebra may be decomposed into three parts, as a triangular decomposition, ${\cal G}\equiv\Pi^+\oplus\Pi^0\oplus\Pi^-$. As discussed once we have identified the positive root and their generators $E_a$, the negative roots follow straightforwardly. The trivial missing generators are all contained in the Cartan sub-algebra, $\Pi^0$, which we may determine from the Dynkin diagram. For the case of $E_{11}$ we find the elements of the Cartan sub-algebra are,
\begin{align}
\nonumber H_a&={K^a}_a-{K^{a+1}}_{a+1} \qquad a=1\ldots 10 \\
H_{11}&=-\frac{1}{3}({K^1}_1+\ldots +{K^8}_8)+\frac{2}{3}({K^9}_9+{K^{10}}_{10}+{K^{11}}_{11})
\end{align}
The complexification\footnote{This means simply allowing the coefficients of the generators to take complex values instead of just real values. In our example in chapter two we already took this as the standard when discussing the representations of $su(2)$.} of the algebra of $su(D)$ is isomorphic to $sl(D)$, for a review see \cite{Hall}. The additional element of the Cartan sub-algebra in the case of $E_{11}$ enlarges $sl(11)$ to $gl(11)$ - it adds an additional scalar to the algebra \cite{EnglertHouartTaorminaWest}. In the case of $K_{27}$ we find the Cartan sub-algebra,
\begin{align}
\nonumber H_a&={K^a}_a-{K^{a+1}}_{a+1} \qquad a=1\ldots 25 \\
\nonumber H_{26}&=-\frac{1}{12}({K^1}_1+\ldots +{K^{24}}_{24})+\frac{11}{12}({K^{25}}_{25}+{K^{26}}_{26})+\frac{1}{\sqrt{3}}R\\
H_{27}&=-\frac{11}{12}({K^1}_1+\ldots +{K^4}_4)+\frac{1}{12}({K^{5}}_{5}+\ldots+{K^{26}}_{26})-\frac{1}{\sqrt{3}}R
\end{align}
For $K_{27}$ we deleted two nodes to arrive at the preferred $A_{25}$ sub-algebra, one of the extra elements of the Cartan sub-algebra naturally enlarges the preferred sub-algebra to $gl(26)$, the extra scalar further enlarges the preferred sub-algebra to $gl(26)+R$ where $R$ is an additional scalar field, associated to the dilaton, $\phi$, of string theory.

Having made these comments let us turn our attention to the more serious "missing" generator, that is the translation generator, $P_a$. To address this issue we will concentrate on the $E_{11}$ algebra. Our first reaction ought to be one of surprise since the translation generator, which has $A_{10}$ weights $[1,0,\ldots ,0]$, does actually appear in the algebra of $E_{11}$. Specifically, using our variables from section \ref{Mtheory}, we must solve $A=1$, which has an apparent string of solutions parameterised by,
\begin{equation}
m_{11}=\frac{1}{3}(-1+11k)=7+11m
\end{equation}
To ensure this is a positive integer we have set $k=2+3m$, where $m\in{\mathbb Z}^+$. Applying the root length condition, we have,
\begin{equation}
\beta^2=-2[(3m+2)^2+2m(m+2)]\leq -8
\end{equation}
Subject to the roots having a non-zero multiplicity, which is reported to be the case \cite{FischbacherNicolai}, it appears we have very many translation generators. Incidentally, this kind of repetition of generators at higher levels happens for all representations that we find in the algebra. The content of the algebra may be thought of as being defined by the set of generators which appear at the lowest levels of such a string of generators; these generators are akin to the prime numbers in that they are the basic building blocks of the algebra, and one might refer to such generators as prime generators. The problem with the translation generator is that its prime generator occurs at a level in the algebra which is above the dual gravity field. Since we are dealing with a physical theory the content that we are concerned with might be reasonably expected to occur before the dual gravity field, which would be expected to have the largest index structure that we may be able to interpret as a form in a $D$ dimensional spacetime theory. Indeed if we truncate the theory to the generators occurring at levels below that of the dual gravity field we find all the fields of bosonic supergravity that occurred in the nonlinear realisation of chapter three. It is natural, in terms of the index structure of the generator, to expect the translation generator to occur at low levels. Nevertheless it has been suggested that the level 7 translation generator may still be used to generate spacetime translations \cite{EnglertHouart}. Later in this thesis we will argue for the relevancy of some of the higher level fields of $E_{11}$ and consequently it is difficult here to dismiss the level 7 translation generator. 

\begin{figure}[cth]
\hspace{180pt} \includegraphics[viewport=0 150 80 200,angle=90]{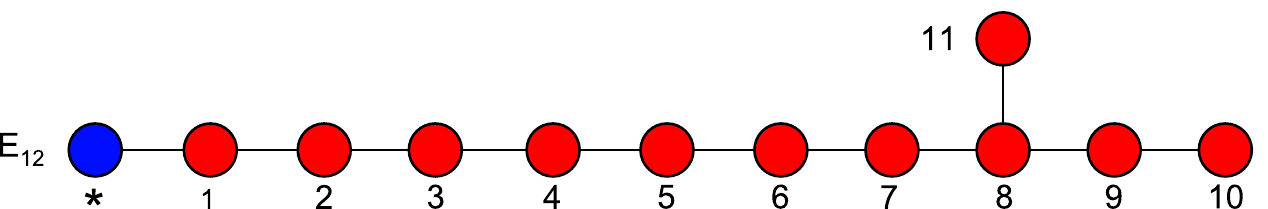}
\caption{The Dynkin Diagram of $E_{12}$} \label{E12}
\end{figure}

However, we will adopt the proposal of \cite{West3}, which is to enlarge the algebra to include its $l_1$ representation as well as adjoint representation of $E_{11}$ which we constructed above. The associations between the $l_1$ representation and the adjoint representation of $E_{11}$ as well as other very-extended algebras has been studied in detail in \cite{KleinschmidtWest}. The $l_1$ representation of $E_{11}$ has highest weight $l_1$, which is our notation for the first fundamental weight of $E_{11}$ and hence when decomposed to representations of $A_{10}$ corresponds to the translation generator. It is found by attaching another node, which we denote with a $*$, to node $1$ of the Dynkin diagram of $E_{11}$ by a single line, to give $E_{12}$, see figure (\ref{E12}), and then restricting the full $E_{12}$ to just those roots with the Dynkin coefficient of $\alpha_*$ set to one. 
This is clearly a representation of $E_{11}$ and if we were to label the fundamental weights of $E_{11}$ as $l_i$ where $1\leq i \leq 11$, then we would find $\alpha_*=x-l_1$, so that a generic weight in the $l_1$ representation therefore takes the form, 
\begin{equation}
l_1-\sum_{i=1}^{11}m_i\alpha_i
\end{equation}
Hence we see the origin of the name "$l_1$ representation". If we had set $m_*=0$ we would have found the adjoint representation of $E_{11}$ that was discussed earlier. The advantage of this representation is that we find the translation generator for free at level zero. But that is not all, it turns out \cite{West3,KleinschmidtWest} that the content of the $l_1$ representation may be interpreted as the central charges derived from the gauge fields of $E_{11}$.
\subsection{The Central Charges of M-theory}
Let us now explicitly construct the low level generators of the $l_1$ representation of $E_{11}$. We extend the notation of equation (\ref{E11simpleroots}) so that the $i$ index runs over $\{*,1,\ldots 10\}$ and we use a similar extension of the components in the inner product of equation (\ref{innerproduct1}). The deleted roots are,
\begin{align}
\nonumber \alpha_*&=x+\frac{3}{2}z-\nu_1 \\
\alpha_{11}&=z-\nu_8
\end{align}
Where $\nu_i$ are fundamental weights of $A_{10}$ given in our vector space basis by equation (\ref{Anweights}), and the orthogonal vectors are,
\begin{align}
\nonumber x&=e_*-\frac{1}{2}(e_1+\ldots +e_{11}), \qquad x^2=\frac{3}{2}\\
z&=\frac{3}{11}(e_1+\ldots +e_{11}), \qquad z^2=-\frac{2}{11}
\end{align}
A generic root in the $l_1$ representation, having $m_*=1$, takes the form,
\begin{equation}
\beta=x+\frac{3}{2}z+m_{11}z-\Lambda
\end{equation}
Where $\Lambda$ is,
\begin{equation}
\Lambda=\nu_1+m_{11}\nu_8-\sum_{i=1}^{10}m_i\alpha_i\equiv\sum_{i=1}^{10}p_i\nu_i
\end{equation}
We find as Dynkin coefficients,
\begin{equation}
m_j=
\begin{cases}
\frac{j}{11}(3m_{11}+A-1)-B_j+1,  \qquad j\leq 8 \\
\frac{j}{11}(-8m_{11}+A-1)-B_j+8m_{11}+1,\qquad j>8
\end{cases} \label{l1mj}
\end{equation}
We find a set of solutions given by 
\begin{equation}
m_{11}=\frac{1}{3}(-A+1+11k)\label{m11l1}
\end{equation}
Such that,
\begin{equation}
m_j=
\begin{cases}
jk-B_j+1,  \qquad j\leq 8 \\
\frac{j}{3}(A-1-8k)-B_j+1+\frac{8}{3}(-A+1+11k),\qquad j>8
\end{cases} 
\end{equation}
We find that $A\rightarrow A+1$, $B_j\rightarrow B_j+1$ maps our $l_1$ representations into those of the adjoint found earlier. This is to be expected since we have simply shifted our representations by the $A_{10}$ weight $[1,0,\ldots 0]$. The question of when representations of the $A_n$ sub-algebra occur in both the adjoint and $l_1$ representations of ${\cal G}^{+++}$ with this one-to-one relationship is addressed in \cite{KleinschmidtWest}. The generic root of the $l_1$ representation is,
\begin{equation}
\beta=e_*+\sum_{n=1}^{10}(k-\sum_{i=n}^{10}p_i)e_n+ke_{11}
\end{equation}
Its length squared is,
\begin{equation}
\beta^2=1+\sum_{n=1}^{10}(k-\sum_{n}^{10}p_i)^2+k^2-(m_{11})^2 
\end{equation}
The roots of the $l_1$ representation of $E_{11}$ are given up to level 12 in table \ref{l1E11roots} in appendix A, where we have dropped the $\alpha_*$ or the $e_*$ in each basis. 

Let us now offer some interpretation of the content of this representation. By comparing the tables of roots of the $l_1$ representation and the adjoint representation one may observe that for every $A_{10}$ weight occurring in the $l_1$ representation one can find a weight in the adjoint representation whose $A_{10}$ weight, $(p_1,p_2,\ldots ,p_{10})$, is simply shifted one place to the left in our notation to give $(p_2,\ldots ,p_{10},0)$. Let us denote the variables referring to the $l_1$ representation with a hat. Specifically, if we denote the weights appearing in the adjoint by $p_i$, one is led to conjecture that if $\hat{p}_i$ appears in the $l_1$ representation then $p_i=\hat{p}_{i+1},p_{10}=0$ occurs in the adjoint. Let us check. 

First we look to see if $m_{11}$ is an integer,
\begin{align}
\nonumber m_{11}&=\frac{1}{3}(-A+11k)\\
\nonumber &=\frac{1}{3}(-\sum_{i=1}^{10}(i-1)\hat{p}_i+11k)\\
&=\frac{1}{3}(-\hat{A}+\sum_{i=1}^{10}\hat{p}_i+11k)
\end{align}
Comparing this expression with equation (\ref{m11l1}), we conclude that $m_{11}$ will be an integer if it is possible to express $\sum_{i=1}^{10}\hat{p}_i$ in the following form,
\begin{equation}
\sum_{i=1}^{10}\hat{p}_i=1+3n+11m
\end{equation}
Where $n,m\in\mathbb{Z}$. This is true since the $\hat{p}_i$ could sum to any integer and so we must be able to find a unit grading on the right hand side, and we can see that by taking $n=4r$ and $m=-r$, where $r\in\mathbb{Z}$, this is satisfied, as,
\begin{equation}
\sum_{i=1}^{10}\hat{p}_i=1+r
\end{equation}
If we now substitute this back into our expression for $m_{11}$ we find,
\begin{align}
\nonumber m_{11}&=\frac{1}{3}(-\hat{A}+1+11(k-r))+4r\\
&=\hat{m}_{11}+4r
\end{align}
In the last line we have shifted our variable such that $k-r=\hat{k}$, and we see that we find a putative root in the adjoint representation corresponding to every root of the $l_1$ representation, but occurring at a shifted, higher level. One can work through the tables of the $l_1$ representation and check this relation explicitly. We note in particular that if $r=0$, that is if we are considering a fundamental weight of $A_{10}$ then the corresponding roots occur at the same level. We have not checked the root length condition here, but such a check can be found in \cite{KleinschmidtWest}, where one can also  finds results of this nature, although presented somewhat differently than here, applied to very-extended Kac-Moody algebras other than $E_{11}$. 

The consequence of this injection, relating $l_1$ roots to adjoint roots, by shifting the $A_{10}$ weights one place lower in the weight vector allows to say that for every field arising in the $l_1$ there is a related field in the adjoint whose index structure is enlarged by one index for each representation of $A_{10}$ that it carries. For the simple cases we find that $[0,\ldots ,0,1,0]$, corresponding to a two-form field $Z^{a_1a_2}$, in the $l_1$ representation is mapped to $[0,\ldots 0,1,0,0]$ in the adjoint, corresponding to the three-form gauge field of the M2-brane, $A^{a_1\ldots a_3}$. The natural interpretation is that $Z^{a_1a_2}$ is the charge associated to the M2-brane. The same is true for $[0,\ldots, 0,1,0,0,0,0]$, $Z^{a_1\ldots a_5}$ which we interpret as the central charge of the M5-brane. For these simple cases it is as if the $l_1$ representation simply adds an index to the gauge fields of the adjoint representation. The change in the index structure is more complex for the non-fundamental representations. Nevertheless, we are able to interpret the fields introduced to our algebra from the $l_1$ representation as being associated to the central charges of M-theory \cite{KleinschmidtWest}, should the $E_{11}$ conjecture be established beyond doubt. It is quite a surprising consequence of simply introducing the translation generator to the fields of the theory that one finds a full set of central charges, and one must take it to imply that the coordinates of spacetime are on an equal footing with all the fields associated to the central charges. In fact due to the level shift in relating roots of the $l_1$ representation and the adjoint, one can hope to learn about the higher level fields of $E_{11}$ from studying the central charges at low levels. The full algebra constructed from both the adjoint representation of $E_{11}$ and its $l_1$ representation is the semidirect product, $l_1\ltimes E_{11}$.

The reader encouraged by aesthetic considerations might take the construction employed for the $l_1$ representation as hinting at an underlying $E_{12}$ symmetry, simply because its construction would be much more elegant than the semidirect product of $l_1\ltimes E_{11}$, and one would find the desired representations occurring at levels $m_*=0,1$. If one were to specify the extra coordinate to be timelike one would also be able to recognise the Lorentz group $SO(2,10)$, which is isomorphic to the conformal group in a Minkowski spacetime of signature $(1,9)$, immediately. 
\subsection{The $l_1$ Representation of the Bosonic String}
We finish this chapter by constructing the $l_1$ algebra of $K_{27}$, for which explicit tables of roots have not been provided in the literature.
\begin{figure}[hct]
\hspace{180pt} \includegraphics[viewport=0 150 80 200,angle=90]{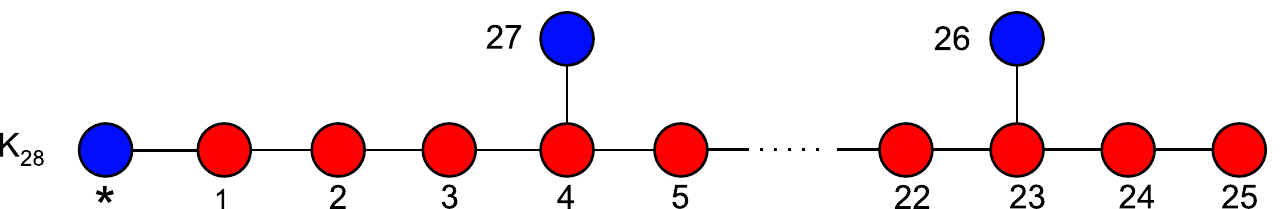}
\caption{The Dynkin Diagram of $K_{28}$} \label{K28}
\end{figure}
We use the extension of the inner product of equation (\ref{innerproduct2}), so that the simple roots are represented by the following vectors,
\begin{align}
\nonumber \alpha_*&=e_*-e_1 \\
\nonumber \alpha_{26}&=e_{25}-e_{26} \\
\alpha_{27}&=-(e_*+e_1+\ldots +e_4)+\sqrt{3}e_{27}
\end{align}
The algebra is decomposed into representations of $A_{25}$ by deleting the nodes labelled $*$, $26$ and $27$ of the Dynkin diagram in figure \ref{K28}. The deleted roots may be expressed as,
\begin{align}
\nonumber \alpha_{*}&=x-\nu_1 \\
\nonumber \alpha_{26}&=y-\frac{2}{27}x-\nu_{24}\\
\alpha_{27}&=z-\frac{5}{27}y-\frac{22}{27}x-\nu_{4}
\end{align}
Where $\nu_i$ are the fundamental weights of $A_{25}$ and $x$, $y$ and $z$ are vectors orthogonal to the root spaces of $A_{25}$, $A_{26}$ and $D_{27}$ encountered in the decomposition. They are explicitly,
\begin{align}
\nonumber x&=e_*-\frac{1}{26}(e_1+\ldots +e_{26}), \qquad x^2=\frac{27}{26}\\
\nonumber y&=\frac{2}{27}(e_*+e_1+\ldots +e_{26}), \qquad y^2=\frac{4}{27}\\
z&=\sqrt{3}e_{27}, \qquad z^2=-3
\end{align}
A general root of the $l_1$ representation, having $m_*=1$ is therefore,
\begin{equation}
\beta=x+m_{26}(y-\frac{2}{27}x)+m_{27}(z-\frac{5}{27}y-\frac{22}{27}x)-\Lambda
\end{equation}
Where,
\begin{equation}
\Lambda=\nu_1+m_{26}\nu_{24}+m_{27}\nu_{4}-\sum_{i=1}^{25}m_i\alpha_i\equiv \sum_{i=1}^{25}p_i\nu_i
\end{equation}
As usual, we look for solutions so that the following are non-negative integers,
\begin{align}
m_j=\begin{cases}
\frac{j}{26}(-1+2m_{26}+22m_{27}+A)-B_j+1 \qquad &j\leq4\\
\frac{j}{26}(-1+2m_{26}-4m_{27}+A)+4m_{27}-B_j+1 \qquad &5\leq j\leq24\\
\frac{1}{26}(1+24m_{26}+4m_{27}-A) &j=25\end{cases}
\end{align}
A general set of solutions is given by
\begin{equation}
m_{26}=-11m_{27}+\frac{1}{2}(1-A)+13k
\end{equation}
In the case of the adjoint representation of $K_{27}$ we were restricted to representations of $A_{25}$ such that $A$ was even, here, we find the complementary representations since $A$ must be an odd number. The other Dynkin coefficients become,
\begin{align}
m_j=\begin{cases}
jk-B_j+1 \qquad &j\leq4\\
j(k-m_{27})+4m_{27}-B_j+1 \qquad &5\leq j\leq24\\
\frac{1}{2}(1-A)-10m_{27}+12k &j=25\end{cases}
\end{align}
These are integers with the proviso that $A$ is odd, and give the general root,
\begin{equation}
\beta=e_*+\sum_{n=1}^{25}(k-m_{27}-\sum_{i=n}^{25}p_i)e_n+(k-m_{27})e_{26}+\sqrt{2}m_{27}e_{27}
\end{equation}
Which has length squared,
\begin{equation}
\beta^2=1+\sum_{n=1}^{25}(k-m_{27}-\sum_{i=n}^{25}p_i)^2+(k-m_{27})^2-2m_{27}^2
\end{equation}
Using these expressions we may generate the root system with the help of a computer program, and we list the low level content of the $l_1$ representation of $K_{27}$ in table \ref{l1K27roots} of appendix A.

As for the $E_{11}$ algebra we can check to see if we can interpret the fields of the $l_1$ representation as central charges related to the fields of the $K_{27}$ theory. We switch to using hatted indices to indicate variables associated to the $l_1$ representation, and repeat the analysis of the previous section. We are curious to find if the following $m_{26}$ is integer valued, for some root of the $l_1$ representation, encoded in $\hat{A}$ and $\hat{p}_i$,
\begin{equation}
m_{26}=-11m_{27}-\frac{1}{2}(\sum_{i=1}^{25}\hat{p}_i-\hat{A})+13k
\end{equation}
Allowing for shifts in the parameters we may require,
\begin{equation}
\sum_{i=1}^{25}\hat{p}_i=1+22n+26m+2q=1+2r
\end{equation}
We see that we can only find associated representations in the adjoint if $\sum_{i=1}^{25}p_i$ is odd. This is to be expected due to the complementary even-odd parity of the sum, $A$, that was used to construct the two algebras.
The associated roots in the adjoint occur at level,
\begin{align}
\nonumber m_{27}&=\hat{m}_{27} \\
m_{26}&=\hat{m}_{26}+r
\end{align}
For example in the tables of appendix A, we may identify a number of such associations which occur at the same level. There are some examples where a shift in the level $m_{26}$ can be seen even in our low order tables. For example the weight $[2,0,0,0,1,0,\ldots ,0]$ at level $(\hat{m}_{26},\hat{m}_{27})=(1,2)$, so that $r=1$ and indeed we do find the $A_{25}$ representation $[0,0,0,1,0,\ldots ,0]$ in the adjoint at level $(m_{26},m_{27})=(2,2)$ of our tables. Between the same levels we can also find the related roots with $A_{25}$ weights $[0,1,0,0,0,1,0,\ldots ,0,1]\rightarrow [1,0,0,0,1,0,\ldots ,0,1,0]$ as expected from the above analysis.

We leave open the question of how to interpret the remaining roots of the $l_1$ representation, those with even values for $\sum_{i=1}^{25}p_i$, and indeed do not address the interesting question of how to physically interpret the higher order gauge fields in the $K_{27}$ algebra.
\newpage
\section{A Solution Generating Group Element}
In this chapter we discover a group element of a special form that encodes the algebra of $E_{11}$ as well as other so-called very-extended Kac-Moody algebras \cite{GaberdielOliveWest}, in such a way that the roots of the adjoint representation can be readily identified with brane solutions from $M$-theory. The reader is referred to appendix \ref{Dynkindiagrams} for the Dynkin diagrams of the very-extended algebras considered in this chapter. We will describe the results published originally in \cite{West4} and \cite{CookWest}.

In section \ref{scalars} we demonstrated the process of dimensional reduction and observed that in a reduction to three dimensions one obtains a generic theory containing gravity, gauge fields of various rank and possibly a dilaton that is described by the properties of its scalars. In the cases we looked at the scalars belonged to a nonlinear realisation of $A_{D-3}$ and $E_8$. It is sensible to ask the reverse question, that is if we start with a theory in three dimensions and allow the scalars to parameterise a coset with the symmetry of an arbitrary semisimple Lie group, what is the related higher dimensional theory? Such a set of theories whose dimensional reduction to three dimensions yield every possible semisimple Lie group, $\cal G$, was found in \cite{CremmerJuliaLuPope}. The process of reversing the dimensional reduction takes its name from chemistry and is called oxidation. For a detailed discussion of oxidation in the context of group theory the reader is referred to \cite{Keurentjes}. Furthermore a correspondence between the low-level fields in the adjoint representation of a very-extended Lie group, $\cal G^{+++}$, and the maximally oxidised theory associated with $\cal G$ has been demonstrated in reference \cite{KleinschmidtSchnakenburgWest}. There is a substantial literature deriving single brane solutions in generic supergravity theories \cite{BPSsolutions}, for a review see \cite{Argurio, Stelle}. The solutions for the oxidised theories mentioned above were considered in \cite{EnglertHouartWest}. In this chapter we will prove by construction that there exists a group element, which is a modification of that found in \cite{West4}, which encodes brane solutions of the corresponding maximally oxidised supergravity theories in a universal manner. We commence by reviewing the results of \cite{West4}.

\subsection{M-theory, IIA and IIB brane solutions}
It was shown in \cite{West4} that the usual half-BPS solutions of the type IIA and type IIB ten dimensional supergravity theories and the eleven dimensional theory possess a general formulation in terms of $E_{11}$ group elements, namely, 
\begin{equation}
g=\exp(-\frac{1}{2}\ln N \beta \cdot H)\exp((1-N)E_\beta) \label{groupelement2}
\end{equation}
Where, $\beta$ is a root in the adjoint representation of $E_{11}$, $E_{\beta}$ is its corresponding generator, $H$ is a vector consisting of the elements of the Cartan sub-algebra and $N$ is a harmonic function related to the gauge field, $A$, such that $A=N^{-1}-1$. In the literature concerning brane solutions in supergravity \cite{Argurio,Stelle} one finds harmonic functions, $N$, having the form,
\begin{equation}
N=1+\frac{k}{r^{D-p-3}}\label{harmonicfunctionabridged}
\end{equation}
Where $k$ is a constant dependent upon the space-time dimension, the dimension of the brane, dilaton coupling (if a dilaton is present) and the charge carried by the brane, an explicit form is given in equation (\ref{harmonicfunction}) later in this thesis. The variable $r$ is the usual notion of distance defined over the coordinates transverse to the brane worldvolume, i.e. $r^2=g_{ab}x^ax^b$ where the indices $a,b$ are transverse to the brane worldvolume. It is possible to write the harmonic function of equation (\ref{harmonicfunctionabridged}) in terms of group theoretical quantities. The assertion, that we will demonstrate by construction, is that for simply-laced groups,
\begin{equation}
p+1=\sum_{i=1}^{D}(i-D)<\alpha_i,\beta>=D\sum_{i=1}^{D}p_i-\tilde{A}
\end{equation}
Where we have used the notation from chapter 4, that is $\tilde{A}\equiv\sum_{i=1}^{D}ip_i$ (as oppose to A which we have used in this chapter to denote a gauge field) and $p_i$ indicate the weights along the gravity line, $A_{D-1}$. We may write the harmonic function $N$ in the group theoretic notation,
\begin{equation}
N=1+\frac{k}{r^{D-2-\sum_{i=1}^{D}(i-D)<\alpha_i,\beta>}}
\end{equation}
In fact the results of \cite{CookWest} demonstrated that the group element of equation (\ref{groupelement2}) should be generalised to take account of the varying root lengths of $E_{11}$ to,
\begin{equation}
g=\exp(-\frac{1}{\beta^2}\ln N \beta \cdot H)\exp((1-N)E_\beta) \label{groupelement}
\end{equation}
And similarly the expression for the harmonic function becomes,
\begin{equation}
N=1+\frac{k}{r^{D-2-\sum_{i=1}^{D}(i-D)<\alpha_i,\beta>\frac{<\beta,\beta>}{<\alpha_i,\alpha_i>}}} \label{realharmonicfunction}
\end{equation}
We will use these latter expressions throughout. 

Let us consider the well-known solutions of M-theory, the $M2$-brane, the $M5$-brane and the $pp$-wave, we demonstrate that these are solutions of the Einstein equations of a truncated form of the eleven dimensional supergravity action in appendix B, and now we turn our attention to demonstrating that they are contained in the $E_{11}$ algebra by a group element of the form of equation (\ref{groupelement}). Before we commence it will be helpful to write down the elements of the Cartan sub-algebra of $E_{11}$ which we find using equation (\ref{Serrerelations}),
\begin{align}
\nonumber H_i&={K^i}_i-{K^{i+1}}_{i+1}, \qquad i=1\ldots 10 \\
H_{11}&=-\frac{1}{3}({K^1}_1+\ldots+{K^8}_8)+\frac{2}{3}({K^9}_9+{K^{10}}_{10}+{K^{11}}_{11})
\end{align}
Now, using table \ref{E11roots} as our reference, we take the first few roots and substitute them into our group element to see how the brane solutions are encoded. Taking $\beta=\alpha_{11}$ we find that, 
\begin{align}
\nonumber g&=\exp(-\frac{1}{2}\ln N H_{11})\exp((1-N)R^{91011})\\
&=\exp(-\frac{1}{2}\ln N [-\frac{1}{3}({K^1}_1+\ldots +{K^8}_8)+\frac{2}{3}({K^9}_9+{K^{10}}_{10}+{K^{11}}_{11})])\exp((1-N)R^{91011})
\end{align}
Where $R^{91011}$ is the generator associated to the root $\alpha_{11}$. Recalling from chapter three that the coefficients, ${h_a}^b$, of the generators ${K^a}_b$ give the vielbein as ${{(e^h)}_\mu}^a$ we can read off the line element corresponding to this element, it is,
\begin{equation}
ds^2=N^{\frac{1}{3}}({dt_1}^2+\ldots+dy_8^2)+N^{-\frac{2}{3}}(dx_9^2+dx_{10}^2-dt^2) 
\end{equation}
Where we have imposed the signature of spacetime upon our solution by Wick rotating $x^{11}\rightarrow -it$ we will not worry about the consequences of this process at this stage since it is the subject of chapter six, but we will mention in passing that it is simply a matter of choosing a particular real form of the $A_{10}$ sub-algebra that we are working with. We read off our gauge field too,
\begin{equation}
A^T_{91011}=1-N
\end{equation}
Where the "$T$" is used to indicate that its indices are the flat tangent space indices of the group. To find a familiar gauge field we must transform the tangent space indices to world volume indices using the vielbein on the relevant coordinates, i.e. ${e_9}^9={e_{10}}^{10}={e_{11}}^{11}=N^{-\frac{1}{3}}$. We find the familiar relation between the gauge field and harmonic function, namely,
\begin{equation}
A_{91011}=N^{-1}-1
\end{equation}
From equation (\ref{realharmonicfunction}) we find that,
\begin{equation}
N=1+\frac{k}{r^6}
\end{equation}
In other words we have reproduced the familiar $M2$ brane solution. We can repeat the process for the next root in table \ref{E11roots} $\beta=(0,\ldots ,0,1,2,3,2,1,2)$, we find that,
\begin{align}
\nonumber \beta\cdot H&=H_6+2H_7+3H_8+2H_9+H_{10}+2H_{11} \\
&=-\frac{2}{3}({K^1}_1+\ldots +{K^5}_5)+\frac{1}{3}({K^6}_6+\ldots +{K^{11}}_{11})
\end{align}
Giving the line element of the $M5$-brane,
\begin{equation}
ds^2=N^{\frac{2}{3}}(dt_1^2+\ldots +dy_5^2)+N^{-\frac{1}{3}}(dx_6^2+\ldots +dx_{10}^2-dt^2)
\end{equation}
The gauge field and harmonic function are those of the $M5$ solution. The $pp$-wave solution follows similarly as can be see in \cite{West4}. 

An interesting property of the $E_{11}$ algebra, that was reported earlier in \cite{KleinschmidtSchnakenburgWest}, is also demonstrated in \cite{West4} using the group element of equation (\ref{groupelement}). This is the interesting observation that upon dimensional reduction to ten dimensions, which corresponds in the algebra decomposition to deleting a node on the gravity line so that the diagram remains connected, there are two different ways to decompose to an $A_9$ sub-algebra. One way gives rise to all the bosonic fields of IIA supergravity and the second choice leads to the bosonic fields of IIB supergravity. The group element then encodes the well-known brane solutions of each theory. Dimensional reduction of eleven dimensional supergravity gives rise to a non-chiral IIA theory, and to construct the IIB chiral theory requires a little premeditation. Given that much of the motivation for an $E_{11}$ symmetry has been based upon the symmetries that appear in dimensional reduction, but that the IIB theory is not derived in such a straightforward manner, it must be viewed as a minor success that the three supergravity theories in eleven and ten dimensions may be united in such a satisfying scheme to the reader sympathetic to the $E_{11}$ conjecture.

\subsection{The Commutators}
In order to work with the nonlinear realisation of a particular very-extended group, $\cal{G}^{+++}$ we will need to construct the commutators of the algebra at low orders. While this will be a useful process to go through it will also transpire that there is a delicate aspect to finding brane solutions encoded in the group element of equation (\ref{groupelement}), for which knowledge of the algebras commutators will be crucial. The delicate point is that ordering of the generators in the group element is important. We will see this most clearly in the diverse examples that occur when finding the solution content of the maximally oxidised supergravity theories, and in particular with the role of the dilaton, whose generator $R_0$ will occur in the first exponential of the group element, coming from $\beta\cdot H$, but we will find that we only find the expected brane solutions when we group together the dilaton fields into their own exponential. This aspect of the dilaton in the non-linear realisation is discussed in section \ref{dilatongenerator}. In order to manoeuvre the dilaton generator we will use the identities of section \ref{identities}, so that knowledge of the commutators of the dilaton will be crucial. 

By definition a Kac-Moody algebra is the multiple commutators of the simple root generators, $E_a$ and separately those of $F_a$, subject to the Serre relations (\ref{Serrerelations}). However as a matter of practise this is very difficult to carry out, in this section we explain how to construct the commutators of the $\cal G^{+++}$ algebra using the tables of low-order generators in \cite{KleinschmidtSchnakenburgWest}. If we consider two $A_n$ representations with root coefficients $(a_1,a_2,\ldots ,a_n)$ and $(b_1,b_2,\ldots ,b_n)$ then their commutator has a root which is the sum, i.e. it has root coefficients $((a_1+b_1),(a_2+b_2),\ldots ,(a_n+b_n))$.

The Borel sub-algebra is formed from the Cartan sub-algebra generators, $H_a$, where $a=1\ldots n$, the positive root generators, ${K^a}_b$, where $a<b$ and $a,b=1\ldots n $, and the generators, ${R^{a_1\ldots a_n}}_{b_1\ldots b_m}$ which arise from our decomposition of $\cal G^{+++}$. The generators of the Cartan sub-algebra are constructed from the ${K^a}_a$ generators where $a=1\ldots n$ and a dilaton generator, which we label $R_0$, although some of the theories we are interested in have no dilaton. The ${K^a}_b$ and ${R^{a_1\ldots a_n}}_{b_1\ldots b_m}$ obey the commutation rules
\begin{align}
\nonumber[{K^a}_b,{K^c}_d]=& \delta^b_c{K^a}_d-\delta^a_d{K^b}_c\\
[{K^a}_b,{R^{c_1\ldots c_n}}_{d_1\ldots d_m}]=& \delta^{c_1}_b{R^{ac_2\ldots c_n}}_{d_1\ldots d_m}+\ldots +\delta^{c_n}_b{R^{c_1\ldots c_{n-1}a}}_{d_1\ldots d_m}\\
\nonumber &-\delta^a_{d_1}{R^{c_1\ldots c_n}}_{bd_2\ldots d_m}-\ldots -\delta^a_{d_m}{R^{c_1\ldots c_n}}_{d_1\ldots d_{m-1}b}
\end{align}
Where the second equation includes a contribution from the trace of ${K^a}_b$, which is really a $GL(n)$ generator. Let us make use of the following notation for a generic commutator
\begin{equation}
[R^{n_1n_2\ldots n_a}_{(s_1)},R^{n_1n_2\ldots n_b}_{(s_2)}]=c^{s_1,s_2}_{a,b}R^{n_1n_2\ldots n_{a+b}}_{(s_1+s_2)}
\label{generalcommutator}
\end{equation}
The $s$ labels are the levels associated with the elimination of one of the simple roots in theories where the decomposition involves the elimination of more than one simple root. It is used to differentiate different types of generators, and can be simply read off from the root. We define which root is associated with the $s$ labels in each of the $\cal G^{+++}$ that we consider. In $\cal G^{+++}$ theories where only one root is eliminated the $s$ label is redundant and will be dropped.

Some general restrictions on $c^{s_1,s_2}_{a,b}$ are determined from the general Jacobi identity, where we simplify our notation so that the figure in square brackets is the number of antisymmetrised indices on each operator
\begin{align}
\nonumber &[R^{[a]}_{s_1},[R^{[b]}_{s_2},R^{[c]}_{s_3}]]+[R^{[b]}_{s_2},[R^{[c]}_{s_3},R^{[a]}_{s_1}]]+[R^{[c]}_{s_3},[R^{[a]}_{s_1},R^{[b]}_{s_2}]]=0\\
&\Rightarrow c^{s_1,s_2+s_3}_{a,b+c}c^{s_2,s_3}_{b,c}+c^{s_2,s_3+s_1}_{b,c+a}c^{s_3,s_1}_{c,a}+c^{s_3,s_1+s_2}_{c,a+b}c^{s_1,s_2}_{a,b}=0
\label{jacobiidentity}
\end{align}
Having commenced with the generators associated with the simple roots, we construct new generators by taking their commutators as we proceed. Subsequently by taking a commutator with a third generator we find restrictions on our commutator coefficients, $c^{s_1,s_2}_{a,b}$, using the Jacobi identity given above.

We will be particularly interested in finding how the dilaton commutes with other generators, which is denoted by $R^{[a]}_{s_1}=R_0$ to be the dilaton generator. The dilaton generator appears as a member of the Cartan sub-algebra, so it doesn't change the $A_n$ representation of the operator it commutes with. Of particular use are the following specific cases of (\ref{jacobiidentity}) which give the commutator coefficients for the general commutator $[R_0, R^{[m]}_s]=c_{0,m}^{0,s}R^{[m]}_s$ in a recursive form which we have found by setting $b=1$ and $c=m-1$ for the following values of $s_2$ and $s_3$ in (\ref{jacobiidentity})
\begin{align}
s_2=0, s_3=0 \qquad\Rightarrow\qquad c_{0,m}^{0,0}&=c_{0,m-1}^{0,0}+c_{0,1}^{0,0} \label{jicoefficients1}\\
s_2=0, s_3=1 \qquad\Rightarrow\qquad c_{0,m}^{0,1}&=c_{0,m-1}^{0,1}+c_{0,1}^{0,0}\\
s_2=1, s_3=0 \qquad\Rightarrow\qquad c_{0,m}^{0,1}&=c_{0,m-1}^{0,0}+c_{0,1}^{0,1}\\
s_2=1, s_3=1 \qquad\Rightarrow\qquad c_{0,m}^{0,2}&=c_{0,m-1}^{0,1}+c_{0,1}^{0,1} \label{jicoefficients4}
\end{align}
These relations allow us to find all the commutators with the dilaton, $R_0$, up to low orders, and we only need to specify one unknown which we choose to be ($c_{0,1}^{0,0}$).\footnote{The relation between $c_{0,1}^{0,1}$ and $c_{0,1}^{0,0}$ can be found by considering the commutation any of the Cartan elements, e.g. $H_n$, that contain $R_0$ with the generator for the simple root corresponding the second eliminated root, e.g. for the $F_4^{+++}$ algebra this is $E_7$, which can be seen in appendix \ref{Dynkindiagrams} on the Dynkin diagram.} Similar recursive commutator relations can be found for higher order generators by considering $s$ labels which are greater than one in equation (\ref{jacobiidentity}).\\

Let us consider the example of $F_4^{+++}$ to demonstrate how we build the algebra. The low order generators of $F_4^{+++}$ are given in reference \cite{KleinschmidtSchnakenburgWest} and we list them here with their associated root coefficients in brackets after them. 
\begin{alignat}{4}
\nonumber & &\qquad &R_0 (0000000)& \qquad &R_1 (0000001)& \qquad & \\ 
\nonumber &&\qquad &R_0^6 (0000010) & \qquad & R_1^6 (0000011) & \qquad & \\ 
& &\qquad &R_0^{56} (0000120) & \qquad &R_1^{56} (0000121)& \qquad &R_2^{56} (0000122) \\ 
\nonumber & &\qquad & & \qquad &R_1^{456} (0001231) & \qquad &R_2^{456} (0001232) \\ 
\nonumber & &\qquad & & \qquad &R_1^{3456} (0012341) & \qquad &R_2^{456,6} (0001242)\\
\nonumber & &\qquad & & \qquad & & \qquad &R_2^{3456} (0012342) 
\end{alignat}
The seven generators associated with the simple roots are $E_a={K^a}_{a+1}$ for $a=1\ldots 5$, $E_6=R_0^6$ and $E_7=R_1$. The subscript $s$ labels are the levels corresponding to the simple root generator $R_1$. We read off the following commutation relations for the dilaton $R_0$ with the other generators from the root coefficients
\begin{alignat}{3}
\nonumber &[R_0,R_0^a]= c_{0,1}^{0,0}R_0^a& \qquad &[R_0,R_1^a]= c_{0,1}^{0,1}R_1^a& \qquad &\\
\nonumber &[R_0,R_0^{ab}]= c_{0,2}^{0,0}R_0^{ab}& \qquad &[R_0,R_1^{ab}]= c_{0,2}^{0,1}R_1^{ab} &\qquad &[R_0,R_2^{ab}]= c_{0,2}^{0,2}R_2^{ab} \\
& &\qquad &[R_0,R_1^{abc}]= c_{0,3}^{0,1}R_1^{abc} &\qquad &[R_0,R_2^{abc}]= c_{0,3}^{0,2}R_2^{abc} \\
\nonumber & &\qquad &[R_0,R_1^{abcd}]= c_{0,4}^{0,1}R_1^{abcd} &\qquad &[R_0,R_2^{abc,d}]= c_{0,4}^{0,2}R_2^{abc,d}
\end{alignat}
By considering $[H_7,E_7]=2E_7$ and $[H_7,E_6]=-E_6$, from equation (\ref{Serrerelations}), we find that $c_{0,1}^{0,1}=-c_{0,1}^{0,0}$ and using equations (\ref{jicoefficients1})-(\ref{jicoefficients4}) we find that in terms of $c_{0,1}^{0,0}$ the commutation relations above are
\begin{alignat}{3}
\nonumber &[R_0,R_0^a]= c_{0,1}^{0,0}R_0^a& \qquad &[R_0,R_1^a]= -c_{0,1}^{0,0}R_1^a& \qquad &\\
\nonumber &[R_0,R_0^{ab}]= 2c_{0,1}^{0,0}R_0^{ab}& \qquad &[R_0,R_1^{ab}]= 0 &\qquad &[R_0,R_2^{ab}]= -2c_{0,1}^{0,0}R_2^{ab} \\
& &\qquad &[R_0,R_1^{abc}]= c_{0,1}^{0,0}R_1^{abc} &\qquad &[R_0,R_2^{abc}]= -c_{0,1}^{0,0}R_2^{abc} \\
\nonumber & &\qquad &[R_0,R_1^{abcd}]= 2c_{0,1}^{0,0}R_1^{abcd} &\qquad &[R_0,R_2^{abc,d}]= 0
\end{alignat}
We fix $c_{0,0}^{0,1}$ to be $\frac{1}{\sqrt{8}}$ and obtain the relations given later in equation (\ref{Fcommutators}). Other commutators can easily be found using the table of roots given in \cite{KleinschmidtSchnakenburgWest} together with the Jacobi identity (\ref{jacobiidentity}) and the Serre relations (\ref{Serrerelations}).

\subsection{The Dilaton Generator} \label{dilatongenerator}
Equation (\ref{groupelement}) plays a central role in our work and may contain the dilaton generator, $R_0$, through $\beta \cdot H$ for particular algebras. The non-linear realisation allows us to write down the most generally covariant field equations, as carried out in chapter two, for the field content arising from a particular $\cal G^{+++}$ in which the dilaton field, $A$ appears within a factor in the field strength. The procedure for finding the field equations is given in \cite{SchnakenburgWest, West1, West2} and, using the above commutator relations, in the non-linear realisation of $\cal G^{+++}$ the field strengths take the form
\begin{equation}
{F_{a_1a_2\ldots a_p}}_{s_1}=pe^{-c_{0,p-1}^{0,s_1}A}(\partial_{[a_1}{A_{a_2\ldots a_p]}}_{s_1}+\ldots)
\label{generalcovariantfieldstrength}
\end{equation}
Where "$+\ldots$" indicates terms which are not total derivatives but have the correct number of indices and within which the sum of $s$ labels matches the $s$ label of the field strength. The required dual field strength in the non-linear realisation is found to take the form
\begin{equation}
{F_{a_1a_2\ldots a_{(D-p)}}}_{s_2}=(D-p)e^{-c_{0,D-(p-1)}^{0,s_2}A}(\partial_{[a_1}{A_{a_2\ldots a_{D-p}]}}_{s_2}+\ldots)
\end{equation}
Where $D$ is the number of background spacetime dimensions, and we construct the remaining field strengths similarly. Having constructed the field strengths for a particular theory, we follow the process in \cite{SchnakenburgWest, West1, West2} and write down the equations relating a field strength and its Hodge dual, which take the general form
\begin{equation}
*{F_{a_1a_2\ldots a_{(D-p)}}}_{s_1}\equiv{F_{a_1a_2\ldots a_p}}_{s_1}=\frac{1}{p!}\epsilon_{a_1a_2\ldots a_D}F_{s_2}^{a_{(p+1)}\ldots a_D}
\label{Hodgedual}
\end{equation}
Where the Hodge dual is indicated by $*$, and $\epsilon_{a_1a_2\ldots a_D}$ is the usual completely antisymmetric tensor. We then substitute our expressions for ${F_{a_1a_2\ldots a_p}}_{s_1}$ and ${F_{a_1a_2\ldots a_{D-p}}}_{s_2}$ into this equation. The second order field equations are obtained by bringing all factors containing the dilaton field $e^{kA}$ together as a coefficient of the field strength appearing in the Lagrangian and differentiating the equation. Any total derivatives vanish by the Bianchi identity and we obtain an equation of the form
\begin{equation}
\partial_{[a_{m}}*(pe^{(-c_{0,p-1}^{0,s_1}+c_{0,D-(p-1)}^{0,s_2})A}({\partial_{a_1}A_{a_2\ldots a_p]}}_{s_1})+\ldots)=\partial_{[a_{m}}(+\ldots)_]
\end{equation}
The "$+\ldots$" correspond to the non-total derivative terms coming from our non-linear formulation of the field strengths. The coefficient of the total derivative on the left-hand-side is now exactly that which appears in a Lagrangian formulation of this system, where the field strength is just the total derivative, $p\partial_{[a_1}A_{a_2\ldots a_p]}$. These considerations are useful in making comparisons with the literature, in particular with \cite{CremmerJuliaLuPope}. The reader may see a fully worked example of this method for $D_{n-3}^{+++}$ in section \ref{D}.

\subsection{Solutions in Generic Gravity Theories}
We use the notation of \cite{Argurio} to express the general equations of motion, line element and dilaton that arise from a theory containing gravity, a gauge field and a dilaton, which is the truncation of the full supergravity action, and allows us to investigate single brane solutions. The generic action integral is
\begin{equation}
A=\frac{1}{16\pi G_D}\int {d^Dx\sqrt{-g}(R-\frac{1}{2}\partial_\mu\phi\partial^\mu\phi-\frac{1}{2.n_i!}e^{a_i\phi}F_{a_1\ldots a_{n_i}}F^{a_1\ldots a_{n_i}})}
\end{equation}
Where $D$ is the dimension of the background spacetime, $a_i$ is the dilaton coupling constant, $F_{a_1\ldots a_{n_i}}$ is a general $n_i$-form field strength. We note here that the equations of motion derived from a truncated action, as above, is always consistent with the single electric brane solutions presented here. Chern-Simons terms in the full action will alter the equation of motion obtained from varying the gauge potentials. For our solutions the only non-zero gauge potentials possess a timelike index, any remaining Chern-Simons-like terms in the gauge equation  will be a wedge product of such gauge fields and so are identically zero. Consequently the Chern-Simons terms do not effect our single electric brane solutions as they vanish at the level of the equations of motion. The equations of motion coming from the general action above are
\begin{align}
\nonumber {R^\mu}_\nu \nonumber &= \frac{1}{2}\partial^\mu\phi\partial_\nu\phi+\frac{1}{2n_i!}e^{a_i\phi}(n_iF^{\mu a_2\ldots a_{n_i}}F_{\nu a_2\ldots a_{n_i}} -\frac{n_i-1}{D-2}{\delta^\mu}_\nu F_{a_1\ldots a_{n_i}}F^{a_1\ldots a_{n_i}})\\
\nonumber \partial_\mu(\sqrt{-g}\partial^\mu\phi)&=\frac{\sqrt{-g}.a_i}{2.n_i!}e^{a_i\phi}F_{a_1\ldots a_{n_i}}F^{a_1\ldots a_{n_i}} \\
\partial_\mu(\sqrt{-g}e^{a_i\phi}F^{\mu a_2 \ldots a_{n_i}})&= 0
\end{align}
A theory containing such field strengths, $F_{a_1\ldots a_{n_i}}$, has conserved charges which are associated with $(n_i-2)$-branes, and their dual field strengths are associated to $(D-n_i-2)$-branes. A BPS $p$-brane solution with arbitrary dilaton coupling has line element
\begin{equation}
ds^2=N_p^{-2(\frac{D-p-3}{\Delta})}(-dt_1^2+dx_2^2+\ldots dx_{p+1}^2)+N_p^{2(\frac{p+1}{\Delta})}(dy_{p+2}^2+\ldots dy_D^2) \label{Lineelement}
\end{equation}
Where $\Delta=(p+1)(D-p-3)+\frac{1}{2}a_i^2(D-2)$, $a_i$ is the dilaton coupling constant of the associated field strength, $F_{a_1\ldots a_{n_i}} 
$, and $N_p$ is an harmonic function, equal to (\ref{realharmonicfunction}), and taking the general form
\begin{equation}
N_p=1+\frac{1}{D-p-3}\sqrt{\frac{\Delta}{2(D-2)}}\frac{\|\bf Q\|}{r^{(D-p-3)}} 
\end{equation}
Where $\bf Q$ is the conserved charge associated with the $p$-brane solution, and $r^2={y_{p+2}^2+\ldots y_D^2}$. We are able to read from any given line element the value of the dilaton coupling constant, $a_i$, to within a minus sign. The associated dilaton, for coupling constant $a_i$, is
\begin{equation}
e^{\phi}=N_p^{a_i\frac{D-2}{\Delta}} \label{Dilaton}
\end{equation}
A second brane solution is associated with the dual of the field strength that gives rise to the $p$-brane solution. This is a $(D-p-4)$-brane, where the dilaton coupling constant, $-a_i$, is now the negative of that for the $p$-brane. We find that for the $(D-p-4)$-brane
\begin{equation}
\Delta'=(D-p-3)(p+1)+\frac{1}{2}a_i^2(D-2)=\Delta
\end{equation}
And the line element for the brane associated with the dual field strength is
\begin{equation}
ds^2=N_{D-p-4}^{-2(\frac{p+1}{\Delta})}(-dt_1^2+dx_2^2+\ldots dx_{D-p-3}^2)+N_{D-p-4}^{2(\frac{D-p-3}{\Delta})}(dy_{D-p-2}^2+\ldots dy_D^2) \label{Duallineelement}
\end{equation}
The associated dilaton is
\begin{equation}
e^{\phi}=N_{D-p-4}^{-a_i\frac{D-2}{\Delta}} \label{Dualdilaton}
\end{equation}
Ensuring that the line elements are related as (\ref{Lineelement}) is to (\ref{Duallineelement}) and confirmation of a change of sign in the power of the harmonic function $N$ in the dilaton, as between (\ref{Dilaton}) and (\ref{Dualdilaton}), enables us to recognise which field strengths are dual to each other in the $\cal G^{+++}$ considered in the remainder of this chapter. 

\subsection{Simply Laced Groups}
Having established the background material and tools we now proceed to check that our group element of equation (\ref{groupelement}) does indeed encode the electric brane solutions of $\cal{G}^{+++}$. We commence with simply-laced groups. A simply laced group has simple roots which all have the same length, here our roots are normalised such that $\alpha_a^2=2$.

We also change our convention for picking out the timelike coordinate in our spacetime. In the earlier examples from $E_{11}$ we chose $x^{11}$ to be our timelike coordinate. As we will see in chapter six, all such choices of timelike coordinate are equivalent for the algebra, and are related by Weyl reflections in roots along the gravity line. Consequently in the remainder of this chapter we stick to the more conventional choice of singling out $x^1$ as a time coordinate. In terms of the generators, those that appear in the decompositions of chapter four and in the literature \cite{KleinschmidtSchnakenburgWest}, are highest weight generators, whose indices are the highest sequential coordinates of the spacetime, one of which is the time coordinate. We may lower the indices of such generators using commutators with the generators, ${K^a}_b$ of the $A_n$ sub-algebra, or by applying Weyl reflections in the roots of the $A_n$ sub-algebra. We will highlight such a shift to a lowest weight generator in the following examples.

\subsubsection{Very Extended $D_{n-3}$} \label{D}
$D_{24}^{+++}\equiv K_{27}$ is the conjectured symmetry underlying the effective action of the bosonic string theory \cite{West1}, and the analogous n-dimensional generalisation is $D_{n-3}^{+++}$ \cite{West1, EnglertHouartTaorminaWest}, these are maximally oxidised theories. The Dynkin diagram for $D_{n-3}^{+++}$ is shown in appendix \ref{Dynkindiagrams}, where the red nodes indicate the gravity line we shall consider, which is an $A_{n-2}$ sub-algebra.

We decompose the $D_{n-3}^{+++}$ algebra with respect to its $A_{n-2}$ sub-algebra, the simple roots whose nodes we delete are $\alpha_{n-1}$ and $\alpha_n$, as enumerated in appendix \ref{Dynkindiagrams} and we associate the levels $l_1$ and $l_2$ with these respectively. Our $s$ labels are chosen to be the $l_2$ level that the generator appears at. We find the generators ${K^a}_b$ at level $(0,0)$, corresponding to the $A_{n-2}$ sub-algebra of the gravity line, and the following other generators up to level $(1,1)$
\begin{alignat}{3}
\nonumber l_1 \longrightarrow \qquad &\qquad &\qquad &0 &\qquad &1\\ 
\nonumber l_2 \downarrow \qquad \qquad &0 &\qquad & R_0 & \qquad & R_1^{a_1a_2\ldots a_{(n-5)}}\\ 
\nonumber &1 &\qquad &R_0^{ab} & \qquad &R_1^{a_1a_2\ldots a_{(n-4)},b}\label{Dgenerators}\\
& & & & &R_1^{a_1a_2\ldots a_{(n-3)}}
\end{alignat}
The so-called dual to gravity $R_1^{a_1a_2\ldots a_{(n-4)},b}$ is listed amongst our low-level generators for completeness but it is not as well understood as the other listed generators and we will not consider it here, nor throughout this chapter in any of the $\cal G^{+++}$ theories that we consider, as a starting point generator for finding the encoded brane solutions via equation (\ref{groupelement}). Each of the generators is associated with a root, $\beta$, such that $\beta^2=2$ except for the dilaton generator $R_0$ and the generator $R_1^{a_1a_2\ldots a_{(n-3)}}$, for which $\beta^2=0$. That $\beta^2=0$ for these generators means that we cannot commence with the generators $R_0$ and $R_1^{a_1a_2\ldots a_{(n-3)}}$ and use the group element in equation (\ref{groupelement}) to deduce an electric brane associated with them, as we would have a singularity coming from the $\frac{1}{\beta^2}$ for these generators. As such they are discarded as starting points for our method, as are all such generators which possess such a singularity. It would be interesting to understand this obstruction to finding a brane solution by the method advocated in this chapter, as it would certainly provide information relating the root length to the physical spectrum. Since in the nonlinear realisations of Kac-Moody algebras we have very many more roots than physical branes, any information that can be gleamed about such a property of the algebra could be invaluable.

We make use of the following commutators, where we have chosen the coefficient $\frac{24}{D-2}$ in keeping with \cite{West1} and the other commutator has been determined from the Serre relations (\ref{Serrerelations})
\begin{align}
\nonumber [R_0,R_0^{ab}]&={\frac{24}{D-2}}R_0^{ab}\\
[R_0,R_1^{a_1a_2\ldots a_{n-5}}]&=-{\frac{24}{D-2}}R_1^{a_1a_2\ldots a_{n-5}}
\label{Dcommutators}
\end{align}
The simple root generators of $D_{n-3}^{+++}$ are
\begin{align}
\nonumber E_a&={K^a}_{a+1}, \qquad a=1,\ldots (n-2)\\
\nonumber E_{(n-1)}&=R_0^{(n-2)(n-1)}\\
E_n&=R_1^{56\ldots(n-1)}
\end{align}
and the Cartan sub-algebra generators, $H_a$, are given by
\begin{align}
\nonumber H_a&={K^a}_a-{K^{a+1}}_{a+1}, a=1,\ldots(n-2),\\
\nonumber H_{(n-1)}&=-\frac{2}{D-2}({K^1}_1+\ldots{K^{n-3}}_{n-3})+\frac{D-4}{D-2}({K^{n-2}}_{n-2}+{K^{n-1}}_{n-1})+{\frac{1}{6}}R_0 \\
H_n&=-\frac{D-4}{D-2}({K^1}_1+\ldots{K^4}_4)+\frac{2}{D-2}({K^5}_5+\ldots{K^{n-1}}_{n-1})-{\frac{1}{6}}R_0 \label{Dcartansub-algebra}
\end{align}
The low-level field content \cite{KleinschmidtSchnakenburgWest} is ${\hat{h}}_a\hspace{0pt}^b$, $A$, $A_{a_1a_2}$ and their field strengths have duals derived from  $A_{a_1\ldots a_{n-4},b}$, $A_{a_1\ldots a_{n-3}}$, $A_{a_1\ldots a_{n-5}}$ respectively. Our choice of local sub-algebra for the non-linear realisation allows us to write the group element as
\begin{align}
\nonumber g&=\exp(\sum_{a\leq b}{\hat{h}}_a\hspace{0pt}^b{K^a}_b)\exp(\frac{1}{(n-3)!}A_{a_1a_2\ldots a_{n-3}}R_1^{a_1a_2\ldots a_{n-3}})\exp(\frac{1}{(n-4)!}A_{a_1a_2\ldots a_{n-4},b}R_1^{a_1a_2\ldots a_{n-4},b})\\
&\qquad \exp(\frac{1}{(n-5)!}A_{a_1a_2\ldots a_{n-5}}R_1^{a_1a_2\ldots a_{n-5}})\exp(\frac{1}{2!}A_{ab}R_0^{ab})\exp(AR_0) \label{Dgroupelement}
\end{align}
The field content of the associated maximally oxidised theory given in \cite{CremmerJuliaLuPope} agrees with the low-order $D_{n-3}^{+++}$ content given above. The Lagrangian for the oxidised theory, contains gravity, a dilaton, $A$, and a 3-form field strength, $F_{\mu\nu\rho}=3\partial_{[\mu}A_{\nu\rho]}$, and is given by
\begin{equation}
A=\frac{1}{16\pi G_{n-1}}\int d^{n-1}x\sqrt{-g}(R-\frac{1}{2}\partial_\mu\phi\partial^\mu\phi-\frac{1}{2.3!}e^{\sqrt{\frac{8}{D-2}}\phi}{F_{\mu\nu\rho}}{F^{\mu\nu\rho}})
\label{Dlagrangian}
\end{equation}
Using the group element of equation (\ref{groupelement}) applied to the roots and generators listed in equation (\ref{Dgenerators}), we find the following electric branes:
\paragraph{A String}
We commence by setting $\beta$ in (\ref{groupelement}) to be the root associated with the generator, $R_0^{(n-2)(n-1)}$ $(0 0\ldots 0 1 0)$, the highest weight in the $R^{a_1a_2}$ representation, corresponding to the field $A_{a_1a_2}$ in \cite{KleinschmidtSchnakenburgWest}. In order to use conventional notation we choose to associate time-like coordinate with $x^1$. This means finding the lowest weight of $R_0^{(n-2)(n-1)}$ using multiple commutation with the appropriate ${K^a}_b$ generators such that all the indices on our operator are lowered to the lowest consecutive sequence of indices. Our corresponding lowest weight in this representation is $R_0^{12} (1 2\ldots 2 1 1 0)$. The new root is found by adding a simple root $\alpha_a$ for each ${K^a}_{a+1}$ that we commute with $R_0^{(n-2)(n-1)}$ to lower its indices. We find (taking $\beta=(1 2\ldots 2 1 1 0)$),
\begin{equation}
\nonumber \beta \cdot H=\frac{D-4}{D-2}({K^1}_1+{K^2}_2)-\frac{2}{D-2}({K^3}_3+\ldots +{K^{n-1}}_{n-1})+{\frac{1}{6}}R_0 
\label{Dstringexpansion}
\end{equation}
We now substitute equation (\ref{Dstringexpansion}) into our group element given in equation (\ref{groupelement}), 
\begin{align}
\nonumber g&=\exp(-\frac{1}{2}\ln N_1(\frac{D-4}{D-2}({K^1}_1+{K^2}_2)-\frac{2}{D-2}({K^3}_3+\ldots +{K^{n-1}}_{n-1})+{\frac{1}{6}}R_0))\exp((1-N_1)E_{\beta})\\
\nonumber &=\exp(-\frac{1}{2}\ln N_1(\frac{D-4}{D-2}({K^1}_1+{K^2}_2)-\frac{2}{D-2}({K^3}_3+\ldots +{K^{n-1}}_{n-1})))\\
&\qquad \exp(N_1^{-\frac{2}{D-2}}(1-N_1)E_{\beta})\exp(-{\frac{1}{12}}\ln N_1R_0) 
\label{Dstringelement}
\end{align}
Where we have moved the generator associated with the dilaton, $R_0$, to the right so that it agrees with the structure of the group element from which the non-linear realisation is constructed in equation (\ref{Dgroupelement}). We make use of $[R_0,R_0^{ab}]$ in equation (\ref{Dcommutators}) to do this\footnote{We note that in this case the commutator takes the form $[X,Y]=kY$ where $k$ is some constant so that $e^Xe^Y=e^Ye^Xe^{[X,Y]}$ appears to have defeated the object of moving the generator $Y$ past $X$. But by the Serre relations, $[X,[X,Y]]=[Y,[X,Y]]=0$ so that $e^Xe^Y=e^Ye^{[X,Y]}e^X$. Furthermore we also make use of the approximation $1+\ln {N^m}\approx N^m$.} and by examining the resulting $R_0$ term we find a dilaton which is given by
\begin{equation}
e^A=N_1^{-\frac{1}{12}}
\end{equation}
By reading off the coefficients of the ${K^a}_a$ in the group element we find a line element corresponding to a string
\begin{equation}
ds^2=N_1^{-\frac{D-4}{D-2}}(-dt_1^2+dx_2^2)+N_1^{\frac{2}{D-2}}(dy_3^2+\ldots+dy_{n-1}^2)
\label{Dstringlineelement}
\end{equation}
The brane is derived from a gauge potential which we can also read off from equation (\ref{Dstringelement}) as the coefficient of $E_\beta$
\begin{equation}
{A_{12}^T}_{(0)}=N_1^{-\frac{2}{D-2}}(1-N_1)
\end{equation}
To make the change to world volume indices we use the appropriate vielbein components which we read from equation (\ref{Dstringlineelement}), $({e^{\hat{h}})^1}_1=({e^{\hat{h}})^2}_2=N_1^{-\frac{D-4}{2(D-2)}}$ and find
\begin{equation}
{A_{12}}_{(0)}=N_1^{-1}-1
\end{equation}
This gives rise to a 3-form field strength, which we label ${F_{\mu\nu\rho}}_0$.
\paragraph{An $(n-6)$ Brane}
Let us consider the representation $R^{a_1a_2\ldots a_{n-5}}$, corresponding to the field $A_{a_1a_2\ldots a_{n-5}}$ in \cite{KleinschmidtSchnakenburgWest} whose highest weight generator is $R_1^{5\ldots(n-1)}$ $(0 0\ldots 0 1)$. The lowest weight generator in the representation is $R_1^{1\ldots(n-5)}$ $(1 2 3 4\ldots 4 3 2 1 0 1)$ and the corresponding element in the Cartan sub-algebra is
\begin{equation}
\beta \cdot H=\frac{2}{D-2}({K^1}_1+\ldots +{K^{n-5}}_{n-5})-\frac{D-4}{D-2}({K^{n-4}}_{n-4}+\ldots +{K^{n-1}}_{n-1})-{\frac{1}{6}}R_0
\end{equation}
Substituting this into equation (\ref{groupelement}), recalling that $\beta^2=2$, we find the corresponding group element is
\begin{align}
\nonumber g&=\exp(-\frac{1}{2}\ln N_{n-6}(\frac{2}{D-2}({K^1}_1+\ldots +{K^{n-5}}_{n-5})\\
\nonumber &\qquad -\frac{D-4}{D-2}({K^{n-4}}_{n-4}+\ldots +{K^{n-1}}_{n-1})))\exp(N_{n-6}^{-\frac{2}{D-2}}(1-N_{n-6})E_{\beta})\\
\nonumber &\qquad \exp({\frac{1}{12}}\ln N_{n-6}R_0)
\end{align}
Where in the last line we have made use of $[R_0,R_1^{a_1\ldots a_{n-5}}]$, given in equation (\ref{Dcommutators}) to move the dilaton generator, $R_0$, to the far right of the expression in agreement with the group element from which the non-linear realisation is constructed (\ref{Dgroupelement}). By examining the $R_0$ term we find a dilaton given by
\begin{equation}
e^A=N_{n-6}^{\frac{1}{12}}
\end{equation}
And a line element corresponding to an electric $(n-6)$-brane
\begin{equation}
ds^2=N_{n-6}^{-\frac{2}{D-2}}(-dt_1^2+\ldots dx_{n-5}^2)+N_{n-6}^{\frac{D-4}{D-2}}(dy_{n-4}^2+\ldots+dy_{n-1}^2)
\end{equation}
The brane is derived from a gauge potential
\begin{equation}
{A_{12\ldots (n-5)}^T}_{(1)}=N_{n-6}^{-\frac{2}{D-2}}(1-N_{n-6})
\end{equation}
We use $({e^{\hat{h}})^1}_1=\ldots ({e^{\hat{h}})^{n-5}}_{n-5}=N_{n-6}^{-\frac{1}{D-2}}$ and make the change to world volume indices
\begin{align}
\nonumber {A_{12\ldots (n-5)}}_{(1)}=&N_{n-6}^{-\frac{2}{D-2}}N_{n-6}^{-\frac{n-5}{D-2}}(1-N_{n-6})\\
=&N_{n-6}^{-1}-1
\end{align}
Where we have used $D=n-1$ to get to the second line. This gauge potential is associated with an $(n-4)$-form field strength which we conclude is the dual of ${F_{\mu\nu\rho}}_0$.

We have reproduced all the usual BPS electric branes using the group element given in equation (\ref{groupelement}). The formulae we have found above do indeed correspond to the solutions of the Lagrangian (\ref{Dlagrangian}). Let us now consider a fully worked example of the method outlined in section \ref{dilatongenerator} to find a relation between our dilaton generator $A$ and $\phi$. We have found the gauge potentials $A^{(0)}, A_{a_1a_2}^{(0)}$ and $A_{a_1\ldots a_{n-5}}^{(1)}$ and we have one 3-form field strength appearing in the Lagrangian of equation (\ref{Dlagrangian}). Using equation (\ref{generalcovariantfieldstrength}) we write down the most general equation for the 3-form field strength
\begin{equation}
{F_{a_1a_2a_3}}_{(0)}=3e^{-c_{0,2}^{0,0}A}\partial_{[a_1}{A_{a_2a_3]}}_{(0)} \label{D3form}
\end{equation}
Its dual is an $(n-4)$-form field strength which is most generally
\begin{equation}
{F_{a_1\ldots a_{n-4}}}_{(1)}=(n-4)e^{-c_{0,n-5}^{0,1}A}\partial_{[a_1}{A_{a_2\ldots a_{n-5}]}}_{(1)} \label{D(n-4)form}
\end{equation}
These are related by duality giving
\begin{equation}
*{F_{a_1a_2a_3}}_{(0)}=\frac{1}{(n-4)!}\epsilon_{a_1a_2a_3b_1\ldots b_{n-4}}{F^{a_1\ldots a_3}}_{(0)}={F_{b_1\ldots b_{n-4}}}_{(1)} \label{Ddualityeq}
\end{equation}
Substituting equations (\ref{D3form}) and (\ref{D(n-4)form}) into equation (\ref{Ddualityeq}) yields
\begin{equation}
*(3e^{-c_{0,2}^{0,0}A}\partial_{[a_1}{A_{a_2a_3]}}_{(0)})=(n-4)e^{-c_{0,n-5}^{0,1}A}\partial_{[a_1}{A_{a_2\ldots a_{n-4}]}}_{(1)}
\end{equation}
If we bring the exponentials of $A$ together on the left-hand-side with the 3-form and differentiate we obtain
\begin{equation}
\partial_{[a_m}*(3e^{(-c_{0,2}^{0,0}+c_{0,n-5}^{0,1})A}\partial_{a_1}{A_{a_2a_3]}}_{(0)})=0 \label{Dfieldeq}
\end{equation}
Where the right-hand-side has vanished under differentiation by the Bianchi identity. The commutation of $H_{n-1}$ given in equation (\ref{Dcartansub-algebra}) with the positive root generators $E_n$ and $E_{n-1}$ enables us to find a relation between $c_{0,n-5}^{0,1}$ and $c_{0,2}^{0,0}$, we take $l={\frac{1}{12}}$ and we find
\begin{align}
\nonumber[H_{n-1},E_n]&=[H_{n-1},R^{56\ldots (n-1)}_1]=0 \qquad \Rightarrow l c_{0,n-5}^{0,1}=-\frac{4}{D-2}\\
\nonumber[H_{n-1},E_{n-1}]&=[H_{n-1},R^{(n-2)(n-1)}_0]=2R^{(n-2)(n-1)}_0 \qquad \Rightarrow l c_{0,2}^{0,0}=\frac{4}{D-2}\\
& \Rightarrow c_{0,2}^{0,0}=-c_{0,n-5}^{0,1}
\end{align}
Substituting this identity into our field equation (\ref{Dfieldeq}) gives
\begin{equation}
\partial_{[a_m}*(3e^{(-2c_{0,2}^{0,0})A}\partial_{a_1}{A_{a_2a_3]}}_{(0)})=0
\end{equation}
We have fixed our structure constants such that $c_{0,2}^{0,0}=\frac{24}{D-2}$ in accordance with reference \cite{West1} and by comparison with (\ref{Dlagrangian}) we find that
\begin{equation}
A=-\frac{1}{12}\sqrt{\frac{D-2}{2}}\phi
\end{equation}
We find the electric branes from the group element to $E_6^{+++}$ and $E_7^{+++}$ in appendix \ref{oxidisedgravity} and show similarly that the low order field content derived from the group element (\ref{groupelement}) matches the field content given in reference \cite{CremmerJuliaLuPope} for the corresponding oxidised theories.
\subsection{Non-Simply Laced Groups}
The general method is slightly less straightforward for non-simply laced groups in that it becomes possible for the roots we find from the group theory to have a norm squared that differs from two. This difference manifests itself in two places. The first arises when we find the element in the Cartan sub-algebra corresponding to the lowest weight in a representation, at this point we must be wary that we don't have an identity between the root coefficients and the Cartan sub-algebra coefficients but a proportionality factor of $\frac{<\alpha, \alpha>}{2}$. So that, for example, an element of the Cartan sub-algebra corresponding to a root with norm squared half that of the gravity line has an additional factor of $\frac{1}{2}$ in front of it. The second place we must be wary of the varying size of the simple roots comes when we make use of the group element where we must remember to put in the correct value of $\beta^2$ for the generator we are considering.
\subsubsection{Very Extended $B_{n-3}$}
The Dynkin diagram for $B_{n-3}^{+++}$ is shown in appendix \ref{Dynkindiagrams}, where the red nodes indicate the gravity line we shall consider, which in this case is an $A_{n-2}$ sub-algebra. We decompose the $B_{n-3}^{+++}$ algebra with respect to its $A_{n-2}$ sub-algebra by deleting the nodes corresponding to the simple roots $\alpha_n$ and $\alpha_{n-1}$ which have generators $R_1^{56\ldots(n-1)}(00\ldots 01)$ and $R_0^{(n-1)}(00\ldots 10)$ respectively. We associate the levels $l_1$ and $l_2$ with these and the $s$ labels on our generators are chosen to be the $l_1$ level of that generator. From reference \cite{KleinschmidtSchnakenburgWest} we find the $B_{n-3}^{+++}$ algebra contains the generators ${K^a}_b$ at level $(0,0)$, corresponding to the $A_{n-2}$ sub-algebra of the gravity line. We have the following other generators up to level $(1,2)$ \cite{KleinschmidtSchnakenburgWest}
\begin{alignat}{3}
\nonumber l_1 \longrightarrow \qquad &\qquad &\qquad &0 &\qquad &1\\
\nonumber l_2 \downarrow \qquad \qquad &0 & & R_0 & &R_1^{a_1a_2\ldots a_{(n-5)}}\\
\nonumber &1 & &R_0^{a} & &R_1^{a_1a_2\ldots a_{(n-4)}} \label{Bgenerators}\\
\nonumber &2 & &R_0^{a_1a_2} & &R_1^{a_1a_2\ldots a_{(n-4)},b}\\
& & & & &R_1^{a_1a_2\ldots a_{(n-3)}}
\end{alignat}
The generators $R_0^{a}$ and $R_1^{a_1a_2\ldots a_{n-4}}$ have associated roots $\beta$ such that $\beta^2=1$, the rest and the gravity line nodes have $\beta^2=2$, with the exception of the dilaton generator $R_0$ and the generator $R_1^{a_1a_2\ldots a_{(n-3)}}$ which have $\beta^2=0$. We cannot apply our method to $R_0$ and $R_1^{a_1a_2\ldots a_{(n-3)}}$ so they are discarded as starting points when we come to find the electric branes from the generators. We make use of the following commutator relations where we have chosen the commutator coefficient of $[R_0,R_0^{a_1}]$ and the rest have followed from the relations (\ref{jicoefficients1})-(\ref{jicoefficients4}) and the Serre relations (\ref{Serrerelations})
\begin{align}
\nonumber [R_0,R_0^{a_1}]&=\frac{1}{4}\sqrt{\frac{8}{D-2}}R_0^{a_1}\\
\nonumber [R_0,R_0^{a_1a_2}]&=\frac{1}{2}\sqrt{\frac{8}{D-2}}R_0^{a_1a_2}\label{Bcommutators}\\
\nonumber [R_0,R_1^{a_1a_2\ldots a_{{n-5}}}]&=-\frac{1}{2}\sqrt{\frac{8}{D-2}}R_1^{a_1a_2\ldots a_{n-5}}\\
[R_0,R_1^{a_1a_2\ldots a_{{n-4}}}]&=-\frac{1}{4}\sqrt{\frac{8}{D-2}}R_1^{a_1a_2\ldots a_{n-4}}
\end{align}
We note from content of the table given in \cite{KleinschmidtSchnakenburgWest} for $B_{n-3}^{+++}$ that no higher order commutators are needed.
The simple root generators of $B_{n-3}^{+++}$ are
\begin{align}
\nonumber E_a&={K^a}_{a+1}, \qquad a=1,\ldots (n-2)\\
\nonumber E_{(n-1)}&=R_0^{(n-1)},\\
E_n&=R_1^{56\ldots(n-1)}
\end{align}
and the Cartan sub-algebra generators, $H_a$ are given by
\begin{align}
\nonumber H_a&={K^a}_a-{K^{a+1}}_{a+1}, a=1,\ldots(n-2),\\
\nonumber H_{(n-1)}&=-\frac{2}{D-2}({K^1}_1+\ldots +{K^{n-2}}_{n-2})+2\frac{D-3}{D-2}({K^{n-1}}_{n-1})+\sqrt{\frac{8}{D-2}}R_0,\\
H_n&=-\frac{D-4}{D-2}({K^1}_1+\ldots +{K^4}_4)+\frac{2}{D-2}({K^5}_5+\ldots{K^{n-1}}_{n-1})-\sqrt{\frac{8}{D-2}}R_0
\end{align}
The low-level field content \cite{KleinschmidtSchnakenburgWest} is ${\hat{h}}_a\hspace{0pt}^b$, $A$, $A_{a_1}$, $A_{a_1a_2}$ and their field strengths have duals derived from  $A_{a_1\ldots a_{n-4},b}$, $A_{a_1\ldots a_{n-3}}$, $A_{a_1\ldots a_{n-4}}$, $A_{a_1\ldots a_{n-5}}$ respectively. Our choice of local sub-algebra for the non-linear realisation allows us to write the group element of $B_{n-3}^{+++}$ as
\begin{align}
\nonumber g&=\exp(\sum_{a\leq b}{\hat{h}}_a\hspace{0pt}^b{K^a}_b)\exp(\frac{1}{(n-3)!}A_{a_1a_2\ldots a_{n-3}}R_1^{a_1a_2\ldots a_{n-3}})\exp(\frac{1}{(n-4)!}A_{a_1a_2\ldots a_{n-4},b}R_1^{a_1a_2\ldots a_{n-4},b})\\
\nonumber &\qquad \exp(\frac{1}{(n-4)!}A_{a_1a_2\ldots a_{n-4}}R_1^{a_1a_2\ldots a_{n-4}})\exp(\frac{1}{(n-5)!}A_{a_1a_2\ldots a_{n-5}}R_1^{a_1a_2\ldots a_{n-5}})\\
&\qquad \exp(\frac{1}{2!}A_{ab}R_0^{ab})\exp(A_aR_0^a)\exp(AR_0)
\end{align}
The field content of the associated maximally oxidised theory given in \cite{CremmerJuliaLuPope} agrees with the low-order $B_{n-3}^{+++}$ content given above. The Lagrangian for the oxidised theory, containing gravity, a 2-form field strength, ${F_{\mu\nu}}_0=2\partial_{[\mu}{A_{\nu]}}_0$, a 3-form field strength, ${F_{\mu\nu\rho}}_0=3(\partial_{[\mu}{A_{\nu\rho]}}_0+{A_{[\mu}}_0\partial_{\nu}{A_{\rho]}}_0)$ and a dilaton, $A$, is
\begin{equation}
A=\frac{1}{16\pi G_{n-1}}\int d^{n-1}x\sqrt{-g}(R-\frac{1}{2}\partial_\mu\phi\partial^\mu\phi-\frac{1}{2.2!}e^{\frac{1}{2}\sqrt{\frac{8}{D-2}}\phi}{{F_{\mu\nu}}_0}{F^{\mu\nu}_0}-\frac{1}{2.3!}e^{\sqrt{\frac{8}{D-2}}\phi}{F_{\mu\nu\rho}}_0F^{\mu\nu\rho}_0)\label{Blagrangian}
\end{equation}
We now demonstrate the use of the group element (\ref{groupelement}) to generate all of the usual BPS electric branes of $B_{n-3}^{+++}$ starting from the generators given in equation (\ref{Bgenerators}). We find the following electric branes
\paragraph{A Particle}
We wish to find the electric brane associated with the generator $R_0^a$ that has highest weight $R_0^{(n-1)} (0 0\ldots 0 1 0)$ and we note that $\beta^2=1$. The lowest weight generator is $R_0^{1} (1 1\ldots 1 1 0)$. To find the corresponding element in the Cartan sub-algebra we observe from appendix \ref{Dynkindiagrams} that only the root corresponding to the $(n-1)$th node does not have norm squared equal to two. Indeed $\alpha_{n-1}^2=1$, so because we are working in the Cartan basis we have to effectively halve the $H_{n-1}$ generator contribution when we find $\beta\cdot H$ from the root coefficients, $\alpha^i$, so that
\begin{align}
\nonumber \beta \cdot H=&H_1+\ldots +H_{n-2}+\frac{1}{2}H_{n-1}\\
=&\frac{D-3}{D-2}({K^1}_1)-\frac{1}{D-2}({K^2}_2+\ldots+{K^{n-1}}_{n-1})+\sqrt{\frac{2}{D-2}}R_0
\label{Bparticleexpansion}
\end{align}
By substituting (\ref{Bparticleexpansion}) into equation (\ref{groupelement}), and bearing in mind that $\beta^2=1$ for $R_0^1$, we find the corresponding group element is
\begin{align}
\nonumber g&=\exp(-\ln N_0(\frac{D-3}{D-2}({K^1}_1)-\frac{1}{D-2}({K^2}_2+\ldots+{K^{n-1}}_{n-1})))\\
&\qquad \exp(N_0^{-\frac{1}{D-2}}(1-N_0)E_{\beta})\exp(-\sqrt{\frac{2}{D-2}}\ln N_0R_0) 
\end{align}
In the last line we have made use of $[R_0, R_0^{a_1}]$ from equation (\ref{Bcommutators}) to move the dilaton generator to the far right so that it agrees with the group element from which the non-linear realisation is constructed. By examining the $R_0$ term we find a dilaton given by
\begin{equation}
e^A=N_0^{-\sqrt{\frac{2}{D-2}}}
\end{equation}
And a line element corresponding to a particle
\begin{equation}
ds^2=N_0^{-\frac{2(D-3)}{D-2}}(-dt_1^2)+N_0^{\frac{2}{D-2}}(dy_2^2+\ldots+dy_{n-1}^2)
\end{equation}
The particle is derived from a gauge potential
\begin{equation}
{A_{1}^T}_{(0)}=N_0^{-\frac{1}{D-2}}(1-N_0)
\end{equation}
We complete the change to world volume indices using $({e^{\hat{h}})^1}_1=N_0^{-\frac{D-3}{D-2}}$
\begin{equation}
{A_{1}}_{(0)}=N_1^{-1}-1
\end{equation}
This gives rise to a 2-form field strength, which we label ${F_{\mu\nu}}_0$.
\paragraph{An $(n-5)$ Brane}
Proceeding in the usual manner we find the electric brane associated with the generator $R_1^{a_1a_2\ldots a_{(n-4)}}$ whose lowest weight is $R_1^{12\ldots (n-4)}$ $(1234\ldots 43211)$ and the corresponding element in the Cartan sub-algebra is, (recalling that the root associated with the $(n-1)$th node of the Dynkin diagram is short, being half the length of the other roots),
\begin{equation}
\beta \cdot H=\frac{1}{D-2}({K^1}_1+\ldots {K^{n-4}}_{n-4})-\frac{D-3}{D-2}({K^{n-3}}_{n-3}+\ldots{K^{n-1}}_{n-1})-\frac{1}{2}\sqrt{\frac{8}{D-2}}R_0
\end{equation}
We note that $\beta^2=1$ and substitute this expression into equation (\ref{groupelement}) to find that the group element for the generator $R_1^{45\ldots (n-1)}$ is
\begin{align}
\nonumber g&=\exp(-\ln N_{n-5}(\frac{1}{D-2}({K^1}_1+\ldots {K^{n-4}}_{n-4})-\frac{D-3}{D-2}({K^{n-3}}_{n-3}+\ldots{K^{n-1}}_{n-1})))\\
& \qquad \exp(N_{n-5}^{-\frac{1}{D-2}}(1-N_{n-5})E_{\beta})\exp(\sqrt{\frac{2}{D-2}}\ln N_{n-5}R_0) 
\end{align}
We find a dilaton given by
\begin{equation}
e^A=N_{n-5}^{\sqrt{\frac{2}{D-2}}}
\end{equation}
And a line element corresponding to an $(n-5)$-brane
\begin{equation}
ds^2=N_{n-5}^{-\frac{2}{D-2}}(-dt_1^2+dx_2^2+\ldots +dx_{n-4}^2)+N_{n-5}^{\frac{2(D-3)}{D-2}}(dy_{n-3}^2+\ldots+dy_{n-1}^2)
\end{equation}
The particle is derived from a gauge potential
\begin{equation}
{A_{12\ldots (n-4)}^T}_{(1)}=N_{n-5}^{-\frac{1}{D-2}}(1-N_{n-5})
\end{equation}
We use $({e^{\hat{h}})^1}_1=\ldots ({e^{\hat{h}})^{n-4}}_{n-4}= N_{n-5}^{\frac{-1}{D-2}}$ to complete the change to world volume indices
\begin{equation}
{A_{12\ldots (n-4)}}_{(1)}=N_{n-5}^{-1}-1
\end{equation}
This gives rise to an $(n-3)$ form field strength, which we interpret to be the dual of ${F_{\mu\nu}}_0$.
\paragraph{A String}
We wish to find the electric brane associated with the generator $R_0^{a_1a_2}$ whose lowest weight generator is $R_0^{12} (1 2\ldots 2 2 0)$ and the corresponding element in the Cartan sub-algebra is
\begin{equation}
\beta \cdot H=\frac{D-4}{D-2}({K^1}_1+{K^2}_2)-\frac{2}{D-2}({K^3}_3+\ldots {K^{n-1}}_{n-1})+\sqrt{\frac{8}{D-2}}R_0
\end{equation}
The corresponding group element is
\begin{align}
\nonumber g=&\exp(-\frac{1}{2}\ln N_1(\frac{D-4}{D-2}({K^1}_1+{K^2}_2)-\frac{2}{D-2}({K^3}_3+\ldots {K^{n-1}}_{n-1})))\\
&\exp(N_1^{-\frac{2}{D-2}}(1-N_1)E_{\beta})\exp(-\sqrt{\frac{2}{D-2}}\ln N_1R_0) 
\end{align}
We find a dilaton given by
\begin{equation}
e^A=N_0^{-\sqrt{\frac{2}{D-2}}}
\end{equation}
And a line element corresponding to a string
\begin{equation}
ds^2=N_1^{-\frac{D-4}{D-2}}(-dt_1^2+dx_2^2)+N_1^{\frac{2}{D-2}}(dy_3^2+\ldots+dy_{n-1}^2)
\end{equation}
The string is derived from a gauge potential
\begin{equation}
{A_{12}^T}_{(0)}=N_1^{-\frac{2}{D-2}}(1-N_1)
\end{equation}
We complete the change to world volume indices using $({e^{\hat{h}})^1}_1=({e^{\hat{h}})^2}_2=N_1^{-\frac{D-4}{2(D-2)}}$
\begin{equation}
{A_{12}}_{(0)}=N_1^{-1}-1
\end{equation}
From this we derive a 3-form field strength, which we label ${F_{\mu\nu\rho}}_0$.
\paragraph{An $(n-6)$ Brane}
We wish to find the electric brane associated with the generator $R_1^{a_1a_2\ldots a_{n-5}}$ whose lowest weight generator is $R_1^{12\ldots (n-5)}$ $(1 2 3 4\ldots 4 3 2 1 0 1)$ and the corresponding element in the Cartan sub-algebra is
\begin{equation}
\beta \cdot H=\frac{2}{D-2}({K^1}_1+\ldots +{K^{n-5}}_{n-5})-\frac{D-4}{D-2}({K^{n-4}}_{n-4}+\ldots +{K^{n-1}}_{n-1})-\sqrt{\frac{8}{D-2}}R_0
\end{equation}
The corresponding group element is
\begin{align}
\nonumber g&=\exp(-\frac{1}{2}\ln N_{n-6}(\frac{2}{D-2}({K^1}_1+\ldots +{K^{n-5}}_{n-5})-\frac{D-4}{D-2}({K^{n-4}}_{n-4}+\ldots +{K^{n-1}}_{n-1})))\\
& \qquad \exp(N_{n-6}^{-\frac{2}{D-2}}(1-N_{n-6})E_{\beta})\exp(\sqrt{\frac{2}{D-2}}\ln N_{n-6}R_0) 
\end{align}
We find a dilaton given by
\begin{equation}
e^A=N_{n-6}^{\sqrt{\frac{2}{D-2}}}
\end{equation}
And a line element corresponding to an $(n-6)$-brane
\begin{equation}
ds^2=N_{n-6}^{-\frac{2}{D-2}}(-dt_1^2+dx_2^2+\ldots +dx_{n-5}^2)+N_{n-6}^{\frac{D-4}{D-2}}(dy_{n-4}^2+\ldots+dy_{n-1}^2)
\end{equation}
The brane is derived from a gauge potential
\begin{equation}
{A_{12\ldots (n-5)}^T}_{(1)}=N_{n-6}^{-\frac{2}{D-2}}(1-N_{n-6})
\end{equation}
We use $({e^{\hat{h}})^1}_1=\ldots ({e^{\hat{h}})^{n-5}}_{n-5}= N_{n-6}^{-\frac{1}{D-2}}$ to complete the change to world volume indices
\begin{equation}
{A_{12\ldots (n-5)}}_{(1)}=N_{n-6}^{-1}-1
\end{equation}
This gives rise to an $(n-4)$-form field strength, which we interpret to be the dual of ${F_{\mu\nu\rho}}_0$. We have reproduced all the usual BPS electric branes using the group element given in equation (\ref{groupelement}). The formulae we have found above do indeed correspond to the solutions of the Lagrangian (\ref{Blagrangian}). Our dilaton field $A$ to be related to $\phi$ by
\begin{equation}
A = -\phi
\end{equation}

\subsubsection{Very Extended $F_4$}
The ${F_4}^{+++}$ Dynkin diagram is shown in appendix \ref{Dynkindiagrams}, where the red nodes represent the gravity line with respect to which we decompose our Kac-Moody algebra. The shorter roots, enumerated as nodes 6 and 7 have $\alpha^2=1$ where the gravity line roots are normalised to have norm squared of two.

The $F_4^{+++}$ algebra is decomposed with respect to its $A_5$ sub-algebra, the simple roots whose nodes we delete are $\alpha_7$ and $\alpha_6$ and their generators are $R_1 (0000001)$ and $R_0^6(0000010)$. We associate the levels $l_1$ and $l_2$ with these respectively. The $s$ labels on our generators are chosen to be the $l_1$ level of the generator. From reference \cite{KleinschmidtSchnakenburgWest} we find the ${F_4}^{+++}$ algebra decomposed with respect to an $A_5$ sub-algebra contains the generators ${K^a}_b$ at level (0,0) and the following generators at levels $(l_1, l_2)$
\begin{alignat}{4}
\nonumber l_1 \longrightarrow \qquad &\qquad &\qquad &0 &\qquad &1 &\qquad &2 \\ 
\nonumber l_2 \downarrow \qquad \qquad &0 &\qquad &R_0 & \qquad &R_1 & \qquad & \\ 
\nonumber &1 &\qquad &R_0^a & \qquad & R_1^a & \qquad & \\ 
\nonumber &2 &\qquad &R_0^{ab} & \qquad &R_1^{ab} & \qquad &{R_2}^{ab}\\ 
\nonumber &3 &\qquad & & \qquad &R_1^{abc} & \qquad &{R_2}^{abc} \\ 
\nonumber &4 &\qquad & & \qquad &R_1^{abcd} & \qquad &{R_2}^{abc,d}\\
& &\qquad & & \qquad & & \qquad &{R_2}^{abcd}  \label{Fgenerators}
\end{alignat}
The generators $R_0$ and ${R_2}^{abcd}$ have associated roots $\beta$ such that $\beta^2=0$ so we discard them as starting points for our method in this section, and ${R_2}^{abc,d}$, ${R_2}^{ab}$, $R_0^{ab}$ have $\beta^2=2$, the other generators all have $\beta^2=1$. We make use of the following commutation relations where we have chosen the commutator coefficient for $[R_0,R_0^a]$ and the rest have been deduced from equations (\ref{jicoefficients1})-(\ref{jicoefficients4}) and the Serre relations (\ref{Serrerelations})
\begin{alignat}{3}
\nonumber &[R_0,R_0^a]= \frac{1}{\sqrt{8}}R_0^a &\qquad &[R_0,R_1^a]=-\frac{1}{\sqrt{8}}R_1^a &\qquad & \\
\nonumber &[R_0,R_0^{ab}]= \frac{1}{\sqrt{2}}R_0^{ab} &\qquad &[R_0,R_1^{ab}]=0 &\qquad &[R_0,{R_2}^{ab}]=-\frac{1}{\sqrt{2}}{R_2}^{ab} \label{Fcommutators}\\
\nonumber & &\qquad &[R_0,R_1^{abc}]= \frac{1}{\sqrt{8}}R_1^{abc} &\qquad &[R_0,{R_2}^{abc}]=-\frac{1}{\sqrt{8}}{R_2}^{abc} \\
& &\qquad &[R_0,R_1^{abcd}]= \frac{1}{\sqrt{2}}R_1^{abcd} &\qquad &[R_0,{R_2}^{abc,d}]=0 
\end{alignat}
The simple root generators of ${F_4}^{+++}$ are
\begin{align}
\nonumber E_a&={K^a}_{a+1}, \qquad a=1,\ldots5\\
\nonumber E_6&=R_0^6\\
E_7&=R_1
\end{align}
The Cartan sub-algebra generators, $H_a$, are given by \cite{EnglertHouart}
\begin{align}
\nonumber H_a&={K^a}_a-{K^{a+1}}_{a+1}, a=1,\ldots5,\\
\nonumber H_6&=-\frac{1}{2}({K^1}_1+\ldots{K^5}_5)+\frac{3}{2}{K^6}_6+\sqrt{2}R_0,\\
H_7&=-\sqrt{8}R_0
\end{align}
The low-level field content \cite{KleinschmidtSchnakenburgWest} is ${\hat{h}}_a\hspace{0pt}^b$, $A$, ${A_{a_1}}_0$, ${A_{a_1}}_1$, ${A_{a_1a_2}}_2$, ${A_{a_1a_2a_3a_4}}_1$ and their field strengths have duals derived from  ${A_{a_1a_2a_3,b}}_2$, ${A_{a_1a_2a_3a_4}}_2$ ${A_{a_1a_2a_3}}_2$, ${A_{a_1a_2a_3}}_1$, ${A_{a_1a_2}}_0$, $A_1$, respectively, we also have an ${A_{a_1a_2}}_1$ field without a dual listed above, so we treat it as a self-dual field. Our choice of local sub-algebra for the non-linear realisation allows us to write the group element of ${F_4}^{+++}$ as
\begin{align}
\nonumber g=\exp(\sum_{a\leq b}{\hat{h}}_a\hspace{0pt}^b{K^a}_b)&\exp(\frac{1}{4!}{A_{a_1a_2a_3a_4}}_1R_1^{a_1a_2a_3a_4})\exp(\frac{1}{4!}{A_{a_1a_2a_3a_4}}_2{R_2}^{a_1a_2a_3a_4})\\
\nonumber &\exp(\frac{1}{3!}{A_{a_1a_2a_3,b}}_2{R_2}^{a_1a_2a_3,b})\exp(\frac{1}{3!}{A_{a_1a_2a_3}}_2{R_2}^{a_1a_2a_3})\\
\nonumber &\exp(\frac{1}{3!}{A_{a_1a_2a_3}}_1R_1^{a_1a_2a_3})\exp(\frac{1}{2!}{A_{a_1a_2}}_2{R_2}^{a_1a_2})\\
\nonumber &\exp(\frac{1}{2!}{{A_{a_1a_2}}_1R_1^{a_1a_2})\exp(\frac{1}{2!}A_{a_1a_2}}_0R_0^{a_1a_2})\\
&\exp({A_{a_1}}_1R_1^a)\exp({A_{a_1}}_0R_0^{a_1})\exp(A_1R_1)\exp(AR_0)
\end{align}
The field content of the associated maximally oxidised theory given in \cite{CremmerJuliaLuPope} agrees with the low-order $F_4^{+++}$ content given above. The Lagrangian for the oxidised theory, contains two 2-form field strengths ${F_{\mu\nu}}_0=2\partial_{[\mu}{A_{\nu]}}_0+\sqrt{2}A_1\partial_{[\mu}{A_{\nu]}}_1$ and ${F_{\mu\nu}}_1=2\partial_{[\mu}{A_{\nu]}}_1$, a 3-form field strength ${F_{\mu\nu\rho}}_2=3(\partial_{[\mu}{A_{\nu\rho]}}_2+{A_{[\mu}}_1\partial_{\nu}{A_{\rho]}}_1)$, a self-dual 3-form field strength ${F_{\mu\nu\rho}}_1=3(\partial_{[\mu}{A_{\nu\rho]}}_1-{A_{[\mu}}_0\partial_{\nu}{A_{\rho]}}_1)-\frac{1}{\sqrt{2}}{A_1}{F_{\mu\nu\rho}}_2$, a 1-form field strength ${F_\mu}_1=\partial_{\mu}A_1$, and the dilaton $A$.
\begin{align}
\nonumber A=\frac{1}{16\pi G_6}\int d^6x\sqrt{-g}(&R-\frac{1}{2}\partial_\mu\phi\partial^\mu\phi-\frac{1}{2.2!}e^{{\frac{1}{\sqrt{2}}}\phi}{F_{\mu\nu}}_0F^{\mu\nu}_0-\frac{1}{2.2!}e^{{-\frac{1}{\sqrt{2}}}\phi}{F_{\mu\nu}}_1F^{\mu\nu}_1\\
\nonumber &-\frac{1}{2.3!}e^{-\sqrt{2}\phi}{{F_{\mu\nu\rho}}_2}F^{\mu\nu\rho}_2-\frac{1}{2}e^{\sqrt{2}\phi}{{F_{\mu}}_1}F^{\mu}_1)\\
\nonumber -\int_{C.S.} d^6x\epsilon^{a_1\ldots a_6}&(\frac{1}{\sqrt{2}.3!3!}A_1{F_{a_1a_2a_3}}_2{F_{a_4a_5a_6}}_1+\frac{1}{2.2!3!}{A_{a_1}}_0{F_{a_2a_3}}_0{F_{a_4a_5a_6}}_2\\
&+\frac{1}{2.2!3!}{A_{a_1}}_0{F_{a_2a_3}}_1{F_{a_4a_5a_6}}_1)
\label{Flagrangian}
\end{align}
The reader must remember that ${F_{\mu\nu\rho}}_1$ is self-dual and its kinetic term in the action integral vanishes. We again use the group element (\ref{groupelement}) to generate all of the electric branes of $F_4^{+++}$ starting from the generators given in equation (\ref{Fgenerators}). We find the following electric branes
\paragraph{A Particle}
Commencing by setting $\beta$ to be the root lowest weight generator $R_0^1$ $(1 1 1 1 1 1 0)$ and taking account of the shorter root on the sixth node of the ${F_4}^{+++}$ Dynkin diagram, we find the element corresponding to $R_0^1$ in the Cartan sub-algebra
\begin{equation}
\beta \cdot H=\frac{3}{4}({K^1}_1)-\frac{1}{4}({K^2}_2+\ldots +{K^6}_6)+\frac{1}{\sqrt{2}}R_0
\end{equation}
And from equation (\ref{groupelement}) we find the corresponding group element is
\begin{equation}
g = \exp(-\ln N_0(\frac{3}{4}({K^1}_1)-\frac{1}{4}({K^2}_2+\ldots +{K^6}_6)))\exp(N_0^{-\frac{1}{4}}(1-N_0)E_\beta)\exp(-\frac{1}{\sqrt{2}}\ln N_0R_0)
\end{equation} 
So we find a dilaton given by 
\begin{equation}
e^A = N_0^{-\frac{1}{\sqrt{2}}}
\end{equation}
And a line element corresponding to a particle
\begin{equation}
ds^2=N_0^{-\frac{3}{2}}(-dt_1^2)+N_0^{\frac{1}{2}}(dy_2^2+\ldots+dy_6^2)
\label{Fparticlelinelement}
\end{equation}
We have a gauge field given by
\begin{equation}
{A_1^T}_{(0)} = N_0^{-\frac{1}{4}}(1-N_0)
\end{equation}
The change to world volume indices uses the vielbein component ${(e^h)^1}_1=N_0^{-\frac{3}{4}}$
\begin{equation}
{A_1}_{(0)}=N_0^{-1}-1
\end{equation}
We associate this gauge potential with a 2-form field strength, ${F_{\mu\nu}}_0$.
\paragraph{A 2-Brane}
We wish to find the electric brane associated with the lowest weight generator ${R_2}^{123} (1 2 3 3 3 3 2)$. Noting that $\beta^2=1$, then the corresponding element in the Cartan sub-algebra is
\begin{equation}
\beta \cdot H=\frac{1}{4}({K^1}_1+\ldots{K^3}_3)-\frac{3}{4}({K^4}_4+\ldots{K^6}_6)-\frac{1}{\sqrt{2}}R_0
\end{equation}
Substituting this into equation (\ref{groupelement}) we find the appropriately arranged element is
\begin{equation}
g=\exp(-\ln N_2(\frac{1}{4}({K^1}_1+\ldots +{K^3}_3)-\frac{3}{4}({K^4}_4+\ldots +{K^6}_6)\exp(N_2^{-\frac{1}{4}}(1-N_2)E_{\beta})\exp(+\frac{1}{\sqrt{2}}\ln N_2R_0)
\end{equation}
We find a dilaton given by
\begin{equation}
e^A=N_2^{+\frac{1}{\sqrt{2}}}
\end{equation}
And a line element corresponding to a 2-brane
\begin{equation}
ds^2=N_2^{-\frac{1}{2}}(-dt_1^2+\ldots dx_3^2)+N_2^{\frac{3}{2}}(dy_4^2+\ldots+dy_6^2) \label{F2branelineeelement}
\end{equation}
The brane is derived from the gauge field given by
\begin{equation}
{A_{123}^T}_{(2)}=N_2^{-\frac{1}{4}}(1-N_2)
\end{equation}
We complete the change to world volume indices using $({e^{\hat{h}})^1}_1=\ldots({e^{\hat{h}})^3}_3=N_0^{-\frac{1}{4}}$
\begin{equation}
{A_{123}}_{(2)}=N_2^{-1}-1
\end{equation}
We conclude that ${R_2}^{456}$ is the generator associated with the dual of ${F_{\mu\nu}}_0$.
\paragraph{A Second Particle}
The calculation for the generator which has highest weight $R_1^6$ is the same as that for $R_0^6$ except that the expansion of $\beta \cdot H$ has an extra $\frac{1}{2}H_7$ added on. That is,
\begin{equation}
\beta \cdot H=\frac{3}{4}({K^1}_1)-\frac{1}{4}({K^2}_2+\ldots +{K^6}_6)-\frac{1}{\sqrt{2}}R_0
\end{equation}
And from equation (\ref{groupelement}) we find the corresponding group element is
\begin{align}
\nonumber g &= \exp(-\ln N_0(\frac{3}{4}({K^1}_1)-\frac{1}{4}({K^2}_2+\ldots +{K^6}_6)-\frac{1}{\sqrt{2}}R_0))\exp((1-N_0)E_\beta)\\
& \qquad = \exp(-\ln N_0(\frac{3}{4}({K^1}_1)-\frac{1}{4}({K^2}_2+\ldots +{K^6}_6)))\exp(N_0^{-\frac{1}{4}}(1-N_0)E_\beta)\exp(\frac{1}{\sqrt{2}}\ln N_0R_0)
\end{align} 
We find a line element identical to (\ref{Fparticlelinelement}) for a particle and a dilaton given by
\begin{equation}
e^A = N_0^{\frac{1}{\sqrt{2}}}
\end{equation}
We have a gauge field given by
\begin{equation}
{A_1^T}_{(1)} = N_0^{-\frac{1}{4}}(1-N_0)
\end{equation}
We change to world volume indices via the vielbein component ${(e^h)^1}_1=N_0^{-\frac{3}{4}}$ and find
\begin{equation}
{A_1}_{(1)}=N_0^{-1}-1
\end{equation}
We associate this gauge potential with a 2-form field strength which we label ${F_{\mu\nu}}_1$.
\paragraph{A Second 2-Brane}
Similarly, the derivation of the electric brane associated with the highest weight generator $R_1^{456}$ is the same as that for ${R_2}^{456}$ but with a $\frac{1}{2}H_7$ taken off of expansion of the $\beta \cdot H$ expansion. That is,
\begin{equation}
\beta \cdot H=\frac{1}{4}({K^1}_1+\ldots +{K^3}_3)-\frac{3}{4}({K^4}_4+\ldots +{K^6}_6)+\frac{1}{\sqrt{2}}R_0
\end{equation}
Substituting this into equation (\ref{groupelement}) we find the appropriately arranged element is
\begin{align}
\nonumber g&=\exp(-\ln N_2(\frac{1}{4}({K^1}_1+\ldots{K^3}_3)-\frac{3}{4}({K^4}_4+\ldots{K^6}_6)\\
&\qquad\exp(N_2^{-\frac{1}{4}}(1-N_2)E_{\beta})\exp(-\frac{1}{\sqrt{2}}\ln N_2R_0)
\end{align}
We find a dilaton given by
\begin{equation}
e^A=N_2^{-\frac{1}{\sqrt{2}}}
\end{equation}
And a line element corresponding to a 2-brane identical to (\ref{F2branelineeelement}). The brane is derived from the gauge field given by
\begin{equation}
{A_{123}^T}_{(1)}=N_2^{-\frac{1}{4}}(1-N_2)
\end{equation}
We complete the change to world volume indices using $({e^{\hat{h}})^1}_1=\ldots({e^{\hat{h}})^3}_3=N_0^{-\frac{1}{4}}$ and find
\begin{equation}
{A_{123}}_{(1)}=N_2^{-1}-1
\end{equation}
We conclude that ${R_1}^{456}$ is the generator associated with the dual of ${F_{\mu\nu}}_1$.
\paragraph{A String}
We wish to find the electric brane associated with the ${R^{a_1a_2}}_2$ generator whose lowest weight generator is ${R_2}^{12} (1 2 2 2 2 2 2)$, and noting that $\beta^2=2$, the corresponding element in the Cartan sub-algebra is
\begin{equation}
\beta \cdot H=\frac{1}{2}({K^1}_1+{K^2}_2)-\frac{1}{2}({K^3}_3+\ldots +{K^6}_6)-\sqrt{2}R_0
\end{equation}
The corresponding group element (\ref{groupelement}) is
\begin{align}
g=\exp(-\frac{1}{2}\ln N_1(\frac{1}{2}({K^1}_1+{K^2}_2)-\frac{1}{2}({K^3}_3+\ldots +{K^6}_6)\exp(N_1^{-\frac{1}{2}}(1-N_1)E_{\beta})\exp(\frac{1}{\sqrt{2}}\ln N_1R_0)
\end{align}
We find a dilaton given by
\begin{equation}
e^A=N_1^{\frac{1}{\sqrt{2}}}
\end{equation}
And a line element corresponding to a string
\begin{equation}
ds^2=N_1^{-\frac{1}{2}}(-dt_1^2+dx_2^2)+N_1^{\frac{1}{2}}(dy_3^2+\ldots+dy_6^2)
\end{equation}
The brane is derived from the gauge field given by
\begin{equation}
{A_{12}^T}_{(2)}=N_1^{-\frac{1}{2}}(1-N_1)
\end{equation}
We complete the change to world volume indices using $({e^{\hat{h}})^1}_1=({e^{\hat{h}})^2}_2=N_1^{-\frac{1}{4}}$
\begin{equation}
{A_{12}}_{(2)}=N_1^{-1}-1
\end{equation}
This gauge potential gives rise to a 3-form field strength, which we label ${F_{\mu\nu\rho}}_2$.
\paragraph{A Second String}
Starting with the lowest weight generator $R_0^{12} (1 2 2 2 2 2 0)$, with $\beta^2=2$. We find that the component in the Cartan sub-algebra is,
\begin{equation}
\beta \cdot H=\frac{1}{2}({K^1}_1+{K^2}_2)-\frac{1}{2}({K^3}_3+\ldots +{K^6}_6)+{\sqrt{2}}R_0
\end{equation}
The corresponding group element (\ref{groupelement}) is
\begin{align}
g=\exp(-\frac{1}{2}\ln N_1(\frac{1}{2}({K^1}_1+{K^2}_2)-\frac{1}{2}({K^3}_3+\ldots +{K^6}_6)\exp(N_1^{-\frac{1}{2}}(1-N_1)E_{\beta})\exp(-\frac{1}{\sqrt{2}}\ln N_1R_0)
\end{align}
We find a dilaton given by
\begin{equation}
e^A=N_1^{-\frac{1}{\sqrt{2}}}
\end{equation}
And a line element corresponding to a string
\begin{equation}
ds^2=N_1^{-\frac{1}{2}}(-dt_1^2+dx_2^2)+N_1^{\frac{1}{2}}(dy_3^2+\ldots+dy_6^2)
\end{equation}
The brane is derived from the gauge field given by
\begin{equation}
{A_{12}^T}_{(0)}=N_1^{-\frac{1}{2}}(1-N_1)
\end{equation}
We complete the change to world volume indices using $({e^{\hat{h}})^1}_1=({e^{\hat{h}})^2}_2=N_1^{-\frac{1}{4}}$
\begin{equation}
{A_{12}}_{(0)}=N_1^{-1}-1
\end{equation}
We conclude that this is the group element associated with the dual of ${F_{\mu\nu\rho}}_2$.
\paragraph{A "Gooseberry" String} \label{Gooseberrystring}
We find the electric brane associated with the lowest weight generator $R_1^{12} (1 2 2 2 2 2 1)$, and ,noting that $\beta^2=1$, this has component in the Cartan sub-algebra,
\begin{equation}
\beta \cdot H=\frac{1}{2}({K^1}_1+{K^2}_2)-\frac{1}{2}({K^3}_3+\ldots{K^6}_6)
\end{equation}
Substituting this expression into equation (\ref{groupelement}) gives the corresponding element
\begin{align}
g=&\exp(-\ln N_1(\frac{1}{2}({K^1}_1+{K^2}_2)-\frac{1}{2}({K^3}_3+\ldots +{K^6}_6)))\exp((1-N_1)E_{\beta})
\end{align}
There is no dilaton and we find a line element corresponding to a string
\begin{equation}
ds^2=N_1^{-1}(-dt_1^2+dx_2^2)+N_1(dy_3^2+\ldots+dy_6^2)
\end{equation}
The brane is derived from the gauge field given by
\begin{equation}
{A_{12}^T}_{(1)}=(1-N_1)
\end{equation}
We complete the change to world volume indices using $({e^{\hat{h}})^1}_1=({e^{\hat{h}})^2}_2=N_1^{-\frac{1}{2}}$
\begin{equation}
{A_{12}}_{(1)}=N_1^{-1}-1
\end{equation}
This gauge field is associated with a second 3-form field strength which we label, ${F_{\mu\nu\rho}}_1$. We deduce that this field strength is self-dual as we have no more two-form generators in (\ref{Fgenerators}) that could produce an electric brane that would be associated with a dual to ${F_{\mu\nu\rho}}_1$. We note that the difference in the line element for the string here and those for the other strings derived from $F_4^{+++}$ is due to the dilaton coupling constant being zero here. And due to its singular nature we refer to it as the "gooseberry" string amongst the triplet of strings occurring in the $F_4^{+++}$ theory.
\paragraph{A 3-Brane} \label{3-brane}
We consider the lowest weight generator $R_1^{1234} (1234441)$, for which we note that $\beta^2=1$. This corresponds to an element in the Cartan sub-algebra given by
\begin{equation}
\beta \cdot H=-({K^5}_5+{K^6}_6)+{\sqrt{2}}R_0
\end{equation}
The group element given by equation (\ref{groupelement}) is
\begin{equation}
g=\exp(-\ln N_3(0({K^1}_1+\ldots{K^4}_4)-({K^5}_5+{K^6}_6)))\exp(N_3^{-1}(1-N_3)E_\beta)\exp(-{\sqrt{2}}\ln N_3R_0)
\end{equation}
We find a dilaton given by
\begin{equation}
e^A=N_3^{-{\sqrt{2}}}
\end{equation}
And a line element corresponding to a 3-brane
\begin{equation}
ds^2=(-dt_1^2+\ldots +dx_4^2)+N_3^{2}(dy_5^2+dy_6^2)
\end{equation}
The associated gauge potential is
\begin{equation}
{A_{1234}}^T_{(1)}=N_3^{-1}-1
\end{equation}
The expression is the same in world volume indices as $({e^{\hat{h}})^1}_1=\ldots ({e^{\hat{h}})^4}_4=1$ that is
\begin{equation}
{A_{1234}}_{(1)}=N_3^{-1}-1
\end{equation}
This gauge potential gives rise to a 5 form field strength and we conclude this is the dual to the one form field strength formed from $R_1$, ${F_\mu}_1=\partial_\mu R_1$. We have reproduced all the BPS electric branes in the oxidised theory using the group element given in equation (\ref{groupelement}). The formulae we have found above do indeed correspond to the solutions of the Lagrangian (\ref{Flagrangian}). Our dilaton field $A$ is related to $\phi$ by
\begin{equation}
A = -\phi
\end{equation}

\subsection{Higher Level Branes}
In this chapter we generalised the result of \cite{West}, namely that the group element of equation (\ref{groupelement}) encodes the usual electric BPS brane solutions of the $\cal G^{+++}$ theories at low levels. So far we have only considered the low-order field content, up to the level of the dual gravity field. We have found precise agreement with the solutions of the known actions of the oxidised theories \cite{CremmerJuliaLuPope}. The $pp$-wave for each $\cal G^{+++}$ theory is also present in the low order theory and can be found as advocated in \cite{West} from the generators of the $A_{D-1}$ sub-algebra used to form the gravity line. The elegance and generality of the solution generating group element leads us to expect that we can also apply equation (\ref{groupelement}) to the higher order generators of any $\cal G^{+++}$ to find putative solutions to the nonlinearly realised theory. This possibility was considered in \cite{West4}.

We consider the example of the $F_4^{+++}$ theory. The field content at higher levels in $F_4^{+++}$ is given in the tables of \cite{KleinschmidtSchnakenburgWest},  and in particular we find the field ${A_{a_1a_2a_3a_4}}_3$ at level $(3,4)$. This representation has a highest weight generator $R^{3456}_3$ $(0012343)$ and commutes with the dilaton via 
\begin{equation}
[R_0,R^{a_1a_2a_3a_4}_3]=-\frac{1}{\sqrt{2}}R^{a_1a_2a_3a_4}_3
\end{equation}
Applying our method for the root $\beta$ associated with this generator, for which we note that $\beta^2=1$, we find a group element from (\ref{groupelement}) given by
\begin{equation}
g=\exp(-\ln N_3(0({K^1}_1+\ldots {K^4}_4)-({K^5}_5+{K^6}_6)))\exp(N_3^{-1}(1-N_3)E_\beta)\exp(\sqrt{2}\ln N_3R_0)
\end{equation}
We find a dilaton given by
\begin{equation}
e^A=N_3^{\sqrt{2}}
\end{equation}
A line element corresponding to a 3-brane
\begin{equation}
ds^2=(-dt_1^2+\ldots dx_4^2)+N_3(dy_5^2+dy_6^2)
\end{equation}
The brane is derived from a gauge potential
\begin{equation}
{A_{1234}}_{(3)}^T=N_3^{-1}-1={A_{1234}}_{(3)}
\end{equation}
This gives rise to a 5-form field strength, ${F_{\mu\nu\rho\sigma\tau}}_3$. If the dual field strength were to exist we would expect it to be a 1-form formed from a scalar, which we label $R_{-1}$. Such a scalar does not appear in the field content tables of \cite{KleinschmidtSchnakenburgWest} and does not exist but had it done it would commute with the dilaton via $[R_0, R_{-1}]=\frac{1}{\sqrt{8}}R_{-1}$.

As is well known, the eleven dimensional supergravity theory has a $pp$-wave, a 2-brane and a 5-brane whose corresponding conserved charges occur in the eleven dimensional supersymmetry algebra. This is consistent with the results of \cite{AczarragaGauntlettIzquierdoTownsend} which showed that each of the brane solutions had a topological charge that appeared as a central charge in the supersymmetry algebra. In reference \cite{West3}, and in chapter 4 of this thesis, it was argued that the brane solutions of the non-linearly realised $\cal G^{+++}$ theory possessed charges which belonged to the $l_1$ representation of $\cal G^{+++}$. The $l_1$ multiplet of charges for $E_{11}$ in table \ref{l1E11roots} begins with the generators of spacetime translations, $P^a$, and then contains the 2-form and the 5-form central charges of the supersymmetry algebra at the lowest levels, as well as an infinite number of other charges in the higher orders. Let us consider the example of $F_4^{+++}$ and list the fields, $A_{[n]}$, with their corresponding conserved charges, $Z^{[n-1]}$, associated with the brane solution we have found in each case, including the higher level field, ${A_{a_1a_2a_3a_4}}_{(s=3)}$, which we have just considered
\begin{center}
\begin{tabular}{|c|c|c|}
\hline
Levels, $(l_1,l_2)$ & Fields, $A_{[n]}$ & Associated Charges, $Z^{[n-1]}$\\
\hline
$(0,0)$ & ${\hat{h}}_a\hspace{0pt}^b$ & $P^a$\\
$(0,0)$ & $A$ & $-$\\
$(1,0)$ & $A_1$ & $-$\\
$(0,1), (1,1)$ & $A_{a_1}^i$ & $Z^i$\\
$(0,2),(1,2),(2,2)$ & $A_{a_1a_2}^{(ij)}$ & $Z^{a_1(ij)}$\\
$(1,3),(2,3)$ & $A_{a_1a_2a_3}^i$ & $Z^{a_1a_2i}$\\
$(2,4)$ & $A_{a_1a_2a_3,b}$ & $Z^{a_1a_2,b}$\\
$(1,4),(2,4),(3,4)$ & $A_{a_1a_2a_3a_4}^{(ij)}$ & $Z^{a_1a_2a_3(ij)}$\\
\hline
\end{tabular}
\end{center}
\vspace{5pt}
The $F_4^{+++}$ theory possesses an $SL(2)$ symmetry associated with the seventh node of its Dynkin diagram, which has the generator $R_1$, which gives rise to the assignment of the $i,j$ indices above.
We note that while we could not use our method to find the brane solution associated to ${A_{a_1a_2a_3a_4}}_{(s=2)}$, as its generator had associated root, $\beta$, such that $\beta^2=0$, we are able to deduce that its field strength is dual to a 1-form field strength formed from the dilaton, $A$. Indeed we expect its brane solution to have a third rank tensor conserved charge, as listed above.

It is interesting to compare these with the supersymmetric central charges of the $(1,0)$ supersymmetry algebra
\begin{equation}
\left\{Q_\alpha^i,Q_\beta^j\right\}=\epsilon^{ij}(\Gamma_m)_{\alpha\beta}P^m+(\Gamma_{m_1m_2m_3})_{\alpha\beta}Z^{m_1m_2m_3(ij)} \label{SuSy}
\end{equation}
The supersymmetric generator of spacetime translations, $P^m$, is equivalent to the conserved charge arising from the $pp$-wave solution corresponding to the graviton, ${\hat{h}}_a\hspace{0pt}^b$, in the $F_4^{+++}$ field content. Similarly we can make an association between the central charge $Z^{m_1m_2m_3(ij)}$ of the supersymmetry algebra and the conserved charges coming from brane solutions which couple to the fields $A_{a_1a_2a_3a_4}^i$. The 5-form field strengths constructed from ${A_{a_1a_2a_3a_4}}_{(s=1)}$ and ${A_{a_1a_2a_3a_4}}_{(s=2)}$ are dual to the 1-form field strengths formed from the axion, $A_1$, and the dilaton, $A$, respectively, which can be seen by referring to the 3-brane solution found on page \pageref{3-brane} and from considering the low-level content. We have just considered, above in this section, the field, ${A_{a_1a_2a_3a_4}}_{(s=3)}$, whose field strength is a 5-form and its associated brane solution is also a 3-brane. Altogether we have a triplet of conserved charges coming from the non-linear realisation of $F_4^{+++}$ which are equivalent to the central charges, $Z^{m_1m_2m_3(ij)}$, of the supersymmetry algebra. Including the generators of spacetime translations, $P^m$, the conserved charges, $Z^{m_1m_2m_3(ij)}$, of $F_4^{+++}$ account for 36 degrees of freedom which is compatible with the anticommutator of the two $Sp(2)$ Majorana-Weyl spinors, $Q_\alpha^i$. Crucially in order to complete the supersymmetric degrees of freedom we have had to include generators that occur in the algebra at levels above that of the dual to gravity. It is not clear how to interpret this, except to say that the higher level generators must be pertinent to a physical theory.
 
We have constructed branes, even at a lower level than those mentioned above, whose conserved charges in the above table are not present amongst the supersymmetry algebra's central charges; charges such as $Z^i$ and the part of $Z^{a_1(ij)}$ triplet associated with the so-called "gooseberry string", listed in our analysis of $F_4^{+++}$ on page \pageref{Gooseberrystring}. Indeed we can carry on and consider the brane solutions associated with the infinite number of higher-order fields in $F_4^{+++}$, or any $\cal G^{+++}$, and find a conserved charge for each one. We are lead to conclude that the central charges in the supersymmetry algebra only account for a few of the brane charges, all of which belong to the $l_1$ representation.
\newpage
\section{Multiple Signatures of M-theory}
Each of the Kac-Moody algebras we have considered so far has a natural euclidean inner product over the $A_{D-1}$ sub-algebra that we have used to incorporate spacetime in our constructions. Hitherto we have implicitly Wick rotated the generators of this sub-algebra to generate a theory in a familiar $(1,D-1)$ signature. In this chapter we look at the natural consequences of this construction, concentrating on the $E_{11}$ case and the effect on the solution generating group element of equation (\ref{groupelement}) that we discussed in chapter five. Using the results of \cite{Keurentjes} we will reach a surprising conclusion, namely that if $E_{11}$ is the symmetry algebra of $M$-theory then $M$-theory solutions exist in not one spacetime signature but six different signatures \cite{CookWest2}.

The chapter begins by looking at the solutions of a generic gravity theory coupled to an n-form field strength in an arbitrary spacetime signature, and we discover a useful shorthand that determines whether a solution in one signature is a solution in alternative signatures. The method for introducing spacetime signature into the $E_{11}$ formulation is then discussed and the results of \cite{Keurentjes}, where it was observed that the signature of the local subgroup is altered under general Weyl reflections, are reviewed. The Weyl transformations of $E_{11}$ were found to correspond in the dimensionally reduced theory to the U-duality transformations \cite{EnglertHouartTaorminaWest}. We apply the Weyl reflection to the brane solutions uncovered using the group element of chapter five, and make use of our shorthand to discover whether the group element still encodes solutions to the Einstein equations in the new signatures. An exhaustive list of "Weyl orbits" of the brane solutions is given and whether or not they remain solutions is indicated using the results of section \ref{solution_preserving}. The Weyl transformations of some solutions in eleven dimensions were discussed in \cite{West3,EnglertHouart,EnglertHouartTaorminaWest,EnglertHenneauxHouart}. Finally in the remainder of the chapter we observe that the "Weyl orbits" of the brane solutions partition the theory into electric solutions and magnetic, or spacelike, solutions. Using this observation we uncover the known spacelike brane solutions of supergravity using the group element of chapter five. Spacelike brane solutions, or $S$-branes, are branes which map out a spacelike world-volume \cite{GutperleStrominger}. The convention for naming spacelike branes is that an S$p$-brane has a $(p+1)$ Euclidean world-volume. Consequently an $S$-brane only exists for an instant in time. 

\subsection{General Signature Formulation of the Einstein Equations} \label{general_signature_einstein_equations}
In this chapter we will be working beyond the usual signatures of supergravity, and it will be useful to get an appreciation of brane solutions in alternative signatures, for this we will need to express the Einstein equations in a form that is readily applicable to different signatures. The Einstein equations and the gauge equations for a single brane solution can be derived by varying the truncated form of a gravity action, where the truncation is the restriction to the kinetic term for one of the theory's n-form field strengths, the dilaton term and the Ricci scalar, e.g.
\begin{equation}
A=\frac{1}{16\pi G_{D}}\int {d^{D}x\sqrt{-g}(R-\frac{1}{2}\partial^\mu\phi\partial_\mu\phi-\frac{1}{2.n!}e^{a_i\phi}F_{\mu_1\ldots \mu_n}F^{\mu_1\ldots \mu_n})} \label{action}
\end{equation}
The usual Chern-Simons term appearing in the eleven dimensional supergravity action has been omitted here since for the class of extremal branes we will consider it plays no role in the dynamics, and will not affect our discussions. The equations of motion determined by varying with respect to the metric, $g_{\mu\nu}$, are the Einstein equations and the equation that comes from varying the gauge field is the gauge equation. There is also an equation of motion coming from the variation of the dilaton field. For the generic truncated action these are
\begin{align}
\nonumber &{R^\mu}_\nu \nonumber = \frac{1}{2}\partial^\mu\phi\partial_\nu\phi+\frac{1}{2n!}e^{a_i\phi}(nF^{\mu \lambda_2\ldots \lambda_{n}}F_{\nu \lambda_2\ldots \lambda_{n}} -\frac{n-1}{D-2}{\delta^\mu}_\nu F_{\lambda_1\ldots \lambda_{n}}F^{\lambda_1\ldots \lambda_{n}})\\
\nonumber &\partial_\mu(\sqrt{-g}e^{a_i\phi}F^{\mu \lambda_2 \ldots \lambda_{n}})= 0 \\
&\frac{1}{\sqrt{-g}}\partial_\mu(\sqrt{-g}\partial^\mu\phi)-\frac{a_i}{2.n!}e^{a_i\phi}F_{\mu \lambda_2 \ldots \lambda_{n}}F^{\mu \lambda_2 \ldots \lambda_{n}}= 0 \label{EinsteinEq}
\end{align}
We compute the curvature components in the spin-connection formalism as described in \cite{Argurio}, but we commence with a line element for a brane solution in an arbitrary signature. Our solution ansatz is,
\begin{equation}
ds^2=A^2(\sum_{i=1}^{i=q}-dt_{i}^2+\sum_{j=1}^{j=p}dx_{j}^2)+B^2(\sum_{a=1}^{a=c}-du_{a}^2+\sum_{b=1}^{b=d}dy_{b}^2) \label{ansatz}
\end{equation}
The coordinates are split into two groups, those that are longitudinal to the brane, $t_i$ and $x_i$, we indicate with indices $\{i,j,k, \ldots \}$, and those that are transverse, $u_a$ and $y_a$ with $\{a,b,c, \ldots \}$. The given line element is the world-volume of a brane with signature $(q,p)$ on the brane and signature $(c,d)$ in the bulk; the corresponding global spacetime signature is $(q+c,p+d)$. We adopt the notation $[(q,p),(c,d)]$ to express a single signature for our ansatz in terms of its longitudinal, $(q,p)$ and transverse components $(c,d)$. For a global signature $(t,s)$ then $q+c=t$ and $p+d=s$. The case when $c=0$ and $q=1$, corresponds to the usual single brane solution ansatz. In eleven-dimensional supergravity there is no dilaton, if we set the dilaton coupling to zero, the coefficients $A$ and $B$, functions of the transverse coordinates $(u_a,y_a)$, take the form in the extremal case \cite{Argurio, Stelle, Ohta1} 
\begin{equation}
A=N_{(c,d)}^{-(\frac{1}{p+1})},\qquad B=N_{(c,d)}^{(\frac{1}{D-p-3})} \label{singlebranecoefficients}
\end{equation}
Where, 
\begin{equation}
N_{(c,d)}=1+\frac{1}{D-p-3}\sqrt{\frac{\Delta}{2(D-2)}}\frac{\|\bf Q\|}{r^{(D-p-3)}} \label{harmonicfunction}
\end{equation}
Where $\bf Q$ is the conserved charge associated with the $p$-brane solution, $\Delta=(p+1)(d-2)+\frac{1}{2}a_i^2(D-2)$ and $r$ is the radial distance in the transverse coordinates such that $r^2=-u_au_a+y_by_b$. That is, $N_{(c,d)}$ are independent of the longitudinal coordinates, and are harmonic functions in the transverse coordinates, $u_a$ and $y_b$, so that $\partial^\mu\partial_\mu N_{(c,d)}=0$.

The full non-zero curvature components are,
\begin{align}
\nonumber {R^{t_i}}_{t_i}=B^{-2} \{ &\partial_{u_a}\partial_{u_a}\ln{A}{\hat{\delta}^{u_a}}_{\hspace{7pt} u_a}-\partial_{y_a}\partial_{y_a}\ln{A}{\hat{\delta}^{y_a}}_{\hspace{7pt} y_a}\\
\nonumber &+\partial_{u_a}\ln{A}\partial_{u_a}\Psi{\hat{\delta}^{u_a}}_{\hspace{7pt} u_a}-\partial_{y_a}\ln{A}\partial_{y_a}\Psi{\hat{\delta}^{y_a}}_{\hspace{7pt} y_a} \} {\hat{\delta}^{t_i}}_{\hspace{7pt} t_i} \\
\nonumber {R^{x_i}}_{x_i}=B^{-2} \{ &\partial_{u_a}\partial_{u_a}\ln{A}{\hat{\delta}^{u_a}}_{\hspace{7pt} u_a}-\partial_{y_a}\partial_{y_a}\ln{A}{\hat{\delta}^{y_a}}_{\hspace{7pt} y_a}\\
\nonumber&+\partial_{u_a}\ln{A}\partial_{u_a}\Psi{\hat{\delta}^{u_a}}_{\hspace{7pt} u_a}-\partial_{y_a}\ln{A}\partial_{y_a}\Psi{\hat{\delta}^{y_a}}_{\hspace{7pt} y_a} \} {\hat{\delta}^{x_i}}_{\hspace{7pt} x_i} \\
\nonumber {R^{u_a}}_{u_a}=B^{-2} \{&\partial_{u_a}\partial_{u_a}\ln{B}{\hat{\delta}^{u_a}}_{\hspace{7pt} u_a}-\partial_{y_a}\partial_{y_a}\ln{B}{\hat{\delta}^{y_a}}_{\hspace{7pt} y_a}\\
 &+\partial_{u_a}\ln{B}\partial_{u_a}\ln{B}({\hat{\delta}^{y_a}}_{\hspace{7pt}y_a}+{\hat{\delta}^{u_a}}_{\hspace{7pt} u_a}-2)\\
\nonumber &+\partial_{u_a}\ln{A}\partial_{u_a}\ln{A}({\hat{\delta}^{t_i}}_{\hspace{7pt}t_i}+{\hat{\delta}^{x_i}}_{\hspace{7pt} x_i})
-\partial_{y_a}\ln{B}\partial_{y_a}\Psi{\hat{\delta}^{y_a}}_{\hspace{7pt} y_a}\\
\nonumber &+\partial_{u_a}\partial_{u_a}\Psi+\partial_{u_a}\ln{B}\partial_{u_a}\Psi({\hat{\delta}^{u_a}}_{\hspace{7pt} u_a}-2)\}{\hat{\delta}^{u_a}}_{\hspace{7pt} u_a}\\
\nonumber {R^{y_a}}_{y_a}=B^{-2} \{&\partial_{u_a}\partial_{u_a}\ln{B}{\hat{\delta}^{u_a}}_{\hspace{7pt} u_a}-\partial_{y_a}\partial_{y_a}\ln{B}{\hat{\delta}^{y_a}}_{\hspace{7pt} y_a}\\
\nonumber &-\partial_{y_a}\ln{B}\partial_{y_a}\ln{B}({\hat{\delta}^{y_a}}_{\hspace{7pt}y_a}+{\hat{\delta}^{u_a}}_{\hspace{7pt} u_a}-2)\\
\nonumber &-\partial_{y_a}\ln{A}\partial_{y_a}\ln{A}({\hat{\delta}^{t_i}}_{\hspace{7pt}t_i}+{\hat{\delta}^{x_i}}_{\hspace{7pt} x_i})
+\partial_{u_a}\ln{B}\partial_{u_a}\Psi{\hat{\delta}^{u_a}}_{\hspace{7pt} u_a}\\
\nonumber &-\partial_{y_a}\partial_{y_a}\Psi-\partial_{y_a}\ln{B}\partial_{y_a}\Psi({\hat{\delta}^{y_a}}_{\hspace{7pt} y_a}-2)\}{\hat{\delta}^{y_a}}_{\hspace{7pt} y_a}
\end{align}
Where ${\hat{\delta}^{x_i}}_{\hspace{7pt} x_i}$ counts the number of $x_i$ coordinates in the line element (e.g. for the $M2$-brane, ${\hat{\delta}^{t_i}}_{\hspace{7pt} t_i}=1$, ${\hat{\delta}^{x_i}}_{\hspace{7pt} x_i}=2$, ${\hat{\delta}^{u_a}}_{\hspace{7pt} u_a}=0$ and ${\hat{\delta}^{y_a}}_{\hspace{7pt} y_a}=8$); repeated lowered indices a, b, i and j are not summed - sums are taken care of via the counting symbols $\hat{\delta}$; and, for the ansatz (\ref{ansatz}) with extremal coefficients A and B (\ref{singlebranecoefficients}),
\begin{align}
\nonumber \Psi\equiv &(p+1)\ln{A}+(D-p-3)\ln{B}\\
=&(p+1)\ln{N_p^{-2(\frac{1}{p+1})}}+(D-p-3)\ln{N_p^{2(\frac{1}{D-p-3})}}\\
\nonumber =&0
\end{align}
The curvature terms reduce to
\begin{align}
\nonumber {R^{t_i}}_{t_i}=B^{-2} \{ &\partial_{u_a}\partial_{u_a}\ln{A}{\hat{\delta}^{u_a}}_{\hspace{7pt} u_a}-\partial_{y_a}\partial_{y_a}\ln{A}{\hat{\delta}^{y_a}}_{\hspace{7pt} y_a} \} {\hat{\delta}^{t_i}}_{\hspace{7pt} t_i} \\
\nonumber {R^{x_i}}_{x_i}=B^{-2} \{ &\partial_{u_a}\partial_{u_a}\ln{A}{\hat{\delta}^{u_a}}_{\hspace{7pt} u_a}-\partial_{y_a}\partial_{y_a}\ln{A}{\hat{\delta}^{y_a}}_{\hspace{7pt} y_a} \} {\hat{\delta}^{x_i}}_{\hspace{7pt} x_i} \\
\nonumber {R^{u_a}}_{u_a}=B^{-2} \{&\partial_{u_a}\partial_{u_a}\ln{B}{\hat{\delta}^{u_a}}_{\hspace{7pt} u_a}-\partial_{y_a}\partial_{y_a}\ln{B}{\hat{\delta}^{y_a}}_{\hspace{7pt} y_a}\\
&+\partial_{u_a}\ln{B}\partial_{u_a}\ln{B}({\hat{\delta}^{y_a}}_{\hspace{7pt}y_a}+{\hat{\delta}^{u_a}}_{\hspace{7pt} u_a}-2) \label{curvatureterms}\\
\nonumber &+\partial_{u_a}\ln{A}\partial_{u_a}\ln{A}({\hat{\delta}^{t_i}}_{\hspace{7pt}t_i}+{\hat{\delta}^{x_i}}_{\hspace{7pt} x_i})\}{\hat{\delta}^{u_a}}_{\hspace{7pt} u_a}\\
\nonumber {R^{y_a}}_{y_a}=B^{-2} \{&\partial_{u_a}\partial_{u_a}\ln{B}{\hat{\delta}^{u_a}}_{\hspace{7pt} u_a}-\partial_{y_a}\partial_{y_a}\ln{B}{\hat{\delta}^{y_a}}_{\hspace{7pt} y_a}\\
\nonumber &-\partial_{y_a}\ln{B}\partial_{y_a}\ln{B}({\hat{\delta}^{y_a}}_{\hspace{7pt}y_a}+{\hat{\delta}^{u_a}}_{\hspace{7pt} u_a}-2)\\
\nonumber &-\partial_{y_a}\ln{A}\partial_{y_a}\ln{A}({\hat{\delta}^{t_i}}_{\hspace{7pt}t_i}+{\hat{\delta}^{x_i}}_{\hspace{7pt} x_i})
\}{\hat{\delta}^{y_a}}_{\hspace{7pt} y_a}
\end{align}
As demonstrated in chapter five, the $M2$, $M5$, and $pp$-wave solutions \cite{BPSsolutions} of $M$-theory are encoded in a group element of the non-linear realisation of $E_{11}$ \cite{West}. For reference and comparison with later solutions, we demonstrate that these electric cases are solutions of the Einstein equations in appendix \ref{electricbranesolutions}.

\subsection{New Solutions from Signature Change} \label{solution_preserving}
A change of signature has the potential to alter both the Einstein equations and the field content of a theory. In preparation for the next section, we pose a question: given a solution to the Einstein equations in one signature, are there any other signatures which would also carry a related version of that solution in the new signature? We note that the harmonic function, $N_{(c,d)}$, is a function of the transverse coordinates and may be transformed by a signature change. By a 'related solution' we specifically mean that if the Einstein equations for a given solution were re-expressed in terms of the functions carrying the new signature then they would remain balanced and we would find a new solution.

Signature change can be brought about in two equivalent ways, the first is as a mapping of a coordinate, or a subset of the coordinates, $x^\mu\rightarrow ix^\mu, x_\mu\rightarrow -i x_\mu$, leaving the metric unaltered. For example, a Lorentzian signature can be made Euclidean by making the change on the temporal coordinate, $t^1\rightarrow ix^1$, having the effects,
\begin{align}
\nonumber ds^2 &= g_{\mu\nu}x^\mu x^\nu \\
\nonumber &=g_{t_1t_1}dt^2_1+g_{x_2x_2}dx^2_2+\ldots +g_{x_Dx_D}dx^2_D \\
\nonumber &=-f_1(N)dt^2_1+f_2(N)dx^2_2+\ldots +f_D(N)dx^2_D \\
& \downarrow \label{euclideanisation}\\
\nonumber ds^2&=f_1(N)dx^2_1+f_2(N)dx^2_2+\ldots +f_D(N)dx^2_D \\
&\nonumber = g_{x_1x_1}dx^2_1+g_{x_2x_2}dx^2_2+\ldots +g_{x_Dx_D}dx^2_D
\end{align}
Where $\pm f_i(N)$, some function of $N$, is the metric component in each case. An electric field $A_\mu$ transforms as
\begin{equation}
A_\mu dx^\mu = A_{t_1}dt^1 \rightarrow iA_{x_1}dx^1 \label{Wickrotation}
\end{equation}
Equivalently, signature change at the quadratic level of the line element can be thought of as a transformation of the metric components, or subset of the metric components, where appropriate, $g_{\mu\nu}\rightarrow -g_{\mu\nu}$ as opposed to the coordinates. Correspondingly we can view the example above (\ref{euclideanisation}), as the transformation $g_{t_1t_1}\rightarrow -g_{x_1x_1}$. Both methods realise the change of line element but only when applied independently.

In equation (\ref{curvatureterms}) we have written out the curvature coefficients for our ansatz, (\ref{ansatz}). We observe that the expressions for ${R^{t_i}}_{t_i}$ and ${R^{x_i}}_{x_i}$ satisfying our explicit ansatz of (\ref{ansatz}) are interchanged under the interchange of longitudinal temporal and spatial coordinates, given by $t^i\rightarrow ix^i$ and $x^j\rightarrow it^j$. These transformations corresponds to the notational swap $t_i\leftrightarrow x_i$ in the equations (\ref{curvatureterms}) and the set of curvature terms as a whole is unaffected. In terms of signature this corresponds to a signature inversion on only the longitudinal coordinates. Alternatively, the swap $u_a$ for $y_a$ and vice-versa, given by $u^a\rightarrow iy^a$ and $y^b\rightarrow iu^b$ and corresponding to a signature inversion on only the transverse coordinates, interchanges the expressions for ${R^{u_a}}_{u_a}$ and ${R^{y_a}}_{y_a}$ and introduces a minus into all the curvature components, ${R^\mu}_\nu\rightarrow -{R^\mu}_\nu$. One could also achieve these signature changes by transforming the metric components: a signature inversion on all of the longitudinal coordinates $g^{t_it_i}\rightarrow -g^{x_ix_i}$ and $g^{x_ix_i}\rightarrow -g^{t_it_i}$ leaves the set of curvature terms unaltered, whereas $g^{u_au_a}\rightarrow -g^{y_ay_a}$ and $g^{y_ay_a}\rightarrow -g^{u_au_a}$ introduces a minus sign for all the curvature terms.

To find a new solution under longitudinal and transverse signature inversions the signs induced in the curvature components must match the sign changes in the remaining terms of the Einstein equations, those derived from the field strength and the dilaton. For the eleven dimensional case, which we consider in this paper, there is no dilaton, $\phi\rightarrow 0$, $a_i\rightarrow 0$, so we shall disregard it in the following discussion.

The field strength terms in each of the Einstein equations for the usual single brane solutions, with only one temporal coordinate longitudinal to the brane and none transverse, given in appendix \ref{electricbranesolutions}, are proportional to 
\begin{equation}
g^{tt'}g^{x_{1}x_{1}'}\ldots g^{x_{p}x_{p}'}g^{y_ay_a'}(F_{t'x_{1}'\ldots x_{p}'y_a'})(F_{tx_{1}\ldots x_{p}y_a})
\end{equation}
With the generalisation to our ansatz (\ref{ansatz}) to include multiple time coordinates longitudinal and transverse to the brane, the equivalent proportional term is 
\begin{equation}
g^{t_{1}t_{1}'}\ldots g^{t_{q}t_{q}'}g^{x_{1}x_{1}'}\ldots g^{x_{p}x_{p}'}g^{\mu \mu'}(F_{t_{1}'\ldots t_{q}'x_{1}'\ldots x_{p}'\mu'})(F_{t_{1}\ldots t_{q}x_{1}\ldots x_{p}\mu}) \label{rhs}
\end{equation}
Where the radial coordinate $\mu$ may now be a spatial or temporal transverse coordinate. An inversion of the longitudinal coordinates only, causes a sign change $(-1)^{p+q}$ in this term. The effect of inverting only the transverse coordinates $g^{\mu\mu'} \rightarrow -g^{\mu\mu'}$ introduces a minus sign as there is only one occurrence of the metric component with transverse coordinates in (\ref{rhs}).

A new solution is found under a signature inversion when the sign changes induced in the Riemann curvature components match the sign changes in the term (\ref{rhs}). For example, since we have observed that a signature inversion on only the transverse coordinates introduces a minus sign in both the curvature components and the remaining terms in the Einstein equations (\ref{rhs}), then a solution with signature components $[(q,p),(c,d)]$, will always find a new solution under an inversion of the transverse signature, taking the signature to $[(q,p),(d,c)]$. Additionally, if $p+q$ is even we may invert just the longitudinal coordinates and find another new solution $[(q,p),(c,d)]\rightarrow[(p,q),(c,d)]$, and furthermore in this case we may invert the full signature, an inversion of both longitudinal and transverse coordinates together,  $[(q,p),(c,d)]\rightarrow[(p,q),(c,d)]$ and find yet another new solution. However if $p+q$ is odd an inversion of the longitudinal signature introduces an unbalanced minus sign and no new solution is found.\footnote{However, we shall observe later that the $F^2$ term in the action may change its sign and in such cases the 'lost' solution of a $-F^2$ theory is a new solution of a $+F^2$ theory, but for the time being we continue to consider the usual $-F^2$ theory} For example, a solution in $(1,D-1)$ with longitudinal and transverse signature components $[(1,p),(0,D-p-1)]$ is the familiar $p$-brane solution. We find that for even $p+q$, we have the following set of signature components that carry a related solution, $[(1,p),(D-p-1,0)]$, $[(p,1),(0,D-p-1)]$ and $[(p,1),(D-p-1,0)]$ which give spacetime signatures $(D-p,p)$, $(p, D-p)$ and $(1,D-1)$ respectively. For the case of odd $(p+q)$ we have only one alternative signature, coming from an inversion of only the transverse coordinates, which gives a new solution, namely $(D-p,p)$ with longitudinal and transverse components $[(1,p),(D-p-1,0)]$.

Following the preceding considerations we are in a position to write down a set of signatures in which the $M2$ and $M5$ branes remain solutions. The $M2$ brane has $p+q=3$ and hence has related solutions in $(1,10)$ and $(9,2)$, with the following parameters,
\begin{center}
\begin{tabular}{|c|c|c|}
\hline
Global & Longitudinal & Transverse \\
Signature & Signature & Signature \\
\hline
$(1,10)$ & $(1,2)$ & $(0,8)$\\
$(9,2)$ & $(1,2)$ & $(8,0)$\\
\hline
\end{tabular}
\end{center}
Equivalently, the $M5$ brane has $p+q=6$ and has related solutions in $(1,10)$, $(6,5)$, $(5,6)$ and $(10,1)$,
\begin{center}
\begin{tabular}{|c|c|c|}
\hline
Global & Longitudinal & Transverse \\
Signature& Signature & Signature \\
\hline
$(1,10)$ & $(1,5)$ & $(0,5)$\\
$(6,5)$ & $(1,5)$ & $(5,0)$\\
$(5,6)$ & $(5,1)$ & $(0,5)$\\
$(10,1)$ & $(5,1)$ & $(5,0)$\\
\hline
\end{tabular}
\end{center}
In addition to inverting components of the signature we may also transform individual temporal coordinates into spacelike coordinates and vice versa. It is observed by following the computations in appendix \ref{electricbranesolutions} and the term in equation (\ref{rhs}), that a given solution will give a new solution by converting an even number of longitudinal temporal coordinates into longitudinal spatial coordinates, while leaving the transverse coordinates unaltered. Such a transformation introduces a sign change $(-1)^{2m}=1$ into term (\ref{rhs}), where $m$ is an integer such that $q\pm2m,p\mp2m\geq 0$, and only alters the counting symbols ${\hat{\delta}^{t_i}}_{\hspace{7pt} t_i}$ and ${\hat{\delta}^{x_i}}_{\hspace{7pt} x_i}$ in the curvature components so that a new solution is found. That is, given a solution in a signature with components $[(q,p),(c,d)]$ then functions carrying the signature $[(q\pm2m,p\mp2m),(c,d)]$ will give a new solution. Applying this to each of the signatures containing solutions related to the $M2$ brane, we find the additional solutions,
\begin{center}
\begin{tabular}{|c|c|c|}
\hline
Global& Longitudinal & Transverse \\
Signature & Signature & Signature \\
\hline
$(3,8)$ & $(3,0)$ & $(0,8)$\\
$(11,0)$ & $(3,0)$ & $(8,0)$\\
\hline
\end{tabular}
\end{center}
And similarly, additional signatures for the $M5$ solutions are,
\begin{center}
\begin{tabular}{|c|c|c|}
\hline
Global & Longitudinal & Transverse \\
Signature& Signature & Signature \\
\hline
$(3,8)$ & $(3,3)$ & $(0,5)$\\
$(8,3)$ & $(3,3)$ & $(5,0)$\\
\hline
\end{tabular}
\end{center}
We can carry this argument to the transverse coordinates, but we are no longer restricted to transforming even multiples of temporal coordinates into spatial coordinates, indeed any integer is possible giving a range of new solutions in signatures $[(q,p),(c\pm m, d \mp m)]$ where m is an integer such that $c\pm m, d \mp m \geq 0$. For the solutions related to the $M2$-brane we find further solutions,
\begin{center}
\begin{tabular}{|c|c|c|}
\hline
Global& Longitudinal & Transverse \\
Signature & Signature & Signature \\
\hline
$(2,9)$ & $(1,2)$ & $(1,7)$\\
$(3,8)$ & $(1,2)$ & $(2,6)$\\
$(4,7)$ & $(1,2)$ & $(3,5)$\\
$(5,6)$ & $(1,2)$ & $(4,4)$\\
$(6,5)$ & $(1,2)$ & $(5,3)$\\
$(7,4)$ & $(1,2)$ & $(6,2)$\\
$(8,3)$ & $(1,2)$ & $(7,1)$\\
$(4,7)$ & $(3,0)$ & $(1,7)$\\
$(5,6)$ & $(3,0)$ & $(2,6)$\\
$(6,5)$ & $(3,0)$ & $(3,5)$\\
$(7,4)$ & $(3,0)$ & $(4,4)$\\
$(8,3)$ & $(3,0)$ & $(5,3)$\\
$(9,2)$ & $(3,0)$ & $(6,2)$\\
$(10,1)$ & $(3,0)$ & $(7,1)$\\
\hline
\end{tabular}
\end{center}
And for the $M5$ solution we find the further related solutions,
\begin{center}
\begin{tabular}{|c|c|c|}
\hline
Global & Longitudinal & Transverse \\
Signature& Signature & Signature \\
\hline
$(2,9)$ & $(1,5)$ & $(1,4)$\\
$(3,8)$ & $(1,5)$ & $(2,3)$\\
$(4,7)$ & $(1,5)$ & $(3,2)$\\
$(5,6)$ & $(1,5)$ & $(4,1)$\\
$(4,7)$ & $(3,3)$ & $(1,4)$\\
$(5,6)$ & $(3,3)$ & $(2,3)$\\
$(6,5)$ & $(3,3)$ & $(3,2)$\\
$(7,4)$ & $(3,3)$ & $(4,1)$\\
$(5,6)$ & $(5,1)$ & $(1,4)$\\
$(7,4)$ & $(5,1)$ & $(2,3)$\\
$(8,3)$ & $(5,1)$ & $(3,2)$\\
$(9,2)$ & $(5,1)$ & $(4,1)$\\
\hline
\end{tabular}
\end{center}
This discussion is exhaustive, we have found all signatures that give solutions in $-F^2$ theories related to the $M2$ and $M5$-brane solutions of $M$-theory. We note that a universal shorthand for assessing whether or not a given signature contains a solution is to count the number of temporal longitudinal coordinates and if this is odd we have a solution to $-F^2$ theories. 

Later in this chapter we will consider theories constructed from an action under the double Wick rotation transforming $A_{\mu_1\ldots\mu_n}\rightarrow -iA_{\mu_1\ldots\mu_n}$. Such an action, upto Chern-Simons terms, looks identical to that of equation (\ref{action}) except that the sign of the kinetic $F^2$ term has changed from "$-$" to "$+$". Such a theory can be imagined as originating with an imaginary brane charge and will be relevant to our later consideration of spacelike branes. Let us now apply the reasoning of this section to such a $+F^2$ action. Adjusting the considerations of this section to $+F^2$ theories introduces an extra minus sign in front of the field strength terms in the Einstein equations (\ref{EinsteinEq}) and as a consequence into the term proportional to the non-curvature components given in equation (\ref{rhs}). In this case we will have a solution if the number of longitudinal time coordinates is even, meaning that exactly all the signatures not listed in this section, from the set of all signatures in 11-dimensions, will admit solutions to a $+F^2$ theory. 

Let us introduce a new term $\kappa$ to keep track of solutions in both $+F^2$ and $-F^2$ theories. For reasons that will become clear we use $f(\alpha_{11})$ to indicate the sign in front of the $F^2$ term, if $f(\alpha_{11})=0$ we have a $-F^2$ theory and if $f(\alpha_{11})=1$ we have a $+F^2$ theory. We define $\kappa$ to be
\begin{equation}
\kappa\equiv(-1)^{({\hat{\delta}^{t_i}}_{\hspace{7pt}t_i}+f(\alpha_{11}))} \label{kappa}
\end{equation}
If $\kappa=-1$ we have a brane solution, otherwise we do not; this criterion will be used to check for solutions throughout this paper. We note that this implies that there are no extremal $S$-branes (${\hat{\delta}^{t_i}}_{\hspace{7pt}t_i}=0$) in $-F^2$ theories ($f(\alpha_{11})=0$) indicating the known result \cite{BhattacharyaRoy,GutperleSabra} that $S$-branes in $M$-theory have an associated imaginary charge $||{\bf Q}||$ in equation (\ref{harmonicfunction}). More simply, the transformation $||{\bf Q}||\rightarrow i||{\bf Q}||$ induces the transformation $-F^2\rightarrow +F^2$.

\subsection{Spacetime Signature and Weyl Reflections} \label{local_sub_algebra_and_weyl}
So far in this thesis we have motivated the idea of forming a nonlinear theory on a coset space $\frac{\cal{G}}{\cal{H}}$ and we have looked at several particular choices of local sub-algebra, $\cal{H}$ in chapter three when motivating the $E_{11}$ conjecture. We have not mentioned how one may determine the local sub-algebra in any general way, and here we simply state that the local denominator sub-algebra, $H$, may be chosen to be Cartan involution invariant. Making this choice yields the maximal compact sub-algebra. It was understood that a Wick rotation would then give a Lorentz invariant non-compact sub-algebra, relevant to spacetime. The Cartan involution, $\Omega$, takes the generators of the positive roots, $E_i={K^i}_{i+1}$, to the negative of the generators of the negative roots, $-F_{i}=-{K^{i+1}}_i$ and vice-versa,
\begin{equation}
\Omega{(E_i)}=-F_i \qquad \text{and} \qquad \Omega{(F_i)}=-E_i
\end{equation}
Such that the set of  generators, $E_i-F_i$, is invariant under the Cartan involution, $\Omega(E_i-F_i)=-F_i-(-E_i)=E_i-F_i$, and form a basis for the local denominator sub-algebra, $H$. It was noted \cite{EnglertHouart} that the Cartan involution can be generalised to what has been called the temporal involution, $\hat{\Omega}$, whose action is:
\begin{equation}
\hat{\Omega}({K^i}_{i+1})=-\epsilon_i{K^{i+1}}_i
\end{equation}
Where $\epsilon_i=\pm 1$. The generalisation redefines $H$ to be the sub-algebra left invariant under the temporal involution, as opposed to the Cartan involution; we denote the local sub-algebra invariant under the temporal involution as $\hat H$. This redefinition allows $\hat H$ to include non-compact generators, and in this way to differentiate between temporal and spatial coordinates, imposing a signature on the sub-algebra and a Lorentzian invariance. The information about which coordinates are timelike is carried by the new variable $\epsilon_i$. For example, we may impose a $(1,10)$ signature where $x^1$ is the temporal coordinate, by taking $\epsilon_1=-1$ and $\epsilon_i=1$ for the remaining spatial coordinates, giving:
\begin{equation}
\hat{\Omega}({K^1}_2)={K^2}_1, \quad \hat{\Omega}({K^i}_{i+1})=-{K^{i+1}}_i \qquad \text{For} \qquad 2\leq i \leq 10
\end{equation}
We find a basis for the local denominator algebra consisting of both compact and non-compact generators that is invariant under the temporal involution:
\begin{align}
\nonumber \hat{\Omega}(E_1+F_1)&=-\epsilon_1F_1-\epsilon_1E_1=E_1+F_1 \\
\hat{\Omega}(E_i-F_i)&=-\epsilon_iF_i+\epsilon_iE_i=E_i-F_i \qquad 2\leq i \leq 10
\end{align}
It was observed in \cite{Keurentjes, Keurentjes2} that the temporal involution does not commute with the Weyl reflections, $S_i$, which are defined in equation (\ref{Weylreflection}). Under the action of the Weyl group, the choice of local sub-algebra is not preserved and we obtain a new set of $\epsilon_i$'s, corresponding to a different temporal involution $\hat\Omega'$. This idea is described in detail elsewhere \cite{Keurentjes, Keurentjes2, EnglertHouart} and may be summarised as
\begin{align}
S_i\hat\Omega({K^j}_{j+1})=S_i(\epsilon_j{K^j}_{j+1})=\epsilon_j\rho_j{K^{j+1}}_j\equiv\hat\rho_j\Omega'({K^j}_{j+1})
\end{align}
Where $S_i{K^j}_{j+1}=\rho_j{K^{j+1}}_j$ and $\rho_j=\pm1$ arises because ${K^j}_{j+1}$ are representations of the Weyl group up to a sign. A consequence of the new temporal involution is that a new set of compact and non-compact generators form the basis of the local denominator sub-algebra $\hat H'_{11}$, potentially corresponding to a new set of temporal and spatial coordinates and signature. 

The Weyl reflections corresponding to nodes on the the gravity line of $E_{11}$ preserve the signature of spacetime while the reflection in the exceptional root $\alpha_{11}$ can change it. Keurentjes makes use of a ${\mathbb Z}_2$ valued function on the root lattice, $f$, which encodes the values of the $\epsilon_i$'s. We may regard the function $f$ as a member of the weight space, it may be written 
\begin{equation}
f=\sum_{i=1}^{11} n_i\lambda_i
\end{equation}
Where $\lambda_i$ are the fundamental weights of $E_{11}$. Its action on the simple roots, via the inner product, is 
\begin{equation}
f(\alpha_i)\equiv<f,\alpha_i>=n_i
\end{equation}
Where $n_i$ take the values $0$ or $1$, a value $f(\alpha_i)=1$ corresponds to a Chevalley generator ${K^i}_{i+1}$ which has $\epsilon_i=-1$. Put more simply, one of $x^i$ or $x^{i+1}$ is timelike and the other spacelike. Alternatively, $f(\alpha_i)=0$ implies that the consecutive coordinates, $x^i$ and $x^{i+1}$, are of the same type, either both timelike or both spacelike. It is worth highlighting that the root associated with the group element of chapter five (\ref{groupelement}) does not determine $f$, to obtain a putative solution we must specify both a position in the root lattice (a root) and a vector ($f$) encoding the signature; hitherto the group element has been used to find solutions in signature $(1,10)$.

Keurentjes offers a useful shorthand notation for following the effect of Weyl reflections upon the function $f$, which we will describe here before making some use of it. The values of the weight, $f$, are written out on the Dynkin diagram, with the value $f(\alpha_i)$ written in the position of $\alpha_i$ on the diagram. As an example a $(1,10)$ signature might have the diagram:
\begin{align}
\nonumber &0\\
\nonumber 0\hspace{5pt} 0\hspace{5pt} 0\hspace{5pt} 1\hspace{5pt} 1\hspace{5pt} 0\hspace{5pt} 0\hspace{5pt} &0\hspace{5pt} 0\hspace{5pt} 0\\
\nonumber s\hspace{6pt} s\hspace{6pt} s\hspace{6pt} s\hspace{6pt} t\hspace{6pt} s\hspace{6pt} s\hspace{6pt} s&\hspace{6pt} s\hspace{6pt} s\hspace{6pt} s
\end{align}
We have indicated beneath the diagram with a series of $s$ (spatial) and $t$'s (temporal) the nature of the coordinates obtained by commencing with a spacelike $x^1$ on the far left of the gravity line. But we may also consider the case where $x^1$ is a temporal coordinate, giving a mostly timelike set of coordinates. In general each signature diagram is ambiguous, representing both $(t,s,\pm)$ and $(s,t,\pm)$, but as we shall see these signatures do not always contain related solutions, so some care must be taken to specify the nature of one of the coordinates so that a signature diagram is not ambiguous. 

The ten values of $f$ on the gravity line gives the nature of all eleven dimensions of spacetime. The value of $f(\alpha_{11})$ plays no part in this, but it is argued in \cite{Keurentjes} that it determines the sign in front of the kinetic term in the action derived from the non-linear realisation. Specifically, we impose the choice that $f(\alpha_{11})=0$ corresponds to the usual minus sign in front of $F^2$, while the alternative implies a plus sign in front of the $F^2$ term. It will be useful to follow the convention and label our signatures as $(t,s,\pm)$, where we will always write the number of timelike coordinates first and where, a little confusingly, $'+'$ implies that the sign of $F^2$ is negative, and $'-'$ that our action has a positive $F^2$ term. 
Let us find the effect of a Weyl reflection, $S_i$, on the function $f$, we have,
\begin{align}
\nonumber S_i(f)&=\sum_jn_j\lambda_j -\sum_jn_j(\alpha_i,\lambda_j)\alpha_i\\
&\equiv \sum_jm_j\lambda_j
\end{align}
Where $m_j$ are the components of $S_i(f)$. Taking the inner product with $\alpha_k$ gives the relation between $n_k$ and $m_k$
\begin{align}
m_k=n_k-n_iA_{ki}
\end{align}
Thus we find Keurentjes' diagrammatic prescription for following signature change: to apply $S_i$ to $f$ we simply add the value of $f(\alpha_i)$ to all the nodes it is connected to, and subsequently reduce modulo two. The reduction modulo two comes from the size of the fundamental lattice which has edge length $\left|-\alpha_i\right|+\left|+\alpha_i\right|=2\left|\alpha_i\right|$, so the sub-algebra repeats with this unit and we need only consider a version of it upto $\pm n2\left|\alpha_i\right|, n \in {\textbf{Z}}$. For example in the signature diagram above, $S_i$ where $ i=\{ 1\ldots 3,6\ldots 11 \}$ have no effect upon the signature, whereas $S_4$ and $S_5$ bring about the following two signature diagrams respectively.
\begin{align}
\nonumber &0\\
\nonumber 0\hspace{5pt} 0\hspace{5pt} 1\hspace{5pt} 1\hspace{5pt} 0\hspace{5pt} 0\hspace{5pt} 0\hspace{5pt} &0\hspace{5pt} 0\hspace{5pt} 0\\
\nonumber s\hspace{6pt} s\hspace{6pt} s\hspace{6pt} t\hspace{6pt} s\hspace{6pt} s\hspace{6pt} s\hspace{6pt} s&\hspace{6pt} s\hspace{6pt} s\hspace{6pt} s\\
\nonumber &0\\
\nonumber 0\hspace{5pt} 0\hspace{5pt} 0\hspace{5pt} 0\hspace{5pt} 1\hspace{5pt} 1\hspace{5pt} 0\hspace{5pt} &0\hspace{5pt} 0\hspace{5pt} 0\\
\nonumber s\hspace{6pt} s\hspace{6pt} s\hspace{6pt} s\hspace{6pt} s\hspace{6pt} t\hspace{6pt} s\hspace{6pt} s&\hspace{6pt} s\hspace{6pt} s\hspace{6pt} s
\end{align}
\subsection{Invariant Roots and Signature Orbits}
In chapter five we described a method for finding single brane solutions from the decomposition of $E_{11}$ with respect to its gravity line, encoded within the group element (\ref{groupelement}). The prescription for finding the brane solution in the original literature \cite{West,CookWest} made use of the lowest weight generator in to find a brane solution, so that the coordinate $x^1$ could be identified with time. It may have been imagined that it would be possible to raise the weight of a generator and remove the timelike index from the associated gauge field so as to uncover spacelike brane solutions. However, other weights associated to a particular generator could be reached from the lowest weight by a series of Weyl reflections, which in the light of the previous discussion implies a potential signature change. If we commence with the $R^{123}$ and $R^{123456}$ generators in $(1,10,+)$, where $x^1$ is the time coordinate, and raise them to their highest weights with the Weyl reflections $S^{-1}_0=(S_1)(S_2S_1)\ldots(S_{10}S_9\ldots S_1)$\footnote{$S_0=(S_{1}\ldots S_{10})\ldots(S_1S_2)(S_1)$ is the series of Weyl reflections that takes the highest weight to the lowest weight for all representations of $A_{10}$, so that $S_0R^{91011}=R^{123}$ and $S_0R^{67891011}=R^{123456}$.}, we find that the effect on the signature diagram is,
\begin{alignat}{2}
\nonumber &0& &1\\
 1\hspace{5pt} 0\hspace{5pt} 0\hspace{5pt} 0\hspace{5pt} 0\hspace{5pt} 0\hspace{5pt} 0\hspace{5pt} &0\hspace{5pt} 0\hspace{5pt} 0 &\qquad \stackrel{S^{-1}_0}{\longrightarrow} \qquad 0\hspace{5pt} 0\hspace{5pt} 0\hspace{5pt} 0\hspace{5pt} 0\hspace{5pt} 0\hspace{5pt} 0\hspace{5pt} &0\hspace{5pt} 0\hspace{5pt} 1 \label{signaturediagram}\\
\nonumber t\hspace{6pt} s\hspace{6pt} s\hspace{6pt} s\hspace{6pt} s\hspace{6pt} s\hspace{6pt} s\hspace{6pt} s&\hspace{6pt} s\hspace{6pt} s\hspace{6pt} s& t\hspace{7pt} t\hspace{7pt} t\hspace{7pt} t\hspace{7pt} t\hspace{7pt} t\hspace{7pt} t\hspace{7pt} t&\hspace{7pt} t\hspace{7pt} t\hspace{7pt} s
\end{alignat}
Up to insisting that $x^1$ is preserved as a time coordinate, we obtain the signature $(10,1,-)$, where the singled out spatial coordinate is in the longitudinal sector, specifically the spatial coordinate here is $x^{11}$, and the gauge fields in each case are $A_{91011}$ and $A_{67891011}$. From the observations of section \ref{solution_preserving}, it is known that the $M2$ brane has a related solution in $(10,1,-)$ with signature components $[(2,1),(8,0)]$, alternatively the $M5$ brane does not have a related solution $[(5,1),(5,0)]$ in $(10,1,-)$. We note that in chapter five we used the highest weights of the low-level generators of $E_{11}$ to find the $M2$ and $M5$ brane solutions. This may seem in contradiction to the above statements but we also used a different choice of the signature determining weight $f$. We will discuss the alchemy of transforming an electric gauge field into a magnetic one in detail in section \ref{spacelikeinvolution}.

Some Weyl reflections may change the signature diagram, and even the signature, without changing the root, so there is an ambiguity about which signature the root and its associated solution exist in. We now consider the example of the exceptional root $\alpha_{11}$, associated with the generator $R^{91011}$, the highest weight of the `$M2$ representation' and find what we shall refer to as its signature orbit.

\subsubsection{Membrane Solution Signatures} \label{membranesolutionsignatures}

Let us first consider the trivial Weyl reflections of the gravity line on the root $\alpha_{11}$. It is noted that $\alpha_{11}$ is invariant under the reflections $\{S_1,\ldots S_7\}$ and $\{S_9,S_{10}\}$ and we may apply any number of these reflections, without altering the root, although we may trivially change the signature diagram but not the signature. Furthermore, we can observe from the signature diagram of the highest weight, shown on the right of (\ref{signaturediagram}), that as only $f(\alpha_{10})=1$ along the gravity line, only a series of reflections composed of $\{S_9,S_{10}\}$ may have an effect on the signature diagram without effecting the root. Explicitly, the only possible different signature diagrams that may be reached without changing the root are,
\begin{alignat}{2}
\nonumber &1& &1\\
\nonumber 0\hspace{5pt} 0\hspace{5pt} 0\hspace{5pt} 0\hspace{5pt} 0\hspace{5pt} 0\hspace{5pt} 0\hspace{5pt} &0\hspace{5pt} 0\hspace{5pt} 1 &\qquad \stackrel{S_{10}}{\longrightarrow} \qquad 0\hspace{5pt} 0\hspace{5pt} 0\hspace{5pt} 0\hspace{5pt} 0\hspace{5pt} 0\hspace{5pt} 0\hspace{5pt} &0\hspace{5pt} 1\hspace{5pt} 1 \\
\nonumber t\hspace{7pt} t\hspace{7pt} t\hspace{7pt} t\hspace{7pt} t\hspace{7pt} t\hspace{7pt} t\hspace{7pt} t&\hspace{7pt} t\hspace{7pt} t\hspace{7pt} s& t\hspace{7pt} t\hspace{7pt} t\hspace{7pt} t\hspace{7pt} t\hspace{7pt} t\hspace{7pt} t\hspace{7pt} t&\hspace{7pt} t\hspace{7pt} s\hspace{7pt} t\\
&1& &1\\
\nonumber 0\hspace{5pt} 0\hspace{5pt} 0\hspace{5pt} 0\hspace{5pt} 0\hspace{5pt} 0\hspace{5pt} 0\hspace{5pt} &0\hspace{5pt} 0\hspace{5pt} 1 &\qquad \stackrel{S_9S_{10}}{\longrightarrow} \qquad 0\hspace{5pt} 0\hspace{5pt} 0\hspace{5pt} 0\hspace{5pt} 0\hspace{5pt} 0\hspace{5pt} 0\hspace{5pt} &1\hspace{5pt} 1\hspace{5pt} 0 \\
\nonumber t\hspace{7pt} t\hspace{7pt} t\hspace{7pt} t\hspace{7pt} t\hspace{7pt} t\hspace{7pt} t\hspace{7pt} t&\hspace{7pt} t\hspace{7pt} t\hspace{7pt} s& t\hspace{7pt} t\hspace{7pt} t\hspace{7pt} t\hspace{7pt} t\hspace{7pt} t\hspace{7pt} t\hspace{7pt} t&\hspace{7pt} s\hspace{7pt} t\hspace{7pt} t
\end{alignat}
Here $x^1$ has been held as a temporal coordinate. The interpretation is that the trivial Weyl reflection in the roots of the gravity line shift the singled-out coordinate between $\{x^9,x^{10},x^{11}\}$, the longitudinal brane coordinates. While it is always true that the gravity line Weyl reflections do not alter the signature, it is not generally true that they do not alter the value of $f(\alpha_{11})$. For example, consider the root, $\alpha_8+\alpha_9+\alpha_{10}+\alpha_{11}$, associated to the generator $R^{8910}$, for which only a series of reflections composed of $\{S_8,S_9\}$ may alter the signature diagram without changing the root. In this example the reflection $S_8$ may change the value of $f(\alpha_{11})$ in addition to shifting the singled-out coordinate amongst the longitudinal coordinates. In general, the gravity line, or trivial, Weyl reflections preserve a signature $(t,s)$ but do not necessarily preserve whether we are working with a $-F^2$ or a $+F^2$ theory.

We now turn our attention to the non-trivial signature changes that may be applied to a root, $\beta$, without altering it and we outline here a prescription for finding alternative signatures without altering the root. In order to consider the effects of the reflection $S_{11}$ on the signature we first transform our root to a new root that is invariant under $S_{11}$, we call the series of Weyl reflections applied to achieve this $U$. Furthermore, we restrict ourselves to using only trivial Weyl reflections in this transformation, $U$, so that we only effect a non-trivial signature change after we have transformed to an $S_{11}$ invariant root. These restrictions identify a unique $S_{11}$ invariant root for a given non-zero coefficient of $\alpha_{11}$ in the simple root expansion of the root, $\beta$, or the level of $\beta$. At this stage $S_{11}$ may be applied without changing the root, but with the potential of altering the signature. Our original root in the new signature may be re-obtained by applying $U^{-1}$. These steps allow an algorithmic exploration of the related signatures for a specific root. 

For $\alpha_{11}$, at level one, the $S_{11}$ invariant root is $\alpha_7+2\alpha_8+\alpha_9+\alpha_{11}$. It is obtained from $\alpha_{11}$ by reflections $S_8S_7S_9S_8\equiv U$, so that $\alpha_{11}=U^{-1}S_{11}U\alpha_{11}$, and a new class of signature diagrams is obtained that is not trivially related to the first class. This process is repeated for every trivially related signature diagram and in this way all possible Weyl reflections preserving $\alpha_{11}$ are applied and the associated set of signature diagrams including $(10,1,-)$ is obtained, we call this set the signature orbit of $\alpha_{11}$. An equivalent approach would be to apply all possible trivial reflections to $S_{11}U\alpha_{11}$, before transforming back to $\alpha_{11}$ with $U^{-1}$. This procedure is simply completed by a computer program, with the results shown in table \ref{sigorbitshighm2}, where we have only listed the cases where we have taken $x^1$ to be a temporal coordinate. The equivalent signature orbit where $x^1$ is taken to be spacelike is found by inverting all signatures, while keeping $f(\alpha_{11})$ constant. The signature orbits for the lowest weight, associated with generator $R^{123}$, are found by applying $(S_1\ldots S_{10})(S_1\ldots S_9)(S_1\ldots S_8)$ and the results are shown in table \ref{sigorbitslowm2}. 

\begin{table}[htpb]
\centering
\begin{tabular}{|c|c|c|c|c|}
\hline
Global & Longitudinal & Transverse & Trivially& \\
Signature & Signature & Signature & Related & $\kappa$\\
(temporal $x^1$)&&& Signatures&\\
\hline
$(10,1,-)$ & $(2,1)$ & $(8,0)$ & $3$ & $-1$\\
$(9,2,-)$ & $(3,0)$ & $(6,2)$ & $21$ & $+1$\\
$(2,9,-)$ & $(0,3)$ & $(2,6)$ & $7$ & $-1$\\
$(6,5,-)$ & $(2,1)$ & $(4,4)$ & $105$ & $-1$\\
$(5,6,-)$ & $(1,2)$ & $(4,4)$ & $105$ & $+1$\\
$(6,5,-)$ & $(0,3)$ & $(6,2)$ & $21$ & $-1$\\
$(5,6,-)$ & $(3,0)$ & $(2,6)$ & $7$ & $+1$\\
$(9,2,-)$ & $(1,2)$ & $(8,0)$ & $3$ & $+1$\\
\hline
&&&$272$&\\
\hline
\end{tabular}
\caption{The signature orbit of the root associated to $R^{91011}$}\label{sigorbitshighm2}
\end{table}
\begin{table}[htpb]
\centering
\begin{tabular}{|c|c|c|c|c|}
\hline
Global & Longitudinal & Transverse & Trivially& \\
Signature& Signature & Signature & Related & $\kappa$\\
(temporal $x^1$)&&& Signatures&\\
\hline
$(1,10,+)$ & $(1,2)$ & $(0,8)$ & $1$ & $-1$\\
$(10,1,+)$ & $(2,1)$ & $(8,0)$ & $2$ & $+1$\\
$(9,2,+)$ & $(3,0)$ & $(6,2)$ & $15$ & $-1$\\
$(9,2,-)$ & $(3,0)$ & $(6,2)$ & $13$ & $+1$\\
$(6,5,+)$ & $(2,1)$ & $(4,4)$ & $70$ & $+1$\\
$(6,5,-)$ & $(2,1)$ & $(4,4)$ & $70$ & $-1$\\
$(5,6,+)$ & $(1,2)$ & $(4,4)$ & $35$ & $-1$\\
$(5,6,-)$ & $(1,2)$ & $(4,4)$ & $35$ & $+1$\\
$(5,6,+)$ & $(3,0)$ & $(2,6)$ & $13$ & $-1$\\
$(5,6,-)$ & $(3,0)$ & $(2,6)$ & $15$ & $+1$\\
$(2,9,-)$ & $(2,1)$ & $(0,8)$ & $2$ & $-1$\\
$(9,2,-)$ & $(1,2)$ & $(8,0)$ & $1$ & $+1$\\
\hline
&&&$272$&\\
\hline
\end{tabular}
\caption{The signature orbit of the root associated to $R^{123}$} \label{sigorbitslowm2}
\end{table}
We have found the signature orbits of the highest and lowest weights in the membrane representation, but we have not checked whether each prescribed signature offers a solution to the Einstein and gauge equations. Using our observations of section \ref{solution_preserving} it is, in fact, a quick exercise to check all signatures and see if they offer a solution. We have evaluated $\kappa$, defined in equation (\ref{kappa}) for each putative solution given in the tables, and wherever we find $\kappa=-1$ we have a solution of the Einstein equations. If we count the number of trivially related signatures for these cases we notice that exactly half of the total signature orbit are solutions, that is $3+7+105+21=136$ solutions associated to the generator $R^{91011}$. For spacelike $x^1$ we find $21+105+7+3=136$ solutions too. For the lowest weight generator $R^{123}$ considered in table \ref{sigorbitslowm2}, we again find $1+15+70+35+13+2=136$ solutions for timelike $x^1$, and similarly $2+13+70+35+15+1=136$ solutions for spacelike $x^1$.

\subsubsection{Fivebrane Solution Signatures} \label{fivebranesolutionsignatures}
We now turn our attention to the representation that gives the $M5$ solution. Its highest weight generator is $R^{67891011}$ which is associated with the root $\beta\equiv\alpha_6+2\alpha_7+3\alpha_8+2\alpha_9+\alpha_{10}+2\alpha_{11}$. Only the Weyl reflections $\{S_5,S_{11}\}$ alter the root. The level two $S_{11}$-invariant root is obtained from $\beta$ by acting upon it with the series of reflections given by $U\equiv S_{10}S_9S_8S_7S_6S_5$ and the lowest weight representation, with generator $R^{123456}$, is obtained under the reflection $(S_1\ldots S_{10})\ldots(S_1\ldots S_5)$. The signature orbits are shown in tables \ref{highm5sigorbits} and \ref{lowm5sigorbits} respectively. Again the cases where $\kappa=-1$ give solutions, but we note that there is a difference to the $M2$ case when we consider the solutions for spacelike $x^1$. As before the spacelike $x^1$ case is found by a global signature inversion, however for the $M5$ case this does not bring about a change of sign in $\kappa$. The solutions for spacelike and timelike $x^1$ are no longer complementary but identical. For the highest weight generator we find $3+3+12+60+40+3+12+3=136$ solutions, the same number of solutions as for the $M2$ case, but for the lowest weight we find $5+1+3+15+40+40+15+3+5+1=128$ solutions.

\begin{table}[htpb]
\centering
\begin{tabular}{|c|c|c|c|c|}
\hline
Global & Longitudinal & Transverse & Trivially& \\
Signature& Signature & Signature & Related & $\kappa$\\
(temporal $x^1$)&&& Signatures&\\
\hline
$(10,1,+)$ & $(5,1)$ & $(5,0)$ & $3$ & $-1$\\
$(10,1,-)$ & $(5,1)$ & $(5,0)$ & $3$ & $+1$\\
$(2,9,+)$ & $(1,5)$ & $(1,4)$ & $3$ & $-1$\\
$(2,9,-)$ & $(1,5)$ & $(1,4)$ & $3$ & $+1$\\
$(9,2,+)$ & $(5,1)$ & $(4,1)$ & $12$ & $-1$\\
$(9,2,-)$ & $(5,1)$ & $(4,1)$ & $12$ & $+1$\\
$(6,5,+)$ & $(3,3)$ & $(3,2)$ & $60$ & $-1$\\
$(6,5,-)$ & $(3,3)$ & $(3,2)$ & $60$ & $+1$\\
$(5,6,+)$ & $(3,3)$ & $(2,3)$ & $40$ & $-1$\\
$(5,6,-)$ & $(3,3)$ & $(2,3)$ & $40$ & $+1$\\
$(6,5,+)$ & $(5,1)$ & $(1,4)$ & $3$ & $-1$\\
$(6,5,-)$ & $(5,1)$ & $(1,4)$ & $3$ & $+1$\\
$(5,6,+)$ & $(1,5)$ & $(4,1)$ & $12$ & $-1$\\
$(5,6,-)$ & $(1,5)$ & $(4,1)$ & $12$ & $+1$\\
$(6,5,+)$ & $(1,5)$ & $(5,0)$ & $3$ & $-1$\\
$(6,5,-)$ & $(1,5)$ & $(5,0)$ & $3$ & $+1$\\
\hline
&&&$272$&\\
\hline
\end{tabular}
\caption{The signature orbit of the root associated to $R^{67891011}$} \label{highm5sigorbits}
\end{table}
\begin{table}[htpb]
\centering
\begin{tabular}{|c|c|c|c|c|}
\hline
Global & Longitudinal & Transverse & Trivially& \\
Signature& Signature & Signature & Related & $\kappa$\\
(temporal $x^1$)&&& Signatures&\\
\hline
$(10,1,+)$ & $(5,1)$ & $(5,0)$ & $5$ & $-1$\\
$(1,10,+)$ & $(1,5)$ & $(0,5)$ & $1$ & $-1$\\
$(2,9,+)$ & $(1,5)$ & $(1,4)$ & $3$ & $-1$\\
$(2,9,-)$ & $(1,5)$ & $(1,4)$ & $2$ & $+1$\\
$(9,2,+)$ & $(5,1)$ & $(4,1)$ & $15$ & $-1$\\
$(9,2,-)$ & $(5,1)$ & $(4,1)$ & $10$ & $+1$\\
$(6,5,+)$ & $(3,3)$ & $(3,2)$ & $40$ & $-1$\\
$(6,5,-)$ & $(3,3)$ & $(3,2)$ & $60$ & $+1$\\
$(5,6,+)$ & $(3,3)$ & $(2,3)$ & $40$ & $-1$\\
$(5,6,-)$ & $(3,3)$ & $(2,3)$ & $60$ & $+1$\\
$(6,5,+)$ & $(5,1)$ & $(1,4)$ & $15$ & $-1$\\
$(6,5,-)$ & $(5,1)$ & $(1,4)$ & $10$ & $+1$\\
$(5,6,+)$ & $(1,5)$ & $(4,1)$ & $3$ & $-1$\\
$(5,6,-)$ & $(1,5)$ & $(4,1)$ & $2$ & $+1$\\
$(5,6,+)$ & $(5,1)$ & $(0,5)$ & $5$ & $-1$\\
$(6,5,+)$ & $(1,5)$ & $(5,0)$ & $1$ & $-1$\\
\hline
&&&$272$&\\
\hline
\end{tabular}
\caption{The signature orbit of the root associated to $R^{123456}$} \label{lowm5sigorbits}
\end{table}

\subsubsection{$pp$-Wave Solution Signatures} \label{ppWave}
The $pp$-wave is treated in the same manner. Its highest weight is associated to the root $\beta=\alpha_1+\ldots \alpha_{10}$ and its lowest weight is associated to the root $S_0\beta=-\beta$. In both cases only the Weyl reflections $\{S_1,S_{10},S_{11}\}$ alter the root. However, the negative roots have generators of the form ${K^a}_b$, where $a>b$, which are projected out of the general group element of $E_{11}$, see equation (2.24) in \cite{West}, so the lowest weight representation has generator ${K^1}_2$, and associated root $\alpha_1$. We note that $\alpha_1$ is only altered by the Weyl reflections $\{S_1,S_2\}$, and is related to the highest weight by $\alpha_1=S_2S_3\ldots S_{10}\beta$. There are a number of possible level 0 $S_{11}$-invariant roots, we make use of $U\equiv S_8S_7S_6S_5S_4S_3S_2S_1$ to transform $\beta$ into $\alpha_9+\alpha_{10}$ and then effect the signature-changing Weyl reflection, $S_{11}$, before transforming back to $\beta$. The signature orbits containing the $M$-theory $pp$-wave coming from the root associated to the highest weight and the lowest weight with a positive root generator are listed in tables \ref{highppwave} and \ref{lowppwave} respectively. 

Our analysis of solutions to the Einstein equations from $\kappa$ is not appropriate for the $pp$-wave, instead we have a $pp$-wave solution if the ansatz given in appendix \ref{electricbranesolutions} is satisfied \cite{Argurio, Stelle}. From the tables we obtain $1+2+7=10$ solutions for the $pp$-wave for each weight of the representation and, in addition, we note that there is no associated $M'$-theory $pp$-wave, in our signature orbits.

\begin{table}[htpb]
\centering
\begin{tabular}{|c|c|c|c|c|}
\hline
Global & Longitudinal & Transverse & Signature & Trivially\\
Signature& Signature & Signature & of $\Omega_9$ & Related \\
(temporal $x^1$)&&&& Signatures\\
\hline
$(10,1,-)$ & $(1,0)$ & $(0,1)$ & $(9,0)$ & $1$ \\
$(2,9,+)$ & $(1,0)$ & $(0,1)$ & $(1,8)$ & $2$ \\
$(2,9,-)$ & $(1,0)$ & $(0,1)$ & $(1,8)$ & $7$ \\
\hline
&&&&$10$\\
\hline
\end{tabular}
\caption{The signature orbit of the root associated to highest weight $pp$-wave} \label{highppwave}
\end{table}
\begin{table}[htpb]
\centering
\begin{tabular}{|c|c|c|c|c|}
\hline
Global & Longitudinal & Transverse & Signature & Trivially\\
Signature & Signature & Signature & of $\Omega_9$ & Related \\
(temporal $x^1$)&&&& Signatures\\
\hline
$(1,10,+)$ & $(1,0)$ & $(0,1)$ & $(0,9)$ & $1$\\
$(9,2,-)$ & $(1,0)$ & $(0,1)$ & $(8,1)$ & $2$\\
$(9,2,+)$ & $(1,0)$ & $(0,1)$ & $(8,1)$ & $7$\\
\hline
&&&&$10$ \\
\hline
\end{tabular}
\caption{The signature orbit of the root associated to lowest weight $pp$-wave} \label{lowppwave}
\end{table}
\subsection{$S$-Branes from a Choice of Local Sub-Algebra} \label{spacelikeinvolution}
Spacelike branes or $S$-branes were discovered as a constituent of string theory by Gutperle and Strominger \cite{GutperleStrominger}, who argued that they were a timelike kink in the tachyon field on the world volume of an unstable D-brane, or D-brane anti-D-brane pair. There is a wealth of literature on the rolling tachyon \cite{rollingtachyon} whose association with $S$-branes was first highlighted by Sen. Supergravity $S$-branes were found by Chen, Gal'tsov and Gutperle in arbitrary dimension, D, in \cite{ChenGaltsovGutperle} and in D=10 by Kruczenski, Myers and Peet in \cite{KruczenskiMyersPeet}. These solutions were shown to be equivalent under a coordinate transformation by Bhattacharya and Roy in \cite{BhattacharyaRoy}. General $S$-brane solutions in eleven dimensions were also found in reference \cite{Ohta2}, where intersection rules are also considered. We concentrate here on simply identifying the spacelike branes of $M$-theory and the related solutions in other theories that may be constructed from the brane spectrum of $E_{11}$. 

The group element (\ref{groupelement}) has been used to find brane solutions in exotic signatures by Weyl reflecting the known electric brane solutions of $M$-theory. The group element itself does not know which signature its associated solution exists in, indeed signature information comes from the choice of local sub-algebra. In our solutions we have singled out electric field strengths, those which always have a temporal coordinate, and used these as a starting point for the signature orbits of the previous section. It was observed in section \ref{membranesolutionsignatures} that the Weyl reflections that did not alter the root kept the temporal coordinate on the brane world-volume for the $M$-theory solutions, presenting an obstacle to finding $S$-branes from the electric solutions by Weyl reflecting the group element (\ref{groupelement}). Let's look at this in more detail. If one rotates the coordinates, using a Weyl reflection, to obtain a new root and associated gauge field, the effect of $S_i$ on the expansion of $\beta\cdot H$ in \ref{groupelement} is to interchange ${K^i}_i$ and ${K^{i+1}}_{i+1}$, where $i=1\ldots 10$, and in terms of the coordinate indices on the gauge potential the indices $x^i$ and $x^{i+1}$ are swapped. For example consider the lowest weight of the $M2$ representation with gauge field $A_{123}$, whose indices are transformed in the following manner,
\begin{align}
\nonumber A_{123}&\stackrel{S_1}{\longrightarrow}A_{123}\\
A_{123}&\stackrel{S_2}{\longrightarrow}A_{123}\\
\nonumber A_{123}&\stackrel{S_3}{\longrightarrow}A_{124}
\end{align}
If we pick $x^3$ to be the timelike coordinate we might think that to remove it from the gauge field would require a Weyl reflection in $S_3$. The effect of $S_3$ on the signature diagram is to change the timelike coordinate from  $x^3$ to $x^4$,
\begin{alignat}{2}
\nonumber &0& &0\\
\nonumber 0\hspace{5pt} 1\hspace{5pt} 1\hspace{5pt} 0\hspace{5pt} 0\hspace{5pt} 0\hspace{5pt} 0\hspace{5pt} &0\hspace{5pt} 0\hspace{5pt} 0 &\qquad \stackrel{S_{3}}{\longrightarrow} \qquad 0\hspace{5pt} 0\hspace{5pt} 1\hspace{5pt} 1\hspace{5pt} 0\hspace{5pt} 0\hspace{5pt} 0\hspace{5pt} &0\hspace{5pt} 0\hspace{5pt} 0 \\
\nonumber s\hspace{6pt} s\hspace{6pt} t\hspace{6pt} s\hspace{6pt} s\hspace{6pt} s\hspace{6pt} s\hspace{6pt} s&\hspace{6pt} s\hspace{6pt} s\hspace{6pt} s& s\hspace{6pt} s\hspace{6pt} s\hspace{6pt} t\hspace{6pt} s\hspace{6pt} s\hspace{6pt} s\hspace{6pt} s&\hspace{6pt} s\hspace{6pt} s\hspace{6pt} s
\end{alignat}
Consequently the new gauge field $A_{124}$ remains electric and similar considerations for each possible choice of timelike coordinate show that an electric gauge field remains electric under Weyl reflections. Importantly the $S$-brane solutions of $M$-theory are not related to the electric solutions by Weyl reflections, and are not found in the signature orbits of the usual electric solutions. 

However, given any real form of an algebra that leaves a Lorentzian form, e.g. $-t^2+x_1^2+\ldots x_{10}^2$, invariant we may consider the complex extension of the algebra such that a Euclidean form is left invariant. A specific example of how the generators transform is ${K^1}_2\rightarrow i{K^1}_2$ where $x^1$ is the temporal coordinate. To reintroduce a Lorentzian symmetry we apply the inverse transformation. For example to make $x^{10}$ temporal, we transform ${K^{10}}_{11}\rightarrow -i{K^{10}}_{11}$, and obtain a complexified version of the original set of generators preserving a real Lorentzian form, $t^2+x_1^2+\ldots +x_9^2-x_{10}^2$. The result is that the generators, $A_{123}, A_{123456}$ and ${K^1}_2$ used to find the electric solutions of appendix \ref{electricbranesolutions} become $iA_{123}, iA_{123456}$ and $i{K^1}_2$, as would be expected by the Wick rotations as in equation \ref{Wickrotation}, and all their indices are now spacelike. 

Equivalently, there is a different choice of local sub-algebra with a different set of generators, all real, that preserve the same Lorentzian form, that give an identical theory but with a different sign in front of $F^2$ in the action.  For example the $S2$ and $S5$-branes are solutions in signature $(1,10,-)$ with a real set of generators. Our Weyl reflections lead us to pick out the real form of the sub-algebra, and the sign of $F^2$. The $pp$-wave has a null field strength, hence we make use of the complex generators to find its spacelike solution. These spacelike solutions are verified in appendix \ref{spacelikebranesolutions}.

We may commence with the $S2$ and $S5$-brane solutions of $M$-theory given in appendix \ref{spacelikebranesolutions}, encoded in the group element and find their signature orbits. This solution is identical to commencing using a local subgroup, $H$, whose temporal coordinate with respect to our ansatz (\ref{ansatz}) is not part of the brane world-volume. The metric for the $M$-theory spacelike solution takes the same form as our ansatz (\ref{ansatz}), explicitly,
\begin{equation}
ds^2=A^2(\sum_{j=1}^{j=p}dx_{j}^2)+B^2(-du^2+\sum_{b=1}^{b=d}dy_{b}^2)
\end{equation}
Where $A$ and $B$, are as defined in (\ref{singlebranecoefficients}), using the harmonic function,
\begin{equation}
N_{(1,D-p-2)}=1+\frac{1}{D-p-3}\sqrt{\frac{\Delta}{2(D-2)}}\frac{\|\bf Q\|}{\hat{r}^{(D-p-3)}} 
\end{equation}
Where $\hat{r}^2=-(u^1)^2+(y^1)^2+\ldots (y^{D-p-2})^2$. For the $S2$ and $S5$ branes we have the associated field strengths
\begin{align}
\nonumber F_{x_1x_2x_3\hat{r}}=&4\partial_{[\hat{r}}A_{x_1x_2x_3]}=\partial_{\hat{r}}A_{x_1x_2x_3}=\partial_{\hat{r}}{N_{(1,7)}^{-1}} \\ F_{x_1x_2x_3x_4x_5x_6\hat{r}}=&7\partial_{[\hat{r}}A_{x_1x_2x_3x_4x_5x_6]}=\partial_{\hat{r}}A_{x_1x_2x_3x_4x_5x_6}=\partial_{\hat{r}}{N_{(1,4)}^{-1}}\label{spacelikefieldstrengths}
\end{align}
Weyl reflections of these solutions then give new orbits of possible solutions and completes the range of signature configurations related by $E_{11}$, in that we find solutions of $M$-theory where the temporal coordinate may be any of $\{x^1\ldots x^{11}\}$ for each weight. We list these signature orbits for the case of the highest and lowest weights of the $S2$-brane in the tables \ref{highs2}, \ref{lows2} and similarly for the $S5$-brane in tables \ref{highs5}, \ref{lows5}. 
\begin{table}[htpb]
\centering
\begin{tabular}{|c|c|c|c|c|}
\hline
Global & Longitudinal & Transverse & Trivially& \\
Signature & Signature & Signature & Related & $\kappa$\\
(temporal $x^1$)&&& Signatures&\\
\hline
$(1,10,-)$ & $(0,3)$ & $(1,7)$ & $1$ & $-1$\\
$(10,1,-)$ & $(3,0)$ & $(7,1)$ & $7$ & $+1$\\
$(7,4,-)$ & $(2,1)$ & $(5,3)$ & $105$ & $-1$\\
$(4,7,-)$ & $(1,2)$ & $(3,5)$ & $63$ & $+1$\\
$(5,6,-)$ & $(2,1)$ & $(3,5)$ & $63$ & $-1$\\
$(6,5,-)$ & $(1,2)$ & $(5,3)$ & $105$ & $+1$\\
$(5,6,-)$ & $(0,3)$ & $(5,3)$ & $35$ & $-1$\\
$(6,5,-)$ & $(3,0)$ & $(3,5)$ & $21$ & $+1$\\
$(7,4,-)$ & $(0,3)$ & $(7,1)$ & $7$ & $-1$\\
$(4,7,-)$ & $(3,0)$ & $(1,7)$ & $1$ & $+1$\\
$(3,8,-)$ & $(0,3)$ & $(3,5)$ & $35$ & $-1$\\
$(8,3,-)$ & $(3,0)$ & $(5,3)$ & $21$ & $+1$\\
$(3,8,-)$ & $(2,1)$ & $(1,7)$ & $21$ & $-1$\\
$(8,3,-)$ & $(1,2)$ & $(7,1)$ & $3$ & $+1$\\
$(9,2,-)$ & $(2,1)$ & $(7,1)$ & $21$ & $-1$\\
$(2,9,-)$ & $(1,2)$ & $(1,7)$ & $3$ & $+1$\\
\hline
&&&$512$&\\
\hline
\end{tabular}
\caption{The signature orbit of $S2$-brane from $E_{11}$ from $R^{91011}$} \label{highs2}
\end{table}
\begin{table}[htpb]
\centering
\begin{tabular}{|c|c|c|c|c|}
\hline
Global & Longitudinal & Transverse & Trivially& \\
Signature & Signature & Signature & Related & $\kappa$\\
(temporal $x^1$)&&& Signatures&\\
\hline
$(10,1,+)$ & $(3,0)$ & $(7,1)$ & $3$ & $-1$\\
$(10,1,-)$ & $(3,0)$ & $(7,1)$ & $5$ & $+1$\\
$(7,4,+)$ & $(2,1)$ & $(5,3)$ & $50$ & $+1$\\
$(7,4,-)$ & $(2,1)$ & $(5,3)$ & $62$ & $-1$\\
$(4,7,+)$ & $(1,2)$ & $(3,5)$ & $25$ & $-1$\\
$(4,7,-)$ & $(1,2)$ & $(3,5)$ & $31$ & $+1$\\
$(5,6,+)$ & $(2,1)$ & $(3,5)$ & $62$ & $+1$\\
$(5,6,-)$ & $(2,1)$ & $(3,5)$ & $50$ & $-1$\\
$(6,5,+)$ & $(1,2)$ & $(5,3)$ & $31$ & $-1$\\
$(6,5,-)$ & $(1,2)$ & $(5,3)$ & $25$ & $+1$\\
$(8,3,+)$ & $(3,0)$ & $(5,3)$ & $31$ & $-1$\\
$(8,3,-)$ & $(3,0)$ & $(5,3)$ & $25$ & $+1$\\
$(4,7,+)$ & $(3,0)$ & $(1,7)$ & $5$ & $-1$\\
$(4,7,-)$ & $(3,0)$ & $(1,7)$ & $3$ & $+1$\\
$(9,2,+)$ & $(2,1)$ & $(7,1)$ & $10$ & $+1$\\
$(9,2,-)$ & $(2,1)$ & $(7,1)$ & $6$ & $-1$\\
$(2,9,+)$ & $(1,2)$ & $(1,7)$ & $5$ & $-1$\\
$(2,9,-)$ & $(1,2)$ & $(1,7)$ & $3$ & $+1$\\
$(6,5,+)$ & $(3,0)$ & $(3,5)$ & $25$ & $-1$\\
$(6,5,-)$ & $(3,0)$ & $(3,5)$ & $31$ & $+1$\\
$(3,8,+)$ & $(2,1)$ & $(1,7)$ & $6$ & $+1$\\
$(3,8,-)$ & $(2,1)$ & $(1,7)$ & $10$ & $-1$\\
$(8,3,+)$ & $(1,2)$ & $(7,1)$ & $3$ & $-1$\\
$(8,3,-)$ & $(1,2)$ & $(7,1)$ & $5$ & $+1$\\
\hline
&&&$512$&\\
\hline
\end{tabular}
\caption{The signature orbit of $S2$-brane from $E_{11}$ from $R^{123}$} \label{lows2}
\end{table}

\begin{table}[htpb]
\centering
\begin{tabular}{|c|c|c|c|c|}
\hline
Global & Longitudinal & Transverse & Trivially& \\
Signature & Signature & Signature & Related & $\kappa$\\
(temporal $x^1$)&&& Signatures&\\
\hline
$(1,10,-)$ & $(0,6)$ & $(1,4)$ & $4$ & $-1$\\
$(10,1,-)$ & $(6,0)$ & $(4,1)$ & $1$ & $-1$\\
$(7,4,+)$ & $(4,2)$ & $(3,2)$ & $36$ & $+1$\\
$(7,4,-)$ & $(4,2)$ & $(3,2)$ & $54$ & $-1$\\
$(4,7,+)$ & $(2,4)$ & $(2,3)$ & $24$ & $+1$\\
$(4,7,-)$ & $(2,4)$ & $(2,3)$ & $36$ & $-1$\\
$(5,6,+)$ & $(4,2)$ & $(1,4)$ & $6$ & $+1$\\
$(5,6,-)$ & $(4,2)$ & $(1,4)$ & $9$ & $-1$\\
$(6,5,+)$ & $(2,4)$ & $(4,1)$ & $24$ & $+1$\\
$(6,5,-)$ & $(2,4)$ & $(4,1)$ & $36$ & $-1$\\
$(5,6,+)$ & $(2,4)$ & $(3,2)$ & $54$ & $+1$\\
$(5,6,-)$ & $(2,4)$ & $(3,2)$ & $36$ & $-1$\\
$(6,5,+)$ & $(4,2)$ & $(2,3)$ & $36$ & $+1$\\
$(6,5,-)$ & $(4,2)$ & $(2,3)$ & $24$ & $-1$\\
$(7,4,+)$ & $(2,4)$ & $(5,0)$ & $9$ & $+1$\\
$(7,4,-)$ & $(2,4)$ & $(5,0)$ & $6$ & $-1$\\
$(3,8,-)$ & $(0,6)$ & $(3,2)$ & $6$ & $-1$\\
$(8,3,-)$ & $(6,0)$ & $(2,3)$ & $4$ & $-1$\\
$(3,8,+)$ & $(2,4)$ & $(1,4)$ & $9$ & $+1$\\
$(3,8,-)$ & $(2,4)$ & $(1,4)$ & $6$ & $-1$\\
$(8,3,+)$ & $(4,2)$ & $(4,1)$ & $36$ & $+1$\\
$(8,3,-)$ & $(4,2)$ & $(4,1)$ & $24$ & $-1$\\
$(9,2,+)$ & $(4,2)$ & $(5,0)$ & $6$ & $+1$\\
$(9,2,-)$ & $(4,2)$ & $(5,0)$ & $9$ & $-1$\\
$(2,9,+)$ & $(0,6)$ & $(2,3)$ & $4$ & $+1$\\
$(9,2,+)$ & $(6,0)$ & $(3,2)$ & $6$ & $+1$\\
$(4,7,+)$ & $(0,6)$ & $(4,1)$ & $4$ & $+1$\\
$(7,4,+)$ & $(6,0)$ & $(1,4)$ & $1$ & $+1$\\
\hline
&&&$510$&\\
\hline
\end{tabular}
\caption{The signature orbit of $S5$-brane from $E_{11}$ generator $R^{67891011}$} \label{highs5}
\end{table}
\begin{table}[htpb]
\centering
\begin{tabular}{|c|c|c|c|c|}
\hline
Global & Longitudinal & Transverse & Trivially& \\
Signature & Signature & Signature & Related & $\kappa$\\
(temporal $x^1$)&&& Signatures&\\
\hline
$(10,1,+)$ & $(6,0)$ & $(4,1)$ & $3$ & $+1$\\
$(10,1,-)$ & $(6,0)$ & $(4,1)$ & $2$ & $-1$\\
$(7,4,+)$ & $(4,2)$ & $(3,2)$ & $40$ & $+1$\\
$(7,4,-)$ & $(4,2)$ & $(3,2)$ & $60$ & $-1$\\
$(4,7,+)$ & $(2,4)$ & $(2,3)$ & $20$ & $+1$\\
$(4,7,-)$ & $(2,4)$ & $(2,3)$ & $30$ & $-1$\\
$(5,6,+)$ & $(2,4)$ & $(3,2)$ & $20$ & $+1$\\
$(5,6,-)$ & $(2,4)$ & $(3,2)$ & $30$ & $-1$\\
$(6,5,+)$ & $(4,2)$ & $(2,3)$ & $40$ & $+1$\\
$(6,5,-)$ & $(4,2)$ & $(2,3)$ & $60$ & $-1$\\
$(3,8,+)$ & $(2,4)$ & $(1,4)$ & $15$ & $+1$\\
$(3,8,-)$ & $(2,4)$ & $(1,4)$ & $10$ & $-1$\\
$(8,3,+)$ & $(4,2)$ & $(4,1)$ & $30$ & $+1$\\
$(8,3,-)$ & $(4,2)$ & $(4,1)$ & $20$ & $-1$\\
$(5,6,+)$ & $(4,2)$ & $(1,4)$ & $30$ & $+1$\\
$(5,6,-)$ & $(4,2)$ & $(1,4)$ & $20$ & $-1$\\
$(6,5,+)$ & $(2,4)$ & $(4,1)$ & $15$ & $+1$\\
$(6,5,-)$ & $(2,4)$ & $(4,1)$ & $10$ & $-1$\\
$(7,4,+)$ & $(6,0)$ & $(1,4)$ & $3$ & $+1$\\
$(7,4,-)$ & $(6,0)$ & $(1,4)$ & $2$ & $-1$\\
$(8,3,+)$ & $(6,0)$ & $(2,3)$ & $4$ & $+1$\\
$(8,3,-)$ & $(6,0)$ & $(2,3)$ & $6$ & $-1$\\
$(9,2,+)$ & $(6,0)$ & $(3,2)$ & $4$ & $+1$\\
$(9,2,-)$ & $(6,0)$ & $(3,2)$ & $6$ & $-1$\\
$(2,9,+)$ & $(2,4)$ & $(0,5)$ & $5$ & $+1$\\
$(9,2,+)$ & $(4,2)$ & $(5,0)$ & $10$ & $+1$\\
$(4,7,+)$ & $(4,2)$ & $(0,5)$ & $10$ & $+1$\\
$(7,4,+)$ & $(2,4)$ & $(5,0)$ & $5$ & $+1$\\
\hline
&&&$510$&\\
\hline
\end{tabular}
\caption{The signature orbit of $S5$-brane from $E_{11}$ generator $R^{123456}$} \label{lows5}
\end{table}

The $pp$-wave solution distinguishes three sets of coordinates, namely a longitudinal coordinate, a transverse coordinate and the nine remaining coordinates in 11-dimensions, $\Omega_9$. Consequently there are two alternative choices of local sub-algebra that may be made, the first introduces a transverse time coordinate and the second introduces a time coordinate into $\Omega_9$, which we then relabel $\Omega_{(1,8)}$. The former case is similar to the original $pp$-wave solution under an interchange of $K\rightarrow -K$, having a line element:
\begin{equation}
ds^2=-(1+K){dt_1}^2+(1-K){dx_1}^2+2Kdt_1dx_1+d\Omega_9^2
\end{equation}
Consequently this second choice of local sub-algebra leads to the same solution as the usual sub-algebra, but the $pp$-wave in this case is completely out of phase with the original $pp$-wave. This solution is still dependent on a static harmonic function.

There is a time-dependent solution, which we label the $Spp$-wave, coming from the choice of local sub-algebra that introduces a time coordinate into $\Omega_9$, which we indicate by $\Omega_{(1,8)}$. The solution is given in appendix \ref{spacelikebranesolutions}. We list the signature orbit of the highest weight case in table \ref{hSppwave}, as in the case of the $pp$-wave the lowest weight signature orbit is identical.
\begin{table}[htpb]
\centering
\begin{tabular}{|c|c|c|c|c|}
\hline
Signature & Longitudinal & Transverse & Signature & Trivially \\
(temporal $x^1$)& Signature & Signature & of $\Omega_9$ & Related \\
&&&& Signatures\\
\hline
$(10,1,-)$ & $(1,0)$ & $(1,0)$ & $(8,1)$ & $7$ \\
$(10,1,+)$ & $(1,0)$ & $(1,0)$ & $(8,1)$ & $2$ \\
$(4,7,-)$ & $(1,0)$ & $(1,0)$ & $(2,7)$ & $22$ \\
$(4,7,+)$ & $(1,0)$ & $(1,0)$ & $(2,7)$ & $14$ \\
\hline
&&&&$45$\\
\hline
\end{tabular}
\caption{The signature orbits of the root associated to highest weight $Spp$-wave} \label{hSppwave}
\end{table}

Let us count the solutions in the $S2$-brane signature orbit from the M, $M*$ and $M'$-theories\footnote{That is, only those with signature $(1,10,\pm)$,$(10,1,\pm)$,$(2,9,\pm)$,$(9,2,\pm)$,$(5,6,\pm)$ and $(6,5,\pm)$}. From the highest weight signature orbit in table \ref{highs2}, we find $1+63+35+21=120$ solutions, where $x^1$ is timelike and $7+105+21+3=136$ where $x^1$ is spacelike. Recollect that we found $136$ solutions related to the equivalent highest weight $M2$ signature orbit, giving a total of $256$ solutions with timelike $x^1$ and $272$ with spacelike $x^1$; in all we have $528$ solutions from the root associated to the $R^{91011}$ generator. Similarly for the lowest weight associated to the $S2$-brane we find $3+50+31+6+5+25=120$ solutions with timelike $x^1$ and $5+62+25+10+3+31=136$ with spacelike $x^1$, giving a total of $528$ solutions from the root associated to the $R^{123}$ generator.

The $S5$-brane signature orbit coming from the highest weight given in table \ref{highs5} has $4+1+9+36+36+24+9=119$ solutions of the three $M$-theories for both timelike and spacelike $x^1$. If we include the $136$ solutions from the highest weight of the $M5$ representation for both timelike and spacelike $x^1$, we find a total of $510$ solutions associated to the $R^{67891011}$ generator. Perhaps most interesting are the results from the lowest weight generator $R^{123456}$; from the $S5$ signature orbit in table \ref{lows2} we find $2+30+60+20+10+6=128$ solutions to the three $M$-theories for both timelike and spacelike $x^1$. Recalling that we earlier counted $128$ solutions from the lowest weight signature orbit containing the $M5$ brane for each choice of $x^1$ giving a total of $512$ solutions associated to the $R^{123456}$ generator. We note that there is a difference between both the total number and type of solutions associated the generator $R^{a_1\ldots a_6}$ at different weights, in contrast to the $R^{a_1a_2a_3}$ generator to which a consistent set of solutions is associated at each weight.
\subsection{A Naive Interpretation for Signature Change}
In this chapter we have strayed into what would seem to be non-physical territory, and certainly unfamiliar, by simply following through the consequences of imposing a Lorentzian symmetry on spacetime as it occurs in the nonlinear realisation. However it has been observed \cite{Keurentjes} that there are some physically appealing aspects to this construction, namely, that one is able to obtain the $M*$ and $M'$ theories without the need for a compactification on a closed timelike loop \cite{Hull}. Instead such theories arise as a consequence of U-duality transformations applied in the form of a Weyl reflection of our algebra, this certainly makes the $M*$ and $M'$ theories physically more viable. In this chapter we also saw that the consequence of signature changes in the algebra of $E_{11}$ led us to the signature orbits of the known theories. The signature orbits of the known electric branes revealed a naturally complementary set of brane signature orbits and we saw that these were the S-branes of supergravity. It may be that the notion of multiple signatures goes hand-in-hand with the concept of spacelike branes. 

Indeed there is a familiar mathematical interpretation for introducing signature changes into ordinary quantum field theory in Minkowski space where one does require solutions in Euclidean space, namely instantons, which occur when the particle enters a region forbidden by energy considerations. The famous example is the particle in a square-well potential, moving in one dimension, where in ordinary quantum mechanics one must solve the Schrodinger equation,
\begin{equation}
-\frac{\hbar}{2m}\frac{\partial^2\psi}{\partial x^2}=(E-V(x))\psi
\end{equation}
Where $V(x)$ is the potential function, $E$ is the energy of the particle. We find the wave-function is given by,
\begin{equation}
\psi=Ae^{i\frac{\sqrt{2m}}{\hbar}\sqrt{E-V(x)}t}
\end{equation}
While $V(x)<E$ the wavefunction is oscillatory, and when $V(x)>E$ the wavefunction decays exponentially. One could consider this situation as being similar to a signature change, $t\rightarrow it$ on the time coordinate, which is the only longitudinal coordinate of the particle. Of course it is not simply that the wavefunction in the allowed region is mapped to that of the forbidden region by such a Wick rotation, there is additional information that is given by the magnitude of $E-V(x)$, but the calculation could at least be interpreted as if a signature change occurred, rather than a change in the potential function. One could imagine a similar scenario for branes and, in particular, one brane in the potential of another. We note that for every brane solution in the signature orbits given in this paper there is an exact pairing between putative solutions in $[(p,q),(c,d)]$ and $[(q,p),(c,d)]$, that is, under an signature inversion of the brane coordinates. For the fivebrane case either both or none of these signatures will carry solutions, while for the membrane case only one of the two signatures will carry a solution in a theory with a particular sign of $F^2$. Specifically the M5-brane, $[(1,5),(0,5)]=(1,10)$, is paired with a solution in $[(5,1),(0,5)]=(5,6)$ which is an $M'$-theory solution; and the M2-brane, $[(1,2),(0,8)]=(1,10)$, has a solution in $[(2,1),(0,8)]=(2,9)$, with a '$+F^2$' term, which is an $M*$-theory solution. From this viewpoint, we are always considering the theory in a $(1,10)$ signature but certain calculations would require theories in other signatures, namely $(2,9)$ and $(5,6)$.

We will finish this chapter by writing down a systematic set of rules for applying the kind of Wick rotations present in the quantum mechanics of particles to extended objects. We will find that we can obtain all the signatures of branes in the signature orbits of the $M2$ and the $M5$ described in this chapter. These comments are not intended to be scientific, but are merely a passing observation that the quite complex set of branes and signatures coming from $E_{11}$ may be reproduced in a much more simple manner. Indeed we do not imagine that the Schrodinger equation in one dimension can be equally well applied to the objects of a quantum field theory of gravity, we simply wonder if we may find any results consistent with the findings of this chapter by commencing with a set of rules whose prototype is the tunneling effect observed in the calculations of the one-dimensional potential well.

The rules of the game are derived from the Wick rotations discussed above. We commence with the prototype electric brane solutions of $M$-theory, the $M2$ brane with signature $[(1,2),(0,8)]$ and the $M5$ brane with signature $[(1,5),(0,5)]$. Then we imagine that such solutions may just as easily define the function $V(x_1,\ldots x_p)$ as the energy function. Indeed our usual notion of kinetic and potential energy is given by the sign of the associated terms appearing in the Lagrangian, and throughout this chapter we have discussed changing the sign of the $F^2$ term - we now offer the interpretation that the different signed terms be associated with the kinetic energy and the potential associated to the brane. Our guiding rule is that should two branes' coordinates overlap we imagine these coordinates to be Wick rotated. 

We are allowed to combine any two branes in a given signature to produce a third subject to the following two provisions:
\begin{enumerate}
	\item two branes may only be combined if they exist in the same global signature
\item no spacelike brane may be combined with any other brane
\end{enumerate}
To visualise the process described here we use the following notation: to combine two branes one writes out the two signatures using t's (temporal) and s's (spatial) to indicate the coordinate type. The temporal coordinates in the two objects must be aligned. For one of the objects discard all the transverse spatial coordinates - this object we are imagining to define the region containing a different potential function to the background. 
We note that had the basic object in this game been just the particle in $(1,3)$ we would have generated a second solution signature which we would associate with the instanton, as indicated in table \ref{Twoparticles}.
\begin{table}[tbp]
\centering
\begin{tabular}{|c|c|c|c|c|}
\hline
t&&&& \\
\hline
t&&s&s&s\\
\hline
s&&s&s&s\\
\hline
\end{tabular}
	\caption{The composition of two particle signatures}
	\label{Twoparticles}
\end{table}
For the $M2$ and the $M5$ brane we must write down all the possible spatial overlaps of the two objects, e.g. for two $M2$ branes we would consider the different alignments shown in table \ref{TwoM2s}.
\begin{table}[htbp]
\centering
\begin{tabular}{|c|c|c|c|c|c|c|c|c|c|c|c|}
\hline
t&s&s&&&&&&&&& \\
\hline
t&s&s&&s&s&s&s&s&s&s&s\\
\hline
s&t&t&&s&s&s&s&s&s&s&s\\
\hline
\hline
t&s& &&s&&&&&&& \\
\hline
t&s&s&&s&s&s&s&s&s&s&s\\
\hline
s&t&s&&t&s&s&s&s&s&s&s\\
\hline
\hline
t&&&&s&s&&&&&& \\
\hline
t&s&s&&s&s&s&s&s&s&s&s\\
\hline
s&s&s&&t&t&s&s&s&s&s&s\\
\hline
\end{tabular}
	\caption{The various compositions of two M2 branes signatures}
	\label{TwoM2s}
\end{table}
The rule for finding the combined object is that all coordinate types with a t or an s written above them are switched, while the rest remain unchanged. In the above example one finds three objects in signature $(2,9)$ coming from the overlap of two $M2$ branes. These have signatures $[(2,1),(0,8)]$, $[(1,2),(1,7)]$ and $[(0,3),(2,6)]$. Now one simply starts combining as many objects as possible and finds all possible different new objects. It is not a very exciting game. One soon realises that the sets of objects are just those branes which are potential solutions of the $M$, $M*$ and $M'$ theories - no other signatures ever arise. Compared to finding the signature orbits of a brane solution arising from $E_{11}$ this convoluted game is relatively simple. One also thinks the rules we have suggested are also quite physical, namely that only objects in the same signature may be combined and that spacelike branes may not take part in this dynamical process. It may be no more than a triviality, but we think it is worth noting. Of course, within the closure of such a set, we also find the $S$-branes of $M$-theory in this manner even though we started with only electric branes. We emphasise that this "game" is intended as no more than a light-hearted remark. 
\newpage
\section{Kac-Moody Algebras and U-duality Charge Multiplets}
It was recognised \cite{LuPopeStelle, CremmerLuPopeStelle,ObersPioline2,ElitzurGiveonKutasovRabinovici,ObersPiolineRabinovici,ObersPioline} that the action of the Weyl group of $E_d$ $d=1\ldots 8$ corresponded to the U-duality transformations of M-theory compactified on a torus. By encoding the tension, or mass, of BPS states appearing in string theory in a weight vector, the authors of \cite{ObersPioline2,ElitzurGiveonKutasovRabinovici,ObersPiolineRabinovici,ObersPioline} were able to apply Weyl reflections to the tensions of known solutions of M-theory and demonstrate that they replicated the symmetries induced by the U-dualities of M-theory. Furthermore the weight vectors encoding the tensions of the particle, string and brane states were the fundamental weights of a semisimple Lie algebra. For M-theory compactified on a torus to three dimensions the relevant Lie group was recognised as $E_8$. The orbits of the highest weights of the algebra under the Weyl group gave rise to U-duality multiplets labelled by the solution corresponding to the highest weight. This work is discussed in detail in the original papers, but especially in the review \cite{ObersPioline}.

The conjecture that $E_{11}$ is a symmetry of M-theory \cite{West1} gave an eleven-dimensional origin to these observations. The reduction to three dimensions of the algebra of the $l_1$ representation of $E_{11}$ was shown in \cite{West4} to give perfect agreement with the U-duality multiplet of charges, and incorporate the Weyl group of $E_8$ implicitly from the outset.

In this chapter we extend the work of \cite{West4} and outline the decomposition of the $l_1$ algebra to arbitrary dimensions, $d<11$, on a torus. 

\subsection{General Decomposition}
In chapter four we reviewed the proposal that the full set of brane charges of M-theory are contained in the $l_1$ representation of $E_{11}$ \cite{West3}. Following this proposition the U-duality multiplet of charges has been found for the reduction to three dimensions on an 8-torus as the $l_1$ representation of an $A_2\otimes E_8$ sub-algebra \cite{West4}.

In this section we wish to find the decomposition of the $l_1$ fundamental representation of $E_{11}$ in terms of its $A_{d-1}$ and $E_{11-d}$. The details of this decomposition can be found in \cite{West4}, but may be straightforwardly extrapolated from the decompositions of $E_{12}$ outlined in chapter four. 

\begin{figure}[cth]
\hspace{190pt} \includegraphics[viewport=0 150 80 200,angle=90]{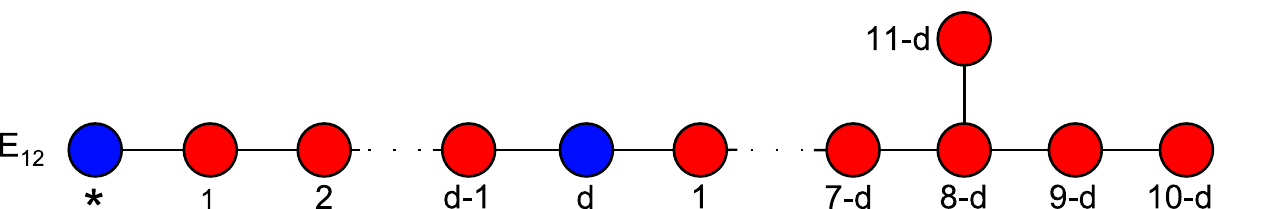}
\caption{The Dynkin diagram of $E_{12}$ decomposed into $A_{d-1}\otimes E_{11-d}$} \label{ae}
\end{figure}

Recall that the $l_1$ representation of $E_{11}$ takes the first fundamental weight of $E_{11}$, $\hat{\lambda}_1$, whose associated generator is the translation generator, and treats it as the highest weight of a representation in the $E_{11}$ lattice. A generic weight in the $l_1$ representation therefore takes the form:
\begin{equation}
\hat{\lambda}_1-\sum_{i=1}^{11}m_i\alpha_i
\end{equation}
Where $\sum_{i=1}^{11}m_i\alpha_i$ is a root in the $E_{11}$ root lattice and so has length $2-2n$ where $n\in \mathbb{Z^{0+}}$. As we saw in chapter four, deletion of the component of $E_{12}$ orthogonal to the $E_{11}$ root lattice yields the $l_1$ representation of $E_{11}$ if we restrict to roots in $E_{12}$, $\beta=\sum_{i=*}^{11}m_i\alpha_i$ such that $m_*=1$. To find the $l_1$ representation of $A_{d-1}\otimes E_{11-d}$ we delete, in addition, the node labelled d, in figure \ref{ae}, to obtain,
For example, to find representations of $A_4\otimes E_6$ we delete the fifth node. The two deleted roots may be expressed in terms of vectors in the root space of $E_{11}$ with components orthogonal to the vectors in the remaining root space after deletion of nodes $*$ and $d$. 

Let us denote the simple roots of $E_{11}$ by $\alpha_i$ and its fundamental weights by $\lambda_i$ and commence by writing the root associated to the $*$ node $\alpha_*$ in terms of the fundamental weights of $E_{11}$ and a vector, $x$, in the root space of $E_{12}$ orthogonal to the roots of $E_{11}$,
\begin{equation}
\alpha_* = -{\hat\lambda_1} + x 
\end{equation}
We recall that 
\begin{equation}
<{\hat\lambda_i},{\hat\lambda_j}>=(A_{(E_{11})}^{-1})_{ij}
\end{equation}
So that $\lambda_1^2 = {1\over 2}$. Since $\alpha_*^2=2$ then $x^2= {3\over 2}$, exactly the same as in chapter four. Consequently we see the reason this construction leads us naturally to the $l_1$ representation, since a root in $E_{12}$ with $m_*=1$ becomes,
$$\beta=y-\hat{\lambda}_1+\sum_{i=1}^{11}m_i\alpha_i\equiv y-\Lambda $$
Where $\Lambda$ is a weight in the $l_1$ representation. 

Similarly we carry out the decomposition to $A_{d-1}\otimes E_{11-d}$ by expressing the d'th root, $\alpha_d$, in the form,
\begin{equation}
\alpha_d = -\nu + y, \qquad \nu = -\sum_{i=1, i\neq d}^{i=11} \lambda_{i}(A_{(E_{11})})_{id}
\end{equation}
Where $(A_{(E_{11})})_{id}$ is the $i,d$'th component in the Cartan matrix of $E_{11}$, $x$ is orthogonal to all the remaining roots after deletion, and now $\lambda_i$ for $i=1\ldots d-1$ are fundamental weights of $A_{d-1}$, and when $i=(d+1) \ldots 11$ they are fundamental weights of $E_{11-d}$. Calling to mind the defining relation of the fundamental weights to the simple roots,
\begin{equation}
<\alpha_i, \lambda_j>=\delta_{ij}
\end{equation}
We are able to denote the fundamental weights of the decomposed $E_{11}$ by $l_i$ and have,
\begin{align}
l_d&={1\over y^2}y\\
l_i^{(r)}&=\lambda_i^{(r)} + {<\nu,\lambda_i^{(r)}> \over y^2}y
\end{align}
Where $r$ is either 1, referring to the $A_{d-1}$ sub-algebra, or 2, referring to the $E_{d-11}$ sub-algebra. In this notation, 
\begin{equation}
\alpha_*=-l_1^{(1)}+x
\end{equation}
We now return to our consideration of $E_{12}$, a general root of which is given by, 
\begin{equation}
\beta = m_*\alpha_* + m_d\alpha_d + \sum_{i=1,i\neq d}^{11} m_i\alpha_i
\end{equation}
For a positive root, $m_*$, $m_d$ and $m_i$ are positive integers. Substituting our expressions for $\alpha_*$ and $\alpha_c$, we obtain,
\begin{equation}
\beta = m_*x+(m_d-m_*{<\lambda_1^{(1)},\lambda_{d-1}^{(1)}> \over x^2})y-\sum_{(r=1,2)} \Lambda^{(r)}
\end{equation}
Where,
\begin{equation}
\Lambda^{(r)}=-\sum_{i=1,i\neq d}^{11} m_i^{(r)}\alpha_i^{(r)}+m_d\nu^{(r)}+m_*\lambda_1^{(r)}\delta_{(r,1)}
\end{equation}
Adopting the notation $m_i^{(1)}=m_i$, $m_i^{(2)}=n_i$, $\lambda_i^{(1)}=\mu_i$ and $\lambda_i^{(2)}=\lambda_i$, we have,
\begin{align}
\nonumber \Lambda^{(1)}&=-\sum_{i=1}^{d-1} m_i\alpha_i^{(1)}+m_d\mu_{d-1}+m_*\mu_1\equiv \sum_i^{d-1}q_i\mu_i\\
\Lambda^{(2)}&=-\sum_{i=1}^{11-d} n_i\alpha_i^{(2)}+m_d\lambda_1\equiv \sum_i^{11-d}p_i\lambda_i
\end{align}
Taking the inner product with $\mu_j$ and $\lambda_j$ respectively we obtain expressions for the Dynkin labels, $n_i$ and $m_i$,
\begin{align}
\nonumber A_{d-1}:&\quad -m_j=\sum_i q_i<\mu_i,\mu_j>-m_d<\mu_{d-1},\mu_j>-m_*<\mu_1,\mu_j> \\
E_{11-d}:&\quad -n_j=\sum_i p_i<\lambda_i,\lambda_j>-m_d<\lambda_1,\lambda_j> \label{dynkinlabels}
\end{align}
The inner product of fundamental weights of $A_{d-1}$ are defined to give the inverse of the Cartan matrix of the group, and are specified by,
\begin{equation}
<\mu_i,\mu_j>={1\over d}i(d-j), \qquad i\leq j
\end{equation}
This formula is arrived at by use of $(\det{A})A^{-1}=adj{A}$, remembering that $A$ is symmetric. It is shown in \cite{GaberdielOliveWest} that $\det{A_{d-1}}=d$ and we recall that the i'th row and j'th column entry of the adjugate of a matrix is proportional to the determinant of the matrix once it's i'th row and j'th column have been removed, we denote this reduced Cartan matrix ${\hat A(ij)}$. Then,
\begin{equation}
\det{\hat A(ij)}=(-1)^{j-i}(\det{A_{(i-1)}})(\det{A_{(d-1-j)}})=(-1)^{j-i}i(d-j), \qquad i\leq j
\end{equation}
The adjugate matrix is defined by,
\begin{equation}
(adj{A})_{ij}=(-1)^{i+j}\det{\hat{A}(ij)}=i(d-j), \qquad i\leq j
\end{equation}
Similar formulae for the weights of $E_{11-d}$ are derived in appendix A of reference \cite{West4} and are,
\begin{align}
\nonumber \lambda_i&=\hat{\mu_i}+{3i\over d-2}z, \qquad i=1,\ldots, 8-d\\
\nonumber &=\hat{\mu}_i+{(8-d)(11-d-i)\over d-2}z, \qquad  i=9-d,10-d \\
&={(11-d)z\over (d-2)},\qquad i=11-d \label{weightsEn}
\end{align}
These weights are derived by deleting the n'th node with respect to an $E_n$ diagram, $z$ is the vector in the root space corresponding to the linear independence of the n'th node and $\hat{\mu}_i$ are the weights of the $A_{n-1}$ subalgebra, where $n=11-d$. We note that $z^2={d-2\over 11-d}$ and,
\begin{equation}
\lambda_1^2=<\hat{\mu}_1,\hat{\mu}_1>+{9\over (d-2)(11-d)}={d-1\over d-2}
\end{equation}
Consequently,
\begin{equation}
\alpha_c^2=x^2+{\mu_{d-1}^2}+{\lambda_1^2}=x^2+{d-1\over d}+{d-1\over d-2}
\end{equation}
Normalising $\alpha_c^2=2$ gives,
\begin{equation}
x^2={-2\over d(d-2)}
\end{equation}
\subsection{Rank $p$ topological charges}
We commence by looking for solutions that have a single $\mu_{d-p}$ weight in the decomposition i.e. $\sum_i q_i\mu_i =\mu_{d-p}$. Such a weight in the $l_1$ representation of $A_{d-1}$ corresponds to a rank $p$ charge, $Z^{a_1\ldots a_p}$. It was demonstrated in \cite{KleinschmidtWest} and also in chapter four of this thesis, that for this weight there exists a corresponding object in the adjoint representation of $A_{d-1}$ with $p+1$ antisymmetric indices, $R^{a_1\ldots a_{p+1}}$, and a corresponding gauge field, $A_{a_1 \ldots a_{p+1}}$, coupling to a p-brane. 

A $\mu_{d-p}$ weight in the $A_{d-1}$ sub-algebra of our decomposition leads to constraints upon the values to be taken by $m_d$ in equation (\ref{mdAn}). We are interested in the decomposition of the $l_1$ representation of $E_{11}$ and as such we take $m_*=1$ and from the $A_{d-1}$ equation in (\ref{dynkinlabels}) we obtain,
\begin{align}
\nonumber -m_j&=<\mu_{(d-p)},\mu_j>-m_d<\mu_{(d-1)},\mu_j>-<\mu_1,\mu_j>\\
&=\begin{cases}(d-p-1)-{j\over d}(d-p-1+m_d),\qquad  j \geq (d-p)\\
-1+{j\over d}(p+1-m_d),\qquad  j \leq (d-p) \label{Andynkinlabels}
\end{cases}
\end{align}
Since $-m_j$ must be integer valued and negative (we may restrict our interest to only the positive Dynkin labels as the negative roots have the negative of the Dynkin labels of the positive roots) we find a simple set of solutions for $m_d$ having the form,
\begin{equation}
m_d=p+1+kd, \qquad k\in {\mathbb Z} \label{mdAn}
\end{equation}
\subsection{Representations of the fundamental weights of $E_{11-d}$}
Having found a criterion for $m_d$ from a particularly interesting weight of $A_{d-1}$, we now turn our attention to restrictions on $m_d$ coming from specific weights of the $E_{11-d}$ sub-algebra. We commence by finding conditions for representations of the fundamental weights of $E_{11-d}$, $\lambda_i$, for which we set $\sum_i p_i\lambda_i =\lambda_i$ in (\ref{dynkinlabels}) and making use of equation (\ref{weightsEn}) we find,
\begin{align}
-n_j&=<\lambda_i,\lambda_j>-m_d<\lambda_1,\lambda_j> \\
\nonumber &=\begin{cases}i\leq 8-d
\begin{cases}
i-m_d+{j \over d-2}(i-m_d), \qquad &j\leq 8-d, i\leq j \\
j-m_d+{j\over d-2}(i-m_d), &j\leq 8-d, i\geq j \\
{2(11-d-j)\over d-2}(i-m_d), &j=9-d, 10-d \\
{3\over d-2}(i-m_d), &j=11-d 
\end{cases} \\
i=9-d,10-d
\begin{cases}
-m_d+ {j\over d-2}(2(11-d-i)-m_d), \qquad &j\leq 8-d \\
{11-d-j\over d-2}((8-d)^2-i(6-d)-2m_d), &j=9-d,10-d, i\leq j \\
\frac{1}{d-2}(4(11-d-j)-2m_d), &j=9-d,10-d, i\geq j \\
{1\over d-2}((8-d)(11-d-i)-3m_d), &j=11-d 
\end{cases}\\
i=11-d 
\begin{cases}
-m_d+{j\over d-2}(3-m_d), \qquad &j\leq 8-d \\
{11-d-j\over d-2}(8-d-2m_d), &j=9-d,10-d \\
{1\over d-2}(11-d-3m_d), &j=11-d
\end{cases}
\end{cases}
\end{align}
The simplest case with a solution is dependent upon the choice of fundamental weight $\lambda_i$ and is
\begin{align}
m_d=\begin{cases}
i+l(d-2), \qquad  &i\leq 8-d \\
2(11-d-i)+l(d-2), \qquad &i=9-d,10-d \\
3+l(d-2), \qquad &i=11-d 
\end{cases} \label{mdEn}
\end{align}
Where $l$ is a positive integer or zero. 
\subsection{Representations of $A_{d-1}\otimes E_{11-d}$ with fundamental $E_{11-d}$ weights}
We may find which rank $p$ charges in the $A_{d-1}$ sub-algebra of the $l_1$ representation are compatible with each of the fundamental weights of $E_{11-d}$ by equating our two conditions for $m_d$, equations (\ref{mdAn}) and (\ref{mdEn}), 
\begin{align}
p=
\begin{cases}
i+l(d-2)-kd-1, \qquad & i\leq 8-d \\
2(11-d-i)+l(d-2)-kd-1, \qquad & i=9-d,10-d\\
2+l(d-2)-kd, & i=11-d 
\end{cases}
\end{align}
In particular for $l=k=0$ we find the content indicated in table \ref{AEreps}, where we use the index $a_i$ to indicate an index in the non-compact spacetime associated to the weight of $A_{d-1}$ being considered, and the index $j_i$ to indicate compactified coordinates coming from the weights of $E_{11-d}$.
\begin{table}[h]
	\centering
		\begin{tabular}{|c|c|c|}
	\hline
			Index of $\lambda_i$&Weight of $E_{11-d}\otimes A_{d-1}$ &Central Charge\\
		\hline
			$i\leq 8-d$&$\lambda_i\otimes\mu_{d-i}$&$Z^{a_1\ldots a_{i-1}j_1\ldots j_{n-i}}$\\
\hline
$i=9-d$&$\lambda_{9-d} \otimes\mu_{d-3}$&$Z^{a_1a_2a_3j_1j_2}$\\
\hline
$i=10-d$&$\lambda_{10-d}\otimes\mu_{d-1}$&$Z^{a_1j_1}$\\
\hline
$i=11-d$&$\lambda_{11-d}\otimes\mu_{d-2}$&$Z^{a_1a_2}$\\
\hline
		\end{tabular}
			\caption{Representations of $A_{d-1}\otimes E_{11-d}$ with fundamental $E_{11-d}$ weights}
	\label{AEreps}
\end{table}
For the case of $d=3$ \cite{West4}, the charges associated to the fundamental weights of nodes $11-d, 10-d$ and $1$ of $E_{11-d}$ are the highest weights of the membrane, string and particle multiplets of \cite{LuPopeStelle,CremmerLuPopeStelle,ObersPioline2,ElitzurGiveonKutasovRabinovici,ObersPiolineRabinovici,ObersPioline}. Let us look to see the content of the various multiplets appears in the $l_1$ representation of $E_{11}$.
\subsubsection{The Particle Multiplet}
Let us identify the roots of the $l_1$ representation associated to the weight $\lambda_1\otimes \mu_{d-1}$. Recall that this is a solution with Dynkin coefficient $m_d=1$. From equation (\ref{Andynkinlabels}), with $p=0$ for the particle we find that,
\begin{equation}
m_j=1 \qquad j=1,\ldots d-1
\end{equation}
Since we are considering the $l_1$ representation $m_*=1$ then the particle multiplet contains roots with Dynkin labels $(1^{d+1},m_{d+1},\ldots m_{11})$. For the reduction to $d$ dimensions we should find the particle multiplet being made up of roots whose first $d+1$ Dynkin labels are $1$'s. The most complicated case is that of the reduction to $d=3$ which has been studied already in \cite{West4}, and the $l_1$ representations have been shown to form a multiplet of $E_8$. Let us look specifically at the next most complex case, that of the reduction to $d=4$ where the particle multiplet should belong to representations of $E_7$. In particular we hope to find the $56=7+21+21+7$ of $E_7$. From our table \ref{l1E11roots}, we find the $l_1$ content listed in the first three lines of table \ref{particlemultiplet}.
\begin{table}[h]
	\centering
		\begin{tabular}{|c|c|c|}
		\hline
 $l_1$ Root&Central Charge&Dimension of charge\\ 
\hline
$(1^9,0,0,1)$&$Z^{j_1j_2}$&$21$ \\
\hline
$(1^7,2,3,2,1,2)$&$Z^{j_1\ldots j_5}$&$21$ \\
\hline
$(1^5,2,3,4,5,3,1,3)$&$Z^{j_1\ldots j_7,k}$&$7$ \\
\hline
$(1^5,0^7)$&$Z^{j_1\ldots j_6}$&$7$ \\
\hline
		\end{tabular}
			\caption{The particle charge multiplet in $d=4$} \label{particlemultiplet}
\end{table}
In the fourth line of table \ref{particlemultiplet} we have included the root ${K^1}_5$ coming from the $A_{10}$ subalgebra of the $l_1$ representation, whose positive roots, being well understood, were not included in the table \ref{l1E11roots} as discussed in chapter four. Thus in $d=4$ the particle mutiplet is complete. In the reduction to $d=5$ we look for the $27=6+15+6$ of $E_6$, and the corresponding roots are listed in table \ref{particlemultiplet5d}.
\begin{table}[h]
	\centering
		\begin{tabular}{|c|c|c|}
		\hline
$l_1$ Root&Central Charge&Dimension of charge\\ 
\hline
$(1^9,0,0,1)$&$Z^{j_1j_2}$&$15$ \\
\hline
$(1^7,2,3,2,1,2)$&$Z^{j_1\ldots j_5}$&$6$ \\
\hline
$(1^6,0,0,1)$&$Z^{j_1\ldots j_5}$&$6$ \\
\hline
		\end{tabular}
			\caption{The particle charge multiplet in $d=5$} \label{particlemultiplet5d}
\end{table}
For the reduction to $d=6$ we look for the $16=10+5+1$ of $SO(5)$. From the $l_1$ we find the states of table \ref{particlemultiplet6d}.
\begin{table}[h]
	\centering
		\begin{tabular}{|c|c|c|}
		\hline
$l_1$ Root&Central Charge&Dimension of charge\\ 
\hline
$(1^9,0,0,1)$&$Z^{j_1j_2}$&$10$ \\
\hline
$(1^7,2,3,2,1,2)$&$Z^{j_1\ldots j_5}$&$1$ \\
\hline
$(1^7,0,0,1)$&$Z^{j_1\ldots j_4}$&$5$ \\
\hline
		\end{tabular}
			\caption{The particle charge multiplet in $d=6$} \label{particlemultiplet6d}
\end{table}
We repeat the process for $d=7$ where we look for the $10=6+4$ of $Sl(5)$, the appropriate roots are listed in table \ref{particlemultiplet7d}.
\begin{table}[h]
	\centering
		\begin{tabular}{|c|c|c|}
		\hline
$l_1$ Root&Central Charge&Dimension of charge\\ 
\hline
$(1^9,0,0,1)$&$Z^{j_1j_2}$&$6$ \\
\hline
$(1^8,0,0,1)$&$Z^{j_1\ldots j_3}$&$4$ \\
\hline
\end{tabular}
\caption{The particle charge multiplet in $d=7$} \label{particlemultiplet7d}
\end{table}
The considerations giving rise to the formulae used in this chapter are even robust up to $d=8$ where we find the $6=3+3$ of $Sl(3)\times Sl(2)$ contained in the $l_1$ representation. We find one from the root $(1^9,0,0,1)$ and another copy of the same root from the $A_{10}$ subalgebra, both of which give rise to a charge $Z^{j_1j_2}$ of dimension 3. Thus the particle multiplet of charges appears to have a higher dimensional origin in the $l_1$ representation containing the charges of the eleven dimensional theory. 
\subsubsection{The String Multiplet}
Let us identify the roots of the $l_1$ representation associated to the weight $\lambda_{10-d}\otimes \mu_{d-1}$. Recall that this is a solution with Dynkin coefficient $m_d=2$. From equation (\ref{Andynkinlabels}), with $p=1$ for the string we find that,
\begin{equation}
m_j=1 \qquad j=1,\ldots d-1
\end{equation}
Using equation (\ref{mdEn}) we find, the corresponding root is $(1^{d},2^{9-d},0,0,1)$, which, as discussed in \cite{West4} is related to the root $(1^9,0,0,1)$, with charge $Z^{8.11}$ by a series commutators with the generators ${K^d}_{d+1}$, ${K^{d+1}}_{d+2}$, $\ldots$ ${K^7}_8$, giving the component $Z^{d.11}$. This charge has one compact index (j=11) and one non-compact index (a=d), and so is a component of a string charge $Z^{aj}$. We can also apply the generators ${K^a}_{a+1}$ to the roots of the particle multiplet and find string charges in a similar manner. 
We find the string multiplet has $l_1$ roots with Dynkin labels $(1^d,2,m_5\ldots m_{11})$. In $d=4$ we look for the $133=7+35+49+35+7$ of $E_7$ and the suitable roots from the $l_1$ are listed in table \ref{stringmultiplet4d} together with the roots derived from the particle multiplet under the action of the $A_{10}$ subalgebra.
\begin{table}[h]
	\centering
		\begin{tabular}{|c|c|c|}
		\hline
$l_1$ Root&Central Charge&Dimension of charge\\ 
\hline
$(1^4,2^5,0,0,1)$&$Z^{a_1,j_1}$&$7$ \\
\hline
$(1^4,2^4,3,2,1,2)$&$Z^{a_1,j_1\ldots j_4}$&$35$ \\
\hline
$(1^4,2^2,3,4,5,3,1,3)$&$Z^{a_1,j_1\ldots j_6,k}$&$49$ \\
\hline
$(1^4,2,3,4,5,6,4,2,4)$&$Z^{a_1,j_1\ldots j_7,k_1k_2k_3}$&$35$ \\
\hline
$(1^4,2,3,5,7,9,6,3,5)$&$Z^{a_1,j_1\ldots j_7,k_1\ldots k_6}$&$7$ \\
\hline
		\end{tabular}
			\caption{The string charge multiplet in $d=4$} \label{stringmultiplet4d}
\end{table}
We note that there is a root in the table \ref{l1E11roots} having Dynkin labels $(1^4,2,3,4,5,6,4,2,3)$ and a charge $Z^{a_1,j_1\ldots j_7}$ contributing one extra degree of freedom, however this root is contained in the representation $Z^{a_1,j_1\ldots j_6,k}$ listed in table \ref{stringmultiplet4d}, differing in terms of Dynkin labels by $(0^5,1^6,0)$ which is a root of the $A_{10}$ subalgebra having generator ${K^4}_{11}$, and so has already been accounted for in the table.

In $d=5$ we look for the $27=6+15+6$ of $E_6$. This corresponds to roots of the form $(1^5,2, m_6 \ldots m_{11})$, which we list in table \ref{stringmultiplet5d}.
\begin{table}[h]
	\centering
		\begin{tabular}{|c|c|c|}
		\hline
$l_1$ Root&Central Charge&Dimension of charge\\ 
\hline
$(1^5,2^4,0,0,1)$&$Z^{a_1,j_1}$&$6$ \\
\hline
$(1^5,2^3,3,2,1,2)$&$Z^{a_1,j_1\ldots j_4}$&$15$ \\
\hline
$(1^5,2,3,4,5,3,1,3)$&$Z^{a_1,j_1\ldots j_6,k}$&$6$ \\
\hline
		\end{tabular}
			\caption{The string charge multiplet in $d=5$} \label{stringmultiplet5d}
\end{table}

In $d=6$ we look for the $10=5+5$ of $SO(5,5)$. This corresponds to roots of the form $(1^6,2, m_7 \ldots m_{11})$, which we list in table \ref{stringmultiplet6d}.
\begin{table}[h]
	\centering
		\begin{tabular}{|c|c|c|}
		\hline
$l_1$ Root&Central Charge&Dimension of charge\\ 
\hline
$(1^6,2^3,0,0,1)$&$Z^{a_1,j_1}$&$5$ \\
\hline
$(1^6,2^2,3,2,1,2)$&$Z^{a_1,j_1\ldots j_4}$&$5$ \\
\hline
		\end{tabular}
			\caption{The string charge multiplet in $d=6$} \label{stringmultiplet6d}
\end{table}

In $d=7$ we look for the $5=4+1$ of $Sl(5)$. This corresponds to roots of the form $(1^7,2, m_8 \ldots m_{11})$, which we list in table \ref{stringmultiplet7d}.
\begin{table}[h]
	\centering
		\begin{tabular}{|c|c|c|}
		\hline
$l_1$ Root&Central Charge&Dimension of charge\\ 
\hline
$(1^7,2^2,0,0,1)$&$Z^{a_1,j_1}$&$4$ \\
\hline
$(1^7,2,3,2,1,2)$&$Z^{a_1,j_1\ldots j_4}$&$1$ \\
\hline
		\end{tabular}
			\caption{The string charge multiplet in $d=7$} \label{stringmultiplet7d}
\end{table}

In $d=8$ we look for the $3$ of $Sl(3)\times Sl(2)$. This corresponds to roots of the form $(1^8,2, m_9 \ldots m_{11})$, and we find only the root with Dynkin labels $(1^8,2,0,0,1)$, corresponding to a charge $Z^{a_1,j_1}$ of dimension $3$.

Thus the full string multiplet is reproduced from the $l_1$ multiplet in dimensions $d$, such that $3\leq d \leq 8$.

In addition to the string and particle multiplets, we may also find multiplets associated with the membrane charge, a threebrane charge, a fourbrane charge and other charges upto that of a $9-d$-brane, as discussed in \cite{ObersPioline2,ElitzurGiveonKutasovRabinovici,ObersPiolineRabinovici,ObersPioline}. It may be useful to associate the particle, string and membrane multiplets with the Dynkin diagram of $E_{11-d}$,
\begin{figure}[cth]
\hspace{80pt} \includegraphics{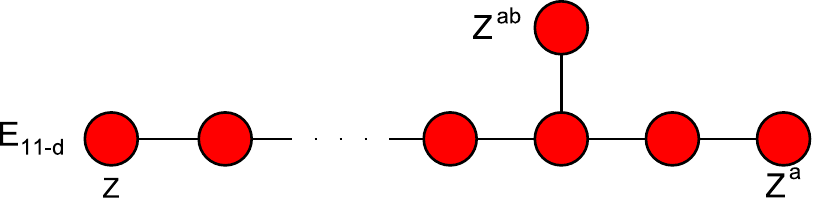}
\caption{Charge multiplets and $E_{11-d}$} \label{chargemultiplets}
\end{figure}
One might also find a fivebrane multiplet whose highest weight is $\mu_{d-5}\times 2\lambda_{11-d}$, indeed one may find multiplets corresponding to a variety of exotic charges whose interpretation in string theory is obscure, but are all a natural consequence of the conjectured eleven dimensional $E_{11}$ symmetry.
\subsection{More Exotic Charges from a general weight of $E_{11-d}$}
Having considered the fundamental representations of $E_{11-d}$ it might be useful to consider a general weight of $E_{11-d}$, i.e. $\sum p_i\lambda_i$, for we shall see that this has a straightforward form. We expand equation (\ref{dynkinlabels}) to find,
\begin{align}
\nonumber -n_j &= \sum p_i <\lambda_i,\lambda_j> - m_d<\lambda_1,\lambda_j>\\
\nonumber &=(p_1-m_d)<\lambda_1,\lambda_j>+p_2<\lambda_2,\lambda_j>+\ldots +p_{11-d}<\lambda_{11-d},\lambda_j> \\
\nonumber &=(p_1-m_d)(1+{j \over d-2})+p_22(1+{j \over d-2})+\ldots p_jj(1+{j \over d-2})+p_{(j+1)}j(1+{j+1 \over d-2})\\
\nonumber &\qquad \ldots + p_{(8-d)}j(1+{8-d \over d-2})+ p_{(9-d)}2j({2 \over d-2})+ p_{(10-d)}2j({1 \over d-2})+ p_{(11-d)}3j({1 \over d-2}) \\
\nonumber &=(p_1-m_d)+2p_2+\ldots jp_j+j(p_{(j+1)}+\ldots +p_{(8-d)})+{j\over d-2}((p_1-m_d)+ \\
&\qquad 2p_2+\ldots jp_j+(j+1)p_{j+1}+\ldots (8-d)p_{(n-3)}+4p_{(9-d)}+2p_{(10-d)}+3p_{(11-d)})
\end{align}
We see that we have a general solution giving positive integer values for $n_j$ when,
\begin{equation}
m_d=\sum_{i=1}^{8-d}ip_i + 4p_{(9-d)}+2p_{(10-d)}+3p_{(11-d)}+h(d-2) \label{generalmdEn}
\end{equation}
Where $h$ is some positive integer, or zero. We could use this equation to find some central charges corresponding to more general weights of $E_{11-d}$ than we have considered so far. We match the value of $m_d$ in equation (\ref{generalmdEn}) with the value of $m_d$ found from considering weights of $A_{d-1}$ as given in equation (\ref{mdAn}).

For example let us consider an $E_{11-d}$ representation carrying the weights $\lambda_i+\lambda_{i+1}$, the values of equation (\ref{generalmdEn}) are listed in table \ref{generalweights1} along with the corresponding $p$-brane derived from equation (\ref{mdAn}), and the central charge.
\begin{table}[h]
	\centering
		\begin{tabular}{|c|c|c|c|}
		\hline
 &$m_d$&$p$&Central Charge\\ 
\hline
$i=1$&$3$&$2$&$Z^{a_1a_2}_{j_1,k_1k_2}$ \\
\hline
$i=i$&$2i+1$&$2i$&$Z^{a_1\ldots a_{2i}}_{j_1\ldots j_i,k_1\ldots k_i}$\\
\hline
$i=6-d$&$13-2d$&$2(6-d)$&$Z^{a_1\ldots a_{2(6-d)}}_{j_1\ldots j_{6-d},k_1\ldots k_{7-d}}$\\
\hline
$i=7-d$&$15-2d$&$2(7-d)$&$Z^{a_1\ldots a_{2(7-d)}}_{j_1\ldots j_{7-d},k_1\ldots k_{8-d}}$\\
\hline
$i=8-d$&$12-d$&$11-d$&$Z^{a_1\ldots a_{12-d},j_1j_2j_3,k_1k_2}$\\
\hline
$i=9-d$&$6$&$5$&$Z^{a_1\ldots a_5,j_1j_2,k_1}$\\
\hline
$i=10-d$&$5$&$4$&$Z^{a_1\ldots a_4,j_1,k_1k_2k_3}$\\
\hline
		\end{tabular}
			\caption{Representations of $A_{d-1}\otimes E_{11-d}$ with with $\lambda_i+\lambda_{i+1}$ $E_{11-d}$ weights} \label{generalweights1}
\end{table}
In particular we note that the representation with $E_{11-d}$ weight $\lambda_{9-d}+\lambda_{10-d}$ corresponds to part of the fivebrane multiplet, and indeed equation (\ref{generalmdEn}) may prove useful in future endeavours to comprehend the content of exotic charge multiplets.

It seems that it is indeed possible to find brane charges with almost arbitrary indices when we compactify to low dimensions. Once again we are faced with the challenge of interpreting the higher level content of the $E_{11}$ algebra before we can hope to understand the plethora of exotic brane charges that seem to arise in dimensional reduction of the $l_1$ representation. However while it may prove difficult to interpret the higher level objects of $E_{11}$ and its $l_1$ representation, the origin of the exotic charges that arose in \cite{ObersPioline2,ElitzurGiveonKutasovRabinovici,ObersPiolineRabinovici,ObersPioline} due to the considerations of the $U$-duality symmetry of string theory do have a natural higher dimensional origin in this setting as reported in \cite{West4}. The results presented in this chapter do lend credence to the notion that the $l_1$ representation of $E_{11}$ contains all the central charges of the eleven dimensional theory that was presented in chapter four, as well as providing further evidence in favour of the $E_{11}$ conjecture.
\newpage
\appendix
\section{Tables of low level roots of $E_{11}$, $K_{27}$ and their $l_1$ representations} \label{roottables}
\end{spacing}
\begin{center}
\begin{longtable}{|c|c|c|c|}

\caption[The roots of $E_{11}$ to level 12]{The roots of $E_{11}$ to level 12} \label{E11roots}\\
\hline \multicolumn{1}{|c|}{\textbf{$A_{10}$ weights}}& 
\multicolumn{1}{|c|}{\textbf{$\beta$ ($\alpha_i$ basis)}}&
\multicolumn{1}{|c|}{\textbf{$\beta$ ($e_i$ basis)}}&
\multicolumn{1}{|c|}{\textbf{$\beta^2$}}\\
\hline
\endfirsthead

\multicolumn{4}{c}
{{\bfseries \tablename\ \thetable{} - continued from the previous page}}\\
\hline \multicolumn{1}{|c|}{\textbf{$A_{10}$ weights}}& 
\multicolumn{1}{|c|}{\textbf{$\beta$ ($\alpha_i$ basis)}}&
\multicolumn{1}{|c|}{\textbf{$\beta$ ($e_i$ basis)}}&
\multicolumn{1}{|c|}{\textbf{$\beta^2$}}\\
\hline
\endhead

\hline \multicolumn{4}{|r|}{Continued on the next page} \\ 
\hline
\endfoot

\hline
\hline
\endlastfoot

$[0,0,0,0,0,0,0,1,0,0]$& $(0,0,0,0,0,0,0,0,0,0,1)$&$(0,0,0,0,0,0,0,0,1,1,1)$&$2$ \\ 
\hline$[0,0,0,0,1,0,0,0,0,0]$& $(0,0,0,0,0,1,2,3,2,1,2)$&$(0,0,0,0,0,1,1,1,1,1,1)$&$2$ \\ 
\hline$[0,1,0,0,0,0,0,0,0,0]$& $(0,0,1,2,3,4,5,6,4,2,3)$&$(0,0,1,1,1,1,1,1,1,1,1)$&$0$ \\ 
$[0,0,1,0,0,0,0,0,0,1]$& $(0,0,0,1,2,3,4,5,3,1,3)$&$(0,0,0,1,1,1,1,1,1,1,2)$&$2$ \\ 
\hline$[0,0,0,0,0,0,0,0,0,1]$& $(1,2,3,4,5,6,7,8,5,2,4)$&$(1,1,1,1,1,1,1,1,1,1,2)$&$-2$ \\ 
$[0,1,0,0,0,0,0,1,0,0]$& $(0,0,1,2,3,4,5,6,4,2,4)$&$(0,0,1,1,1,1,1,1,2,2,2)$&$2$ \\ 
$[1,0,0,0,0,0,0,0,1,0]$& $(0,1,2,3,4,5,6,7,4,2,4)$&$(0,1,1,1,1,1,1,1,1,2,2)$&$0$ \\ 
$[1,0,0,0,0,0,0,0,0,2]$& $(0,1,2,3,4,5,6,7,4,1,4)$&$(0,1,1,1,1,1,1,1,1,1,3)$&$2$ \\
\end{longtable}
\end{center}
\begin{spacing}{1.5}

\end{spacing}
\begin{landscape}
\begin{center}
\begin{longtable}{|c|c|c|}

\caption[The roots of $K_{27}$ to level (3,3)]{The roots of $K_{27}$ to level (3,3)} \label{K27roots}\\
\hline \multicolumn{1}{|c|}{\textbf{$A_{25}$ weights}}& 
\multicolumn{1}{|c|}{\textbf{$\beta$ ($\alpha_i$ basis)}}&
\multicolumn{1}{|c|}{\textbf{$\beta^2$}}\\
\hline
\endfirsthead

\multicolumn{3}{c}
{{\bfseries \tablename\ \thetable{} - continued from the previous page}}\\
\hline \multicolumn{1}{|c|}{\textbf{$A_{10}$ weights}}& 
\multicolumn{1}{|c|}{\textbf{$\beta$ ($\alpha_i$ basis)}}&
\multicolumn{1}{|c|}{\textbf{$\beta^2$}}\\
\hline
\endhead

\hline \multicolumn{3}{|r|}{Continued on the next page} \\ 
\hline
\endfoot

\hline
\hline
\endlastfoot

\hline$[0,0,0,0,0,0,0,0,0,0,0,0,0,0,0,0,0,0,0,0,0,0,0,1,0]$& $(0,0,0,0,0,0,0,0,0,0,0,0,0,0,0,0,0,0,0,0,0,0,0,0,0,1,0)$&$2$ \\ 
\hline$[0,0,0,1,0,0,0,0,0,0,0,0,0,0,0,0,0,0,0,0,0,0,0,0,0]$& $(0,0,0,0,0,0,0,0,0,0,0,0,0,0,0,0,0,0,0,0,0,0,0,0,0,0,1)$&$2$ \\ 
\hline$[0,1,0,0,0,0,0,0,0,0,0,0,0,0,0,0,0,0,0,0,0,0,0,0,0]$& $(0,0,1,2,2,2,2,2,2,2,2,2,2,2,2,2,2,2,2,2,2,2,2,2,1,1,1)$&$0$ \\ 
$[0,0,1,0,0,0,0,0,0,0,0,0,0,0,0,0,0,0,0,0,0,0,0,0,1]$& $(0,0,0,1,1,1,1,1,1,1,1,1,1,1,1,1,1,1,1,1,1,1,1,1,0,1,1)$&$2$ \\ 
\hline$[0,1,0,0,0,0,0,0,0,0,0,0,0,0,0,0,0,0,0,0,0,0,0,1,0]$& $(0,0,1,2,2,2,2,2,2,2,2,2,2,2,2,2,2,2,2,2,2,2,2,2,1,2,1)$&$2$ \\ 
$[1,0,0,0,0,0,0,0,0,0,0,0,0,0,0,0,0,0,0,0,0,0,0,0,1]$& $(0,1,2,3,3,3,3,3,3,3,3,3,3,3,3,3,3,3,3,3,3,3,3,3,1,2,1)$&$0$ \\ 
\hline$[0,0,0,0,0,0,0,0,0,0,0,0,0,0,0,0,0,0,0,0,0,0,0,1,0]$& $(1,2,3,4,4,4,4,4,4,4,4,4,4,4,4,4,4,4,4,4,4,4,4,4,2,3,1)$&$0$ \\ 
$[1,0,0,0,0,0,0,0,0,0,0,0,0,0,0,0,0,0,0,0,0,0,1,0,0]$& $(0,1,2,3,3,3,3,3,3,3,3,3,3,3,3,3,3,3,3,3,3,3,3,4,2,3,1)$&$2$ \\
\end{longtable}
\end{center}
\end{landscape}
\begin{spacing}{1.5}
\end{spacing}
\begin{center}
\begin{longtable}{|c|c|c|c|}

\caption[The roots of the $l_1$ representation of $E_{11}$ to level 12]{The roots of $l_1$ representation of $E_{11}$ to level 12} \label{l1E11roots}\\
\hline \multicolumn{1}{|c|}{\textbf{$A_{10}$ weights}}& 
\multicolumn{1}{|c|}{\textbf{$\beta$ ($\alpha_i$ basis)}}&
\multicolumn{1}{|c|}{\textbf{$\beta$ ($e_i$ basis)}}&
\multicolumn{1}{|c|}{\textbf{$\beta^2$}}\\
\hline
\endfirsthead

\multicolumn{4}{c}
{{\bfseries \tablename\ \thetable{} - continued from the previous page}}\\
\hline \multicolumn{1}{|c|}{\textbf{$A_{10}$ weights}}& 
\multicolumn{1}{|c|}{\textbf{$\beta$ ($\alpha_i$ basis)}}&
\multicolumn{1}{|c|}{\textbf{$\beta$ ($e_i$ basis)}}&
\multicolumn{1}{|c|}{\textbf{$\beta^2$}}\\
\hline
\endhead

\hline \multicolumn{4}{|r|}{Continued on the next page} \\ 
\hline
\endfoot

\hline
\hline
\endlastfoot

$[1,0,0,0,0,0,0,0,0,0]$& $(0,0,0,0,0,0,0,0,0,0,0)$&$(-1,0,0,0,0,0,0,0,0,0,0)$&$2$ \\ 
\hline$[0,0,0,0,0,0,0,0,1,0]$& $(1,1,1,1,1,1,1,1,0,0,1)$&$(0,0,0,0,0,0,0,0,0,1,1)$&$2$ \\ 
\hline$[0,0,0,0,0,1,0,0,0,0]$& $(1,1,1,1,1,1,2,3,2,1,2)$&$(0,0,0,0,0,0,1,1,1,1,1)$&$2$ \\ 
\hline$[0,0,1,0,0,0,0,0,0,0]$& $(1,1,1,2,3,4,5,6,4,2,3)$&$(0,0,0,1,1,1,1,1,1,1,1)$&$0$ \\ 
$[0,0,0,1,0,0,0,0,0,1]$& $(1,1,1,1,2,3,4,5,3,1,3)$&$(0,0,0,0,1,1,1,1,1,1,2)$&$2$ \\ 
\hline$[0,0,1,0,0,0,0,1,0,0]$& $(1,1,1,2,3,4,5,6,4,2,4)$&$(0,0,0,1,1,1,1,1,2,2,2)$&$2$ \\ 
$[0,1,0,0,0,0,0,0,1,0]$& $(1,1,2,3,4,5,6,7,4,2,4)$&$(0,0,1,1,1,1,1,1,1,2,2)$&$0$ \\ 
$[1,0,0,0,0,0,0,0,0,1]$& $(1,2,3,4,5,6,7,8,5,2,4)$&$(0,1,1,1,1,1,1,1,1,1,2)$&$-2$ \\
\end{longtable}
\end{center}
\begin{spacing}{1.5}

\end{spacing}
\begin{landscape}
\begin{center}
\begin{longtable}{|c|c|c|}

\caption[The roots of the $l_1$ representation of $K_{27}$ to level (3,3)]{The roots of the $l_1$ representation of $K_{27}$ to level (3,3)} \label{l1K27roots}\\
\hline \multicolumn{1}{|c|}{\textbf{$A_{25}$ weights}}& 
\multicolumn{1}{|c|}{\textbf{$\beta$ ($\alpha_i$ basis)}}&
\multicolumn{1}{|c|}{\textbf{$\beta^2$}}\\
\hline
\endfirsthead

\multicolumn{3}{c}
{{\bfseries \tablename\ \thetable{} - continued from the previous page}}\\
\hline \multicolumn{1}{|c|}{\textbf{$A_{10}$ weights}}& 
\multicolumn{1}{|c|}{\textbf{$\beta$ ($\alpha_i$ basis)}}&
\multicolumn{1}{|c|}{\textbf{$\beta^2$}}\\
\hline
\endhead

\hline \multicolumn{3}{|r|}{Continued on the next page} \\ 
\hline
\endfoot

\hline
\hline
\endlastfoot

$[1,0,0,0,0,0,0,0,0,0,0,0,0,0,0,0,0,0,0,0,0,0,0,0,0]$& $(0,0,0,0,0,0,0,0,0,0,0,0,0,0,0,0,0,0,0,0,0,0,0,0,0,0,0)$&$2$ \\ 
\hline$[0,0,0,0,0,0,0,0,0,0,0,0,0,0,0,0,0,0,0,0,0,0,0,0,1]$& $(1,1,1,1,1,1,1,1,1,1,1,1,1,1,1,1,1,1,1,1,1,1,1,1,0,1,0)$&$2$ \\ 
$[0,0,1,0,0,0,0,0,0,0,0,0,0,0,0,0,0,0,0,0,0,0,0,0,0]$& $(1,1,1,2,2,2,2,2,2,2,2,2,2,2,2,2,2,2,2,2,2,2,2,2,1,1,1)$&$2$ \\ 
\hline$[1,0,0,0,0,0,0,0,0,0,0,0,0,0,0,0,0,0,0,0,0,0,0,0,0]$& $(1,2,3,4,4,4,4,4,4,4,4,4,4,4,4,4,4,4,4,4,4,4,4,4,2,2,1)$&$0$ \\ 
$[0,1,0,0,0,0,0,0,0,0,0,0,0,0,0,0,0,0,0,0,0,0,0,0,1]$& $(1,1,2,3,3,3,3,3,3,3,3,3,3,3,3,3,3,3,3,3,3,3,3,3,1,2,1)$&$2$ \\ 
\hline$[0,0,0,0,0,0,0,0,0,0,0,0,0,0,0,0,0,0,0,0,0,0,0,0,1]$& $(2,3,4,5,5,5,5,5,5,5,5,5,5,5,5,5,5,5,5,5,5,5,5,5,2,3,1)$&$0$ \\ 
$[1,0,0,0,0,0,0,0,0,0,0,0,0,0,0,0,0,0,0,0,0,0,0,1,0]$& $(1,2,3,4,4,4,4,4,4,4,4,4,4,4,4,4,4,4,4,4,4,4,4,4,2,3,1)$&$2$ \\ 
\hline$[0,0,0,0,0,0,0,0,1,0,0,0,0,0,0,0,0,0,0,0,0,0,0,0,0]$& $(2,3,4,5,4,3,2,1,0,0,0,0,0,0,0,0,0,0,0,0,0,0,0,0,0,0,2)$&$2$ \\
\end{longtable}
\end{center}
\end{landscape}
\begin{spacing}{1.5}

\section{Solutions of M-theory}
\subsection{Electric Branes} \label{electricbranesolutions}
There is a substantial literature deriving single brane solutions in generic supergravity theories \cite{BPSsolutions}, for a review see \cite{Argurio, Stelle}. The $M2$, $M5$ and $pp$-wave solutions of $M$-theory have been derived from $E_{11}$ in \cite{West}.

For $E_{11}$ we have no dilaton and find the single brane solutions to be determined from a truncated action of the form
\begin{equation}
A=\frac{1}{16\pi G_{11}}\int {d^{11}x\sqrt{-g}(R-\frac{1}{2.4!}F_{a_1\ldots a_n}F^{a_1\ldots a_n})}
\end{equation}
$F_{a_1\ldots a_{n}}$ is a general $n$-form field strength formed in the non-linear realisation from the gauge fields. From \cite{West} we have two field strengths, which are dual to each other, which we consider as arising from the a 3-form and a 6-form gauge field
\begin{align}
F_{t_1x_1x_2r}=&4\partial_{[r}A_{t_1x_1x_2]}=\partial_rA_{t_1x_1x_2}=\partial_r{N_{(0,8)}^{-1}}\\
\nonumber F_{t_1x_1x_2x_3x_4x_5r}=&7\partial_{[r}A_{t_1x_1x_2x_3x_4x_5]}=\partial_rA_{t_1x_1x_2x_3x_4x_5}=\partial_r{N_{(0,5)}^{-1}}
\end{align}
The appropriate Einstein and gauge equations may be found by setting $\phi=0$, ${\hat{\delta}^{t_i}}_{\hspace{7pt} t_i}=1$, ${\hat{\delta}^{x_i}}_{\hspace{7pt} x_i}=p$, ${\hat{\delta}^{u_a}}_{\hspace{7pt} u_a}=0$ and ${\hat{\delta}^{y_a}}_{\hspace{7pt} y_a}=D-p-1$ in equations (\ref{EinsteinEq}) and (\ref{curvatureterms}).
The line element of an electric solution is derived from the solution generating group element and coincides with line element for single brane BPS solutions specified by equations (\ref{singlebranecoefficients}) \cite{West, CookWest}.
The non-linear realisation decomposes $E_{11}$ with respect to its longest gravity line $A_{10}$ obtaining a gravitational theory in 11 dimensions and an infinite array of irreducible representations. These representations are classified by level, the coefficient of the exceptional root $\alpha_{11}$ associated to the representation; all the usual solutions of eleven dimensional supergravity appear below level three. In the group element above, $\beta$ is the root associated to the lowest weight of each representation. The following solutions are found,
\subsection{The $M2$-Brane}
The line element of the $M2$-brane solution is \cite{DuffStelle}
\begin{equation}
ds^2=N^{-\frac{2}{3}}_{(0,8)}(-{dt_1}^2+dx_1^2+dx_2^2)+N^{\frac{1}{3}}_{(0,8)}(dy_1^2+\ldots dy_8^2) \label{$M2$metric}
\end{equation}
Giving a metric
\begin{displaymath}
{g_{\mu\nu}}=\bordermatrix{&t_1&x_1&x_2&y_1&\ldots &y_8\cr
t_1&-N_{(0,8)}^{-\frac{2}{3}}&0&0&0&0&0\cr
x_1&0&N_{(0,8)}^{-\frac{2}{3}}&0&0&0&0\cr
x_2&0&0&N_{(0,8)}^{-\frac{2}{3}}&0&0&0\cr
y_1&0&0&0&N_{(0,8)}^{\frac{1}{3}}&0&0\cr
\vdots&0&0&0&0&\ddots&0\cr
y_8&0&0&0&0&0&N_{(0,8)}^{\frac{1}{3}}\cr}
\end{displaymath}
So $\sqrt{-g}=N_{(0,8)}^{\frac{1}{3}}$ and we see that the gauge equation is satisfied by $F_{t_1x_1x_2y_a}=\partial_{y_a}{N_{(0,8)}^{-1}}$ in the following manner,
\begin{align}
\nonumber \partial_{y_a}(\sqrt{-g}F^{t_1x_1x_2y_a})=&\partial_{y_a}(N_{(0,8)}^{\frac{1}{3}}g^{t_1t_1'}g^{x_1x_1'}g^{x_2x_2'}g^{y_ay_a'}F_{t_1'x_1'x_2'y_a'})\\
\nonumber =&\partial_{y_1}(-N_{(0,8)}^2F_{t_1x_1x_2y_1})+\ldots \partial_{y_8}(-N_{(0,8)}^2F_{t_1x_1x_2y_8})\\
\nonumber =&\partial_{y_a}(-N_{(0,8)}^2\partial_{y_a}N_{(0,8)}^{-1}){\hat{\delta}^{y_a}}_{\hspace{7pt} y_a}\\
\nonumber =&\partial_{y_a}\partial_{y_a}N_{(0,8)}{\hat{\delta}^{y_a}}_{\hspace{7pt} y_a}\\
\nonumber =&0
\end{align}
In the last line we have used the fact that $N_{(0,8)}$ is an harmonic function in $y_a$, as can be checked from its definition in equation (\ref{harmonicfunction}). We see that the Einstein equations are satisfied by checking that the right-hand-side of each equation equals the curvature term for the electric solution. A term that appears frequently is $F^{t_1x_1x_2y_a}F_{t_1x_1x_2y_a}$ which we evaluate at the outset
\begin{align}
F^{t_1x_1x_2y_a}F_{t_1x_1x_2y_a}&=F^{t_1x_1x_2y_1}F_{t_1x_1x_2y_1}+\ldots F^{t_1x_1x_2y_8}F_{t_1x_1x_2y_8}\\
\nonumber &=-N_{(0,8)}^{-\frac{1}{3}}N_{(0,8)}^2(F_{t_1x_1x_2y_1})^2-\ldots -N_{(0,8)}^{-\frac{1}{3}}N_{(0,8)}^2(F_{t_1x_1x_2y_8})^2\\
\nonumber &=-8N_{(0,8)}^{-\frac{1}{3}}N_{(0,8)}^{-2}(\partial_{y_a}N_{(0,8)})^2
\end{align}
We proceed to check the Einstein equations,
\begin{align}
\nonumber \left(^{t_i}_{t_i}\right) \qquad \frac{1}{2.4!}(4&F^{t_1\mu_1\mu_2\mu_3}F_{t_1\mu_1\mu_2\mu_3}-\frac{1}{3}F^{\mu_1\mu_2\mu_3\mu_4}F_{\mu_1\mu_2\mu_3\mu_4})\\ 
&=\frac{1}{2.4!}(4.3!-\frac{4!}{3})F^{t_1x_1x_2y_a}F_{t_1x_1x_2y_a}\\
\nonumber &=-\frac{8}{3}N_{(0,8)}^{-\frac{1}{3}}N_{(0,8)}^{-2}(\partial_{y_a}N_{(0,8)})^2\\
\nonumber &=N^{-\frac{1}{3}}_{(0,8)} \{-\partial_{y_a}\partial_{y_a}\ln{N^{-\frac{1}{3}}_{(0,8)}} \}{\hat{\delta}^{y_a}}_{\hspace{7pt} y_a}\\
\nonumber &\equiv {R^{t_i}}_{t_i}\\
\nonumber \left(^{x_i}_{x_i}\right) \qquad 
\frac{1}{2.4!}(4&F^{x_i\mu_1\mu_2\mu_3}F_{x_i\mu_1\mu_2\mu_3}-\frac{1}{3}F^{\mu_1\mu_2\mu_3\mu_4}F_{\mu_1\mu_2\mu_3\mu_4}{\hat{\delta}^{x_i}}_{\hspace{7pt} x_i})\\
&=\frac{1}{2.4!}(4.3!-\frac{4!}{3})F^{t_1x_1x_2y_a}F_{t_1x_1x_2y_a}{\hat{\delta}^{x_i}}_{\hspace{7pt} x_i}\\
\nonumber &=N^{-\frac{1}{3}}_{(0,8)} \{-\partial_{y_a}\partial_{y_a}\ln{N^{-\frac{1}{3}}_{(0,8)}} \}{\hat{\delta}^{y_a}}_{\hspace{7pt} y_a}{\hat{\delta}^{x_i}}_{\hspace{7pt} x_i}\\
\nonumber &\equiv {R^{x_i}}_{x_i}\\
\nonumber \left(^{y_a}_{y_a}\right) \qquad 
\frac{1}{2.4!}(4&F^{y_a\mu_1\mu_2\mu_3}F_{y_a\mu_1\mu_2\mu_3}-\frac{1}{3}F^{\mu_1\mu_2\mu_3\mu_4}F_{\mu_1\mu_2\mu_3\mu_4})\\ 
&=\frac{1}{2.4!}(4.3!-\frac{4!}{3}{\hat{\delta}^{y_a}}_{\hspace{7pt} y_a})F^{t_1x_1x_2y_a}F_{t_1x_1x_2y_a}\\
\nonumber &=\frac{5}{6}N_{(0,8)}^{-\frac{1}{3}}N_{(0,8)}^{-2}{(\partial_{y_a}N_{(0,8)})}^2{\hat{\delta}^{y_a}}_{\hspace{7pt} y_a}\\
\nonumber &=N^{-\frac{1}{3}}_2 \{-8\partial_{y_a}\partial_{y_a}\ln{N^{\frac{1}{6}}_{(0,8)}}-3(\partial_{y_a}\ln{N^{-\frac{1}{3}}_{(0,8)}})^2\\
\nonumber & \qquad \qquad-6(\partial_{y_a}\ln{N^{\frac{1}{6}}_{(0,8)}})^2 \}{\hat{\delta}^{y_a}}_{\hspace{7pt} y_a}\\
\nonumber &\equiv {R^{y_a}}_{y_a}
\end{align}
\subsection{The $M5$-Brane}
The line element of the $M5$-brane solution is \cite{Guven}
\begin{equation}
ds^2=N^{-\frac{1}{3}}_{(0,5)}(-{dt_1}^2+dx_1^2+\ldots dx_5^2)+N^{\frac{2}{3}}_{(0,5)}(dy_1^2+\ldots dy_5^2)
\end{equation}
Giving a metric
\begin{displaymath}
{g_{\mu\nu}}=\bordermatrix{&t_1&x_1&\ldots &x_5&y_1&\ldots &y_5\cr
t_1&-N_{(0,5)}^{-\frac{1}{3}}&0&0&0&0&0&0\cr
x_1&0&N_{(0,5)}^{-\frac{1}{3}}&0&0&0&0&0\cr
\vdots&0&0&\ddots&0&0&0&0\cr
x_5&0&0&0&N_{(0,5)}^{-\frac{1}{3}}&0&0&0\cr
y_1&0&0&0&0&N_{(0,5)}^{\frac{2}{3}}&0&0\cr
\vdots&0&0&0&0&0&\ddots&0\cr
y_5&0&0&0&0&0&0&N_{(0,5)}^{\frac{2}{3}}\cr}
\end{displaymath}
So $\sqrt{-g}=N_{(0,5)}^{\frac{2}{3}}$ and we see that the gauge equation is satisfied by $F_{t_1x_1\ldots x_5y_a}=\partial_{y_a}{N_{(0,5)}^{-1}}$ in the following manner,
\begin{align}
\nonumber \partial_{y_a}(\sqrt{-g}F^{t_1x_1\ldots x_5y_a})=&\partial_{a}(-N_{(0,5)}^2\partial_aN_{(0,5)}^{-1}){\hat{\delta}^{y_a}}_{\hspace{7pt} y_a}\\
\nonumber =&\partial_{y_a}\partial_{y_a}N_{(0,5)}{\hat{\delta}^{y_a}}_{\hspace{7pt} y_a}\\
\nonumber =&0
\end{align}
As $N_{(0,5)}$ is an harmonic function in $y_a$. The Einstein equations are satisfied in the same way as the $M2$-brane solution, but it will be useful to express the equations in terms of the harmonic function $N_{(0,5)}$ for reference.
\begin{align}
\nonumber \left(^{t_i}_{t_i}\right) \hspace{7pt}\frac{1}{2.7!}(7.6!-\frac{2.7!}{3})F^{t_1x_1\ldots x_5y_a}F_{t_1x_1\ldots x_5y_a}=-\frac{1}{6}N_{(0,5)}^{-\frac{2}{3}}N_{(0,5)}^{-2}{(\partial_{y_a}N_{(0,5)})}^2{\hat{\delta}^{y_a}}_{\hspace{7pt} y_a}\\
\nonumber \left(^{x_i}_{x_i}\right) \hspace{7pt}\frac{1}{2.7!}(7.6!-\frac{2.7!}{3})5F^{t_1x_1\ldots x_5y_a}F_{t_1x_1\ldots x_5y_a}=-\frac{5}{6}N_{(0,5)}^{-\frac{2}{3}}N_{(0,5)}^{-2}{(\partial_{y_a}N_{(0,5)})}^2{\hat{\delta}^{y_a}}_{\hspace{7pt} y_a}\\
\nonumber \left(^{y_a}_{y_a}\right) \hspace{7pt}\frac{1}{2.7!}(7.6!-5\frac{2.7!}{3})F^{t_1x_1\ldots x_5y_a}F_{t_1x_1\ldots x_5y_a}=\frac{7}{6}N_{(0,5)}^{-\frac{2}{3}}N_{(0,5)}^{-2}{(\partial_{y_a}N_{(0,5)})}^2{\hat{\delta}^{y_a}}_{\hspace{7pt} y_a}
\end{align}
\subsection{The $pp$-Wave}
The $pp$-wave solution \cite{Hull2} arises from considering the lowest weight generator associated to a positive root, namely ${K^1}_2$, in the weight chain whose highest weight has root $\beta=\alpha_1+\alpha_2+\ldots \alpha_{10}$. ${K^1}_2$ is the generator of the root $\alpha_1$ and we find an associated line element, 
\begin{equation}
ds^2=-(1-K){dt_1}^2+(1+K){dx_2}^2-2Kdt_1dx_2+d\Omega_9^2
\end{equation}
Where we have made the substitution $N_{pp}=1+K$ and we note that $N_{pp}=1+\frac{\bf Q}{7r^7}$, and $r^2=y_1^2+\ldots y_9^2$. We note that $K=K(y_1,\ldots y_9)$, which is less general than the solution in \cite{Hull2}, but fits with the generic harmonic functions we have used for all brane solutions in this paper.

\subsection{Spacelike Brane Solutions} \label{spacelikebranesolutions}
In this appendix we demonstrate that the $S$-brane solutions discussed in section \ref{spacelikeinvolution} satisfy the Einstein equations (\ref{EinsteinEq}) in signature $(1,10,-)$ for our ansatz (\ref{ansatz}). As discussed in section  \ref{spacelikeinvolution} our field strength may be constructed out of a complexified version of the generators that give rise to the usual electric solutions in $(1,10,+)$, such that these solutions are derived from a truncated action with a $+F^2$ term. Equivalently we may use a real form of the sub-algebra generators in signature $(1,10,-)$ to construct our putative solutions. We follow the same approach as in appendix \ref{electricbranesolutions} and have two field strengths derived from a 3-form and a 6-form gauge field both of which have purely spatial indices, given in equation (\ref{spacelikefieldstrengths}). The appropriate Einstein and gauge equations may be found by setting $\phi=0$, ${\hat{\delta}^{t_i}}_{\hspace{7pt} t_i}=0$, ${\hat{\delta}^{x_i}}_{\hspace{7pt} x_i}=p+1$, ${\hat{\delta}^{u_a}}_{\hspace{7pt} u_a}=1$ and ${\hat{\delta}^{y_a}}_{\hspace{7pt} y_a}=D-p-2$ in equations (\ref{EinsteinEq}) and (\ref{curvatureterms}). The line element of a spacelike solution is derived from the solution generating group element in the same way as the electric case but using a choice of local sub-algebra that invokes a non-compact timelike generator in the transverse coordinates. The following solutions are associated with the lowest weights,
\subsection{The $S2$-Brane}
We now demonstrate that there exists an $S2$-brane solution in signature $(1,10,-)$ for our ansatz (\ref{ansatz}). The line element of the $S2$-brane solution is
\begin{equation}
ds^2=N^{-\frac{2}{3}}_{(1,7)}(dx_1^2+\ldots dx_3^2)+N^{\frac{1}{3}}_{(1,7)}(-du_1^2+dy_1^2+\ldots dy_7^2) \label{$S2$metric}
\end{equation}
Giving a metric
\begin{displaymath}
{g_{\mu\nu}}=\bordermatrix{&x_1&x_2&x_3&u_1&y_1&\ldots &y_7\cr
x_1&N_{(1,7)}^{-\frac{2}{3}}&0&0&0&0&0&0\cr
x_2&0&N_{(1,7)}^{-\frac{2}{3}}&0&0&0&0&0\cr
x_3&0&0&N_{(1,7)}^{-\frac{2}{3}}&0&0&0&0\cr
u_1&0&0&0&-N_{(1,7)}^{\frac{1}{3}}&0&0&0\cr
y_1&0&0&0&0&N_{(1,7)}^{\frac{1}{3}}&0&0\cr
\vdots&0&0&0&0&0&\ddots&0\cr
y_7&0&0&0&0&0&0&N_{(1,7)}^{\frac{1}{3}}\cr}
\end{displaymath}
So $\sqrt{-g}=N_{(1,7)}^{\frac{1}{3}}$ and we see that the gauge equation is satisfied by $F_{x_1x_2x_3\hat{r}}=\partial_{\hat{r}}{N_{(1,7)}^{-1}}$ in the following manner,
\begin{align}
\nonumber \partial_{\hat{r}}(\sqrt{-g}F^{x_1x_2x_3\hat{r}})=&\partial_{\hat{r}}(N_{(1,7)}^{\frac{1}{3}}g^{x_1x_1'}g^{x_2x_2'}g^{x_3x_3'}g^{\hat{r}\hat{r}'}F_{x_1'x_2'x_3'\hat{r}'})\\
\nonumber =&\partial_{u_1}(-N_{(1,7)}^2\partial_{u_1}N_{(1,7)}^{-1})+\partial_{y_a}(N_{(1,7)}^2\partial_{y_a}N_{(1,7)}^{-1}){\hat{\delta}^{y_a}}_{\hspace{7pt} y_a}\\
=&\partial_{u_1}\partial_{u_1}N_{(1,7)}-\partial_{y_a}\partial_{y_a}N_{(1,7)}{\hat{\delta}^{y_a}}_{\hspace{7pt} y_a} \label{S2gauge}\\
\nonumber =&0
\end{align}
In the last line we have used the fact that $N_{(1,7)}$ is an harmonic function in $y_a$, as can be checked from its definition in equation (\ref{harmonicfunction}),
\begin{align}
\nonumber -\partial_{u_1}\partial_{u_1}N_{(1,7)}+\partial_{y_a}\partial_{y_a}N_{(1,7)}{\hat{\delta}^{y_a}}_{\hspace{7pt} y_a}&=-\frac{6k}{\hat{r}^8}-\frac{8.6ku_1^2}{\hat{r}^{10}}-\frac{6k}{\hat{r}^8}{\hat{\delta}^{y_a}}_{\hspace{7pt} y_a}+\frac{8.6ky_a^2}{\hat{r}^{10}}{\hat{\delta}^{y_a}}_{\hspace{7pt} y_a}\\
&=-\frac{8.6k}{\hat{r}^8}+\frac{8.6k\hat{r}^2}{\hat{r}^{10}} \label{S2harmonic}\\
\nonumber &=0
\end{align}
Where $k=\pm\frac{\|\bf Q\|}{6}$ for the $S2$-brane, as defined in equation (\ref{harmonicfunction}).

We now check that the Einstein equations are satisfied by verifying that the right-hand-side of each equation equals the curvature term for the spacelike solution. A term that appears frequently is $F^{x_1x_2x_3\hat{r}}F_{x_1x_2x_3\hat{r}}$ which we evaluate at the outset
\begin{align}
\nonumber F^{x_1x_2x_3\hat{r}}F_{x_1x_2x_3\hat{r}}&=F^{x_1x_2x_3u_1}F_{x_1x_2x_3u_1}+F^{t_1x_1x_2y_1}F_{t_1x_1x_2y_1}+\ldots\\
&\qquad F^{t_1x_1x_2y_7}F_{t_1x_1x_2y_7}\\
\nonumber &=-N_{(1,7)}^{-\frac{1}{3}}N_{(1,7)}^2(F_{x_1x_2x_3u_1})^2+7N_{(1,7)}^{-\frac{1}{3}}N_{(1,7)}^2(F_{x_1x_2x_3y_a})^2
\end{align}
We proceed to check the Einstein equations,
\begin{align}
\nonumber \left(^{x_i}_{x_i}\right) \qquad 
-\frac{1}{2.4!}(4&F^{x_i\mu_1\mu_2\mu_3}F_{x_i\mu_1\mu_2\mu_3}-\frac{1}{3}F^{\mu_1\mu_2\mu_3\mu_4}F_{\mu_1\mu_2\mu_3\mu_4}{\hat{\delta}^{x_i}}_{\hspace{7pt} x_i})\\
&=-\frac{1}{2.4!}(4.3!-\frac{4!}{3})F^{x_1x_2x_3\hat{r}}F_{x_1x_2x_3\hat{r}}{\hat{\delta}^{x_i}}_{\hspace{7pt} x_i}\\
\nonumber &=N^{-\frac{1}{3}}_{(1,7)} \{\partial_{u_1}\partial_{u_1}\ln{N^{-\frac{1}{3}}_{(1,7)}}-\partial_{y_a}\partial_{y_a}\ln{N^{-\frac{1}{3}}_{(1,7)}}{\hat{\delta}^{y_a}}_{\hspace{7pt} y_a} \}{\hat{\delta}^{x_i}}_{\hspace{7pt} x_i}\\
\nonumber &\equiv {R^{x_i}}_{x_i}\\
\nonumber \left(^{u_a}_{u_a}\right) \qquad -\frac{1}{2.4!}(4&F^{u_1\mu_1\mu_2\mu_3}F_{u_1\mu_1\mu_2\mu_3}-\frac{1}{3}F^{\mu_1\mu_2\mu_3\mu_4}F_{\mu_1\mu_2\mu_3\mu_4})\\ 
&=-\frac{1}{2.4!}(4.3!F^{x_1x_2x_3u_1}F_{x_1x_2x_3u_1}-\frac{4!}{3}F^{x_1x_2x_3\hat{r}}F_{x_1x_2x_3\hat{r}})\\
\nonumber &=N_{(1,7)}^{-\frac{1}{3}}N_{(1,7)}^{-2}\{\frac{1}{3}(\partial_{u_1}N_{(1,7)})^2+\frac{7}{6}(\partial_{y_a}N_{(1,7)})^2\}\\
\nonumber &\equiv {R^{u_a}}_{u_a}\\
\nonumber \left(^{y_a}_{y_a}\right) \qquad 
-\frac{1}{2.4!}(4&F^{y_a\mu_1\mu_2\mu_3}F_{y_a\mu_1\mu_2\mu_3}-\frac{1}{3}F^{\mu_1\mu_2\mu_3\mu_4}F_{\mu_1\mu_2\mu_3\mu_4})\\ 
&=-\frac{1}{2.4!}(4.3!F^{x_1x_2x_3y_a}F_{x_1x_2x_3y_a}-\frac{4!}{3}{\hat{\delta}^{y_a}}_{\hspace{7pt} y_a}F^{x_1x_2x_3\hat{r}}F_{x_1x_2x_3\hat{r}})\\
\nonumber &=N_{(1,7)}^{-\frac{1}{3}}N_{(1,7)}^{-2}\{-\frac{1}{6}(\partial_{u_1}N_{(1,7)})^2+\frac{2}{3}(\partial_{y_a}N_{(1,7)})^2\}{\hat{\delta}^{y_a}}_{\hspace{7pt} y_a}\\
\nonumber &\equiv {R^{y_a}}_{y_a}
\end{align}

\subsection{The $S5$-Brane}
We now demonstrate that there exists an $S5$-brane solution in signature $(1,10,-)$ for our ansatz (\ref{ansatz}). The line element of the $S5$-brane solution is
\begin{equation}
ds^2=N^{-\frac{1}{3}}_{(1,4)}(dx_1^2+\ldots dx_6^2)+N^{\frac{2}{3}}_{(1,4)}(-du_1^2+dy_1^2+\ldots dy_4^2)
\end{equation}
Giving a metric
\begin{displaymath}
{g_{\mu\nu}}=\bordermatrix{&x_1&\ldots&x_6&u_1&y_1&\ldots &y_4\cr
x_1&N_{(1,4)}^{-\frac{1}{3}}&0&0&0&0&0&0\cr
\vdots&0&\ddots &0&0&0&0&0\cr
x_6&0&0&N_{(1,4)}^{-\frac{1}{3}}&0&0&0&0\cr
u_1&0&0&0&-N_{(1,4)}^{\frac{2}{3}}&0&0&0\cr
y_1&0&0&0&0&N_{(1,4)}^{\frac{2}{3}}&0&0\cr
\vdots&0&0&0&0&0&\ddots&0\cr
y_4&0&0&0&0&0&0&N_{(1,4)}^{\frac{2}{3}}\cr}
\end{displaymath}
So $\sqrt{-g}=N_{(1,4)}^{\frac{2}{3}}$ and we see that the gauge equation is satisfied by $F_{x_1\ldots x_6\hat{r}_a}=\partial_{\hat{r}_a}{N_{(1,4)}^{-1}}$ in the same manner as the field strength associated to the $S2$-brane in equations (\ref{S2gauge}) and (\ref{S2harmonic}). The Einstein equations are also satisfied in the same way as the $S2$-brane solution. We first note that
\begin{equation}
F^{x_1\ldots x_6\hat{r}}F_{x_1\ldots x_6\hat{r}}=-N_{(1,4)}^{-\frac{2}{3}}N_{(1,4)}^{-2}(\partial_{u_1}N_{(1,4)})^2+4N_{(1,4)}^{-\frac{2}{3}}N_{(1,4)}^{-2}(\partial_{y_a}N_{(1,4)})^2
\end{equation}
Let us now confirm that the Einstein equations in $(1,10,-)$ are satisfied for our ansatz (\ref{ansatz}).
\begin{align}
\nonumber \left(^{x_i}_{x_i}\right) \qquad 
-\frac{1}{2.7!}(7&F^{x_i\mu_1\ldots\mu_6}F_{x_i\mu_1\ldots\mu_6}-\frac{2}{3}F^{\mu_1\ldots\mu_7}F_{\mu_1\ldots\mu_7}{\hat{\delta}^{x_i}}_{\hspace{7pt} x_i})\\
&=-\frac{1}{2.7!}(7.6!-\frac{2.7!}{3})F^{x_1\ldots x_6\hat{r}}F_{x_1\ldots x_6\hat{r}}{\hat{\delta}^{x_i}}_{\hspace{7pt} x_i}\\
\nonumber &=N^{-\frac{2}{3}}_{(1,4)} \{\partial_{u_1}\partial_{u_1}\ln{N^{-\frac{1}{6}}_{(1,4)}}-\partial_{y_a}\partial_{y_a}\ln{N^{-\frac{1}{6}}_{(1,4)}}{\hat{\delta}^{y_a}}_{\hspace{7pt} y_a} \}{\hat{\delta}^{x_i}}_{\hspace{7pt} x_i}\\
\nonumber &\equiv {R^{x_i}}_{x_i}\\
\nonumber \left(^{u_a}_{u_a}\right) \qquad -\frac{1}{2.7!}(7&F^{u_1\mu_1\ldots\mu_6}F_{u_1\mu_1\ldots\mu_6}-\frac{2}{3}F^{\mu_1\ldots\mu_7}F_{\mu_1\ldots\mu_7})\\ 
&=-\frac{1}{2.7!}(7.6!F^{x_1\ldots x_6u_1}F_{x_1\ldots x_6u_1}-\frac{2.7!}{3}F^{x_1\ldots x_6\hat{r}}F_{x_1\ldots x_6\hat{r}})\\
\nonumber &=N_{(1,4)}^{-\frac{2}{3}}N_{(1,4)}^{-2}\{\frac{1}{6}(\partial_{u_1}N_{(1,4)})^2+\frac{4}{3}(\partial_{y_a}N_{(1,4)})^2\}\\
\nonumber &\equiv {R^{u_a}}_{u_a}\\
\nonumber \left(^{y_a}_{y_a}\right) \qquad 
-\frac{1}{2.7!}(7&F^{y_a\mu_1\ldots\mu_6}F_{y_a\mu_1\ldots\mu_6}-\frac{2}{3}F^{\mu_1\ldots\mu_7}F_{\mu_1\ldots\mu_7})\\ 
&=-\frac{1}{2.7!}(7.6!F^{x_1\ldots x_6y_a}F_{x_1\ldots x_6y_a}-\frac{2.7!}{3}{\hat{\delta}^{y_a}}_{\hspace{7pt} y_a}F^{x_1\ldots x_6\hat{r}}F_{x_1\ldots x_6\hat{r}})\\
\nonumber &=N_{(1,4)}^{-\frac{2}{3}}N_{(1,4)}^{-2}\{-\frac{1}{3}(\partial_{u_1}N_{(1,4)})^2+\frac{5}{6}(\partial_{y_a}N_{(1,4)})^2\}{\hat{\delta}^{y_a}}_{\hspace{7pt} y_a}\\
\nonumber &\equiv {R^{y_a}}_{y_a}
\end{align}
\subsection{The $Spp$-Wave}
The $Spp$-wave solution arises from considering the lowest weight in the weight chain whose highest weight has root $\beta=\alpha_1+\alpha_2+\ldots \alpha_{10}$, with a choice of local sub-algebra such that the temporal coordinate of $M$-theory is not one of the two distinguished coordinates as it is in the $pp$-wave solution. The line element derived from the group element (\ref{groupelement}) using $E_\beta=i{K^1}_2$ is a solution of the vacuum Einstein equations and is,
\begin{equation}
ds^2=(1-K)dx_1^2+(1+K)dx_2^2-2iKdx_1dx_2+d\Omega_{(1,8)}^2
\end{equation}
Where $d\Omega^2_{(1,8)}=-du_1^2+dy_1^2+\ldots dy_{8}^2$, and $K=K(u_1,y_1,\ldots y_8)$. This $Spp$-wave metric is the line element expected from a double Wick rotation of the $pp$-wave solution. A further Wick rotation would give a $pp$-wave solution of the $M*$-theory in $(2,9)$. It is non-static and has wavefronts that progress in the spacelike directions transverse to $\{x_1,x_2\}$.

\newpage
\section{Electric Branes from Kac-Moody Algebras} \label{oxidisedgravity}
\subsection{Very Extended $E_6$}
The Dynkin diagram for $E_6^{+++}$ is given in appendix \ref{Dynkindiagrams}. The $E_6^{+++}$ algebra may only be uniquely decomposed with respect to an $A_7$ gravity line, giving an 8-dimensional theory. The simple roots that we eliminate are $\alpha_9$ and $\alpha_8$ whose corresponding generators are $R_1 (00\ldots 001)$ and $R_0^{678} (00\ldots 010)$ and we associate the levels $l_1$ and $l_2$ with these respectively. The $s$ labels on our generators are chosen to be the $l_1$ level of the generator. The alternative decomposition along the other branch of the Dynkin diagram is related to the first decomposition via an automorphism.\\

From \cite{KleinschmidtSchnakenburgWest} we find the remaining algebra contains the generators ${K^a}_b$ at level (0,0) and the following generators at up to level (1,2)
\begin{alignat}{3}
\nonumber l_1 \longrightarrow \qquad &\qquad &\qquad &0 &\qquad &1\\ 
\nonumber l_2 \downarrow \qquad \qquad &0 &\qquad & R_0 & \qquad & R_1\\ 
&1 &\qquad &R_0^{a_1a_2a_3} & \qquad &R_1^{a_1a_2a_3} \label{E6generators}\\
\nonumber &2 &\qquad &R_0^{a_1\ldots a_6} & \qquad &R_1^{a_1\ldots a_6}\\
\nonumber & &\qquad & & \qquad &R_1^{a_1\ldots a_5,b}
\end{alignat}
All the generators have $\beta^2=2$, where $\beta$ is the associated root for the generator, with the exceptions of $R_0$ and $R_1^{a_1\ldots a_6}$ which have $\beta^2=0$ and so will not be used as starting points for finding electric branes.\\

We make use of the following commutator relations where we have chosen the commutator coefficient for $[R_0,R_1]$ and the others have followed from equations (\ref{jicoefficients1})-(\ref{jicoefficients4}) and the Serre relations (\ref{Serrerelations})
\begin{alignat}{3}
\nonumber \qquad \qquad & &\qquad & & \qquad & [R_0,R_1]=-R_1\\ 
& &\qquad &[R_0,R_0^{a_1a_2a_3}]=\frac{1}{2}R_0^{a_1a_2a_3} & \qquad &[R_0,R_1^{a_1a_2a_3}]=-\frac{1}{2}{R_1}^{a_1a_2a_3}\\
\nonumber & &\qquad &[R_0,R_0^{a_1\ldots a_6}]=R_0^{a_1\ldots a_6} & \qquad &[R_0,R_1^{a_1\ldots a_6}]=0
\end{alignat}
The simple root generators of $E_6^{+++}$ are
\begin{equation}
E_a={K^a}_{a+1}, a=1,\ldots 7, \qquad E_8=R_0^{678}, \qquad E_9=R_1
\end{equation}
and the Cartan sub-algebra generators, $H_a$, are given by
\begin{eqnarray}
\nonumber &H_a={K^a}_a-{K^{a+1}}_{a+1}, a=1,\ldots 7,\\
&H_8=-\frac{1}{2}({K^1}_1+\ldots{K^5}_5)+\frac{1}{2}({K^6}_6+\ldots{K^8}_8)+R_0,\\
\nonumber&H_9=-2R_0
\end{eqnarray}
The low-level field content \cite{KleinschmidtSchnakenburgWest} is ${\hat{h}}_a\hspace{0pt}^b$, $A$, ${A_{a_1a_2a_3}}_0$, ${A_{a_1\ldots a_6}}_0$ and their field strengths have duals derived from  ${A_{a_1\ldots a_5,b}}_1$, ${A_{a_1\ldots a_6}}_1$, ${A_{a_1a_2a_3}}_1$, $A_1$, respectively. Our choice of local sub-algebra for the non-linear realisation allows us to write the group element of $E_6^{+++}$ as
\begin{align}
\nonumber g=\exp(\sum_{a\leq b}{\hat{h}}_a\hspace{0pt}^b{K^a}_b)&\exp(\frac{1}{6!}{A_{a_1\ldots a_6}}_1R_1^{a_1\ldots a_6})\exp(\frac{1}{6!}{A_{a_1\ldots a_6}}_0R_0^{a_1\ldots a_6})\\
&\exp(\frac{1}{5!}{A_{a_1\ldots a_5,b}}_1R_1^{a_1\ldots a_5,b})\exp(\frac{1}{3!}{A_{a_1a_2a_3}}_1R_1^{a_1\ldots a_3})\\
\nonumber &\exp(\frac{1}{3!}{A_{a_1a_2a_3}}_0R_0^{a_1\ldots a_3})\exp(A_1 R_1)\exp(AR_0)
\end{align}
The field content of the associated maximally oxidised theory given in \cite{CremmerJuliaLuPope} agrees with the low-order $E_6^{+++}$ content given above. The Lagrangian for the oxidised theory, contains contains a graviton, $\phi$, a 4-form field strength, ${F_{\mu\nu\rho\sigma}}_0=4\partial_{[\mu}{A_{\nu\rho\sigma]}}_0$, a 1-form field strength, ${F_{\mu}}_1=\partial_{\mu}A_1$ and a dilaton $A$ and is given by
\begin{align}
\nonumber A=\frac{1}{16\pi G_{n-1}}\int d^8x\sqrt{-g}(&R-\frac{1}{2}\partial_\mu\phi\partial^\mu\phi-\frac{1}{2}e^{2\phi}{F_{\mu}}_1F^{\mu}_1\\
&-\frac{1}{2.4!}e^{-\phi}{F_{\mu\nu\rho\sigma}}_0F^{\mu\nu\rho\sigma}_0) \label{E6lagrangian}\\
\nonumber +\int_{C.S.} d^8x\epsilon^{a_1\ldots a_8}&\frac{1}{4!4!}A_1{F_{a_1\ldots a_4}}_0{F_{a_5\ldots a_8}}_0
\end{align}
Using the group element (\ref{groupelement}) we find the following electric branes
\paragraph{A 2-Brane}
Let us find the electric brane associated with the generator ${R^{a_1a_2a_3}}_0$ whose highest weight is $R_0^{678}$ $(00\ldots010)$. The lowest weight generator is $R_0^{123}$ $(123\ldots32110)$ and the associated Cartan sub-algebra element is
\begin{equation}
\nonumber \beta \cdot H=\frac{1}{2}({K^1}_1+\ldots{K^3}_3)-\frac{1}{2}({K^4}_4+\ldots{K^8}_8)+R_0
\end{equation}
Bearing in mind that $\beta^2=2$, we find that the group element is
\begin{align}
\nonumber g_A=&\exp(-\frac{1}{2}\ln N_2(\frac{1}{2}({K^1}_1+\ldots{K^3}_3)-\frac{1}{2}({K^4}_4+\ldots{K^8}_8)))\\
&\exp(N_2^{-\frac{1}{4}}(1-N_2)E_{\beta})\exp(-\frac{1}{2}\ln N_2R_0)
\end{align}
We have a dilaton given by
\begin{equation}
e^A=N_2^{-\frac{1}{2}}
\end{equation}
And a line element corresponding to a 2-brane
\begin{equation}
ds^2=N_2^{-\frac{1}{2}}(-dy_1^2+dx_2^2+dx_3^2)+N_2^{\frac{1}{2}}(dx_4^2+\ldots+dx_8^2)
\end{equation}
The brane is derived from a gauge potential
\begin{equation}
{A_{123}^T}_{(0)}=N_2^{-\frac{1}{4}}(1-N_2)
\end{equation}
The change to world volume indices is made using $({e^{\hat{h}})^1}_1=\ldots({e^{\hat{h}})^3}_3=N_2^{-\frac{1}{4}}$
\begin{equation}
{A_{123}}_{(0)}=N_2^{-1}-1
\end{equation}
This gauge potential gives rise to a 4-form field strength, ${F_{\mu\nu\rho\sigma}}_0$.
\paragraph{A Second 2-Brane}
Let us find the brane associated with the generator ${R^{a_1}}_1$ whose highest weight is $R_1^{678}$ $(00\ldots011)$. The corresponding lowest weight generator is $R_1^{123}$ $(123\ldots32111)$ element in the Cartan sub-algebra is identical to that for $R_0^{123}$ but with $2R_0$ taken off, that is,
\begin{equation}
\beta \cdot H=\frac{1}{2}({K^1}_1+\ldots{K^3}_3)-\frac{1}{2}({K^4}_4+\ldots{K^8}_8)-R_0
\end{equation}
We find that the group element of equation (\ref{groupelement}) is
\begin{align}
\nonumber g_A=&\exp(-\frac{1}{2}\ln N_2(\frac{1}{2}({K^1}_1+\ldots{K^3}_3)-\frac{1}{2}({K^4}_4+\ldots{K^8}_8)))\\
&\exp(N_2^{-\frac{1}{4}}(1-N_2)E_{\beta})\exp(\frac{1}{2}\ln N_2R_0)
\end{align}
We have a dilaton given by
\begin{equation}
e^A=N_2^{\frac{1}{2}}
\end{equation}
And a line element corresponding to a 2-brane
\begin{equation}
ds^2=N_2^{-\frac{1}{2}}(-dy_1^2+dx_2^2+dx_3^2)+N_2^{\frac{1}{2}}(dx_4^2+\ldots+dx_8^2)
\end{equation}
The brane is derived from a gauge potential
\begin{equation}
{A_{123}^T}_{(1)}=N_2^{-\frac{1}{4}}(1-N_2)
\end{equation}
The change to world volume indices is made using $({e^{\hat{h}})^1}_1=\ldots({e^{\hat{h}})^3}_3=N_2^{-\frac{1}{4}}$
\begin{equation}
{A_{123}}_{(1)}=N_2^{-1}-1
\end{equation}
This gives rise to a 4-form field strength, which we conclude is the dual to ${F_{\mu\nu\rho\sigma}}_0$.
\paragraph{A 5-Brane}
Let us find the electric brane associated with the generator ${R^{a_1a_2\ldots a_6}}_0$ which has highest weight $R_0^{345678}(001232120)$. The lowest weight is $R_0^{123456}(123454220)$ and the corresponding element in the Cartan sub-algebra is
\begin{align}
\nonumber \beta \cdot H=&H_1+2H_2+3H_3+4H_4+5H_5+4H_6+2H_7+2H_8\\
=&0({K^1}_1+\ldots{K^6}_6)-({K^7}_7+{K^8}_8)+2R_0
\end{align}
The group element (\ref{groupelement}) is
\begin{align}
\nonumber g_A=&\exp(\frac{1}{2}\ln N_5({K^7}_7+{K^8}_8))\\
&\exp(N_5^{-1}(1-N_5)E_{\beta})\exp(-\ln N_5R_0)
\end{align}
We have a dilaton given by
\begin{equation}
e^A=N_5^{-1}
\end{equation}
And a line element corresponding to a 5-brane
\begin{equation}
ds^2=(-dy_1^2+dx_2^2+\ldots dx_6^2)+N_5(dx_7^2+dx_8^2)
\end{equation}
The brane is derived from a gauge potential
\begin{equation}
{A_{12\ldots 6}^T}_{(0)}=N_5^{-1}-1
\end{equation}
The change to world volume indices leaves the form of the gauge potential unaltered as $({e^{\hat{h}})^1}_1=\ldots({e^{\hat{h}})^6}_6=1$
\begin{equation}
{A_{12\ldots 6}}_{(0)}=N_5^{-1}-1
\end{equation}
This gauge potential gives rise to a 7-form field strength, which we conclude is the dual to the 1 form ${F_\mu}_1=\partial_\mu R_1$.

We have have successfully reproduced all the usual BPS electric branes using the group element given in equation (\ref{groupelement}). The formulae we have found above do indeed correspond to the solutions of the Lagrangian (\ref{E6lagrangian}). Our dilaton field $A$ is related to $\phi$ by
\begin{equation}
A = \phi
\end{equation}

\subsection{Very Extended $E_7$ - The 10-dimensional theory}
The $E_7^{+++}$ algebra may be decomposed with respect to an $A_9$ gravity line, giving a 10-dimensional theory or equally with respect to its $A_7$ gravity line to give an 8-dimensional theory. We consider the 10-dimensional theory here, and note that in doing so we depart from the considerations of \cite{CremmerJuliaLuPope}, in which the 9-dimensional theory is considered.

The simple root whose node we delete is $\alpha_{10}$ and has generator $R^{78910}$ $(00\ldots 001)$ and we associate the level $l$ with it. Our $s$ labels are all zero, as we eliminate only one simple root, and are discarded in this section. From reference \cite{KleinschmidtSchnakenburgWest} we find the ${E_7}^{+++}$ algebra decomposed with respect to an $A_9$ sub-algebra contains the generators ${K^a}_b$ at level 0 and the following generators up to level 2, and notably no dilaton generator,
\begin{alignat}{3}
\nonumber l \downarrow \qquad & &\qquad & &\qquad & \\
\nonumber 1 \qquad & &\qquad &R^{a_1a_2a_3a_4} &\qquad & \\
 2 \qquad & &\qquad &R^{a_1\ldots a_7,b} &\qquad & \label{E7generators}\\
\nonumber \qquad & &\qquad &R^{a_1\ldots a_8} &\qquad &
\end{alignat}
We note that the generator $R^{a_1\ldots a_8}$ has associated root $\beta$ such that $\beta^2=0$ so we discard this as a starting point for finding an electric brane, all the other generators have $\beta^2=2$.
The simple root generators of $E_7^{+++}$ are
\begin{align}
\nonumber E_a&={K^a}_{a+1}, a=1,\ldots 9
E_{10}&=R^{78910}
\end{align}
and the Cartan sub-algebra generators, $H_a$, are given by
\begin{eqnarray}
\nonumber &H_a={K^a}_a-{K^{a+1}}_{a+1}, a=1,\ldots 9,\\
&H_{10}=-\frac{1}{2}({K^1}_1+\ldots{K^6}_6)+\frac{1}{2}({K^7}_7+\ldots{K^{10}}_{10})
\end{eqnarray}
The low-level field content \cite{KleinschmidtSchnakenburgWest} is ${\hat{h}}_a\hspace{0pt}^b$, and its field strength has a dual derived from  ${A_{a_1\ldots a_7,b}}$, we also have fields ${A_{a_1a_2a_3a_4}}$, ${A_{a_1\ldots a_8}}$ which are not related to each other by a duality condition and we take them to be self-dual in our low order theory. Our choice of local sub-algebra for the non-linear realisation allows us to write the group element of $E_{7}^{+++}$ as
\begin{align}
\nonumber g=\exp(\sum_{a\leq b}{\hat{h}}_a\hspace{0pt}^b{K^a}_b)&\exp(\frac{1}{8!}{A_{a_1\ldots a_8}}R^{a_1\ldots a_8})\exp(\frac{1}{7!}A_{a_1\ldots a_7,b}R^{a_1\ldots a_7,b})\\
&\exp(\frac{1}{4!}A_{a_1\ldots a_4}R^{a_1\ldots a_4})
\end{align}
\paragraph{A 3-Brane}
Let us find the electric brane associated with the generator $R^{a_1a_2a_3a_4}$ whose highest weight is $R^{78910}$ $(0 0\ldots 0 0 1)$. The lowest weight generator in this representation is $R^{1234}$ $(1 2 3 4 4 4 3 2 1 1)$ giving,
\begin{equation}
\beta \cdot H=\frac{1}{2}({K^1}_1+\ldots{K^4}_4)-\frac{1}{2}({K^5}_5+\ldots{K^{10}}_{10})
\end{equation}
We now write down the group element from equation (\ref{groupelement})
\begin{equation}
g_A=\exp(-\frac{1}{2}\ln N_3(\frac{1}{2}({K^1}_1+\ldots{K^4}_4)-\frac{1}{2}({K^5}_5+\ldots{K^{10}}_{10})))\exp((1-N_3)E_{\beta})
\end{equation}
We have a line element corresponding to a 3-brane
\begin{equation}
ds^2=N_3^{-\frac{1}{2}}(-dy_1^2+dx_2^2+\ldots dx_4^2)+N_3^{\frac{1}{2}}(dx_5^2+\ldots+dx_{10}^2)
\end{equation}
The brane is derived from a gauge potential
\begin{equation}
A_{1234}^T=1-N_3
\end{equation}
We complete the change to world volume indices using $({e^{\hat{h}})^1}_1=\ldots({e^{\hat{h}})^4}_4=N_3^{-\frac{1}{4}}$
\begin{equation}
A_{1234}=N_3^{-1}-1
\end{equation}
This gives rise to a 5-form field strength,$F_{\mu\nu\rho\sigma\tau}$ which we conclude is self-dual, as there are no other other generators in (\ref{E7generators}) from which we could derive a dual field strength, and consequently we cannot construct a Lagrangian for this theory.
\subsection{Very Extended $G_2$}
The Dynkin diagram for ${G_2}^{+++}$ is shown in appendix \ref{Dynkindiagrams}, with the darkened nodes indicating the gravity line we consider here. The ${G_2}^{+++}$ algebra is decomposed with respect to the $A_4$ sub-algebra of the gravity line where the deleted node corresponds to simple root is $\alpha_5$ whose generator is $R^5 (00001)$, to which our decomposition level, $l$, is associated. we have no need for the $s$ labels as they are all zero, so we discard them in this section. Reference \cite{KleinschmidtSchnakenburgWest} gives the generators at low levels, we find the ${K^a}_b$ at the zeroth level corresponding to the gravity line step operators and we have the following other generators up to level 3, and notably no dilaton generator,
\begin{alignat}{3}
\nonumber l \downarrow \qquad & &\qquad & &\qquad & \\
\nonumber 1 \qquad & &\qquad &R^a &\qquad & \\
 2 \qquad & &\qquad &R^{ab} &\qquad & \label{G2generators}\\
\nonumber 3 \qquad & &\qquad &R^{ab,c} &\qquad &\\
\nonumber  \qquad & &\qquad &R^{abc} &\qquad &
\end{alignat}
The generators $R^a$, $R^{ab}$ and $R^{ab,c}$ each have an associated root $\beta$ such that $\beta^2=\frac{2}{3}$ but $R^{abc}$ has $\beta^2=0$ so we shall discard it as a starting point for our method of finding electric branes.

The simple root generators of ${G_2}^{+++}$ are
\begin{align}
\nonumber E_a&={K^a}_{a+1}, a=1,\ldots4
E_5&=R^5
\end{align}
and the Cartan sub-algebra generators, $H_a$, are given by
\begin{eqnarray}
\nonumber &H_a={K^a}_a-{K^{a+1}}_{a+1}, a=1,\ldots4,\\
&H_5=-({K^1}_1+\ldots{K^4}_4)+2{K^5}_5
\end{eqnarray}
The low-level field content \cite{KleinschmidtSchnakenburgWest} is ${\hat{h}}_a\hspace{0pt}^b$, $A_{a_1}$ and their field strengths have duals derived from  $A_{a_1a_2,b}$, $A_{a_1a_2}$, respectively. We also find the field $A_{a_1a_2a_3}$ which is not related to any of the other fields by a duality condition and we take it to be self-dual in our low order theory. Our choice of local sub-algebra for the non-linear realisation allows us to write the group element of $G_2^{+++}$ as
\begin{align}
g=\exp(\sum_{a\leq b}{\hat{h}}_a\hspace{0pt}^b{K^a}_b)&\exp(\frac{1}{3!}A_{a_1a_2a_3}R^{a_1a_2a_3})\exp(\frac{1}{2!}A_{a_1a_2,b}R^{a_1a_2,b})\\
\nonumber &\exp(\frac{1}{2!}A_{a_1a_2}{R}^{a_1a_2})\exp(A_{a_1}R^{a_1})
\end{align}
The field content of the associated maximally oxidised theory given in \cite{CremmerJuliaLuPope} agrees with the low-order $G_2^{+++}$ content given above. The Lagrangian for the oxidised theory, contains contains a graviton, $\phi$, and contains a 2-form field strength $F_{\mu\nu}=2\partial_{[\mu}A_{\nu]}$, and is given by
\begin{equation}
A=\frac{1}{16\pi G_5}\int d^5x\sqrt{-g}(R-\frac{1}{2.2!}F_{\mu\nu}F^{\mu\nu})+\int_{C.S.} d^5x\epsilon^{a_1\ldots a_5}\frac{1}{3!2!\sqrt{3}}F_{a_1a_2}F_{a_3a_4}A_{a_5} \label{G2lagrangian}
\end{equation}
We find the following electric branes of $G_2^{+++}$ via our group element (\ref{groupelement})
\paragraph{A Particle}
We consider the brane solution associated with the generator $R^{a_1}$ is $R^5 (00001)$, noting that $\beta^2=\frac{2}{3}$. We find that the lowest weight generator in this representation is $R^1 (11111)$. Taking account that the fifth simple root, as enumerated in appendix \ref{Dynkindiagrams}, of ${G_2}^{+++}$ is shorter than the other four such that $\frac{(\alpha_a,\alpha_a)}{(\alpha_5,\alpha_5)}=3$ where $a=1\ldots 4$ we find that the expansion of $R^1$ in the Cartan sub-algebra is
\begin{align}
\nonumber \beta \cdot H=&H_1+H_2+H_3+H_4+\frac{1}{3}H_5\\
&=\frac{2}{3}({K^1}_1)-\frac{1}{3}({K^2}_2+\ldots{K^5}_5)
\end{align}
The group element in equation (\ref{groupelement}) takes the form
\begin{equation}
g_A=\exp(-\frac{3}{2}\ln N_0(\frac{2}{3}({K^1}_1)-\frac{1}{3}({K^2}_2+\ldots{K^5}_5)))\exp((1-N_0)E_{\beta})
\end{equation}
We read off the line element of a particle
\begin{equation}
ds^2=N_0^{-2}(-dy_1^2)+N_0(dx_2^2+\ldots+dx_5^2)
\end{equation}
The associated gauge potential is read from the group element to be
\begin{equation}
{A_1}^T=(1-N_0)
\end{equation}
We complete the change to world volume indices using $({e^{\hat{h}})^1}_1=N_0^{-1}$
\begin{equation}
A_1=N_0^{-1}-1
\end{equation}
This gauge potential gives rise to a 2-form field strength, $F_{\mu\nu}$.
\paragraph{A String}
We start with the generator $R^{45}$ $(00012)$ whose lowest weight generator is $R^{12} (12222)$ and, noting that $\beta^2=\frac{2}{3}$, we find that the corresponding element in the Cartan sub-algebra is
\begin{align}
\nonumber \beta \cdot H=&H_1+2(H_2+\ldots H_4)+\frac{2}{3}H_5\\
&=\frac{1}{3}({K^1}_1+{K^2}_2)-\frac{2}{3}({K^3}_3+\ldots{K^5}_5)
\end{align}
The group element of equation (\ref{groupelement}) takes the form
\begin{equation}
g_A=\exp(-\frac{3}{2}\ln N_1(\frac{1}{3}({K^1}_1+{K^2}_2)-\frac{2}{3}({K^3}_3+\ldots{K^5}_5)))\exp((1-N_1)E_{\beta})
\end{equation}
We have the line element of a string
\begin{equation}
ds^2=N_1^{-1}(-dy_1^2+dx_2^2)+N_1^{2}(dx_3^2+\ldots+dx_5^2)
\end{equation}
The form of the line element is the same as that of the brane associated with the dual of $F_{\mu\nu}$, the 2-form field strength which we obtained in the previous section. The associated gauge potential is read from the group element to be
\begin{equation}
{A_{12}}^T=(1-N_1)
\end{equation}
We change to world volume indices using $({e^{\hat{h}})^1}_1=({e^{\hat{h}})^2}_2=N_1^{-\frac{1}{2}}$
\begin{equation}
A_{12}=N_1^{-1}-1
\end{equation}

We have have reproduced all the usual BPS electric branes of the oxidised $G_2$ theory using the group element given in equation (\ref{groupelement}). The formulae we have found above do indeed correspond to the solutions of the Lagrangian (\ref{G2lagrangian}).

\newpage
\section{Dynkin Diagrams of the Very-Extended Semisimple Lie Groups} \label{Dynkindiagrams}
\begin{figure}[cth]
\includegraphics[scale=0.9,angle=90]{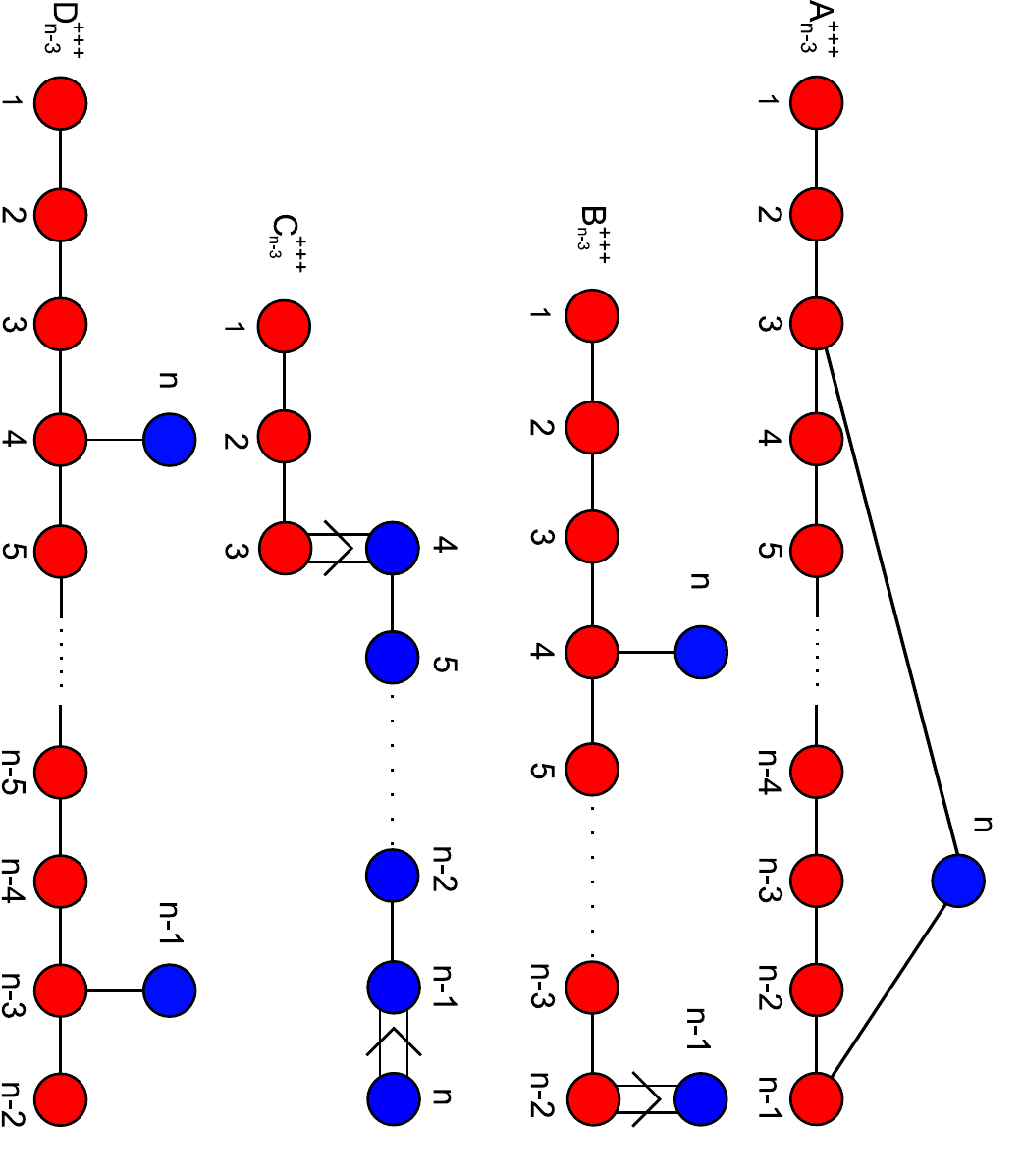}
\caption{The Dynkin diagrams of the very-extended classical groups.} 
\end{figure}
\newpage
\begin{figure}[cth]
\includegraphics[scale=0.9,angle=0]{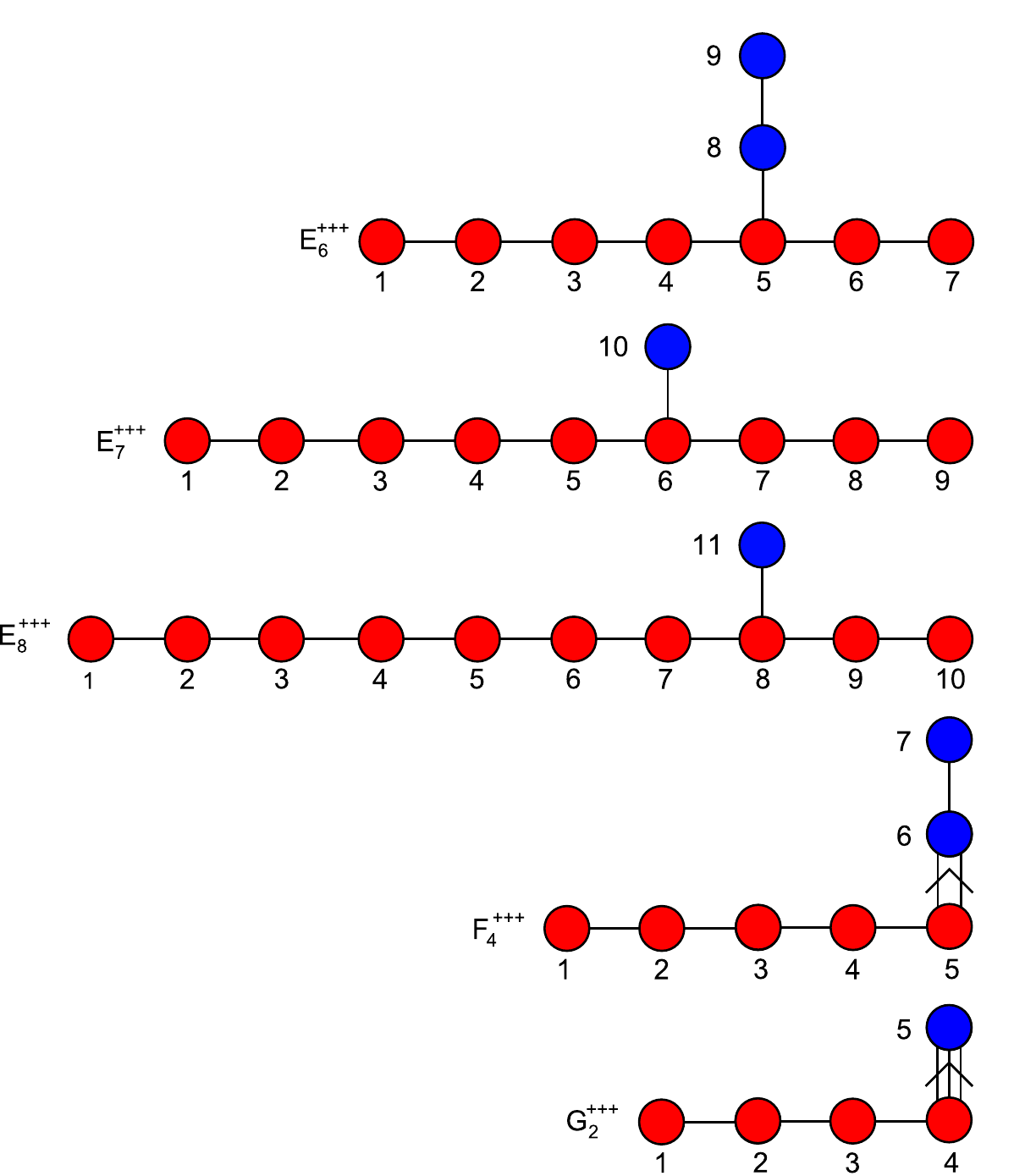}
\caption{The Dynkin diagrams of the very-extended exceptional groups.} 
\end{figure}

\end{spacing}

\newpage


\begin{thebibliography}{widest-label}
\bibitem[1]{Townsend} P. K. Townsend, {\itshape The eleven-dimensional supermembrane revisited}, Phys. Lett. {\bf 350B} (1995) 184-187 \href{http://www.arxiv.org/abs/hep-th/9501068}{\tt hep-th/9501068}
\bibitem[2]{Witten} E. Witten, {\itshape String theory dynamics in various dimensions}, Nucl. Phys. {\bf B443} (1995) 85-126 
\bibitem[3]{West1} P. West, {\itshape $E_{11}$ and M Theory}, Class. Quant.
Grav. {\bf 18} (2001) 4443, \href{http://www.arxiv.org/abs/hep-th/0104081}{{\tt hep-th/0104081}}
\bibitem[4]{CremmerJuliaScherk} E. Cremmer, B. Julia and J. Scherk, {\itshape Supergravity in eleven dimensions}, Phys. Lett. {\bf 76B} (1978) 409-412 
\bibitem[5]{MiemecSchnakenburg} A. Miemec and I. Schnakenburg, {\itshape Basics of M-Theory}, Fortsch.Phys. {\bf 54} (2006) 5-72 \href{http://www.arxiv.org/abs/hep-th/0509137}{\tt hep-th/0509137}
\bibitem[6]{CremmerJulia} E. Cremmer and B. Julia, {\itshape The N=8 supergravity theory - I. The Lagrangian}, Phys. Lett. {\bf B80} (1978) 48 
\bibitem[7]{CremmerJulia1} E. Cremmer and B. Julia, {\itshape The SO(8) supergravity}, Nucl. Phys. {\bf B159} (1979) 141 
\bibitem[8]{FerraraScherkZumino} S. Ferrara, J. Scherk and B. Zumino {\itshape Algebraic Properties of Extended Supersymmetry}, Nucl. Phys. {\bf B121}, (1977) 77
\bibitem[9]{CremmerFerraraScherk} E. Cremmer, S. Ferrara and J. Scherk, {\itshape SU(4) Invariant Supergravity Theory}, Phys. Lett. {\bf 74B}, (1978) 61
\bibitem[10]{Gilmore} R. Gilmore, {\itshape Lie groups, Lie algebras and some of their applications}, Wiley and Sons, New York (1974)
\bibitem[11]{Helgason} S. Helgason, {\itshape Differential geometry and symmetric spaces}, Academic Press, New York, (1962) 
\bibitem[12]{Cremmer} E. Cremmer, {\itshape Supergravities in 5 dimensions}, in Superspace and Supergravities edited by S. Hawking and M. Rocek, Cambridge University Press, (1981) 
\bibitem[13]{Julia} B. Julia, {\itshape Group Disintegrations}, in Superspace and Supergravities edited by S. Hawking and M. Rocek, Cambridge University Press, (1981)
\bibitem[14]{SchwarzWest} J. Schwarz and P. West, {\itshape Symmetries and Transformations of Chiral N=2, D=10 Supergravity}, Phys. Lett. {\bf 126B}, (1983) 301
\bibitem[15]{HullTownsend} C. Hull and P. K. Townsend, {\itshape Enhanced gauge symmetries in superstring theories}, Nucl. Phys. {\bf B534} (1998)  250-260 \href{http://www.arxiv.org/abs/hep-th/9505073}{\tt hep-th/9505073}
\bibitem[16]{DaiLeighPolchinski} J. Dai, R. Leigh and J. Polchinski, {\itshape New connections between string theories}, Mod. Phys. Lett. {\bf A4} (1989) 2073 
\bibitem[17]{DineHuetSeiberg} M. Dine, P. Huet and N. Seiberg, {\itshape Large and small radius in string theory}, Nucl. Phys. {\bf B322} (1989) 301 
\bibitem[18]{Schwarz} J. Schwarz, {\itshape The Power of M-Theory}, Phys. Lett. {\bf B367} (1996) 97-103 \href{http://www.arxiv.org/abs/hep-th/9510086}{\tt hep-th/9510086}
\bibitem[19]{Schwarz1} J. Schwarz, {\itshape An SL(2,Z) Multiplet of Type IIB Superstrings}, Phys. Lett. {\bf B360} (1995) 13-18; Erratum {\bf B364} (1995) 252 \href{http://www.arxiv.org/abs/hep-th/9508134}{\tt hep-th/9508134}
\bibitem[20]{DamourHenneauxNicolai} T. Damour, M. Henneaux and H. Nicolai {\itshape Cosmological Billiards}, Class. Quant. Grav. {\bf 20} (2003) R145-R200 \href{http://www.arxiv.org/abs/hep-th/0212256}{\tt hep-th/0212256}
\bibitem[21]{deBuyl} S. de Buyl, {\itshape Kac-Moody Algebras in M-Theory}, Ph.D. Thesis, Universite Libre de Bruxelles, (2006)  \href{http://www.arxiv.org/abs/hep-th/0608161}{\tt hep-th/0608161}
\bibitem[22]{LuPopeStelle} H. Lu, C. N. Pope and K. S. Stelle, {\itshape Weyl Group Invariance and $p$-brane Multiplets}, Nucl. Phys. {\bf B476} (1996) 89-117 \href{http://www.arxiv.org/abs/hep-th/9602140}{\tt hep-th/9602140}
\bibitem[23]{CremmerLuPopeStelle} E. Cremmer, H. Lu, C. N. Pope and K. S. Stelle, {\itshape Spectrum-generating Symmetries for BPS Solitons}, Nucl. Phys. {\bf B520} (1998) 132-156 \href{http://www.arxiv.org/abs/hep-th/9707207}{\tt hep-th/9707207}
\bibitem[24]{ObersPioline2} N. Obers and B. Pioline, {\it U-duality and M-theory, an algebraic approach}, Proceedings of Quantum Aspects of Gauge Theories, Supersymmetry and Unification, Corfu, September (1998) \href{http://www.arxiv.org/abs/hep-th/9812139}{\tt hep-th/9812139} {\tt hep-th/9812139}
\bibitem[25]{ElitzurGiveonKutasovRabinovici} S. Elitzur, A. Giveon, D. Kutasov, and E. Rabinovici {\it Algebraic aspects of matrix theory on $T^d$}, Nucl. Phys. {\bf B509} (1998) 122-144 \href{http://www.arxiv.org/abs/hep-th/9812139}{\tt hep-th/9707217}{\tt hep-th/9707217}
\bibitem[26]{ObersPiolineRabinovici} N. Obers, B. Pioline and E.  Rabinovici, {\it M-theory and U-duality on $T^d$ with gauge backgrounds}, Nucl. Phys. {\bf B525} (1998) 163-181 \href{http://www.arxiv.org/abs/hep-th/9712084}{\tt hep-th/9712084}
\bibitem[27]{ObersPioline} N. Obers and B. Pioline, {\it U-duality and M-theory}, Phys. Rept. {\bf 318} (1999) 113-225 \href{http://www.arxiv.org/abs/hep-th/9809039}{\tt hep-th/9809039}
\bibitem[28]{CookWest} P. P. Cook and P. West, {\itshape ${\cal G}^{+++}$ and Brane Solutions}, Nucl. Phys. {\bf B705} (2005) 111-151 \href{http://www.arxiv.org/abs/hep-th/0405149}{\tt hep-th/0405149}
\bibitem[29]{CookWest2} P. Cook and P. West, {\itshape $M$-theory solutions in multiple signatures from $E_{11}$}, JHEP {\bf 0512} (2005) 041 \href{http://www.arxiv.org/abs/hep-th/0506122}{\tt hep-th/0506122}
\bibitem[30]{Hall} B. C. Hall, {\itshape Lie Groups, Lie Algebras, and Representations: An Elementary Introduction}, Springer-Verlag (2003) 
\bibitem[31]{Cahn}R. N. Cahn, {\itshape Semi-Simple Lie Algebras and Their Representations}, \href{http://www-physics.lbl.gov/~rncahn/www/liealgebras/book.html}{\tt http://www-physics.lbl.gov/~rncahn/www/liealgebras/book.html}
\bibitem[32]{Humphreys} J. E. Humphreys, {\itshape Introduction to Lie Algebras and Representation Theory}, Springer-Verlag (1972)  
\bibitem[33]{FultonHarris} W. Fulton and J. Harris, {\itshape Representation Theory: A First Course}, Springer-Verlag (1991)  
\bibitem[34]{Cornwell} J. F. Cornwell, {\itshape Group theory in Physics Vol. 2}, Academic Press Inc. (London) Ltd. (1986)  
\bibitem[35]{Varadarajan} V. S. Varadarajan, {\itshape Lie Groups, Lie Algebras, and their Representations}, Springer-Verlag (1974)  
\bibitem[36]{Jacobsen} N. Jacobsen, {\itshape Lie Algebras}, John Wiley and Sons (1962)  
\bibitem[37]{Curtis} M. L. Curtis, {\itshape Matrix Groups}, Springer-Verlag (1979)  
\bibitem[38]{Baker} A. Baker, {\itshape Matrix Groups: An Introduction to Lie Group Theory}, Springer-Verlag (2002)  
\bibitem[39]{BrockertomDieck} T. Brocker and Tammo tom Dieck, {\itshape Representations of Compact Lie Groups}, Springer-Verlag (1985)  
\bibitem[40]{Georgi}H. Georgi, {\itshape Lie Algebras in Particle Physics},Frontiers in Physics Lecture Notes Series 54
\bibitem[41]{Kac} V. Kac, {\itshape Infinite Dimensional Lie Algebras}, Birkhauser (1983)  
\bibitem[42]{Ronan} M. Ronan, {\itshape Symmetry and the Monster}, Oxford University Press (2006)  
\bibitem[43]{Coxeter} H. S. M. Coxeter, {\itshape Regular Polytopes}, Methuen and Co., (1948)  
\bibitem[44]{Stewart} I. Stewart, {\itshape Galois Theory}, Third Edition, Chapman and Hall/CRC (2004)  
\bibitem[45]{Livio} M. Livio, {\itshape The equation that couldn't be solved}, Simon and Schuster, New York (2005) (USA), Souvenir Press Ltd (2006) (UK)
\bibitem[46]{Bao} L. Bao, {\itshape Algebraic Structures in M-theory}, MSc Thesis, Chalmers University of Technology and Goteborg University, (2004) \href{http://www.arxiv.org/abs/hep-th/0604145}{\tt hep-th/0604145}
\bibitem[47]{ColemanWessZumino} S. Coleman, J. Wess and B. Zumino, {\itshape Structure of Phenomenological Lagrangians I}, Phys. Rev. {\bf 177} (1969) 2239; C. G. Callan, S. Coleman, J. Wess and B. Zumino, {\itshape Structure of Phenomenological Lagrangians II}, Phys. Rev. {\bf 177} (1969) 2247
\bibitem[48]{Volkov} D. V. Volkov,  {\itshape Phenomenological Lagrangians}, Sov. J. Part. Nucl. {\bf 4} (1973) 3; D. V. Volkov and V. P. Akulov, {\itshape Possible Universal Neutrino Interaction}, JETP Letters {\bf 16} (1972) 438; D. V. Volkov and V. P. Akulov, {\itshape Is the Neutrino a Goldstone Particle?}, Phys. Lett. {\bf B46} (1973) 109
\bibitem[49]{SalamStrathdee} A. Salam and J. Strathdee, {\itshape Nonlinear Realisations 1: The role of goldstone bosons}, Phys. Rev. {\bf 184} (1969) 1750-1759; A. Salam and J. Strathdee, {\itshape Nonlinear Realisations 2: Conformal Symmetry}, Phys. Rev. {\bf 184} (1969) 1760-1768; C. J. Isham, A. Salam and J. Strathdee, {\itshape Nonlinear realisations of space-time symmetries: Scalar and tensor gravity}, Annals Phys. {\bf 62} (1971) 98-119
\bibitem[50]{Ogievetsky} V. Ogievetsky, {\itshape Infinite-dimensional algebra of general covariance group as the closure of the finite dimensional algebras of conformal and linear groups}, Nuovo. Cimento, {\bf 8} (1973) 988
\bibitem[51]{West2} P. West {\itshape Hidden Superconformal Symmetry in M Theory}, JHEP {\bf 08} (2000) 007, \href{http://www.arxiv.org/abs/hep-th/0005270}{\tt hep-th/0005270}
\bibitem[52]{LambertWest} N. Lambert and P. West, {\itshape Coset Symmetries in Dimensionally Reduced Bosonic String Theory}, Nucl. Phys. {\bf B 615} (2001) 117, \href{http://www.arxiv.org/abs/hep-th/0107209}{\tt hep-th/0107209}
\bibitem[53]{BarwaldWest} O. Barwald and P. West, {\itshape Brane rotating symmetries and the fivebrane equations of motion}, Phys. Lett. {\bf B476} (2000) 157-164 \href{http://www.arxiv.org/abs/hep-th/9912226}{\tt hep-th/9912226}
\bibitem[54]{West5} P. West, {\itshape Automorphisms, Non-linear Realisations and Branes}, JHEP {\bf 0002} (2000) 024 \href{http://www.arxiv.org/abs/hep-th/0001216}{\tt hep-th/0001216}
\bibitem[55]{Ginsparg} P. Ginsparg, {\itshape Applied Conformal Field Theory}, Fields, Strings and Critical Phenomena, (Les Houches, Session XLIX, 1988) ed. by E. Br\'ezin and J. Zinn Justin, 1989 \href{http://www.arxiv.org/abs/hep-th/9108028}{\tt hep-th/9108028}
\bibitem[56]{HoltenProeyen} J. van Holten and A. van Proeyen, {\itshape N=1 supersymmetry algebras in d=2,3,4 mod 8}, J. Phys. A {\bf 15} (1982) 376 
\bibitem[57]{Pope} C. Pope, {\itshape Lectures on Kaluza-Klein Theory}, \\ \href{http://faculty.physics.tamu.edu/pope/ihplec.pdf}{\tt http://faculty.physics.tamu.edu/pope/ihplec.pdf}
\bibitem[58]{Kaluza} T. Kaluza, {\itshape On the problem of unity in physics}, Sitzungsber. Preuss. Akad. Wiss. Berlin (Math. Phys.) (1921) 966-972
\bibitem[59]{Klein} O. Klein, {\itshape Quantum theory and five dimensional theory of relativity}, Z. Phys. {\bf 37} (1926) 895-906
\bibitem[60]{Ehlers} J. Ehlers, {\itshape Dissertation}, Hamburg University (1957)
\bibitem[61]{Geroch} R. Geroch, {\itshape A method for generating solutions of Einstein's equations}, J. Math. Phys. {\bf 12} (1971) 918; R. Geroch, {\itshape A method for generating new solutions of Einstein's equations II}, J. Math. Phys. {\bf 13} (1972) 394
\bibitem[62]{JuliaNicolai} B. Julia and H. Nicolai, {\itshape Conformal internal symmetry of 2d $\sigma$ models coupled to gravity and a dilaton}, Nucl. Phys. {\bf B482} (1996) 431 \href{http://www.arxiv.org/abs/hep-th/9608082}{\tt hep-th/9608082}
\bibitem[63]{FischbacherNicolai} T. Fischbacher and H. Nicolai, {\itshape Low Level representations of E10 and E11},Contribution to the Proceedings of the Ramanujan International Symposium on Kac-Moody Algebras and Applications, ISKMAA-2002, Jan. 28–31, Chennai, India, \href{http://www.arxiv.org/abs/hep-th/0301017}{\tt hep-th/0301017}
\bibitem[64]{KleinschmidtSchnakenburgWest}A. Kleinschmidt, I. Schnakenburg and P. West, {\itshape Very Extended Kac-Moody Algebras and their Interpretation at Low Levels}, Class.Quant.Grav. 21 (2004) 2493-2525 \href{http://www.arxiv.org/abs/hep-th/0309198}{\tt hep-th/0309198}
\bibitem[65]{KleinschmidtWest} A. Kleinschmidt and P. West {\itshape Representations of G+++ and the Role of Space-Time}, JHEP {\bf 0402} (2004) 033, \href{http://www.arxiv.org/abs/hep-th/0312247}{\tt hep-th/0312247}
\bibitem[66]{GaberdielOliveWest} M. R. Gaberdiel, D. I. Olive and P. West {\itshape A Class of Lorentzian Kac-Moody Algebras}, Nucl. Phys. {\bf B 645} (2002) 403-437, \href{http://www.arxiv.org/abs/hep-th/0205068}{\tt hep-th/0205068}
\bibitem[67]{SchnakenburgWest}I. Schnakenburg and P. West {\itshape Kac-Moody Symmetries of IIB Supergravity}, Physics Letters B 517 (2001) 421-428, \href{http://www.arxiv.org/abs/hep-th/0107181}{\tt hep-th/0107181}
\href{http://www.arxiv.org/abs/hep-th/9507121}{\tt hep-th/9507121}
\bibitem[68]{West}Peter West, {\itshape The IIA, IIB and Eleven Dimensional Theories and Their Common $E_{11}$ Origin}, Nucl. Phys. {\bf B693} (2004) 76-102 \href{http://www.arxiv.org/abs/hep-th/0402140}{\tt hep-th/0402140}
\bibitem[69]{SchnakenburgWest1} I. Schnakenburg and P. West, {\itshape Kac-Moody Symmetries of Ten-dimensional Non-maximal Supergravity Theories}, JHEP {\bf 0405} (2004) 019 \href{http://www.arxiv.org/abs/hep-th/0401196}{\tt hep-th/0401196}
\bibitem[70]{EnglertHouartTaorminaWest} F. Englert, L. Houart, A. Taormina and P. West {\itshape The Symmetry of M-Theories}, JHEP {\bf 0309} (2003) 020 \href{http://www.arxiv.org/abs/hep-th/0304206}{\tt hep-th/0304206}
\bibitem[71]{EnglertHouartWest} F. Englert, L. Houart and P. West {\itshape Intersection Rules, Dynamics and Symmetries}, JHEP {\bf 0308} (2003) 025, \href{http://www.arxiv.org/abs/hep-th/0307024}{\tt hep-th/0307024}
\bibitem[72]{EnglertHouart} F. Englert and L. Houart, {\itshape ${\cal G}^{+++}$ Invariant
Formulation of Gravity and M-theories: Exact BPS Solutions}, JHEP {\bf 0401} (2004) 002 \href{http://www.arxiv.org/abs/hep-th/0311255}{\tt hep-th/0311255}
\bibitem[73]{CremmerJuliaLuPope} Cremmer, Julia, Lu and Pope, {\itshape Higher Dimensional Origin of D=3 Coset Symmetries}, \href{http://www.arxiv.org/abs/hep-th/9909099}{\tt hep-th/9909099}
\bibitem[74]{Keurentjes3} A. Keurentjes, {\itshape The Group Theory of Oxidation}, Nucl. Phys. {\bf B658} (2003) 303-347 \href{http://www.arxiv.org/abs/hep-th/0210178}{\tt hep-th/0210178}
\bibitem[75]{DuffLu} M. Duff and J. Lu,, {\itshape Black and super p-branes in diverse
dimensions}, Nucl. Phys. {\bf B416} (1994) 301 \href{http://www.arxiv.org/abs/hep-th/9306052}{\tt hep-th/9306052}; M. Duff, J. Lu and C. Pope, {\itshape The Black Branes of M Theory}, Phys.Lett. {\bf B382} (1996) 73-80 \href{http://www.arxiv.org/abs/hep-th/9604052}{\tt hep-th/9604052}; G. Gibbons and K. Maeda, {\itshape Black holes and Membranes in Higher Dimensional Theories with Dilaton Fields},
 Nucl. Phys. {\bf B298} (1988) 741; P. Breitenlohner, D. Maison and G. Gibbons, {\itshape 4-Dimensional Black Holes from Kaluza-Klein Theories}, Comm. Math. Phys. {\bf 120} (1988) 295; M. Duff and J. Lu, {\itshape The Selfdual TypeIIB Superthreebrane}, Phys. Lett. {\bf 273B} (1991) 409; M. Duff and J. Lu, 
 {\itshape Elementary Five-Brane Solutions of $D=10$ Supergravity}, Nucl. Phys. {\bf B354} (1991) 141;
G. Horowitz and A. Strominger, {\itshape Black Strings and Branes}, Nucl. Phys. {\bf B360} (1991) 197;
C. Callan, J. Harvey and A. Strominger, {\itshape Supersymmetric String Solitons}, \href{http://www.arxiv.org/abs/hep-th/9112030}{\tt hep-th/9112030}; A. Dabholkar, G. Gibbons, J. Harvey and F. Ruiz-Ruiz, {\itshape Superstrings
and Solitons}, Nucl. Phys. {\bf B340} (1990) 33
\bibitem[76]{EnglertHouartArgurio} F. Englert, L. Houart and R. Argurio {\itshape Intersection Rules for p-Branes}, Phys.Lett. B 398 (1997) 61-68 \href{http://www.arxiv.org/abs/hep-th/9701042}{\tt hep-th/9701042}
\bibitem[77]{Argurio} R. Argurio, {\itshape Brane Physics in M-Theory}, PhD thesis, Universite Libre de Bruxelles, \href{http://www.arxiv.org/abs/hep-th/9807171}{\tt hep-th/9807171}
\bibitem[78]{Stelle} K. Stelle, {\itshape BPS Branes in Supergravity}, based on lectures given at the ICTP Summer School in 1996 and 1997 \href{http://www.arxiv.org/abs/hep-th/9803116}{\tt hep-th/9803116}
\bibitem[79]{AczarragaGauntlettIzquierdoTownsend} J. Aczarraga, J. Gauntlett, J. Izquierdo and P Townsend {\itshape Topological Extensions of the Supersymmetry Algebra for Extended Objects}, Phys. Rev. Lett. {\bf 63} 22 (1989) 2443
\bibitem[80]{West3} P. West {\itshape $E_{11}$, SL(32) and Central Charges}, Phys.Lett. B575 (2003) 333-342, \href{http://www.arxiv.org/abs/hep-th/0307098}{\tt hep-th/0307098}
\bibitem[81]{EnglertHouart1} F. Englert and L. Houart, {\itshape ${\cal G}^{+++}$-Invariant
Formulation of Gravity and M-theories: Exact Intersecting Brane Solutions}, \href{http://www.arxiv.org/abs/hep-th/0405082}{\tt hep-th/0405082}
\bibitem[82]{EnglertHenneauxHouart} F. Englert, M. Henneaux, L. Houart, {\itshape From very-extended to overextended gravity and M-theories}, JHEP {\bf 0502} (2005) 070 \href{http://www.arxiv.org/abs/hep-th/0412184}{\tt hep-th/0412184}
\bibitem[83]{deBuylHouartTabti} S. de Buyl, L. Houart, N. Tabti {\itshape Dualities and signatures of G++ invariant theories}, JHEP {\bf 0506} (2005) 084, \href{http://arxiv.org/abs/hep-th/0505199}{\tt hep-th/0505199}
\bibitem[84]{Keurentjes} A. Keurentjes, {\itshape $E_{11}$: Sign of the Times}, Nucl. Phys. {\bf B697} (2004) 302-318 \href{http://www.arxiv.org/abs/hep-th/0402090}{\tt hep-th/0402090}
\bibitem[85]{West4} P. West, {\itshape $E_{11}$ origin of Brane charges and U-duality multiplets}, JHEP {\bf 0408} (2004) 052 \href{http://www.arxiv.org/abs/hep-th/0406150}{\tt hep-th/0406150}
\bibitem[86]{Hull} C. M. Hull, {\itshape Timelike T-duality, de Sitter space, large N gauge theories and topological field theory}, JHEP {\bf 9807} (1998) 021 \href{http://www.arxiv.org/abs/hep-th/9806146}{\tt hep-th/9806146}; C. M. Hull, {\itshape Duality and the signature of spacetime}, JHEP {\bf 9811} (1998) 017 \href{http://www.arxiv.org/abs/hep-th/9807127}{\tt hep-th/9807127}; C. M. Hull and R. R. Khuri, {\itshape Branes, times and dualities}, Nucl. Phys. {\bf B 536} (1998) 219 \href{http://www.arxiv.org/abs/hep-th/9808069}{\tt hep-th/9808069}
\bibitem[87]{GutperleStrominger} M. Gutperle and A. Strominger, {\itshape Spacelike Branes}, JHEP {\bf 0204} (2002) 018 \href{http://www.arxiv.org/abs/hep-th/0202210}{\tt hep-th/0202210}  
\bibitem[88]{ChenGaltsovGutperle} Chiang-Mei Chen, Dmitri V. Gal'tsov and Michael Gutperle {\itshape $S$-brane Solutions in Supergravity Theories}, Phys. Rev. {\bf D66} (2002) 024043 \href{http://www.arxiv.org/abs/hep-th/0204071}{\tt hep-th/0204071}
\bibitem[89]{Ohta1} N. Ohta, {\itshape Intersection Rules for Non-Extreme $p$-Branes}, Phys. Lett. {\bf B403} (1997) 218, \href{http://www.arxiv.org/abs/hep-th/9702164}{\tt hep-th/9702164}
\bibitem[90]{BPSsolutions} M. Duff and J. Lu,, {\itshape Black and super p-branes in diverse
dimensions}, Nucl. Phys. {\bf B416} (1994) 301 \href{http://www.arxiv.org/abs/hep-th/9306052}{\tt hep-th/9306052}; M. Duff, J. Lu and C. Pope, {\itshape The Black Branes of M Theory}, Phys.Lett. {\bf B382} (1996) 73-80 \href{http://www.arxiv.org/abs/hep-th/9604052}{\tt hep-th/9604052}; G. Gibbons and K. Maeda, {\itshape Black holes and Membranes in Higher Dimensional Theories with Dilaton Fields},
 Nucl. Phys. {\bf B298} (1988) 741; P. Breitenlohner, D. Maison and G. Gibbons, {\itshape 4-Dimensional Black Holes from Kaluza-Klein Theories}, Comm. Math. Phys. {\bf 120} (1988) 295; M. Duff and J. Lu, {\itshape The Selfdual TypeIIB Superthreebrane}, Phys. Lett. {\bf 273B} (1991) 409; M. Duff and J. Lu, 
 {\itshape Elementary Five-Brane Solutions of $D=10$ Supergravity}, Nucl. Phys. {\bf B354} (1991) 141;
G. Horowitz and A. Strominger, {\itshape Black Strings and Branes}, Nucl. Phys. {\bf B360} (1991) 197;
C. Callan, J. Harvey and A. Strominger, {\itshape Supersymmetric String Solitons}, \href{http://www.arxiv.org/abs/hep-th/9112030}{\tt hep-th/9112030}; A. Dabholkar, G. Gibbons, J. Harvey and F. Ruiz-Ruiz, {\itshape Superstrings
and Solitons}, Nucl. Phys. {\bf B340} (1990) 33
\bibitem[91]{BhattacharyaRoy} S. Bhattacharya and S. Roy {\itshape Time dependent supergravity solutions in arbitrary dimensions}, JHEP {\bf 0312} (2003) 015 \href{http://www.arxiv.org/abs/hep-th/0309202}{\tt hep-th/0309202}
\bibitem[92]{GutperleSabra} M. Gutperle and W. Sabra {\itshape S-brane solutions in gauged and ungauged supergravities}, Phys.Lett. {\bf B601} (2004) 73-80 \href{http://www.arxiv.org/abs/hep-th/0407147}{\tt hep-th/0407147}
\bibitem[93]{KruczenskiMyersPeet} M. Kruczenski, R. C. Myers and A. W. Peet {\itshape Supergravity S-Branes}, JHEP {\bf 0205} (2002) 039 \href{http://www.arxiv.org/abs/hep-th/0204144}{\tt hep-th/0204144}
\bibitem[94]{Ohta2} N. Ohta {\itshape Intersection Rules for S-Branes}, Phys. Lett. {\bf B558} (2003) 213, \href{http://www.arxiv.org/abs/hep-th/0301095}{\tt hep-th/0301095}
\bibitem[95]{Keurentjes2} A. Keurentjes, {\itshape Time-like T-duality algebra}, JHEP {\bf 0411} (2004) 034 \href{http://www.arxiv.org/abs/hep-th/0404174}{\tt hep-th/0404174}
\bibitem[96]{rollingtachyon} A. Sen, {\itshape Non-BPS states and branes in string theory}, JHEP {\bf 0204} (2002) 048 \href{http://www.arxiv.org/abs/hep-th/9904207}{\tt hep-th/9904207}; A. Sen, {\itshape Rolling Tachyon}, JHEP {\bf 0204} (2002) 018 \href{http://www.arxiv.org/abs/hep-th/0203211}{\tt hep-th/0203211}; A. Sen, {\itshape Tahyon Matter}, JHEP {\bf 0207} (2002) 065 \href{http://www.arxiv.org/abs/hep-th/9904207}{\tt hep-th/9904207}; A. Sen, {\itshape Field Theory of tachyon matter}, Mod. Phys. Lett. A{\bf 17} (2002) 1797 \href{http://www.arxiv.org/abs/hep-th/0204143}{\tt hep-th/0204143}; A. Sen, {\itshape Time evolution in open string theory}, JHEP {\bf 0210} (2002) 003 \href{http://www.arxiv.org/abs/hep-th/0207105}{\tt hep-th/0207105};
N. Lambert, H. Liu and J. Maldacena, {\itshape Closed strings from decaying D-Branes}, \href{http://www.arxiv.org/abs/hep-th/0303139}{\tt hep-th/0303139}
\bibitem[97]{DuffStelle} M. Duff and K. Stelle, {\itshape Multimembrane solutions of d=11 supergravity}, Phys. Lett. {\bf B253} (1991) 113
\bibitem[98]{Guven} R. Guven {\itshape Black p-brane solutions of 11-dimensional supergravity}, Phys. Lett. {\bf B276} (1992) 49
\bibitem[99]{Hull2} C. Hull {\itshape Exact pp wave solutions of eleven dimensional supergravity}, Phys. Lett. {\bf B139} (1984) 39
\end{thebibliography}
\end{document}